%**start of header
\input amstex

\catcode`\@=11
\message{Loading utility definitions,}
\def\identity@#1{#1}
\def\nofrills@@#1{%
 \DN@{#1}%
 \ifx\next\nofrills \let\frills@\eat@
   \expandafter\expandafter\expandafter\next@\expandafter\eat@
  \else \let\frills@\identity@\expandafter\next@\fi}
\def\nofrillscheck#1{\def\nofrills@{\nofrills@@{#1}}%
  \futurelet\next\nofrills@}
\Invalid@\usualspace
\def\addto#1#2{\csname \expandafter\eat@\string#1@\endcsname
  \expandafter{\the\csname \expandafter\eat@\string#1@\endcsname#2}}
\newdimen\bigsize@
\def\big@#1#2{{\hbox{$\left#2\vcenter to#1\bigsize@{}%
  \right.\nulldelimiterspace\z@\m@th$}}}
\def\big{\big@\@ne}
\def\Big{\big@{1.5}}
\def\bigg{\big@\tw@}
\def\Bigg{\big@{2.5}}
\def\raggedcenter@{\leftskip\z@ plus.4\hsize \rightskip\leftskip
  \parfillskip\z@ \parindent\z@ \spaceskip.3333em \xspaceskip.5em
  \pretolerance9999\tolerance9999 \exhyphenpenalty\@M
  \hyphenpenalty\@M \let\\\linebreak}
\def\uppercasetext@#1{%
   {\spaceskip1.3\fontdimen2\the\font plus1.3\fontdimen3\the\font
    \def\ss{SS}\let\i=I\let\j=J\let\ae\AE\let\oe\OE
    \let\o\O\let\aa\AA\let\l\L
    \skipmath@#1$\skipmath@$}}
\def\skipmath@#1$#2${\uppercase{#1}%
  \ifx\skipmath@#2\else$#2$\expandafter\skipmath@\fi}
\def\add@missing#1{\expandafter\ifx\envir@end#1%
  \Err@{You seem to have a missing or misspelled
  \expandafter\string\envir@end ...}%
  \envir@end
\fi}
\newtoks\revert@
\def\envir@stack#1{\toks@\expandafter{\envir@end}%
  \edef\next@{\def\noexpand\envir@end{\the\toks@}%
    \revert@{\the\revert@}}%
  \revert@\expandafter{\next@}%
  \def\envir@end{#1}}
\begingroup
\catcode`\ =11
\gdef\revert@envir#1{\expandafter\ifx\envir@end#1%
\the\revert@%
\else\ifx\envir@end\enddocument \Err@{Extra \string#1}%
\else\expandafter\add@missing\envir@end\revert@envir#1%
\fi\fi}
\xdef\enddocument {\string\enddocument}%
\global\let\envir@end\enddocument %%%%%% don't remove the final space!
\endgroup\relax
\def\first@#1#2\end{#1}
\def\true@{TT}
\def\false@{TF}
\def\empty@{}
\begingroup  \catcode`\-=3
\long\gdef\notempty#1{%
  \expandafter\ifx\first@#1-\end-\empty@ \false@\else \true@\fi}
\endgroup
\message{more fonts,}
\font@\tensmc=cmcsc10 \relax
\let\sevenex=\tenex % needs special handling in \PSAMSFonts
\font@\sevenit=cmti7 \relax
\font@\eightrm=cmr8 \relax % preloaded in plain.tex
\font@\sixrm=cmr6 \relax % preloaded in plain.tex
\font@\eighti=cmmi8 \relax     \skewchar\eighti='177 % preloaded
\font@\sixi=cmmi6 \relax       \skewchar\sixi='177   % preloaded
\font@\eightsy=cmsy8 \relax    \skewchar\eightsy='60 % preloaded
\font@\sixsy=cmsy6 \relax      \skewchar\sixsy='60   % preloaded
\let\eightex=\tenex % needs special handling in \PSAMSFonts
\font@\eightbf=cmbx8 \relax % preloaded in plain.tex
\font@\sixbf=cmbx6 \relax   % preloaded in plain.tex
\font@\eightit=cmti8 \relax % preloaded in plain.tex
\font@\eightsl=cmsl8 \relax % preloaded in plain.tex
\font@\eighttt=cmtt8 \relax % preloaded in plain.tex
\let\eightsmc=\nullfont % needs special handling in \PSAMSFonts
\newtoks\tenpoint@
\def\tenpoint{\normalbaselineskip12\p@
 \abovedisplayskip12\p@ plus3\p@ minus9\p@
 \belowdisplayskip\abovedisplayskip
 \abovedisplayshortskip\z@ plus3\p@
 \belowdisplayshortskip7\p@ plus3\p@ minus4\p@
 \textonlyfont@\rm\tenrm \textonlyfont@\it\tenit
 \textonlyfont@\sl\tensl \textonlyfont@\bf\tenbf
 \textonlyfont@\smc\tensmc \textonlyfont@\tt\tentt
 \ifsyntax@ \def\big##1{{\hbox{$\left##1\right.$}}}%
  \let\Big\big \let\bigg\big \let\Bigg\big
 \else
   \textfont\z@\tenrm  \scriptfont\z@\sevenrm
       \scriptscriptfont\z@\fiverm
   \textfont\@ne\teni  \scriptfont\@ne\seveni
       \scriptscriptfont\@ne\fivei
   \textfont\tw@\tensy \scriptfont\tw@\sevensy
       \scriptscriptfont\tw@\fivesy
   \textfont\thr@@\tenex \scriptfont\thr@@\sevenex
        \scriptscriptfont\thr@@\sevenex
   \textfont\itfam\tenit \scriptfont\itfam\sevenit
        \scriptscriptfont\itfam\sevenit
   \textfont\bffam\tenbf \scriptfont\bffam\sevenbf
        \scriptscriptfont\bffam\fivebf
   \setbox\strutbox\hbox{\vrule height8.5\p@ depth3.5\p@ width\z@}%
   \setbox\strutbox@\hbox{\lower.5\normallineskiplimit\vbox{%
        \kern-\normallineskiplimit\copy\strutbox}}%
   \setbox\z@\vbox{\hbox{$($}\kern\z@}\bigsize@1.2\ht\z@
  \fi
  \normalbaselines\rm\dotsspace@1.5mu\ex@.2326ex\jot3\ex@
  \the\tenpoint@}
\newtoks\eightpoint@
\def\eightpoint{\normalbaselineskip10\p@
 \abovedisplayskip10\p@ plus2.4\p@ minus7.2\p@
 \belowdisplayskip\abovedisplayskip
 \abovedisplayshortskip\z@ plus2.4\p@
 \belowdisplayshortskip5.6\p@ plus2.4\p@ minus3.2\p@
 \textonlyfont@\rm\eightrm \textonlyfont@\it\eightit
 \textonlyfont@\sl\eightsl \textonlyfont@\bf\eightbf
 \textonlyfont@\smc\eightsmc \textonlyfont@\tt\eighttt
 \ifsyntax@\def\big##1{{\hbox{$\left##1\right.$}}}%
  \let\Big\big \let\bigg\big \let\Bigg\big
 \else
  \textfont\z@\eightrm \scriptfont\z@\sixrm
       \scriptscriptfont\z@\fiverm
  \textfont\@ne\eighti \scriptfont\@ne\sixi
       \scriptscriptfont\@ne\fivei
  \textfont\tw@\eightsy \scriptfont\tw@\sixsy
       \scriptscriptfont\tw@\fivesy
  \textfont\thr@@\eightex \scriptfont\thr@@\sevenex
   \scriptscriptfont\thr@@\sevenex
  \textfont\itfam\eightit \scriptfont\itfam\sevenit
   \scriptscriptfont\itfam\sevenit
  \textfont\bffam\eightbf \scriptfont\bffam\sixbf
   \scriptscriptfont\bffam\fivebf
 \setbox\strutbox\hbox{\vrule height7\p@ depth3\p@ width\z@}%
 \setbox\strutbox@\hbox{\raise.5\normallineskiplimit\vbox{%
   \kern-\normallineskiplimit\copy\strutbox}}%
 \setbox\z@\vbox{\hbox{$($}\kern\z@}\bigsize@1.2\ht\z@
 \fi
 \normalbaselines\eightrm\dotsspace@1.5mu\ex@.2326ex\jot3\ex@
 \the\eightpoint@}
\def\linespacing#1{%
  \addto\tenpoint{\normalbaselineskip=#1\normalbaselineskip
    \normalbaselines
    \setbox\strutbox=\hbox{\vrule height.7\normalbaselineskip
      depth.3\normalbaselineskip width\z@}%
    \setbox\strutbox@\hbox{\raise.5\normallineskiplimit
      \vbox{\kern-\normallineskiplimit\copy\strutbox}}%
  }%
  \addto\eightpoint{\normalbaselineskip=#1\normalbaselineskip
    \normalbaselines
    \setbox\strutbox=\hbox{\vrule height.7\normalbaselineskip
      depth.3\normalbaselineskip width\z@}%
    \setbox\strutbox@\hbox{\raise.5\normallineskiplimit
      \vbox{\kern-\normallineskiplimit\copy\strutbox}}%
  }%
}
\def\extrafont@#1#2#3{\font#1=#2#3\relax}
%%%%%%%%%%%%%%%%%%%%%%%%%%%%%%%%%%%%%%%%%%%%%%%%%%%%%%%%%%%%%%%%%%%%%%%%
\newif\ifPSAMSFonts
\def\PSAMSFonts{%
  \def\extrafont@##1##2##3{%
    \font##1=##2%
      \ifnum##3=9 10 at9pt%
      \else\ifnum##3=8 10 at8pt%
      \else\ifnum##3=6 7 at6pt%
              \else ##3\fi\fi\fi\relax}%
  \font@\eightsmc=cmcsc10 at 8pt
  \font@\eightex=cmex10 at 8pt
  \font@\sevenex=cmex10 at 7pt
  \PSAMSFontstrue
}
%%%%%%%%%%%%%%%%%%%%%%%%%%%%%%%%%%%%%%%%%%%%%%%%%%%%%%%%%%%%%%%%%%%%%%%%
\def\loadextrasizes@#1#2#3#4#5#6#7{%
 \ifx\undefined#1%
 \else \extrafont@{#4}{#2}{8}\extrafont@{#6}{#2}{6}%
   \ifsyntax@
   \else
     \addto\tenpoint{\textfont#1#3\scriptfont#1#5%
       \scriptscriptfont#1#7}%
    \addto\eightpoint{\textfont#1#4\scriptfont#1#6%
       \scriptscriptfont#1#7}%
   \fi
 \fi
}
\newtoks\loadextrafonts@@
\def\loadextrafonts@{%
  \loadextrasizes@\msafam{msam}%
    \tenmsa\eightmsa\sevenmsa\sixmsa\fivemsa
  \loadextrasizes@\msbfam{msbm}%
    \tenmsb\eightmsb\sevenmsb\sixmsb\fivemsb
  \loadextrasizes@\eufmfam{eufm}%
    \teneufm\eighteufm\seveneufm\sixeufm\fiveeufm
  \loadextrasizes@\eufbfam{eufb}%
    \teneufb\eighteufb\seveneufb\sixeufb\fiveeufb
  \loadextrasizes@\eusmfam{eusm}%
    \teneusm\eighteusm\seveneusm\sixeusm\fiveeusm
  \loadextrasizes@\eusbfam{eusb}%
    \teneusb\eighteusb\seveneusb\sixeusb\fiveeusb
  \loadextrasizes@\eurmfam{eurm}%
    \teneurm\eighteurm\seveneurm\sixeurm\fiveeurm
  \loadextrasizes@\eurbfam{eurb}%
    \teneurb\eighteurb\seveneurb\sixeurb\fiveeurb
  \loadextrasizes@\cmmibfam{cmmib}%
    \tencmmib\eightcmmib\sevencmmib\sixcmmib\fivecmmib
  \loadextrasizes@\cmbsyfam{cmbsy}%
    \tencmbsy\eightcmbsy\sevencmbsy\sixcmbsy\fivecmbsy
  \let\loadextrafonts@\empty@
  \ifPSAMSFonts
  \else
    \font@\eightsmc=cmcsc8 \relax
    \font@\eightex=cmex8 \relax
    \font@\sevenex=cmex7 \relax
  \fi
  \the\loadextrafonts@@
}
\message{page dimension settings,}
\parindent1pc
\newdimen\normalparindent \normalparindent\parindent
\normallineskiplimit\p@
\newdimen\indenti \indenti=2pc

\topskip10pt \relax
\message{top matter,}
\def\topmatter{\loadextrafonts@ \let\topmatter\relax}
\def\chapterno@{\uppercase\expandafter{\romannumeral\chaptercount@}}
\newcount\chaptercount@
\def\chapter{\let\savedef@\chapter
  \def\chapter##1{\let\chapter\savedef@
  \leavevmode\hskip-\leftskip
   \rlap{\vbox to\z@{\vss\centerline{\eightpoint
   \frills@{CHAPTER\space\afterassignment\chapterno@
       \global\chaptercount@=}%
   ##1\unskip}\baselineskip2pc\null}}\hskip\leftskip}%
 \nofrillscheck\chapter}
\newbox\titlebox@
\def\title{\let\savedef@\title
  \def\title##1\endtitle{\let\title\savedef@
    \global\setbox\titlebox@\vtop{\tenpoint\bf
      \raggedcenter@
      \baselineskip1.3\baselineskip
      \frills@\uppercasetext@{##1}\endgraf}%
    \ifmonograph@
      \edef\next{\the\leftheadtoks}\ifx\next\empty@ \leftheadtext{##1}\fi
    \fi
    \edef\next{\the\rightheadtoks}\ifx\next\empty@ \rightheadtext{##1}\fi
  }%
  \nofrillscheck\title}
\newbox\authorbox@
\def\author#1\endauthor{\global\setbox\authorbox@
  \vbox{\tenpoint\smc\raggedcenter@ #1\endgraf}\relaxnext@
  \edef\next{\the\leftheadtoks}%
  \ifx\next\empty@\leftheadtext{#1}\fi}
\newbox\affilbox@
\def\affil#1\endaffil{\global\setbox\affilbox@
  \vbox{\tenpoint\raggedcenter@#1\endgraf}}
\newcount\addresscount@
\addresscount@\z@
\def\address#1\endaddress{\global\advance\addresscount@\@ne
  \expandafter\gdef\csname address\number\addresscount@\endcsname
  {\nobreak\vskip12\p@ minus6\p@\indent\eightpoint\smc#1\par}}
\def\curraddr{\let\savedef@\curraddr
  \def\curraddr##1\endcurraddr{\let\curraddr\savedef@
  \toks@\expandafter\expandafter\expandafter{%
       \csname address\number\addresscount@\endcsname}%
  \toks@@{##1}%
  \expandafter\xdef\csname address\number\addresscount@\endcsname
  {\the\toks@\endgraf\noexpand\nobreak
    \indent{\noexpand\rm
    \frills@{{\noexpand\it Current address\noexpand\/}:\space}%
    \def\noexpand\usualspace{\space}\the\toks@@\unskip}}}%
  \nofrillscheck\curraddr}
\def\email{\let\savedef@\email
  \def\email##1\endemail{\let\email\savedef@
  \toks@{\def\usualspace{{\it\enspace}}\endgraf\indent\eightpoint}%
  \toks@@{\tt##1\par}%
  \expandafter\xdef\csname email\number\addresscount@\endcsname
  {\the\toks@\frills@{{\noexpand\it E-mail address\noexpand\/}:%
     \noexpand\enspace}\the\toks@@}}%
  \nofrillscheck\email}

\def\urladdr{\let\savedef@\urladdr
  \def\urladdr##1\endurladdr{\let\urladdr\savedef@
  \toks@{\def\usualspace{{\it\enspace}}\endgraf\indent\eightpoint}%
  \toks@@{\tt##1\par}%
  \expandafter\xdef\csname urladdr\number\addresscount@\endcsname
  {\the\toks@\frills@{{\noexpand\it URL\noexpand\/}:%
     \noexpand\enspace}\the\toks@@}}%
  \nofrillscheck\urladdr}
\def\thedate@{}
\def\date#1\enddate{\gdef\thedate@{\tenpoint#1\unskip}}
\def\thethanks@{}
\def\thanks#1\endthanks{%
  \ifx\thethanks@\empty@ \gdef\thethanks@{\eightpoint#1}%
  \else
    \expandafter\gdef\expandafter\thethanks@\expandafter{%
     \thethanks@\endgraf#1}%
  \fi}
\def\thekeywords@{}
\def\keywords{\let\savedef@\keywords
  \def\keywords##1\endkeywords{\let\keywords\savedef@
  \toks@{\def\usualspace{{\it\enspace}}\eightpoint}%
  \toks@@{##1\unskip.}%
  \edef\thekeywords@{\the\toks@\frills@{{\noexpand\it
    Key words and phrases.\noexpand\enspace}}\the\toks@@}}%
 \nofrillscheck\keywords}
\def\thesubjclass@{}
\def\subjclass{\let\savedef@\subjclass
 \def\subjclass##1\endsubjclass{\let\subjclass\savedef@
   \toks@{\def\usualspace{{\rm\enspace}}\eightpoint}%
   \toks@@{##1\unskip.}%
   \edef\thesubjclass@{\the\toks@
     \frills@{{\noexpand\rm1991 {\noexpand\it Mathematics Subject
       Classification}.\noexpand\enspace}}%
     \the\toks@@}}%
  \nofrillscheck\subjclass}
\newbox\abstractbox@
\def\abstract{\let\savedef@\abstract
  \def\abstract{\let\abstract\savedef@
    \setbox\abstractbox@\vbox\bgroup\noindent$$\vbox\bgroup
      \def\envir@end{\endabstract}\advance\hsize-2\indenti
      \def\usualspace{\enspace}\eightpoint \noindent
      \frills@{{\smc Abstract.\enspace}}}%
  \nofrillscheck\abstract}
\def\endabstract{\par\unskip\egroup$$\egroup}
\def\widestnumber{\begingroup \let\head\relax\let\subhead\relax
  \let\subsubhead\relax \expandafter\endgroup\setwidest@}
\def\setwidest@#1#2{%
   \ifx#1\head\setbox\tocheadbox@\hbox{#2.\enspace}%
   \else\ifx#1\subhead\setbox\tocsubheadbox@\hbox{#2.\enspace}%
   \else\ifx#1\subsubhead\setbox\tocsubheadbox@\hbox{#2.\enspace}%
   \else\ifx#1\key
       \if C\refstyle@ \else\refstyle A\fi
       \setboxz@h{\refsfont@\keyformat{#2}}%
       \refindentwd\wd\z@
   \else\ifx#1\no\refstyle C%
       \setboxz@h{\refsfont@\keyformat{#2}}%
       \refindentwd\wd\z@
   \else\ifx#1\page\setbox\z@\hbox{\quad\bf#2}%
       \pagenumwd\wd\z@
   \else\ifx#1\item
       \setboxz@h{(#2)}\rosteritemwd\wdz@
   \else\message{\string\widestnumber\space not defined for this
      option (\string#1)}%
\fi\fi\fi\fi\fi\fi\fi}
\newif\ifmonograph@
\def\Monograph{\monograph@true \let\headmark\rightheadtext
  \let\varindent@\indent \def\headfont@{\bf}\def\proclaimheadfont@{\smc}%
  \def\remarkheadfont@{\smc}}
\let\varindent@\noindent
\newbox\tocheadbox@    \newbox\tocsubheadbox@
\newbox\tocbox@
\newdimen\pagenumwd
\def\toc{\toc@{Contents}}
\def\newtocdefs{%
   \def \title##1\endtitle
       {\penaltyandskip@\z@\smallskipamount
        \hangindent\wd\tocheadbox@\noindent{\bf##1}}%
   \def \chapter##1{%
        Chapter \uppercase\expandafter{%
              \romannumeral##1.\unskip}\enspace}%
   \def \specialhead##1\endspecialhead
       {\par\hangindent\wd\tocheadbox@ \noindent##1\par}%
   \def \Head##1 ##2
       {\par\hangindent\wd\tocheadbox@ \noindent
        \if\notempty{##1}\hbox to\wd\tocheadbox@{\hfil##1\enspace}\fi
        ##2\par}%
   \def \subhead##1 ##2\endsubhead
       {\par\vskip-\parskip {\normalbaselines
        \advance\leftskip\wd\tocheadbox@
        \hangindent\wd\tocsubheadbox@ \noindent
        \if\notempty{##1}%
              \hbox to\wd\tocsubheadbox@{##1\unskip\hfil}\fi
         ##2\par}}%
   \def \subsubhead##1 ##2\endsubsubhead
       {\par\vskip-\parskip {\normalbaselines
        \advance\leftskip\wd\tocheadbox@
        \hangindent\wd\tocsubheadbox@ \noindent
        \if\notempty{##1}%
              \hbox to\wd\tocsubheadbox@{##1\unskip\hfil}\fi
        ##2\par}}}
\def\toc@#1{\relaxnext@
 \DN@{\ifx\next\nofrills\DN@\nofrills{\nextii@}%
      \else\DN@{\nextii@{{#1}}}\fi
      \next@}%
 \DNii@##1{%
\ifmonograph@\bgroup\else\setbox\tocbox@\vbox\bgroup
   \centerline{\headfont@\ignorespaces##1\unskip}\nobreak
   \vskip\belowheadskip \fi
   \def\page####1%
       {\unskip\penalty\z@\null\hfil
        \rlap{\hbox to\pagenumwd{\quad\hfil####1}}%
              \hfilneg\penalty\@M}%
   \setbox\tocheadbox@\hbox{0.\enspace}%
   \setbox\tocsubheadbox@\hbox{0.0.\enspace}%
   \leftskip\indenti \rightskip\leftskip
   \setboxz@h{\bf\quad000}\pagenumwd\wd\z@
   \advance\rightskip\pagenumwd
   \newtocdefs
 }%
 \FN@\next@}
\def\endtoc{\par\egroup}
\let\pretitle\relax
\let\preauthor\relax
\let\preaffil\relax
\let\predate\relax
\let\preabstract\relax
\let\prepaper\relax
\def\dedicatory #1\enddedicatory{\def\preabstract{{\medskip
  \eightpoint\it \raggedcenter@#1\endgraf}}}
\def\thetranslator@{}
\def\translator{%
  \let\savedef@\translator
  \def\translator##1\endtranslator{\let\translator\savedef@
    \edef\thetranslator@{\noexpand\nobreak\noexpand\medskip
      \noexpand\line{\noexpand\eightpoint\hfil
      \frills@{Translated by \uppercase}{##1}\qquad\qquad}%
       \noexpand\nobreak}}%
  \nofrillscheck\translator}
\outer\def\endtopmatter{\add@missing\endabstract
  \edef\next{\the\leftheadtoks}%
  \ifx\next\empty@
    \expandafter\leftheadtext\expandafter{\the\rightheadtoks}%
  \fi
  \ifmonograph@\else
    \ifx\thesubjclass@\empty@\else \makefootnote@{}{\thesubjclass@}\fi
    \ifx\thekeywords@\empty@\else \makefootnote@{}{\thekeywords@}\fi
    \ifx\thethanks@\empty@\else \makefootnote@{}{\thethanks@}\fi
  \fi
  \inslogo@
  \pretitle
  \begingroup % to localize variant topskip
  \ifmonograph@ \topskip7pc \else \topskip4pc \fi
  \box\titlebox@
  \endgroup
  \preauthor
  \ifvoid\authorbox@\else \vskip2.5pcplus1pc\unvbox\authorbox@\fi
  \preaffil
  \ifvoid\affilbox@\else \vskip1pcplus.5pc\unvbox\affilbox@\fi
  \predate
  \ifx\thedate@\empty@\else
    \vskip1pcplus.5pc\line{\hfil\thedate@\hfil}\fi
  \preabstract
  \ifvoid\abstractbox@\else
    \vskip1.5pcplus.5pc\unvbox\abstractbox@ \fi
  \ifvoid\tocbox@\else\vskip1.5pcplus.5pc\unvbox\tocbox@\fi
  \prepaper
  \vskip2pcplus1pc\relax
}
\def\document{%
  \loadextrafonts@
  \let\fontlist@\relax\let\alloclist@\relax
  \tenpoint}
\message{section heads,}
\newskip\aboveheadskip       \aboveheadskip\bigskipamount
\newdimen\belowheadskip      \belowheadskip6\p@
\def\headfont@{\smc}
\def\penaltyandskip@#1#2{\par\skip@#2\relax
  \ifdim\lastskip<\skip@\relax\removelastskip
      \ifnum#1=\z@\else\penalty@#1\relax\fi\vskip\skip@
  \else\ifnum#1=\z@\else\penalty@#1\relax\fi\fi}
\def\nobreak{\penalty\@M
  \ifvmode\gdef\penalty@{\global\let\penalty@\penalty\count@@@}%
  \everypar{\global\let\penalty@\penalty\everypar{}}\fi}
\let\penalty@\penalty
\def\heading#1\endheading{\head#1\endhead}
\def\subheading{\DN@{\ifx\next\nofrills
    \expandafter\subheading@
  \else \expandafter\subheading@\expandafter\empty@
  \fi}%
  \FN@\next@
}
\def\subheading@#1#2{\subhead#1#2\endsubhead}
\newskip\abovespecialheadskip
\abovespecialheadskip=\aboveheadskip
\def\specialheadfont@{\bf}
\outer\def\specialhead{%
  \add@missing\endroster \add@missing\enddefinition
  \add@missing\enddemo \add@missing\endexample
  \add@missing\endproclaim
  \penaltyandskip@{-200}\abovespecialheadskip
  \begingroup\interlinepenalty\@M\rightskip\z@ plus\hsize
  \let\\\linebreak
  \specialheadfont@\noindent}
\def\endspecialhead{\par\endgroup\nobreak\vskip\belowheadskip}
\outer\def\head#1\endhead{%
  \add@missing\endroster \add@missing\enddefinition
  \add@missing\enddemo \add@missing\endexample
  \add@missing\endproclaim
  \penaltyandskip@{-200}\aboveheadskip
  {\headfont@\raggedcenter@\interlinepenalty\@M
  #1\endgraf}\headmark{#1}%
  \nobreak
  \vskip\belowheadskip}
\let\headmark\eat@
\def\restoredef@#1{\relax\let#1\savedef@\let\savedef@\relax}
\newskip\subheadskip       \subheadskip\medskipamount
\def\subheadfont@{\bf}
\outer\def\subhead{%
  \add@missing\endroster \add@missing\enddefinition
  \add@missing\enddemo \add@missing\endexample
  \add@missing\endproclaim
  \let\savedef@\subhead \let\subhead\relax
  \def\subhead##1\endsubhead{\restoredef@\subhead
    \penaltyandskip@{-100}\subheadskip
    {\def\usualspace{\/{\subheadfont@\enspace}}%
     \varindent@\subheadfont@\ignorespaces##1\unskip\frills@{.\enspace}}%
    \ignorespaces}%
  \nofrillscheck\subhead}
\newskip\subsubheadskip       \subsubheadskip\medskipamount
\def\subsubheadfont@{\it}
\outer\def\subsubhead{%
  \add@missing\endroster \add@missing\enddefinition
  \add@missing\enddemo
  \add@missing\endexample \add@missing\endproclaim
  \let\savedef@\subsubhead \let\subsubhead\relax
  \def\subsubhead##1\endsubsubhead{\restoredef@\subsubhead
    \penaltyandskip@{-50}\subsubheadskip
    {\def\usualspace{\/{\subsubheadfont@\enspace}}%
     \subsubheadfont@##1\unskip\frills@{.\enspace}}\ignorespaces}%
  \nofrillscheck\subsubhead}
\message{theorems/proofs/definitions/remarks,}
\def\proclaimheadfont@{\bf}
\def\proclaimfont{\it}
\newskip\preproclaimskip  \preproclaimskip=\medskipamount
\newskip\postproclaimskip \postproclaimskip=\medskipamount
\outer\def\proclaim{%
  \let\savedef@\proclaim \let\proclaim\relax
  \add@missing\endroster \add@missing\enddefinition
  \add@missing\endproclaim \envir@stack\endproclaim
  \def\proclaim##1{\restoredef@\proclaim
    \penaltyandskip@{-100}\preproclaimskip
    {\def\usualspace{\/{\proclaimheadfont@\enspace}}%
     \varindent@\proclaimheadfont@\ignorespaces##1\unskip
     \frills@{.\enspace}}%
    \proclaimfont\ignorespaces}%
  \nofrillscheck\proclaim}
\def\endproclaim{\revert@envir\endproclaim \par\rm
  \penaltyandskip@{55}\postproclaimskip}
\def\remarkheadfont@{\it}
\def\remarkfont{\rm}
\newskip\remarkskip  \remarkskip=\medskipamount
\def\remark{\let\savedef@\remark \let\remark\relax
  \add@missing\endroster \add@missing\endproclaim
  \envir@stack\endremark
  \def\remark##1{\restoredef@\remark
    \penaltyandskip@\z@\remarkskip
    {\def\usualspace{\/{\remarkheadfont@\enspace}}%
     \varindent@\remarkheadfont@\ignorespaces##1\unskip
     \frills@{.\enspace}}%
    \remarkfont\ignorespaces}%
  \nofrillscheck\remark}
\def\endremark{\par\revert@envir\endremark}
\def\qed{\ifhmode\unskip\nobreak\fi\quad
  \ifmmode\square\else$\m@th\square$\fi}
\newskip\postdemoskip  \postdemoskip=\medskipamount
\def\demo{\DN@{\ifx\next\nofrills
    \DN@####1####2{\remark####1{####2}\envir@stack\enddemo
      \ignorespaces}%
  \else
    \DN@####1{\remark{####1}\envir@stack\enddemo\ignorespaces}%
  \fi
  \next@}%
\FN@\next@}
\def\enddemo{\par\revert@envir\enddemo \endremark\vskip\postdemoskip}
\def\definitionfont{\rm}
\newskip\predefinitionskip  \predefinitionskip=\medskipamount
\newskip\postdefinitionskip  \postdefinitionskip=\medskipamount
\def\definition{\let\savedef@\definition \let\definition\relax
  \add@missing\endproclaim \add@missing\endroster
  \add@missing\enddefinition \envir@stack\enddefinition
  \def\definition##1{\restoredef@\definition
    \penaltyandskip@{-100}\predefinitionskip
    {\def\usualspace{\/{\proclaimheadfont@\enspace}}%
     \varindent@\proclaimheadfont@\ignorespaces##1\unskip
     \frills@{.\proclaimheadfont@\enspace}}%
    \definitionfont\ignorespaces}%
  \nofrillscheck\definition}
\def\enddefinition{\revert@envir\enddefinition
  \par\vskip\postdefinitionskip}
\def\example{%
  \DN@{\ifx\next\nofrills
    \DN@####1####2{\definition####1{####2}\envir@stack\endexample
      \ignorespaces}%
  \else
    \DN@####1{\definition{####1}\envir@stack\endexample\ignorespaces}%
  \fi
  \next@}%
\FN@\next@}
\def\endexample{\revert@envir\endexample \enddefinition }
\message{rosters,}
\newdimen\rosteritemwd
\rosteritemwd16pt % approximately the width of (iii) in 10 point text
\newcount\rostercount@
\newif\iffirstitem@
\let\plainitem@\item
\newtoks\everypartoks@
\def\par@{\everypartoks@\expandafter{\the\everypar}\everypar{}}
\def\leftskip@{}
\def\roster{%
  \envir@stack\endroster
  \edef\leftskip@{\leftskip\the\leftskip}%
  \relaxnext@
  \rostercount@\z@% Initialize \rostercount@ to 0.
  \def\item{\FN@\rosteritem@}%      \item, now redefined, has
  \DN@{\ifx\next\runinitem\let\next@\nextii@
    \else\let\next@\nextiii@
    \fi\next@}%
  \DNii@\runinitem% If \runinitem occurs, \nextii@ must kill it off.
    {\unskip% This unskips any space before the original \roster.
     \DN@{\ifx\next[\let\next@\nextii@
       \else\ifx\next"\let\next@\nextiii@\else\let\next@\nextiv@\fi
       \fi\next@}%
     \DNii@[####1]{\rostercount@####1\relax
       \enspace\therosteritem{\number\rostercount@}~\ignorespaces}%
     \def\nextiii@"####1"{\enspace{\rm####1}~\ignorespaces}%
     \def\nextiv@{\enspace\therosteritem1\rostercount@\@ne~}%
     \par@\firstitem@false% Before doing any of this we still change
     \FN@\next@}%      End of definition of \nextii@\runinitem.
  \def\nextiii@{\par\par@% End the present paragraph, change \everypar
    \penalty\@m\smallskip\vskip-\parskip
    \firstitem@true}%
  \FN@\next@}
\def\rosteritem@{\iffirstitem@\firstitem@false
  \else\par\vskip-\parskip\fi
 \leftskip\rosteritemwd \advance\leftskip\normalparindent
 \advance\leftskip.5em \noindent
 \DNii@[##1]{\rostercount@##1\relax\itembox@}%
 \def\nextiii@"##1"{\def\therosteritem@{\rm##1}\itembox@}%
 \def\nextiv@{\advance\rostercount@\@ne\itembox@}%
 \def\therosteritem@{\therosteritem{\number\rostercount@}}%
 \ifx\next[\let\next@\nextii@\else\ifx\next"\let\next@\nextiii@\else
  \let\next@\nextiv@\fi\fi\next@}
\def\itembox@{\llap{\hbox to\rosteritemwd{\hss
  \kern\z@ % kern to thwart \unskip in \rom
  \therosteritem@}\enspace}\ignorespaces}
\def\therosteritem#1{\rom{(\ignorespaces#1\unskip)}}
\newif\ifnextRunin@
\def\endroster{\relaxnext@
 \revert@envir\endroster % restore \envir@end
 \par\leftskip@% End the paragraph, and restore the \leftskip.
 \global\rosteritemwd16\p@ % restore default value
 \penalty-50 \vskip-\parskip\smallskip% Add a good break and
 \DN@{\ifx\next\Runinitem\let\next@\relax
  \else\nextRunin@false\let\item\plainitem@% Otherwise, set
   \ifx\next\par% moreover, if \endroster is followed by \par,
    \DN@\par{\everypar\expandafter{\the\everypartoks@}}%
   \else% but if the \endroster isn't followed by a new paragraph,
    \DN@{\noindent\everypar\expandafter{\the\everypartoks@}}%
  \fi\fi\next@}%
 \FN@\next@}
\newcount\rosterhangafter@
\def\Runinitem#1\roster\runinitem{\relaxnext@
  \envir@stack\endroster
 \rostercount@\z@
 \def\item{\FN@\rosteritem@}%
 \def\runinitem@{#1}%
 \DN@{\ifx\next[\let\next\nextii@\else\ifx\next"\let\next\nextiii@
  \else\let\next\nextiv@\fi\fi\next}%
 \DNii@[##1]{\rostercount@##1\relax
  \def\item@{\therosteritem{\number\rostercount@}}\nextv@}%
 \def\nextiii@"##1"{\def\item@{{\rm##1}}\nextv@}%
 \def\nextiv@{\advance\rostercount@\@ne
  \def\item@{\therosteritem{\number\rostercount@}}\nextv@}%
 \def\nextv@{\setbox\z@\vbox
  {\ifnextRunin@\noindent\fi
  \runinitem@\unskip\enspace\item@~\par
  \global\rosterhangafter@\prevgraf}%
  \firstitem@false% Set \firstitem@false for future \item's.
  \ifnextRunin@\else\par\fi
  \hangafter\rosterhangafter@\hangindent3\normalparindent
  \ifnextRunin@\noindent\fi
  \runinitem@\unskip\enspace%  Put in all the stored stuff
  \item@~\ifnextRunin@\else\par@\fi% and the \item@, and
  \nextRunin@true\ignorespaces}%% Here's where we set \nextRunin@true.
 \FN@\next@}
\message{footnotes,}
\def\footmarkform@#1{$\m@th^{#1}$}
\let\thefootnotemark\footmarkform@
\def\makefootnote@#1#2{\insert\footins
 {\interlinepenalty\interfootnotelinepenalty
 \eightpoint\splittopskip\ht\strutbox\splitmaxdepth\dp\strutbox
 \floatingpenalty\@MM\leftskip\z@skip\rightskip\z@skip
 \spaceskip\z@skip\xspaceskip\z@skip\noindent         % DF (added \noindent)
 \leavevmode{#1}\footstrut\ignorespaces#2\unskip\lower\dp\strutbox 
 \vbox to\dp\strutbox{}}}
\newcount\footmarkcount@
\footmarkcount@\z@
\def\footnotemark{\let\@sf\empty@\relaxnext@
 \ifhmode\edef\@sf{\spacefactor\the\spacefactor}\/\fi
 \DN@{\ifx[\next\let\next@\nextii@\else
  \ifx"\next\let\next@\nextiii@\else
  \let\next@\nextiv@\fi\fi\next@}%
 \DNii@[##1]{\footmarkform@{##1}\@sf}%
 \def\nextiii@"##1"{{##1}\@sf}%
 \def\nextiv@{\iffirstchoice@\global\advance\footmarkcount@\@ne\fi
  \footmarkform@{\number\footmarkcount@}\@sf}%
 \FN@\next@}
\def\footnotetext{\relaxnext@
 \DN@{\ifx[\next\let\next@\nextii@\else
  \ifx"\next\let\next@\nextiii@\else
  \let\next@\nextiv@\fi\fi\next@}%
 \DNii@[##1]##2{\makefootnote@{\footmarkform@{##1}}{##2}}%
 \def\nextiii@"##1"##2{\makefootnote@{##1}{##2}}%
 \def\nextiv@##1{\makefootnote@{\footmarkform@%
  {\number\footmarkcount@}}{##1}}%
 \FN@\next@}
\def\footnote{\let\@sf\empty@\relaxnext@
 \ifhmode\edef\@sf{\spacefactor\the\spacefactor}\/\fi
 \DN@{\ifx[\next\let\next@\nextii@\else
  \ifx"\next\let\next@\nextiii@\else
  \let\next@\nextiv@\fi\fi\next@}%
 \DNii@[##1]##2{\footnotemark[##1]\footnotetext[##1]{##2}}%
 \def\nextiii@"##1"##2{\footnotemark"##1"\footnotetext"##1"{##2}}%
 \def\nextiv@##1{\footnotemark\footnotetext{##1}}%
 \FN@\next@}
\def\adjustfootnotemark#1{\advance\footmarkcount@#1\relax}
\def\footnoterule{\kern-4\p@
  \hrule width5pc\kern 3.6\p@}%      the \hrule is .4pt high
\message{figures and captions,}
\def\captionfont@{\smc}
\def\topcaption#1#2\endcaption{%
  {\dimen@\hsize \advance\dimen@-\captionwidth@
   \rm\raggedcenter@ \advance\leftskip.5\dimen@ \rightskip\leftskip
  {\captionfont@#1}%
  \if\notempty{#2}.\enspace\ignorespaces#2\fi
  \endgraf}\nobreak\bigskip}
\def\botcaption#1#2\endcaption{%
  \nobreak\bigskip
  \setboxz@h{\captionfont@#1\if\notempty{#2}.\enspace\rm\ignorespaces#2\fi}%
  {\dimen@\hsize \advance\dimen@-\captionwidth@
   \leftskip.5\dimen@ \rightskip\leftskip
   \noindent \ifdim\wdz@>\captionwidth@
   \else\hfil\fi
  {\captionfont@#1}%
  \if\notempty{#2}.\enspace\rm\ignorespaces#2\fi\endgraf}}
\def\@ins{\par\begingroup\def\vspace##1{\vskip##1\relax}%
  \def\captionwidth##1{\captionwidth@##1\relax}%
  \setbox\z@\vbox\bgroup} % start a \vbox
\message{miscellaneous,}
\def\block{\RIfMIfI@\nondmatherr@\block\fi
       \else\ifvmode\noindent$$\predisplaysize\hsize
         \else$$\fi
  \def\endblock{\par\egroup$$}\fi
  \vbox\bgroup\advance\hsize-2\indenti\noindent}
\def\endblock{\par\egroup}
\def\cite#1{\rom{[{\citefont@\m@th#1}]}}
\def\citefont@{\rm}
\def\rom#1{\leavevmode
  \edef\prevskip@{\ifdim\lastskip=\z@ \else\hskip\the\lastskip\relax\fi}%
  \unskip
  \edef\prevpenalty@{\ifnum\lastpenalty=\z@ \else
    \penalty\the\lastpenalty\relax\fi}%
  \unpenalty \/\prevpenalty@ \prevskip@ {\rm #1}}
\message{references,}
\def\refsfont@{\eightpoint}
\def\refsheadfont@{\headfont@}
\newdimen\refindentwd
\setboxz@h{\refsfont@ 00.\enspace}
\refindentwd\wdz@
\def\Refsname{References}
\outer\def\Refs{\add@missing\endroster \add@missing\endproclaim
 \let\savedef@\Refs \let\Refs\relax % because of \outer-ness
 \def\Refs##1{\restoredef@\Refs
   \if\notempty{##1}\penaltyandskip@{-200}\aboveheadskip
     \begingroup \raggedcenter@\refsheadfont@
       \ignorespaces##1\endgraf\endgroup
     \penaltyandskip@\@M\belowheadskip
   \fi
   \begingroup\def\envir@end{\endRefs}\refsfont@\sfcode`\.\@m
   }%
 \nofrillscheck{\csname Refs\expandafter\endcsname
  \frills@{{\Refsname}}}}
\def\endRefs{\par % This will check for a missing \endref, also
  \endgroup}
\newif\ifbook@ \newif\ifprocpaper@
\def\nofrills{%
  \expandafter\ifx\envir@end\endref
    \let\do\relax
    \xdef\nofrills@list{\nofrills@list\do\curbox}%
  \else\errmessage{\Invalid@@ \string\nofrills}%
  \fi}%
\def\defaultreftexts{\gdef\edtext{ed.}\gdef\pagestext{pp.}%
  \gdef\voltext{vol.}\gdef\issuetext{no.}}
\defaultreftexts
\def\ref{\par
  \begingroup \def\envir@end{\endref}%
  \noindent\hangindent\refindentwd
  \def\par{\add@missing\endref}%
  \global\let\nofrills@list\empty@
  \refbreaks
  \procpaper@false \book@false \moreref@false
  \def\curbox{\z@}\setbox\z@\vbox\bgroup
}
\let\keyhook@\empty@
\def\endref{%
  \setbox\tw@\box\thr@@
  \makerefbox?\thr@@{\endgraf\egroup}%
  \endref@
  \endgraf
  \endgroup
  \keyhook@
  \global\let\keyhook@\empty@ % \global to conserve save stack
}
\def\key{\gdef\key{\makerefbox\key\keybox@\empty@}\key} \newbox\keybox@
\def\no{\gdef\no{\makerefbox\no\keybox@\empty@}%
  \gdef\keyhook@{\refstyle C}\no}
\def\by{\makerefbox\by\bybox@\empty@} \newbox\bybox@
 % for backward compatibility
\def\bysame{\by\hbox to3em{\hrulefill}\thinspace\kern\z@}
\def\paper{\makerefbox\paper\paperbox@\it} \newbox\paperbox@
\def\paperinfo{\makerefbox\paperinfo\paperinfobox@\empty@}%
  \newbox\paperinfobox@
\def\jour{\makerefbox\jour\jourbox@
  {\aftergroup\book@false \aftergroup\procpaper@false}} \newbox\jourbox@
\def\issue{\makerefbox\issue\issuebox@\empty@} \newbox\issuebox@
\def\yr{\makerefbox\yr\yrbox@\empty@} \newbox\yrbox@
\def\pages{\makerefbox\pages\pagesbox@\empty@} \newbox\pagesbox@
\def\page{\gdef\pagestext{p.}\makerefbox\page\pagesbox@\empty@}
\def\ed{\makerefbox\ed\edbox@\empty@} \newbox\edbox@
\def\eds{\gdef\edtext{eds.}\makerefbox\eds\edbox@\empty@}
\def\book{\makerefbox\book\bookbox@
  {\it\aftergroup\book@true \aftergroup\procpaper@false}}
  \newbox\bookbox@
\def\bookinfo{\makerefbox\bookinfo\bookinfobox@\empty@}%
  \newbox\bookinfobox@
\def\publ{\makerefbox\publ\publbox@\empty@} \newbox\publbox@
\def\publaddr{\makerefbox\publaddr\publaddrbox@\empty@}%
  \newbox\publaddrbox@
\def\inbook{\makerefbox\inbook\bookbox@
  {\aftergroup\procpaper@true \aftergroup\book@false}}
\def\procinfo{\makerefbox\procinfo\procinfobox@\empty@}%
  \newbox\procinfobox@
\def\finalinfo{\makerefbox\finalinfo\finalinfobox@\empty@}%
  \newbox\finalinfobox@
\def\miscnote{\makerefbox\miscnote\miscnotebox@\empty@}%
  \newbox\miscnotebox@
\def\toappear{\miscnote to appear}
\def\lang{\makerefbox\lang\langbox@\empty@} \newbox\langbox@
\newbox\morerefbox@
\def\vol{\makerefbox\vol\volbox@{\ifbook@ \else
  \ifprocpaper@\else\bf\fi\fi}}
\newbox\volbox@
\define\MR#1{\makerefbox\MR\MRbox@\empty@
  \def\next@##1:##2:##3\next@{\ifx @##2\empty@##1\else{\bf##1:}##2\fi}%
  MR \next@#1:@:\next@}
\newbox\MRbox@
\define\AMSPPS#1{\makerefbox\AMSPPS\MRbox@\empty@ AMS\-PPS \##1}
\define\CMP#1{\makerefbox\CMP\MRbox@\empty@ CMP #1}
\newbox\holdoverbox
\def\makerefbox#1#2#3{\endgraf
  \setbox\z@\lastbox
  \global\setbox\@ne\hbox{\unhbox\holdoverbox
    \ifvoid\z@\else\unhbox\z@\unskip\unskip\unpenalty\fi}%
  \egroup
  \setbox\curbox\box\ifdim\wd\@ne>\z@ \@ne \else\voidb@x\fi
  \ifvoid#2\else\Err@{Redundant \string#1; duplicate use, or
     mutually exclusive information already given}\fi
  \def\curbox{#2}\setbox\curbox\vbox\bgroup \hsize\maxdimen \noindent
  #3}
\def\refbreaks{%
  \def\refconcat##1{\setbox\z@\lastbox \setbox\holdoverbox\hbox{%
       \unhbox\holdoverbox \unhbox\z@\unskip\unskip\unpenalty##1}}%
  \def\holdover##1{%
    \RIfM@
      \penalty-\@M\null
      \hfil$\clubpenalty\z@\widowpenalty\z@\interlinepenalty\z@
      \offinterlineskip\endgraf
      \setbox\z@\lastbox\unskip \unpenalty
      \refconcat{##1}%
      \noindent
      $\hfil\penalty-\@M
    \else
      \endgraf\refconcat{##1}\noindent
    \fi}%
  \def\break{\holdover{\penalty-\@M}}%
  \let\vadjust@\vadjust
  \def\vadjust##1{\holdover{\vadjust@{##1}}}%
  \def\newpage{\vadjust{\vfill\break}}%
}
\def\refstyle#1{\uppercase{%
  \gdef\refstyle@{#1}%
  \if#1A\relax \def\keyformat##1{[##1]\enspace\hfil}%
  \else\if#1B\relax
    \refindentwd\parindent
    \def\keyformat##1{\aftergroup\kern
              \aftergroup-\aftergroup\refindentwd}%
 \else\if#1C\relax
   \def\keyformat##1{\hfil##1.\enspace}%
 \fi\fi\fi}% end of \uppercase
}
\refstyle{A}
\def\finalpunct{\ifnum\lastkern=\m@ne\unkern\else.\spacefactor2000 \fi
       \refquotes@\refbreak@}%
\def\continuepunct#1#2#3#4{}%
\def\endref@{%
  \hskip\refindentwd
  \keyhook@
  \def\nofrillscheck##1{%
    \def\do####1{\ifx##1####1\let\frills@\eat@\fi}%
    \let\frills@\identity@ \nofrills@list}%
  \ifvoid\bybox@
    \ifvoid\edbox@
    \else\setbox\bybox@\hbox{\unhbox\edbox@\breakcheck
      \nofrillscheck\edbox@\frills@{\space(\edtext)}\refbreak@}\fi
  \fi
  \ifvoid\keybox@\else\hbox to\refindentwd{%
       \keyformat{\unhbox\keybox@}}\fi
  \ifmoreref@
    \commaunbox@\morerefbox@
  \else
    \kern-\tw@ sp\kern\m@ne sp
  \fi
  \ppunbox@\empty@\empty@\bybox@\empty@
  \ifbook@ % Case 1: \book etc.
    \commaunbox@\bookbox@ \commaunbox@\bookinfobox@
    \ppunbox@\empty@{ (}\procinfobox@)%
    \ppunbox@,{ vol.~}\volbox@\empty@
    \ppunbox@\empty@{ (}\edbox@{, \edtext)}%
    \commaunbox@\publbox@ \commaunbox@\publaddrbox@
    \commaunbox@\yrbox@
    \ppunbox@,{ \pagestext~}\pagesbox@\empty@
  \else
    \commaunbox@\paperbox@ \commaunbox@\paperinfobox@
    \ifprocpaper@ % Case 2: \paper ... \inbook
      \commaunbox@\bookbox@
      \ppunbox@\empty@{ (}\procinfobox@)%
      \ppunbox@\empty@{ (}\edbox@{, \edtext)}%
      \commaunbox@\bookinfobox@
      \ppunbox@,{ \voltext~}\volbox@\empty@
      \commaunbox@\publbox@ \commaunbox@\publaddrbox@
      \commaunbox@\yrbox@
      \ppunbox@,{ \pagestext~}\pagesbox@\empty@
    \else % Case 3: \paper ... \jour
      \commaunbox@\jourbox@
      \ppunbox@\empty@{ }\volbox@\empty@
      \ppunbox@\empty@{ (}\yrbox@)%
      \ppunbox@,{ \issuetext~}\issuebox@\empty@
      \commaunbox@\publbox@ \commaunbox@\publaddrbox@
      \commaunbox@\pagesbox@
    \fi
  \fi
  \commaunbox@\finalinfobox@
  \ppunbox@\empty@{ (}\miscnotebox@)%
  \finalpunct
  \ppunbox@\empty@{ (}\langbox@{)\spacefactor1001 }%
  \ifnum\spacefactor>\@m \ppunbox@{}{ }\MRbox@\empty@
  \else \commaunbox@\MRbox@
  \fi
  \defaultreftexts
}
\def\punct@#1{#1}
\def\ppunbox@#1#2#3#4{\ifvoid#3\else
  \let\prespace@\relax
  \ifnum\lastkern=\m@ne \unkern\let\punct@\eat@
    \ifnum\lastkern=-\tw@ \unkern\let\prespace@\ignorespaces \fi
  \fi
  \nofrillscheck#3%
  \punct@{#1}\refquotes@\refbreak@
  \let\punct@\identity@
  \prespace@
  \frills@{#2\eat@}\space
  \unhbox#3\breakcheck
  \frills@{#4\eat@}{\kern\m@ne sp}\fi}
\def\commaunbox@#1{\ppunbox@,\space{#1}\empty@}
\def\breakcheck{\edef\refbreak@{\ifnum\lastpenalty=\z@\else
  \penalty\the\lastpenalty\relax\fi}\unpenalty}
\def\endquotes{\def\refquotes@{''\let\refquotes@\empty@}}
\let\refquotes@\empty@
\let\refbreak@\empty@
\newif\ifmoreref@
\def\moreref{%
  \setbox\tw@\box\thr@@
  \makerefbox?\thr@@{\endgraf\egroup}%
  \let\savedef@\finalpunct  \let\finalpunct\empty@
  \endref@
  \def\punct@##1##2{##2;}%
  \global\let\nofrills@list\empty@ % global, to conserve save stack
  \let\finalpunct\savedef@
  \moreref@true
  \def\curbox{\morerefbox@}%
  \setbox\morerefbox@\vbox\bgroup \hsize\maxdimen \noindent
}

\message{end of document,}
\ifx\plainend\undefined \let\plainend\end \fi
\outer\def\enddocument{\par% \par will do a runaway check for \endref
  \add@missing\endRefs
  \add@missing\endroster \add@missing\endproclaim
  \add@missing\enddefinition
  \add@missing\enddemo \add@missing\endremark \add@missing\endexample
  \enddocument@text
  \vfill\supereject\plainend}

\def\enddocument@text{%
  \ifmonograph@ % do nothing
  \else
    \nobreak
    \thetranslator@
    \count@\z@
    \loop\ifnum\count@<\addresscount@\advance\count@\@ne
      \csname address\number\count@\endcsname
      \csname email\number\count@\endcsname
      \csname urladdr\number\count@\endcsname
    \repeat
  \fi
}

\message{output routine,}
\def\folio{{\foliofont@\ifnum\pageno<\z@ \romannumeral-\pageno
 \else\number\pageno \fi}}
\def\foliofont@{\eightrm}
\def\headlinefont@{\eightpoint}
\def\leftheadline{\rlap{\folio}\hfill \iftrue\topmark\fi \hfill}
\def\rightheadline{\hfill \expandafter\iffalse\botmark\fi
  \hfill \llap{\folio}}
\newtoks\leftheadtoks
\newtoks\rightheadtoks
\def\leftheadtext{\let\savedef@\leftheadtext
  \def\leftheadtext##1{\let\leftheadtext\savedef@
    \leftheadtoks\expandafter{\frills@\uppercasetext@{##1}}%
    \mark{\the\leftheadtoks\noexpand\else\the\rightheadtoks}
    \ifsyntax@\setboxz@h{\def\\{\unskip\space\ignorespaces}%
        \headlinefont@##1}\fi}%
  \nofrillscheck\leftheadtext}
\def\rightheadtext{\let\savedef@\rightheadtext
  \def\rightheadtext##1{\let\rightheadtext\savedef@
    \rightheadtoks\expandafter{\frills@\uppercasetext@{##1}}%
    \mark{\the\leftheadtoks\noexpand\else\the\rightheadtoks}%
    \ifsyntax@\setboxz@h{\def\\{\unskip\space\ignorespaces}%
        \headlinefont@##1}\fi}%
  \nofrillscheck\rightheadtext}
\headline={\def\\{\unskip\space\ignorespaces}\headlinefont@
  \def\chapter{%
    \def\chapter##1{%
      \frills@{\afterassignment\chapterno@ \chaptercount@=}##1.\space}%
    \nofrillscheck\chapter}%
  \ifodd\pageno \rightheadline \else \leftheadline\fi}
\def\NoRunningHeads{\global\runheads@false\global\let\headmark\eat@}

\newif\iffirstpage@     \firstpage@true
\newif\ifrunheads@      \runheads@true
\output={\output@}
\newdimen\headlineheight \newdimen\headlinespace
\newdimen\dropfoliodepth

\dropfoliodepth=1pc
\headlineheight=5pt
\headlinespace=24pt

\def\pagewidth#1{\hsize#1%
   \captionwidth@\hsize \advance\captionwidth@-2\indenti}

\def\pageheight#1{%
  \vsize=#1 % target height
  \advance\vsize -\headlineheight % subtract height of running head
  \advance\vsize -\headlinespace % subtract space below running head
  \advance\vsize \topskip % but the 24pt is base-to-base, so we need to
                          % compensate for topskip
}

\pagewidth{30pc}\pageheight{50.5pc}

\newinsert\copyins
\skip\copyins=12\p@
\dimen\copyins=40pc
\count\copyins=1000
\def\inslogo@{\insert\copyins{\logo@}}
\def\logo@{\rightline{\eightpoint Typeset by \AmSTeX}}
\def\nologo{\let\logo@\empty@ \let\inslogo@\empty@}
\let\flheadline\hfil \let\frheadline\hfil
\newif\ifplain@  \plain@false
\def\output@{%
  \def\break{\penalty-\@M}\let\par\endgraf
  \shipout\vbox{%
    \ifplain@
      \let\makeheadline\relax \let\makefootline\relax
    \else
      \iffirstpage@ \global\firstpage@false
        \let\rightheadline\frheadline
        \let\leftheadline\flheadline
      \else
        \ifrunheads@ \let\makefootline\relax
        \else \let\makeheadline\relax \fi
      \fi
    \fi
    \makeheadline \pagebody \makefootline
  }%
  \advancepageno \ifnum\outputpenalty>-\@MM\else\dosupereject\fi
}
\def\pagecontents{%
  \ifvoid\topins\else\unvbox\topins\fi
  \dimen@=\dp\@cclv \unvbox\@cclv % open up \box255
  \ifvoid\footins
  \else % footnote info is present
    \vskip\skip\footins
    \footnoterule
    \unvbox\footins
  \fi
  \ifr@ggedbottom \kern-\dimen@ \vfil \fi
  \ifvoid\copyins \else \vskip\skip\copyins \unvbox\copyins \fi
}
\def\makeheadline{%
  \leftskip=\z@
  \vbox{%
    \vbox to\headlineheight{\vss
      \hbox to\hsize{\hskip\z@ plus\hsize\the\headline}%
      \kern-\prevdepth
    }%
    \vskip\headlinespace
    \vskip-\topskip
  }%
  \nointerlineskip
}
\def\makefootline{%
  \relax\ifdim\prevdepth>\z@ \ifdim\prevdepth>\maxdepth \else
    \vskip-\prevdepth \fi\fi
  \nointerlineskip
  \vbox to\z@{\hbox{}%
    \baselineskip\dropfoliodepth
    \hbox to\hsize{\hskip\z@ plus\hsize\the\footline}%
    \vss}}

\message{hyphenation exceptions (U.S. English)}
\hyphenation{acad-e-my acad-e-mies af-ter-thought anom-aly anom-alies
an-ti-deriv-a-tive an-tin-o-my an-tin-o-mies apoth-e-o-ses
apoth-e-o-sis ap-pen-dix ar-che-typ-al as-sign-a-ble as-sist-ant-ship
as-ymp-tot-ic asyn-chro-nous at-trib-uted at-trib-ut-able bank-rupt
bank-rupt-cy bi-dif-fer-en-tial blue-print busier busiest
cat-a-stroph-ic cat-a-stroph-i-cally con-gress cross-hatched data-base
de-fin-i-tive de-riv-a-tive dis-trib-ute dri-ver dri-vers eco-nom-ics
econ-o-mist elit-ist equi-vari-ant ex-quis-ite ex-tra-or-di-nary
flow-chart for-mi-da-ble forth-right friv-o-lous ge-o-des-ic
ge-o-det-ic geo-met-ric griev-ance griev-ous griev-ous-ly
hexa-dec-i-mal ho-lo-no-my ho-mo-thetic ideals idio-syn-crasy
in-fin-ite-ly in-fin-i-tes-i-mal ir-rev-o-ca-ble key-stroke
lam-en-ta-ble light-weight mal-a-prop-ism man-u-script mar-gin-al
meta-bol-ic me-tab-o-lism meta-lan-guage me-trop-o-lis
met-ro-pol-i-tan mi-nut-est mol-e-cule mono-chrome mono-pole
mo-nop-oly mono-spline mo-not-o-nous mul-ti-fac-eted mul-ti-plic-able
non-euclid-ean non-iso-mor-phic non-smooth par-a-digm par-a-bol-ic
pa-rab-o-loid pa-ram-e-trize para-mount pen-ta-gon phe-nom-e-non
post-script pre-am-ble pro-ce-dur-al pro-hib-i-tive pro-hib-i-tive-ly
pseu-do-dif-fer-en-tial pseu-do-fi-nite pseu-do-nym qua-drat-ic
quad-ra-ture qua-si-smooth qua-si-sta-tion-ary qua-si-tri-an-gu-lar
quin-tes-sence quin-tes-sen-tial re-arrange-ment rec-tan-gle
ret-ri-bu-tion retro-fit retro-fit-ted right-eous right-eous-ness
ro-bot ro-bot-ics sched-ul-ing se-mes-ter semi-def-i-nite
semi-ho-mo-thet-ic set-up se-vere-ly side-step sov-er-eign spe-cious
spher-oid spher-oid-al star-tling star-tling-ly sta-tis-tics
sto-chas-tic straight-est strange-ness strat-a-gem strong-hold
sum-ma-ble symp-to-matic syn-chro-nous topo-graph-i-cal tra-vers-a-ble
tra-ver-sal tra-ver-sals treach-ery turn-around un-at-tached
un-err-ing-ly white-space wide-spread wing-spread wretch-ed
wretch-ed-ly Brown-ian Eng-lish Euler-ian Feb-ru-ary Gauss-ian
Grothen-dieck Hamil-ton-ian Her-mit-ian Jan-u-ary Japan-ese Kor-te-weg
Le-gendre Lip-schitz Lip-schitz-ian Mar-kov-ian Noe-ther-ian
No-vem-ber Rie-mann-ian Schwarz-schild Sep-tem-ber}
\loadeufm \loadmsam \loadmsbm
\message{symbol names}\UseAMSsymbols\message{,}

%%   The following definition can be used to provide a \square for
%%   \qed in lieu of the normal \UseAMSsymbols route.
%%\define\square{\vrule width.6em height.5em depth.1em\relax}

\tenpoint

\W@{}
\csname amsppt.sty\endcsname
%% 
%% End of file `amsppt.sty'.

%    Then load specs that are shared between ams-m and ams-p.

\brokenpenalty=10000
\clubpenalty=10000
\widowpenalty=10000

\catcode`\@=11

%    Dummy definition of \keyboarder, for now [mjd,1995/04/03]
\def\keyboarder#1{}%

%    Page dimensions = 30pc x 50.5pc
\def\pagewidth#1{\hsize#1
  \captionwidth@24pc}

\pagewidth{30pc}

\parindent=18\p@
\normalparindent\parindent

\parskip=\z@

\def\foliofont@{\sevenrm}
\def\headlinefont@{\sevenpoint}

%  Heading styles are different from AMSPPT.STY:

\def\specialheadfont@{\elevenpoint\smc}
\def\headfont@{\bf}
\def\subheadfont@{\bf}
\def\refsheadfont@{\bf}
\def\abstractfont@{\smc}
\def\proclaimheadfont@{\smc}
\def\xcaheadfont@{\smc}
\def\captionfont@{\smc}
\def\citefont@{\bf}
\def\refsfont@{\eightpoint}

\font\sixsy=cmsy6

%    Added \msbfam and \eufmfam to all additional sizes so that \Bbb
%    and \frak can be used in titles, running heads.  [bnb, 1996/05/07]

%    Twelvepoint (12/14) used for titles
\font@\twelverm=cmr10 scaled \magstep1
\font@\twelvebf=cmbx10 scaled \magstep1
\font@\twelveit=cmti10 scaled \magstep1
\font@\twelvesl=cmsl10 scaled \magstep1
\font@\twelvesmc=cmcsc10 scaled \magstep1
\font@\twelvett=cmtt10 scaled \magstep1
\font@\twelvei=cmmi10 scaled \magstep1
\font@\twelvesy=cmsy10 scaled \magstep1
\font@\twelveex=cmex10 scaled \magstep1
\font@\twelvemsb=msbm10 scaled \magstep1
\font@\twelveeufm=eufm10 scaled \magstep1

\newtoks\twelvepoint@
\def\twelvepoint{\normalbaselineskip14\p@
  \abovedisplayskip12\p@ plus3\p@ minus9\p@
  \belowdisplayskip\abovedisplayskip
  \abovedisplayshortskip\z@ plus3\p@
  \belowdisplayshortskip7\p@ plus3\p@ minus4\p@
  \textonlyfont@\rm\twelverm \textonlyfont@\it\twelveit
  \textonlyfont@\sl\twelvesl \textonlyfont@\bf\twelvebf
  \textonlyfont@\smc\twelvesmc \textonlyfont@\tt\twelvett
  \ifsyntax@ \def\big##1{{\hbox{$\left##1\right.$}}}%
    \let\Big\big \let\bigg\big \let\Bigg\big
  \else
    \textfont\z@\twelverm  \scriptfont\z@\eightrm
       \scriptscriptfont\z@\sixrm
    \textfont\@ne\twelvei  \scriptfont\@ne\eighti
       \scriptscriptfont\@ne\sixi
    \textfont\tw@\twelvesy \scriptfont\tw@\eightsy
       \scriptscriptfont\tw@\sixsy
    \textfont\thr@@\twelveex \scriptfont\thr@@\eightex
        \scriptscriptfont\thr@@\eightex
    \textfont\itfam\twelveit \scriptfont\itfam\eightit
        \scriptscriptfont\itfam\eightit
    \textfont\bffam\twelvebf \scriptfont\bffam\eightbf
        \scriptscriptfont\bffam\sixbf
    \textfont\msbfam\twelvemsb \scriptfont\msbfam\eightmsb
        \scriptscriptfont\msbfam\sixmsb
    \textfont\eufmfam\twelveeufm \scriptfont\eufmfam\eighteufm
        \scriptscriptfont\eufmfam\sixeufm
    \setbox\strutbox\hbox{\vrule height8.5\p@ depth3.5\p@ width\z@}%
    \setbox\strutbox@\hbox{\lower.5\normallineskiplimit\vbox{%
        \kern-\normallineskiplimit\copy\strutbox}}%
    \setbox\z@\vbox{\hbox{$($}\kern\z@}\bigsize@1.2\ht\z@
  \fi
  \normalbaselines\rm\dotsspace@1.5mu\ex@.2326ex\jot3\ex@
  \the\twelvepoint@}

%    Elevenpoint (11/13) used for authors in proceedings.
\font@\elevenrm=cmr10 scaled \magstephalf
\font@\elevenbf=cmbx10 scaled \magstephalf
\font@\elevenit=cmti10 scaled \magstephalf
\font@\elevensl=cmsl10 scaled \magstephalf
\font@\elevensmc=cmcsc10 scaled \magstephalf
\font@\eleventt=cmtt10 scaled \magstephalf
\font@\eleveni=cmmi10 scaled \magstephalf
\font@\elevensy=cmsy10 scaled \magstephalf
\font@\elevenex=cmex10 scaled \magstephalf
\font@\elevenmsb=msbm10 scaled \magstephalf
\font@\eleveneufm=eufm10 scaled \magstephalf

\newtoks\elevenpoint@
\def\elevenpoint{\normalbaselineskip13\p@
  \abovedisplayskip12\p@ plus3\p@ minus9\p@
  \belowdisplayskip\abovedisplayskip
  \abovedisplayshortskip\z@ plus3\p@
  \belowdisplayshortskip7\p@ plus3\p@ minus4\p@
  \textonlyfont@\rm\elevenrm \textonlyfont@\it\elevenit
  \textonlyfont@\sl\elevensl \textonlyfont@\bf\elevenbf
  \textonlyfont@\smc\elevensmc \textonlyfont@\tt\eleventt
  \ifsyntax@ \def\big##1{{\hbox{$\left##1\right.$}}}%
    \let\Big\big \let\bigg\big \let\Bigg\big
  \else
    \textfont\z@\elevenrm  \scriptfont\z@\eightrm
       \scriptscriptfont\z@\sixrm
    \textfont\@ne\eleveni  \scriptfont\@ne\eighti
       \scriptscriptfont\@ne\sixi
    \textfont\tw@\elevensy \scriptfont\tw@\eightsy
       \scriptscriptfont\tw@\sixsy
    \textfont\thr@@\elevenex \scriptfont\thr@@\eightex
        \scriptscriptfont\thr@@\eightex
    \textfont\itfam\elevenit \scriptfont\itfam\eightit
        \scriptscriptfont\itfam\eightit
    \textfont\bffam\elevenbf \scriptfont\bffam\eightbf
        \scriptscriptfont\bffam\sixbf
    \textfont\msbfam\elevenmsb \scriptfont\msbfam\eightmsb
        \scriptscriptfont\msbfam\sixmsb
    \textfont\eufmfam\eleveneufm \scriptfont\eufmfam\eighteufm
        \scriptscriptfont\eufmfam\sixeufm
    \setbox\strutbox\hbox{\vrule height8.5\p@ depth3.5\p@ width\z@}%
    \setbox\strutbox@\hbox{\lower.5\normallineskiplimit\vbox{%
        \kern-\normallineskiplimit\copy\strutbox}}%
    \setbox\z@\vbox{\hbox{$($}\kern\z@}\bigsize@1.2\ht\z@
  \fi
  \normalbaselines\rm\dotsspace@1.5mu\ex@.2326ex\jot3\ex@
  \the\elevenpoint@}

\addto\tenpoint{\normalbaselineskip12\p@
 \abovedisplayskip6\p@ plus6\p@ minus0\p@
 \belowdisplayskip6\p@ plus6\p@ minus0\p@
 \abovedisplayshortskip0\p@ plus3\p@ minus0\p@
 \belowdisplayshortskip2\p@ plus3\p@ minus0\p@
 \ifsyntax@
 \else
  \setbox\strutbox\hbox{\vrule height9\p@ depth4\p@ width\z@}%
  \setbox\strutbox@\hbox{\vrule height8\p@ depth3\p@ width\z@}%
 \fi
 \normalbaselines\rm}

%    Add sevenpoint for running heads.
\newtoks\sevenpoint@
\def\sevenpoint{\normalbaselineskip9\p@
 \textonlyfont@\rm\sevenrm \textonlyfont@\it\sevenit
 \textonlyfont@\sl\sevensl \textonlyfont@\bf\sevenbf
 \textonlyfont@\smc\sevensmc \textonlyfont@\tt\seventt
  \textfont\z@\sevenrm \scriptfont\z@\sixrm
       \scriptscriptfont\z@\fiverm
  \textfont\@ne\seveni \scriptfont\@ne\sixi
       \scriptscriptfont\@ne\fivei
  \textfont\tw@\sevensy \scriptfont\tw@\sixsy
       \scriptscriptfont\tw@\fivesy
  \textfont\thr@@\sevenex \scriptfont\thr@@\sevenex
   \scriptscriptfont\thr@@\sevenex
  \textfont\itfam\sevenit \scriptfont\itfam\sevenit
   \scriptscriptfont\itfam\sevenit
  \textfont\bffam\sevenbf \scriptfont\bffam\sixbf
   \scriptscriptfont\bffam\fivebf
  \textfont\msbfam\sevenmsb \scriptfont\msbfam\sixmsb
   \scriptscriptfont\msbfam\fivemsb
  \textfont\eufmfam\seveneufm \scriptfont\eufmfam\sixeufm
   \scriptscriptfont\eufmfam\fiveeufm
 \setbox\strutbox\hbox{\vrule height7\p@ depth3\p@ width\z@}%
 \setbox\strutbox@\hbox{\raise.5\normallineskiplimit\vbox{%
   \kern-\normallineskiplimit\copy\strutbox}}%
 \setbox\z@\vbox{\hbox{$($}\kern\z@}\bigsize@1.2\ht\z@
 \normalbaselines\sevenrm\dotsspace@1.5mu\ex@.2326ex\jot3\ex@
 \the\sevenpoint@}

%    Differences from amsppt.sty:
%    - own skips above and below, not the same as \head
%    - centered, not flush left
\newskip\abovespecheadskip   \abovespecheadskip20\p@ plus8\p@ minus2\p@
\newdimen\belowspecheadskip  \belowspecheadskip6\p@
\outer\def\specialhead{%
  \add@missing\endroster \add@missing\enddefinition
  \add@missing\enddemo \add@missing\endexample
  \add@missing\endproclaim
  \penaltyandskip@{-200}\abovespecheadskip
  \begingroup\interlinepenalty\@M\rightskip\z@ plus\hsize
  \let\\\linebreak
  \specialheadfont@\raggedcenter@\noindent}
\def\endspecialhead{\endgraf\endgroup\nobreak\vskip\belowspecheadskip}

\let\varindent@\indent

%    Differences from amsppt.sty:
%    - no \penaltyandskip@ before
\let\subsubhead\relax
\outer\def\subsubhead{%
  \add@missing\endroster \add@missing\enddefinition
  \add@missing\enddemo
  \add@missing\endexample \add@missing\endproclaim
  \let\savedef@\subsubhead \let\subsubhead\relax
  \def\subsubhead##1\endsubsubhead{\restoredef@\subsubhead
      {\def\usualspace{\/{\subsubheadfont@\enspace}}%
    \subsubheadfont@##1\unskip\frills@{.\enspace}}\ignorespaces}%
  \nofrillscheck\subsubhead}

%    Indentation for \proclaim, \demo, etc., are the same as for
%    \subhead, and are taken care of by \varindent.
%    \proclaim head font is small caps, and the text font is italic
%    (not \sl). Vertical space only above proclaim, no added space to
%    other math environments.

\newskip\abstractindent 	\abstractindent=3pc
\long\def\block #1\endblock{\vskip 6pt
	{\leftskip=\abstractindent \rightskip=\abstractindent
	\noindent #1\endgraf}\vskip 6pt}

\long\def\ext #1\endext{\removelastskip\block #1\endblock}

\outer\def\xca{\let\savedef@\xca \let\xca\relax
  \add@missing\endproclaim \add@missing\endroster
  \add@missing\endxca \envir@stack\endxca
  \def\xca##1{\restoredef@\xca
    \penaltyandskip@{-100}\medskipamount
    \bgroup{\def\usualspace{{\xcaheadfont@\enspace}}%
      \varindent@\xcaheadfont@\ignorespaces##1\unskip
      \frills@{.\xcaheadfont@\enspace}}%
      \ignorespaces}%
  \nofrillscheck\xca}
\def\endxca{\egroup\revert@envir\endxca
  \par\medskip}

%    Differences from amsppt.sty:
%    - font is \smc, not \it
%    - good break point before
%    - skip after
\def\remarkheadfont@{\smc}
\def\remark{\let\savedef@\remark \let\remark\relax
  \add@missing\endroster \add@missing\endproclaim
  \envir@stack\endremark
  \def\remark##1{\restoredef@\remark
    \penaltyandskip@{-100}\medskipamount
    {\def\usualspace{{\remarkheadfont@\enspace}}%
     \varindent@\remarkheadfont@\ignorespaces##1\unskip
     \frills@{.\enspace}}\rm
    \ignorespaces}\nofrillscheck\remark}
\def\endremark{\par\revert@envir\endremark\medskip}

%    Differences from amsppt.sty:
%    - square is flush right
\def\qed{\ifhmode\unskip\nobreak\fi\hfill
  \ifmmode\square\else$\m@th\square$\fi}

%%%%%%

%    Expand rosters to three item levels
%    This is identical to code in ams-j.sty except for the width setting
%    of \rosteritemitemitemwd (here "iii"; "ii" in ams-j.sty).
\newdimen\rosteritemsep
\rosteritemsep=.5pc

\newdimen\rosteritemitemwd
\newdimen\rosteritemitemitemwd

\newbox\setwdbox
\setbox\setwdbox\hbox{0.}\rosteritemwd=\wd\setwdbox
\setbox\setwdbox\hbox{0.\hskip.5pc(c)}\rosteritemitemwd=\wd\setwdbox
\setbox\setwdbox\hbox{0.\hskip.5pc(c)\hskip.5pc(iii)}%
  \rosteritemitemitemwd=\wd\setwdbox

%    Differences from amsppt.sty:
%    - adds \itemitem, \itemitemitem
%    - omits \enspace before item label
%    - omits \smallskip following item
\def\roster{%
  \envir@stack\endroster
  \edef\leftskip@{\leftskip\the\leftskip}%
  \relaxnext@
  \rostercount@\z@% Initialize \rostercount@ to 0.
  \def\item{\FN@\rosteritem@}%
  \def\itemitem{\FN@\rosteritemitem@}%
  \def\itemitemitem{\FN@\rosteritemitemitem@}%
  \DN@{\ifx\next\runinitem\let\next@\nextii@
    \else\let\next@\nextiii@
    \fi\next@}%
  \DNii@\runinitem% If \runinitem occurs, \nextii@ must kill it off.
    {\unskip% This unskips any space before the original \roster.
     \DN@{\ifx\next[\let\next@\nextii@
       \else\ifx\next"\let\next@\nextiii@\else\let\next@\nextiv@\fi
       \fi\next@}%
     \DNii@[####1]{\rostercount@####1\relax
       \therosteritem{\number\rostercount@}~\ignorespaces}%
     \def\nextiii@"####1"{{\rm####1}~\ignorespaces}%
     \def\nextiv@{\therosteritem1\rostercount@\@ne~}%
     \par@\firstitem@false% Before doing any of this we still change
     \FN@\next@}%      End of definition of \nextii@\runinitem.
  \def\nextiii@{\par\par@% End the present paragraph, change \everypar
    \penalty\@m\vskip-\parskip
    \firstitem@true}%
  \FN@\next@}

%    Differences from amsppt.sty:
%    - \leftskip = .5pc, not .5em
\def\rosteritem@{\iffirstitem@\firstitem@false
  \else\par\vskip-\parskip
  \fi
  \leftskip\rosteritemwd \advance\leftskip\normalparindent
  \advance\leftskip.5pc \noindent
  \DNii@[##1]{\rostercount@##1\relax\itembox@}%
  \def\nextiii@"##1"{\def\therosteritem@{\rm##1}\itembox@}%
  \def\nextiv@{\advance\rostercount@\@ne\itembox@}%
  \def\therosteritem@{\therosteritem{\number\rostercount@}}%
  \ifx\next[\let\next@\nextii@
  \else\ifx\next"\let\next@\nextiii@\else\let\next@\nextiv@\fi
  \fi\next@}

%    Differences from amsppt.sty:
%    - \skip after label = .5pc, not \enspace
\def\itembox@{\llap{\hbox to\rosteritemwd{\hss
  \kern\z@ % kern to thwart \unskip in \rom
  \therosteritem@}\hskip.5pc}\ignorespaces}

%    Differences from amsppt.sty:
%    - period after, no parentheses around label
\def\therosteritem#1{\rom{\ignorespaces#1.\unskip}}

%    Two new levels.
\def\rosteritemitem@{\iffirstitem@\firstitem@false
  \else\par\vskip-\parskip
  \fi
  \leftskip\rosteritemitemwd \advance\leftskip\normalparindent
  \advance\leftskip.5pc \noindent
  \DNii@[##1]{\rostercount@##1\relax\itemitembox@}%
  \def\nextiii@"##1"{\def\therosteritemitem@{\rm##1}\itemitembox@}%
  \def\nextiv@{\advance\rostercount@\@ne\itemitembox@}%
  \def\therosteritemitem@{\therosteritemitem{\number\rostercount@}}%
  \ifx\next[\let\next@\nextii@
  \else\ifx\next"\let\next@\nextiii@\else\let\next@\nextiv@\fi
  \fi\next@}

\def\itemitembox@{\llap{\hbox to\rosteritemitemwd{\hss
  \kern\z@ % kern to thwart \unskip in \rom
  \therosteritemitem@}\hskip.5pc}\ignorespaces}

\def\therosteritemitem#1{\rom{(\ignorespaces#1\unskip)}}

\def\rosteritemitemitem@{\iffirstitem@\firstitem@false
  \else\par\vskip-\parskip
  \fi
  \leftskip\rosteritemitemitemwd \advance\leftskip\normalparindent
  \advance\leftskip.5pc \noindent
  \DNii@[##1]{\rostercount@##1\relax\itemitemitembox@}%
  \def\nextiii@"##1"{\def\therosteritemitemitem@{\rm##1}\itemitemitembox@}%
  \def\nextiv@{\advance\rostercount@\@ne\itemitemitembox@}%
  \def\therosteritemitemitem@{\therosteritemitemitem{\number\rostercount@}}%
  \ifx\next[\let\next@\nextii@
  \else\ifx\next"\let\next@\nextiii@\else\let\next@\nextiv@\fi
  \fi\next@}

\def\itemitemitembox@{\llap{\hbox to\rosteritemitemitemwd{\hss
  \kern\z@ % kern to thwart \unskip in \rom
  \therosteritemitemitem@}\hskip.5pc}\ignorespaces}

\def\therosteritemitemitem#1{\rom{(\ignorespaces#1\unskip)}}

\def\endroster{\relaxnext@
  \revert@envir\endroster % restore \envir@end
  \par\leftskip@% End the paragraph, and restore the \leftskip.
  \penalty-50 % \vskip-\parskip% Add a good break
  \DN@{\ifx\next\Runinitem\let\next@\relax
    \else\nextRunin@false\let\item\plainitem@% Otherwise, set
      \ifx\next\par% moreover, if \endroster is followed by \par,
        \DN@\par{\everypar\expandafter{\the\everypartoks@}}%
      \else% but if the \endroster isn't followed by a new paragraph,
        \DN@{\noindent\everypar\expandafter{\the\everypartoks@}}%
      \fi
    \fi\next@}%
  \FN@\next@}

%%%%%%%%%%

%    Differences with amsppt.sty (all address-related items):
%    - break not suppressed before \address
%    - \addressfont@ specified
\def\address#1\endaddress{\global\advance\addresscount@\@ne
  \expandafter\gdef\csname address\number\addresscount@\endcsname
  {\vskip12\p@ minus6\p@\indent\addressfont@\smc\ignorespaces#1\par}}

%  Current addresses as well as permanent ones must be accommodated.
%  Check on the skip before the address; it may be a fixed 6pt.
%  \smallskip has been assumed before the current address, as that
%  is what has been used for \email.

\def\curraddr{\let\savedef@\curraddr
  \def\curraddr##1\endcurraddr{\let\curraddr\savedef@
  \toks@\expandafter\expandafter\expandafter{%
       \csname address\number\addresscount@\endcsname}%
  \toks@@{##1}%
  \expandafter\xdef\csname address\number\addresscount@\endcsname
  {\the\toks@\endgraf\noexpand\nobreak
    \indent\noexpand\addressfont@{\noexpand\rm
    \frills@{{\noexpand\it Current address\noexpand\/}:\space}%
    \def\noexpand\usualspace{\space}\the\toks@@\unskip}}}%
  \nofrillscheck\curraddr}

\def\email{\let\savedef@\email
  \def\email##1\endemail{\let\email\savedef@
  \toks@{\def\usualspace{{\it\enspace}}\endgraf\indent\addressfont@}%
  \toks@@{{\tt ##1}\par}%
  \expandafter\xdef\csname email\number\addresscount@\endcsname
  {\the\toks@\frills@{{\noexpand\it E-mail address\noexpand\/}:%
     \noexpand\enspace}\the\toks@@}}%
  \nofrillscheck\email}

\def\rom#1{{\rm #1}}

%    Differences from amsppt.sty:
%    - rule is 2pc, not 3em
\def\bysame{\by\hbox to2pc{\hrulefill}\thinspace\kern\z@}

%    Differences from amsppt.sty:
%    - indent differs in style B
%\def\refstyle#1{\uppercase{%
%  \gdef\refstyle@{#1}%
%  \if#1A\relax \def\keyformat##1{[##1]\enspace\hfil}%
%  \else\if#1B\relax
%    \refindentwd\parindent
%    \def\keyformat##1{\aftergroup\kern
%              \aftergroup-\aftergroup\refindentwd}%
% \else\if#1C\relax
%   \def\keyformat##1{\hfil##1.\enspace}%
% \fi\fi\fi}% end of \uppercase
%}
\def\refstyle#1{\uppercase{%
  \gdef\refstyle@{#1}%
  \if#1A\relax \def\keyformat##1{[##1]\enspace\hfil}%
  \else\if#1B\relax
    \refindentwd2pc
    \def\keyformat##1{\aftergroup\kern
              \aftergroup-\aftergroup\refindentwd}%
  \else\if#1C\relax
    \refindentwd30pt %\refafterindentwd39pt
    \def\keyformat##1{\hfil##1.\enspace}%
  \fi\fi\fi}% end of \uppercase
}

\refstyle{C}

\catcode`\@=11

%    Heading styles are different from AMSPPT.STY:

\def\addressfont@{\tenpoint}

\skip\footins=\bigskipamount % space added when footnote is present

%  \titlefont is 14/18 (in ams-p.sty it is 12/14)
\font@\titlebf=cmbx10 scaled \magstep2	% 14/18
\font@\titlei=cmmi10 scaled \magstep2
\font@\titlesy=cmsy10 scaled \magstep2
\font@\titlemsb=msbm10 scaled \magstep2
\font@\titleeufm=eufm10 scaled \magstep2
\def\titlefont{\normalbaselineskip18\p@
  \textonlyfont@\bf\titlebf
  \ifsyntax@\else
    \textfont\z@\titlebf  \scriptfont\z@\tenbf  \scriptscriptfont\z@\sevenbf
    \textfont\@ne\titlei  \scriptfont\@ne\teni  \scriptscriptfont\@ne\seveni
    \textfont\tw@\titlesy \scriptfont\tw@\tensy \scriptscriptfont\tw@\sevensy
    \textfont\thr@@\tenex \scriptfont\thr@@\tenex \scriptscriptfont\thr@@\tenex
    \textfont\msbfam\titlemsb \scriptfont\msbfam\tenmsb
      \scriptscriptfont\msbfam\sevenmsb
    \textfont\eufmfam\titleeufm \scriptfont\eufmfam\teneufm
      \scriptscriptfont\eufmfam\seveneufm
  \fi
  \normalbaselines\titlebf}

%  added \specialpart to allow suppression of ``Part n'' element; bnb, 30May96
\newif\ifpart	\partfalse
\newif\ifspecialpart  \specialpartfalse

\font\partnofont=cmbx10 scaled \magstep3
%  \partfont is 20/25
%  amplified \partfont to accommodate math; bnb, 21Sep96
%\font\partfont=cmbx10 scaled \magstep4
\font@\partbf=cmbx10 scaled \magstep4
\font@\parti=cmmi10 scaled \magstep4
\font@\partsy=cmsy10 scaled \magstep4
\font@\partmsb=msbm10 scaled \magstep4
\font@\parteufm=eufm10 scaled \magstep4
\def\partfont{\normalbaselineskip25\p@
  \textonlyfont@\bf\partbf
  \ifsyntax@\else
    \textfont\z@\partbf  \scriptfont\z@\titlebf  \scriptscriptfont\z@\tenbf
    \textfont\@ne\parti  \scriptfont\@ne\titlei  \scriptscriptfont\@ne\teni
    \textfont\tw@\partsy \scriptfont\tw@\titlesy \scriptscriptfont\tw@\tensy
    \textfont\thr@@\tenex \scriptfont\thr@@\tenex \scriptscriptfont\thr@@\tenex
    \textfont\bffam\partbf \scriptfont\bffam\titlebf
      \scriptscriptfont\bffam\tenbf
    \textfont\msbfam\partmsb \scriptfont\msbfam\titlemsb
      \scriptscriptfont\msbfam\tenmsb
    \textfont\eufmfam\parteufm \scriptfont\eufmfam\titleeufm
      \scriptscriptfont\eufmfam\teneufm
  \fi
  \normalbaselines\partbf}

%  To get proper inter-word spacing, \raggedcenter@ must follow font change.
\def\part#1\\#2\endpart{\global\parttrue
  \gdef\thepart{\def\\{\hfil\break}%
    {\partnofont\raggedcenter@ \ifspecialpart\null \else Part #1\fi \endgraf}%
    \vskip 20pt
    {\partfont\raggedcenter@ #2\endgraf}}%
  }

\def\specialpart#1\endspecialpart{%
  \global\specialparttrue
  \part\\#1\endpart}

\def\title#1\endtitle{%
  \global\setbox\titlebox@\vtop{\titlefont
   \raggedcenter@
   #1\endgraf}%
 \ifmonograph@ \edef\next{\the\leftheadtoks}%
    \ifx\next\empty@
    \leftheadtext{#1}\fi
 \fi
 \edef\next{\the\rightheadtoks}\ifx\next\empty@ \rightheadtext{#1}\fi
 }

% chapter heading in ten point

\def\romanchapternum{\gdef\chapterno@{\uppercase\expandafter{\romannumeral
    \chaptercount@}}}

\def\chapterno@{\the\chaptercount@}

\def\chapter{\let\savedef@\chapter
  \def\chapter##1{\let\chapter\savedef@
  \leavevmode\hskip-\leftskip
   \rlap{\vbox to\z@{\vss\centerline{\tenpoint
   \frills@{CHAPTER\space\afterassignment\chapterno@
       \global\chaptercount@=}%
   ##1\unskip}\baselineskip36pt\null}}\hskip\leftskip}%
 \nofrillscheck\chapter}

%  authors are set in all caps

\def\author{\let\savedef@\author
  \def\author##1\endauthor{\global\setbox\authorbox@
 \vbox{\tenpoint\raggedcenter@
  \frills@{\ignorespaces##1}\endgraf}
 \edef\next{\the\leftheadtoks}%
 \ifx\next\empty@\expandafter\uppercase{\leftheadtext{##1}}\fi}
\nofrillscheck\author}

%  \abstract differs from the one in AMSPPT.STY by the use of a slightly
%  larger indentation.

\def\abstract{\let\savedef@\abstract
 \def\abstract{\let\abstract\savedef@
  \setbox\abstractbox@\vbox\bgroup\noindent$$\vbox\bgroup\indenti=0pt
  \def\envir@end{\endabstract}\advance\hsize-2\indenti
  \def\usualspace{\enspace}\tenpoint \noindent
  \frills@{{\abstractfont@ Abstract.\enspace}}}%
 \nofrillscheck\abstract}

\def\thanks#1\endthanks{%
  \gdef\thethanks@{{\tenpoint#1\endgraf}}}

\def\keywords{\let\savedef@\keywords
  \def\keywords##1\endkeywords{\let\keywords\savedef@
  \toks@{\def\usualspace{{\it\enspace}}\raggedcenter@\tenpoint}%
  \toks@@{##1\unskip.\endgraf}%
  \edef\thekeywords@{{\the\toks@\frills@{{\noexpand\it
    Key words and phrases.\noexpand\enspace}}\the\toks@@}}}%
 \nofrillscheck\keywords}

\def\subjclass{\let\savedef@\subjclass
 \def\subjclass##1\endsubjclass{\let\subjclass\savedef@
   \toks@{\def\usualspace{{\rm\enspace}}\tenpoint\raggedcenter@}%
   \toks@@{##1\unskip.\endgraf}%
   \edef\thesubjclass@{{\the\toks@
     \frills@{{\noexpand\rm1991 {\noexpand\it Mathematics Subject
       Classification}.\noexpand\enspace}}%
     \the\toks@@}}}%
  \nofrillscheck\subjclass}

%  More generous skips above some section headings, not as much shrink.
\abovespecheadskip=24\p@ plus12\p@ minus\z@
\aboveheadskip=12\p@ plus4\p@ minus2\p@
\subheadskip=6\p@ plus2\p@ minus\z@

\newskip\xcskip   \newskip\afterxcskip
\xcskip=10pt plus2pt minus0pt
\afterxcskip=0pt
\long\def\xcb#1{\par\ifnum\lastskip<\xcskip
  \removelastskip\penalty-100\vskip\xcskip\fi
  \noindent{\bf#1}%
  \nobreak\bgroup
  \xcbrosterdefs}

\def\xcbrosterdefs{\normalparindent=0pt
\setbox\setwdbox\hbox{0.}\rosteritemwd=\wd\setwdbox\relax
  \setbox\setwdbox\hbox{0.\hskip.5pc(c)}\rosteritemitemwd=\wd\setwdbox\relax
  \setbox\setwdbox\hbox{0.\hskip.5pc(c)\hskip.5pc(iii)}%
	\rosteritemitemitemwd=\wd\setwdbox\relax
}

\def\endxcb{\par\egroup}

\outer\def\endtopmatter{\add@missing\endabstract
  \edef\next{\the\leftheadtoks}%
  \ifx\next\empty@
    \expandafter\leftheadtext\expandafter{\the\rightheadtoks}%
  \fi
  \ifpart
%	%  AMS-LaTeX spacing: \null\vfil\thepart\vfil\vfil\newpage
%	%  Fix this at the same distance as a part with 1-line title.
    \global\plain@true
    \null\vskip146\p@
    \thepart
    \vfill\eject
    \null\vfill\eject
  \fi
  \global\plain@false
  \global\firstpage@true
  \begingroup % to localize variant topskip
    \topskip90\p@
    \box\titlebox@
  \endgroup
  \ifvoid\authorbox@\else \vskip2.5pc plus1pc\unvbox\authorbox@\fi
  \ifnum\addresscount@>\z@
    \vfill
    Author address\ifnum\addresscount@>1 es\fi:
    \count@\z@
    \loop\ifnum\count@<\addresscount@\advance\count@\@ne
      \csname address\number\count@\endcsname
      \csname email\number\count@\endcsname
    \repeat
    \vfill\eject
  \fi
  \ifvoid\affilbox@\else \vskip1pcplus\unvbox\affilbox@\fi
  \ifx\thesubjclass@\empty@\else \vfil\thesubjclass@\fi
  \ifx\thekeywords@\empty@\else \vfil\thekeywords@\fi
  \ifx\thethanks@\empty@\else \vfil\thethanks@\fi
  \ifvoid\abstractbox@\else \vfil\unvbox\abstractbox@\vfil\eject \fi
  \ifvoid\tocbox@\else \vskip1.5pcplus.5pc\unvbox\tocbox@\fi
  \prepaper
  \vskip22pt\relax
}

\def\raggedleft@{\leftskip\z@ plus.4\hsize \rightskip\z@
  \parfillskip\z@ \parindent\z@ \spaceskip.3333em \xspaceskip.5em
  \pretolerance9999\tolerance9999 \exhyphenpenalty\@M
  \hyphenpenalty\@M \let\\\linebreak}

\def\aufm #1\endaufm{\vskip-\prevdepth
  \vskip12pt{\raggedleft@ #1\endgraf}}

\widestnumber\key{M} % set default lettered ref style

%%%% table of contents

\begingroup
\let\head\relax \let\specialhead\relax \let\subhead\relax
\let\subsubhead\relax \let\title\relax \let\chapter\relax
\newbox\tocchapbox@

\gdef\newtocdefs{%
  \def\ptitle##1\endptitle
       {\penalty\z@ \vskip8\p@
        \hangindent\wd\tocheadbox@\noindent{\bf ##1}\endgraf}%
  \def\title##1\endtitle
       {\penalty\z@ \vskip8\p@
        \hangindent\wd\tocchapbox@\noindent{##1}\endgraf}%
  \def\chapter##1{\par
        Chapter ##1.\unskip\enspace}%
  \def\part##1{\par
        {\bf Part ##1.}\unskip\enspace}%
  \def\specialhead##1 ##2\endspecialhead{\par
    \begingroup \hangindent.5em \noindent
    \if\notempty{##1}%
      \leftskip\wd\tocheadbox@
      \llap{\hbox to\wd\tocheadbox@{\hfil##1}}\enspace
    \else
      \leftskip\parindent
    \fi
    ##2\endgraf
    \endgroup}%
  \def\head##1 ##2\endhead{\par
    \begingroup \hangindent.5em \noindent
    \if\notempty{##1}%
      \leftskip\wd\tocheadbox@
      \llap{\hbox to\wd\tocheadbox@{\hfil##1}}\enspace
    \else
      \leftskip\parindent
    \fi
    ##2\endgraf
    \endgroup}%
%    Added subhead and subsubhead to prevent \widestnumber error.
%    Style for these in the toc is unspecified; just provide something
%    ad hoc for now. [mjd,1995/03/28]
%    modify sub(sub)heads so that extra lines are hangindented [bnb,1996/07/03]
  \def\subhead##1 ##2\endsubhead{\par
    \begingroup \leftskip4.5pc
    \noindent\llap{##1\enspace}##2\endgraf \endgroup}%
  \def\subsubhead##1 ##2\endsubsubhead{\par
    \begingroup \leftskip7pc
    \noindent\llap{\hss##1\enspace}##2\endgraf \endgroup}%
}%
\gdef\toc@#1{\relaxnext@
  \DN@{\ifx\next\nofrills\DN@\nofrills{\nextii@}%
      \else\DN@{\nextii@{{#1}}}%
      \fi
      \next@}%
  \DNii@##1{%
    \ifmonograph@\bgroup
    \else\setbox\tocbox@\vbox\bgroup
      \centerline{\headfont@\ignorespaces##1\unskip}\nobreak
      \vskip\belowheadskip
    \fi
    \def\page####1%
       {\unskip\penalty\z@\null\hfil
        \rlap{\hbox to\pagenumwd{\quad\hfil\rm ####1}}%
    \global\setbox\tocchapbox@\hbox{Chapter 1.\enspace}%
    \global\setbox\tocheadbox@\hbox{\hskip18pt \S0.0.}%
         \hfilneg\penalty\@M}%
    \leftskip\z@ \rightskip\leftskip
    \setboxz@h{\bf\quad000}\pagenumwd\wd\z@
    \advance\rightskip\pagenumwd
    \newtocdefs
    }%
  \FN@\next@
}%
\endgroup

\def\logo@{}

\def\makefootline{\baselineskip18\p@\line{\the\footline}}

\Monograph

\catcode`\@=13

\catcode`\@=11

\def\pretitle{\vskip84pt}

\aboveheadskip=12\p@ plus 9\p@

\newskip\abovesubheadskip
\abovesubheadskip=9\p@ plus 6\p@

\newskip\abovesubsubheadskip
\abovesubsubheadskip=6\p@ plus 6\p@

\belowheadskip=6\p@

%  Page numbers and running heads are provided by default, as defined
%  in AMSPPT.STY.  Headline text is centered, and page numbers are
%  positioned at the outside corners.  To suppress page numbers and/or
%  running heads, include \NoPageNumbers and/or \NoRunningHeads in the
%  input file, as appropriate.

\def\foliofont@{\eightbf}
\def\headlinefont@{\eightpoint}           %DF

\newif\iflecture@ \lecture@false

\newif\ifLogoOn  \LogoOnfalse
\def\LogoOn{\global\LogoOntrue
  \def\inslogo@{\insert\copyins{\logo@}}%
  }

\def\Monograph{\monograph@true
  \nologo \nojourlogo
  \let\headmark\rightheadtext
  \let\varindent@\noindent
  \def\abstractfont@{\bf}
  \def\addressfont@{\tenpoint}
  \def\specialheadfont@{\twelvepoint\bf}%
  \def\headfont@{\elevenpoint\bf}%
  \def\proclaimheadfont@{\bf}%
  \def\remarkheadfont@{\bf}}

%  Heading styles are different from AMSPPT.STY:

\def\subheadfont@{\bf}
\def\refheadfont@{\headfont@}

\outer\def\specialhead{%
  \add@missing\endroster \add@missing\enddefinition
  \add@missing\enddemo \add@missing\endexample
  \add@missing\endproclaim
  \penaltyandskip@{-200}\abovespecheadskip
  \begingroup\interlinepenalty\@M\rightskip\z@ plus\hsize
  \let\\\linebreak
  \specialheadfont@\raggedright\noindent}
\def\endspecialhead{\endgraf\endgroup\nobreak\vskip\belowspecheadskip\noindent}

\font@\fourteenrm=cmr10 scaled \magstep2
\font@\fourteenbf=cmbx10 scaled \magstep2
\font@\fourteenit=cmti10 scaled \magstep2
\font@\fourteensl=cmsl10 scaled \magstep2
\font@\fourteensmc=cmcsc10 scaled \magstep2
\font@\fourteentt=cmtt10 scaled \magstep2
\font@\fourteeni=cmmi10 scaled \magstep2
\font@\fourteensy=cmsy10 scaled \magstep2
\font@\fourteenex=cmex10 scaled \magstep2

\newtoks\fourteenpoint@
\def\fourteenpoint{\normalbaselineskip16\p@
 \abovedisplayskip12\p@ plus3\p@ minus9\p@
 \belowdisplayskip\abovedisplayskip
 \abovedisplayshortskip\z@ plus3\p@
 \belowdisplayshortskip7\p@ plus3\p@ minus4\p@
 \textonlyfont@\rm\fourteenrm \textonlyfont@\it\fourteenit
 \textonlyfont@\sl\fourteensl \textonlyfont@\bf\fourteenbf
 \textonlyfont@\smc\fourteensmc \textonlyfont@\tt\fourteentt
 \ifsyntax@ \def\big##1{{\hbox{$\left##1\right.$}}}%
  \let\Big\big \let\bigg\big \let\Bigg\big
 \else
   \textfont\z@\fourteenrm  \scriptfont\z@\tenrm
       \scriptscriptfont\z@\eightrm
   \textfont\@ne\fourteeni  \scriptfont\@ne\teni
       \scriptscriptfont\@ne\eighti
   \textfont\tw@\fourteensy \scriptfont\tw@\tensy
       \scriptscriptfont\tw@\eightsy
   \textfont\thr@@\fourteenex \scriptfont\thr@@\tenex
        \scriptscriptfont\thr@@\tenex
   \textfont\itfam\fourteenit \scriptfont\itfam\tenit
        \scriptscriptfont\itfam\tenit
   \textfont\bffam\fourteenbf \scriptfont\bffam\tenbf
        \scriptscriptfont\bffam\eightbf
   \setbox\strutbox\hbox{\vrule height8.5\p@ depth3.5\p@ width\z@}%
   \setbox\strutbox@\hbox{\lower.5\normallineskiplimit\vbox{%
        \kern-\normallineskiplimit\copy\strutbox}}%
   \setbox\z@\vbox{\hbox{$()$}\kern\z@}\bigsize@1.2\ht\z@
  \fi
  \normalbaselines\rm\dotsspace@1.5mu\ex@.2326ex\jot3\ex@
  \the\fourteenpoint@}

\font@\titlebf=cmbx10 scaled \magstep3  % 17/22
\font@\titlei=cmmi10 scaled \magstep3
\font@\titlesy=cmsy10 scaled \magstep3
\def\titlefont{\normalbaselineskip22\p@
 \textonlyfont@\bf\titlebf
 \ifsyntax@\else
  \textfont\z@\titlebf  \scriptfont\z@\tenbf  \scriptscriptfont\z@\sevenbf
  \textfont\@ne\titlei  \scriptfont\@ne\teni  \scriptscriptfont\@ne\seveni
  \textfont\tw@\titlesy \scriptfont\tw@\tensy \scriptscriptfont\tw@\sevensy
  \textfont\thr@@\tenex \scriptfont\thr@@\tenex \scriptscriptfont\thr@@\tenex
 \fi
 \normalbaselines\titlebf}

\def\raggedleft{\leftskip\z@ plus 1fil
\parfillskip\z@ \parindent\z@ \spaceskip\z@ \xspaceskip\z@
 \pretolerance9999\tolerance9999 \exhyphenpenalty\@M}

\def\raggedright{\rightskip\z@ plus 1fil
\parfillskip\z@ \parindent\z@ \spaceskip\z@ \xspaceskip\z@
 \pretolerance9999\tolerance9999 \exhyphenpenalty\@M}

\newif\ifpart   \partfalse

\newif\ifNoNewpage \NoNewpagefalse

\font\partnofont=cmbx10 scaled \magstep1
\font\partfont=cmbx10 scaled \magstep3

\def\part#1\\#2\endpart{\global\parttrue
  \gdef\thepart{\def\\{\break}%
    {\raggedleft
     \noindent\partnofont PART #1\unskip\vskip5pt
     \partfont\baselineskip=25pt
     \noindent#2\unskip\break\endgraf}}}

\def\series #1\endseries{\global\parttrue
  \def\sauth ##1\endsauth{\gdef\thesauth{{%
                \fourteenpoint\bf ##1\unskip\endgraf}}}%
  \gdef\thepart{\def\\{\break}%
    {\raggedleft
     \partfont\baselineskip=25pt
     \noindent#1\unskip\vskip\baselineskip
     \if\notempty{\thesauth}%
       \thesauth
     \fi
     \endgraf}}}

\newif\ifContents \Contentsfalse

\setbox\titlebox@\vbox{}

\def\title#1\endtitle{%
  \def\\{\break}%
  \global\setbox\titlebox@\vtop{\titlefont
   \ifodd\pageno
     \ifContents\raggedright
     \else\raggedleft
     \fi
   \else\raggedright
   \fi
   \noindent#1\unskip\break}%
 \ifmonograph@ \edef\next{\the\leftheadtoks}%
    \ifx\next\empty
    \leftheadtext{#1}\fi
 \fi
 \edef\next{\the\rightheadtoks}\ifx\next\empty \rightheadtext{#1}\fi
 }

\def\shortauthtitle #1\endshortauthtitle{%       %DF 
  \leftheadtext{\uppercase{\ignorespaces#1\unskip}}}

\def\thelecture{}
\def\thelecturelabel{Lecture}                     %DF
\def\thenewlecturelabel{Lecture}                     %DF
\def\lecturelabel#1{\def\thenewlecturelabel{#1}}  %DF   (to get headings right)
\def\lecture#1{%
  \ifNoNewpage\else\vfill\eject\fi               %DF
  \global\lecture@true
  \gdef\lectnomark{#1}%
  \gdef\thelecture{%
    \begingroup\raggedleft
      \twelvepoint\bf
      \expandafter\uppercase\expandafter{\thelecturelabel}
      {\fourteenbf #1\unskip}\break
  }}

\def\lecturename #1\endlecturename{%
  \gdef\lectnamemark{#1}%
  \gdef\shortlectnamemark{\uppercase{\ignorespaces#1\unskip}}%          %DF
  \gdef\thelecturename{%
      \def\\{\break}%
      \baselineskip=16pt #1\unskip\endgraf
    \endgroup}}

\def\shortlecturename #1\endshortlecturename{%  %DF
  \gdef\shortlectnamemark{\uppercase{\ignorespaces#1\unskip}}}          %DF

\def\chapter{\let\savedef@\chapter
  \def\chapter##1{\let\chapter\savedef@
  \leavevmode\hskip-\leftskip
   \rlap{\vbox to\z@{\vss
   \ifodd\pageno
   \rightline{\twelvepoint\bf
   \frills@{CHAPTER\fourteenpoint\bf\space\afterassignment\chapterno@
       \global\chaptercount@=}%
   ##1\unskip}%
   \else
   \leftline{\twelvepoint
   \frills@{CHAPTER\fourteenpoint\bf\space\afterassignment\chapterno@
       \global\chaptercount@=}%
   ##1\unskip}%
   \fi
\baselineskip44pt\null}}\hskip\leftskip}%
 \nofrillscheck\chapter}

\def\author{\let\savedef@\author
  \def\author##1\endauthor{\global\setbox\authorbox@\vbox{%
        \fourteenpoint\bf\raggedleft
  \ignorespaces##1\unskip\break\endgraf}%
 \edef\next{\the\leftheadtoks}%
 \ifx\next\empty{\leftheadtext{##1}}\fi}\author}

\def\address#1\endaddress{%
  \global\advance\addresscount@\@ne
  \expandafter\gdef\csname address\number\addresscount@\endcsname
  {{\ignorespaces#1\unskip}}}

%  Current addresses as well as permanent ones must be accommodated.
%  Check on the skip before the address; it may be a fixed 6pt.

\def\email{\let\savedef@\email
  \def\email##1\endemail{\let\email\savedef@
  \toks@{\def\usualspace{{\bf\enspace}}}%
  \toks@@{{\tt ##1\unskip}}%
  \expandafter\xdef\csname email\number\addresscount@\endcsname
  {\endgraf\noindent\the\toks@\frills@{{\noexpand\bf E-mail address
        \noexpand\/}\rm :%
     \noexpand\enspace}\the\toks@@}}%
  \nofrillscheck\email}

\def\curraddr{\let\savedef@\curraddr
  \def\curraddr##1\endcurraddr{\let\curraddr\savedef@
  \toks@\expandafter\expandafter\expandafter{%
       \csname address\number\addresscount@\endcsname}%
  \toks@@{##1\unskip}%
  \expandafter\xdef\csname address\number\addresscount@\endcsname
  {\the\toks@\endgraf\noindent
    {\noexpand\rm
    \frills@{{\noexpand\bf Current address\noexpand\/}:\space}%
    \def\noexpand\usualspace{\space}\the\toks@@\unskip}}}%
  \nofrillscheck\curraddr}

\def\setaddress@{%
% \iflecture@\else\vfill\eject\fi        %COMMENTED OUT BY DF
{\parskip=0pt\ifnum\addresscount@>\z@
 \count@\z@ \loop\ifnum\count@<\addresscount@\advance\count@\@ne
 \makefootnote@{$^{\number\count@}$}%
        {\csname address\number\count@\endcsname
        \csname email\number\count@\endcsname}%
 \repeat
\fi
}\adjustfootnotemark{\number\addresscount@}
\addresscount@\z@}			%ADDED BY DF

\outer\def\head#1\endhead{%
  \add@missing\endroster \add@missing\enddefinition
  \add@missing\enddemo \add@missing\endexample
  \add@missing\endproclaim
  \penaltyandskip@{-200}\aboveheadskip
  {\headfont@\interlinepenalty\@M
  \noindent #1\unskip\endgraf}\iflecture@\else\headmark{#1}\fi%
  \penalty10000
  \parskip=\belowheadskip\noindent\parskip0pt\relax}

\outer\def\subhead#1\endsubhead{%
  \add@missing\endroster \add@missing\enddefinition
  \add@missing\enddemo \add@missing\endexample
  \add@missing\endproclaim
  \par
  \ifdim\lastskip=\belowheadskip\else
          \penaltyandskip@{-200}\abovesubheadskip\fi
  {\subheadfont@\interlinepenalty\@M
  \noindent #1\unskip\endgraf}%
  \penalty999
  \parskip=\belowheadskip\noindent\parskip0pt\relax}

\let\subsubhead\relax
\def\subsubhead{%
  \add@missing\endroster \add@missing\enddefinition
  \add@missing\enddemo
  \add@missing\endexample \add@missing\endproclaim
  \let\savedef@\subsubhead \let\subsubhead\relax
  \def\subsubhead##1\endsubsubhead{\restoredef@\subsubhead
  \par
  \ifdim\lastskip=\belowheadskip\else
          \penaltyandskip@{-50}\abovesubsubheadskip\fi
      {\def\usualspace{\/{\it\enspace}}%
    \noindent \it##1\unskip\frills@{.\enspace}}}%
  \nofrillscheck\subsubhead}

\def\endtopmatter{\add@missing\endabstract
  \edef\next{\the\leftheadtoks}%
  \ifx\next\empty
    \expandafter\leftheadtext\expandafter{\the\rightheadtoks}%
  \fi
  \ifpart
     \partfalse				% DF
     \global\plain@true
     \null\vskip11pc
     \thepart
     \vfill\eject
     \null\vfill\eject
     \global\firstpage@true	%DF
  \fi
  \global\plain@false
  \ifLogoOn
     \inslogo@ \let\logo@=\relax   %DF
  \fi
  \iflecture@                            %DF
    \ifNoNewpage\NoNewpagefalse\else     %DF
    \global\firstpage@true  %DF
%    \else\plain@true\null\vfill\eject    %DF
%    \global\plain@false\global\firstpage@true    %DF
  \fi  % DF 
  \fi
  \let\thelecturelabel\thenewlecturelabel    %DF      (to get headings right)
  \null\pretitle
  \box\titlebox@
  \topskip10pt
  \ifvoid\authorbox@
  \else
     \ifLogoOn \vskip-.8\baselineskip \else \vskip5pt\fi
     \unvbox\authorbox@
  \fi
  \ifLogoOn
     \vskip13pt
  \else
     \iflecture@ \vskip4pt \fi
  \fi
  \iflecture@
     \ifLogoOn\else\vskip12pt\fi
     \thelecture
     \thelecturename
     \rightheadtext{\uppercase\expandafter{\thelecturelabel\  %   %DF
     \lectnomark.\ \shortlectnamemark}}%  %DF 
  \fi
  \ifnum\addresscount@>\z@
     \setaddress@
  \fi
  \ifvoid\affilbox@
  \else
        \vskip1pcplus\unvbox\affilbox@
  \fi
  \ifx\thesubjclass@\empty\else \vfil\thesubjclass@\fi
  \ifx\thekeywords@\empty\else \vfil\thekeywords@\fi
  \ifx\thethanks@\empty\else \vfil\thethanks@\fi
  \ifvoid\abstractbox@\else \unvbox\abstractbox@\vskip 18pt \fi
  \ifvoid\tocbox@\else\vskip1.5pcplus.5pc\unvbox\tocbox@\fi
  \iflecture@ \vskip2pc\fi
}

\def\xcaheadfont@{\bf}
\def\xcbheadfont@{\headfont@}

% figure
\def\endinsert{\egroup % finish the \vbox
  \if@mid \dimen@\ht\z@ \advance\dimen@\dp\z@ \advance\dimen@12\p@
    \advance\dimen@\pagetotal \advance\dimen@-\pageshrink
    \ifdim\dimen@>\pagegoal\@midfalse\p@gefalse\fi\fi
  \if@mid \bigskip\box\z@\bigbreak
  \else\insert\topins{\penalty100 % floating insertion
    \splittopskip\z@skip
    \splitmaxdepth\maxdimen \floatingpenalty\z@
    \ifp@ge \dimen@\dp\z@
    \vbox to\vsize{\unvbox\z@\kern-\dimen@}% depth is zero
    \else \box\z@\nobreak\vskip 12\p@ plus 12\p@\fi}\fi\endgroup}

\iflecture@
   \def\refsfont@{\eightpoint}%
\else
   \def\refsfont@{\tenpoint}%
   \widestnumber\key{M}
\fi

\def\leftheadtext{\let\savedef@\leftheadtext
  \def\leftheadtext##1{\let\leftheadtext\savedef@
    \leftheadtoks\expandafter{\frills@{##1}}%
    \mark{\the\leftheadtoks\noexpand\else\the\rightheadtoks}
    \ifsyntax@\setboxz@h{\def\\{\unskip\space\ignorespaces}%
        \headlinefont@##1\unskip}\fi}%
  \nofrillscheck\leftheadtext}
\def\rightheadtext{\let\savedef@\rightheadtext
  \def\rightheadtext##1{\let\rightheadtext\savedef@
    \rightheadtoks\expandafter{\frills@{##1}}%
    \mark{\the\leftheadtoks\noexpand\else\the\rightheadtoks}%
    \ifsyntax@\setboxz@h{\def\\{\unskip\space\ignorespaces}%
        \headlinefont@##1\unskip}\fi}%
  \nofrillscheck\rightheadtext}

\def\captionfont@{\bf}

\def\topcaption#1#2\endcaption{%
  {\dimen@\hsize \advance\dimen@-\captionwidth@
   \rm\raggedcenter@ \advance\leftskip.5\dimen@ \rightskip\leftskip
  {\captionfont@#1}%
  \if\notempty{#2}\enspace\ignorespaces#2\unskip\fi
  \endgraf}\nobreak\bigskip}
\def\botcaption#1#2\endcaption{%
  \nobreak\bigskip
  \setboxz@h{\captionfont@#1\if\notempty{#2}.\enspace\rm#2\unskip\fi}%
  {\dimen@\hsize \advance\dimen@-\captionwidth@
   \leftskip.5\dimen@ \rightskip\leftskip
   \noindent \ifdim\wdz@>\captionwidth@
   \else\hfil\fi\eightpoint
  {\captionfont@#1\unskip}%
  \if\notempty{#2}.\enspace\rm#2\fi\endgraf}}

%%%% table of contents

\newif\ifLONG \global\LONGfalse
\newcount\lcount@

\begingroup
\let\head\relax
\let\specialhead\relax
\let\subhead\relax
\let\subsubhead\relax
\let\title\relax
\let\chapter\relax

\newbox\mytocbox@
\setbox\mytocbox@\hbox{Lecture 2.\enspace}
\gdef\tlength#1{\lcount@=0 \testlength#1\end  
         \LONGfalse\ifnum\lcount@=2 \LONGtrue\fi}
\gdef\testlength#1{\ifx#1\end \let\next=\relax \else \advance\lcount@ by 1 
  \let\next=\testlength\fi\next}

\gdef\newtocdefs{%
   \def \Title##1
       {\penalty\z@ \vskip18\p@
	\global\setbox\mytocbox@\hbox{Lecture 2.\enspace}
        \hangindent\wd\mytocbox@\noindent{\bf ##1}\endgraf}%
   \def \title##1\endtitle
       {\penalty\z@ \vskip8\p@
        \hangindent\wd\tocheadbox@\noindent{##1}\endgraf}%
   \def \chapter##1{%
        Chapter ##1.\unskip\enspace}%
   \def \part##1{%
        {\bf PART ##1}\hfill\break}%
   \def\Notetaker##1
	{\penalty\z@ \vskip6\p@\hangindent\wd\mytocbox@ 
        \noindent\hskip\wd\mytocbox@{\bf \llap(Notes by ##1)\endgraf}}%
   \def\Solutions##1 
        {\penalty\z@ \vskip6\p@\hangindent\wd\mytocbox@ 
        \noindent\hskip\wd\mytocbox@{\bf \llap(Solutions by ##1)\endgraf}}%
   \def\Lecture##1##2 
        {\penalty\z@ \vskip9\p@
	\tlength{##1}
	\ifLONG\global\setbox\mytocbox@\hbox{Lecture 12.\enspace} 
	  \else\global\setbox\mytocbox@\hbox{Lecture 2.\enspace}\fi 
 \hangindent\wd\mytocbox@\noindent{Lecture ##1.\enspace##2}\endgraf\vskip3pt}% 
   \def\Headnn##1 
       {\par\hangindent\wd\mytocbox@\parindent=\wd\mytocbox@
	\indent ##1\endgraf}
  \def\ULecture##1 
       {\title ##1\endtitle\vskip3pt} 
   \def \specialhead##1 ##2\endspecialhead
       {\par\hangindent\wd\tocheadbox@\parindent=12pt \indent
        \if\notempty{##1}\hbox to\wd\tocheadbox@{\hfil ##1\unskip\enspace}\fi
        {\bf ##2}\par}%
   \def \Head##1##2 
	{\penalty\z@ \par
        \hangindent\wd\mytocbox@\noindent
        \hbox to\wd\mytocbox@{\hfil\S##1.\unskip\enspace}##2\endgraf}%
   \def \subhead##1 ##2\endsubhead
       {\par\vskip-\parskip {\normalbaselines
        \advance\leftskip\wd\tocheadbox@
        \hangindent\wd\tocsubheadbox@ \parindent=12pt\indent
        \if\notempty{##1}%
              \hbox to\wd\tocsubheadbox@{##1\unskip\hfil}\fi
         ##2\unskip\par}}%
   \def \subsubhead##1 ##2\endsubsubhead{}% not applicable
   \def \Chapter##1##2
       {\penalty\z@ \vskip9\p@
	\tlength{##1}
	\ifLONG\global\setbox\mytocbox@\hbox{Chapter 12.\enspace} 
	  \else\global\setbox\mytocbox@\hbox{Chapter 2.\enspace}\fi 
 \hangindent\wd\mytocbox@\noindent{Chapter ##1.\enspace##2}\endgraf\vskip3pt}%
}%
\gdef\toc@#1{\relaxnext@
 \DN@{\ifx\next\nofrills\DN@\nofrills{\nextii@}%
      \else\DN@{\nextii@{{#1}}}\fi
      \next@}%
 \DNii@##1{%
  \bgroup
   \def\page####1%
       {\unskip\penalty\z@\null\hfil
        \rlap{\hbox to\pagenumwd{\quad\hfil\rm ####1}}%
              \hfilneg\penalty\@M}%
   \setbox\tocheadbox@\hbox{\S99.99.\enspace}%
   \setbox\tocsubheadbox@\hbox{0.0.0.\enspace}%
  \leftskip\z@ \rightskip\leftskip
   \setboxz@h{\bf\quad0000000}\pagenumwd\wd\z@
   \advance\rightskip\pagenumwd
   \vskip-4pt
   \newtocdefs
 }%
 \FN@\next@}%

\endgroup

%%%%%

%       define  a logo for the upper left-hand corner

%\def\nojourlogo{\let\jourlogo\empty@}
\def\nojourlogo{\LogoOnfalse}

\def\jourlogo{%
  \vbox to\headlineheight{%
    \parshape\z@ \leftskip\z@ \rightskip\z@
    \parfillskip\z@ plus1fil\relax
    \eightbf \baselineskip9pt
    \parindent\z@ \frenchspacing
    \vtop{IAS/Park City Mathematics Series\endgraf
      Volume \issuevol@, \issueyear@\par\vss}%
    \vss}%
}

\def\flheadline{\ifLogoOn\jourlogo\global\LogoOnfalse\else\empty@\fi}  % DF
\let\frheadline\flheadline 

%       macros to be put into the \topmatter for the logo

\define\volyear#1#2{\issueinfo{#1}{#1}{}{#2} 
     \let\logo@=\copyrightline@}              % ADDED BY DF

\define\issueinfo#1#2#3#4{%
  \def\issuevol@{#1}\def\issueno@{#2}%
  \def\issuemonth@{#3}\def\issueyear@{#4}}
\issueinfo{00}{0}{}{1997}

\define\copyrightinfo#1#2{\def\cryear@{#1}\def\crholder@{#2}}
\copyrightinfo{\issueyear@}{American Mathematical Society}

\font\sixsy=cmsy6

\def\issn#1{\gdef\issn@{#1}}
\issn{1070-4116}

\skip\copyins=1.5pc
\def\copyrightline@{%
  \rightline{\sixbf \textfont2=\sixsy \copyright\cryear@\ \crholder@}}

\newif\ifplain@  \plain@false

\Monograph

%%%%%%%%%%  Front End to AMS Commands          (DF)        %%%%%%%%%%% 
 
\NoBlackBoxes

\newif\iftajointauthor \tajointauthorfalse 
\newif\iftaassistantauthor \taassistantauthorfalse

\def\SeriesTitle#1{\gdef\seriestitle{#1}%          
		   \gdef\shortseriestitle{\uppercase{#1}}%  
} 
\def\ShortSeriesTitle#1{\gdef\shortseriestitle{\uppercase{#1}}}    %  optional
\def\Author#1{\gdef\authormark{#1}} 
\def\ShortAuthor#1{\gdef\shortauthor{\uppercase{#1}}}

\def\EndTopInfo{ 
%   \volyear{\volumemark}{\yearmark} 
   \LogoOnfalse
   \topmatter 
%   \series\seriestitle
%   \iftajointauthor
%     \iftaassistantauthor
%       \sauth\authormark\\ \\ {\it with the assistance of\/}\\\ta\endsauth 
%     \else 
%       \sauth\authormark\\\ta\endsauth 
%     \fi
%   \else 
%     \sauth\authormark\endsauth 
%   \fi
%   \endseries 
   \title\seriestitle\endtitle 
   \iftajointauthor
     \iftaassistantauthor
       \author\authormark\\ \\ {\it with the assistance of\/}\\\ta\endauthor 
       \shortauthtitle\shortauthor, \shortseriestitle\endshortauthtitle
     \else 
       \author\authormark\\ \ta\endauthor 
       \shortauthtitle\shortauthor, \shortta, %
         \shortseriestitle\endshortauthtitle 
     \fi
   \else 
     \author\authormark\endauthor 
     \shortauthtitle\shortauthor, \shortseriestitle\endshortauthtitle
   \fi
%   \address\authoraddress\endaddress 
%   \email\authoremail\endemail 
   \iftajointauthor 
     \address\taaddress\endaddress 
     \email\taemail\endemail 
   \fi 
   \endtopmatter
}

\def\Lecture#1#2{ 
   \topmatter
   \lecturelabel{Lecture}
   \lecture{#1}
   \lecturename #2\endlecturename
   \endtopmatter
}

\def\Chapter#1#2{ 
   \topmatter
   \lecturelabel{Chapter}
   \lecture{#1}
   \lecturename #2\endlecturename
   \endtopmatter
}

\def\Notetaker#1{
   \footnotetext""{\bf Notes by #1}
}

\def\ShortLectureName#1{ 
  \gdef\shortlectnamemark{\uppercase{\ignorespaces#1\unskip}}
  \rightheadtext{\uppercase\expandafter{\thelecturelabel\  %
  \lectnomark.\ \shortlectnamemark}}%
}

\def\BlankPage{ 
   \plain@true\null\vfill\eject\plain@false
}

\def\Head#1#2{\add@missing\endroster \add@missing\enddefinition
  \add@missing\enddemo \add@missing\endexample
  \add@missing\endproclaim
  \penaltyandskip@{-200}\aboveheadskip
  {\headfont@\interlinepenalty\@M
  \noindent \S#1. #2\unskip\endgraf}%
  \penalty10000
  \parskip=\belowheadskip\flushpar\parskip0pt}

\def\Headnn#1{\add@missing\endroster \add@missing\enddefinition
  \add@missing\enddemo \add@missing\endexample
  \add@missing\endproclaim
  \penaltyandskip@{-200}\aboveheadskip
  {\headfont@\interlinepenalty\@M
  \noindent #1\unskip\endgraf}%
  \penalty10000
  \parskip=\belowheadskip\flushpar\parskip0pt}

\def\Subhead#1#2{%
  \add@missing\endroster \add@missing\enddefinition
  \add@missing\enddemo \add@missing\endexample
  \add@missing\endproclaim
  \par
  \ifdim\lastskip=\belowheadskip\else
          \penaltyandskip@{-200}\abovesubheadskip\fi
  {\subheadfont@\interlinepenalty\@M
  \noindent #1. #2\unskip\endgraf}%
  \penalty999
  \parskip=\belowheadskip\noindent\parskip0pt\relax}

\def\SubHeadnn#1{\add@missing\endroster \add@missing\enddefinition
  \add@missing\enddemo \add@missing\endexample
  \add@missing\endproclaim
  \par
  \ifdim\lastskip=\belowheadskip\else
          \penaltyandskip@{-200}\abovesubheadskip\fi
  {\subheadfont@\interlinepenalty\@M
  \noindent #1\unskip\endgraf}%
  \penalty999
  \parskip=\belowheadskip\flushpar\parskip0pt}

%%%%%%%%%%%%%%%%%%%%%%%%%%%%%%%%%%%%%%%%%%%%%%%%%%%%%%%%%%%%%

%		FOR PROBLEMS 

\setbox0\hbox{(a)\ }  
\newdimen\secwidth \secwidth=\wd0 
\secwidth=0pt
\newcount\probno \probno=0 \newcount\probsecno
\newbox\numbox

\def\prob#1#2{#1.\;\hbox to 40pt{\hrulefill/#2}\hskip25pt}
\long\def\problem#1\endproblem{\bigbreak \global\advance\probno by 1
  \noindent {\parskip2pt plus1pt minus1pt \parindent0pt
  {\bf\problemlabel\number\probno.\enspace}#1}\bigbreak}
\def\probsec{\futurelet\next\probsecc}
\def\probsecc{\bigbreak \global\advance\probno by 1
  \noindent \begingroup  \parskip2pt plus1pt minus1pt \parindent0pt
  \probsecno=0 
  \ifx\next\sec\firstsec\else\firsttext\fi}
\long\def\firsttext#1\sec{{\bf\problemlabel\number\probno.\enspace}#1\sec}
\def\firstsec\else\firsttext\fi\sec#1\sec{\fi\advance\probsecno by 1 
    \leftskip\secwidth\leavevmode{\bf\problemlabel\number\probno.\enspace 
    (\secno\unskip)\ }#1\sec}
\long\def\sec{\par\leftskip\secwidth\smallbreak\advance\probsecno by 1 
  \leavevmode{\bf(\secno\unskip)\ }}
\def\secno{\ifcase\probsecno\or a \or b \or c \or d \or e \or f \or g \or h
  \or i \or j \or k \or l \or m \or n\else\fi}

\def\pretend#1\haswidth#2{\setbox0\hbox{#2}\hbox to \wd0{\hfill #1}}

\def\Solution#1#2{\rightheadtext{SOLUTION TO \uppercase\expandafter{\problemlabel}#1} \Headnn{Solution\footnote{Solution by #2} to %
   \problemlabel#1}}
 
\def\Solutionna#1{\rightheadtext{SOLUTION TO \uppercase\expandafter{\problemlabel}#1}\Headnn{Solution to %
   \problemlabel#1}}

\def\Notes{\begingroup\parindent12pt\eightpoint
\add@missing\endroster \add@missing\enddefinition
  \add@missing\enddemo \add@missing\endexample
  \add@missing\endproclaim
  \penaltyandskip@{-200}\aboveheadskip
  {\tenbf\interlinepenalty\@M
  \noindent References and cross
  references\unskip\endgraf}% 
  \penalty10000
  \parskip=\belowheadskip\flushpar\parskip0pt\refstyle{C}}

%	XREF CODES 

\newcount\commacount@
\def\first@#1#2\end{#1}
\def\true@{TT}
\def\false@{TF}
\def\empty@{}
\begingroup  \catcode`\-=3
\long\gdef\notempty#1{%
  \expandafter\ifx\first@#1-\end-\empty@ \false@\else \true@\fi}
\endgroup
\def\tcomma#1{\commacount@=0 \testcomma#1\end } 
\def\testcomma#1{\ifx#1\end \let\next=\relax 
  \else\if#1,\commacount@=1\fi\let\next=\testcomma\fi\next}
\def\tperiod#1{\commacount@=0 \testperiod#1\end } 
\def\testperiod#1{\ifx#1\end \let\next=\relax 
  \else\if#1.\commacount@=1\fi\let\next=\testperiod\fi\next}

\def\CODE#1#2#3#4{\if\notempty{#2}\tcomma{#2}\ifnum\commacount@=1[#4-{\bf %
   #1}, \S\S#2]\else \tperiod{#2}\ifnum\commacount@=1[#4-{\bf #1},~\S#2]\else%
   #3#2 of [#4-{\bf #1}]\fi\fi\else[#4-{\bf #1}]\fi}

\def\ASH#1{\CODE{Home\-work}{#1}{Problem~ASH}{I}}

\def\CF#1{\CODE{Clas\-sical Fields}{#1}{Chapter~}{I}}
\def\DQFT#1{\CODE{Dynam\-ics of QFT}{#1}{Lecture~}{II}}

\def\DS#1{\CODE{Spin\-ors}{#1}{Chapter~}{I}}

\def\FP#1{\CODE{Home\-work}{#1}{Problem~FP}{I}}

\def\LSS#1{\CODE{Super\-sym\-metry}{#1}{Chapter~}{I}}

\def\SH#1{\CODE{Home\-work}{#1}{Problem~SH}{I}}
\def\SM#1{\CODE{Signs}{#1}{\S }{I}}

\catcode`\@=13

%\endinput

\CenteredTagsOnSplits
\NoBlackBoxes
\def\Schur{\text{\rm Schur}}
\def\sgn{\text{\rm sgn}}
\def\upvee{^{\sssize\vee}}
\define\({\left(}
\define\){\right)}

\define\Ahat{{\hat\scrA}}

\define\CC{{\Bbb C}}

\define\Cbar{\overline{C}}

\define\Crit{\operatorname{Crit}}
\define\DDad{\overline{\Cal D}_{\ad}}
\define\DDa{\DD_a }
\define\DDbar{\overline{\Cal D}}
\define\DDbd{\overline{\Cal D}_{\bd}}
\define\DDbm{\overline{\DD}_{-}}
\define\DDbp{\overline{\DD}_{+}}
\define\DDb{\DD_b }
\define\DDdd{\overline{\Cal D}_{\dd}}
\define\DDd{\DD_d }
\define\DDe{\DD_e }
\define\DDgd{\overline{\Cal D}_{\gd}}
\define\DDg{\DD_c }

\define\DDp{\DD_{+}}
\define\DD{\Cal D}
\define\Dad{\overline{D}_{\ad}}
\define\Dai{D_{a i}}
\define\Da{D_a }
\define\Dbar{\overline{D}}
\define\Dbd{\overline{D}_{\bd}}
\define\Dbm{\overline{D}_{-}}
\define\Dbp{\overline{D}_{+}}
\define\Db{D_b }
\define\Ddd{\overline{D}_{\dd}}
\define\Dd{D_d }
\define\Dens{\operatorname{Dens}}
\define\Dgd{\overline{D}_{\gd}}
\define\Dg{D_c }

\define\Diracp{\Dirac_\phi }
\define\Dirac{D\hskip-.65em /}
\define\Dm{D_- }
\define\Dp{D_+ }

\define\End{\operatorname{End}}
\define\FF{\Bbb F}
\define\Fbar{\overline{F}}
\define\Fb{\overline{F}}

\define\Gam#1#2#3{\Gamma^#1_{#2\dot#3}}
\define\Gamt#1#2#3{\Gamma^{#1\dot#2#3}}
\define\HH{{\Bbb H}}
\define\Hess{\operatorname{Hess}}
\define\Hom{\operatorname{Hom}}
\define\Jch{\check{J}}
\define\Kah{K\"ahler}
\define\La{\Lambda _a }
\define\Lbar{\overline{\Lambda }}
\define\Lb{\Lambda _b }
\define\Lch{\check{L}}

\define\Lie#1{\operatorname{Lie}(#1)}

\define\Mvac{\scrM_{\text{vac}}}
\define\OO{\Bbb{O}}

\define\Phat{\hat\scrP}
\define\Phibar{\overline{\Phi }}
\define\Phitil{\tilde{\Phi }}

\define\Qad{\overline Q_\ad }
\define\Qa{Q_a }
\define\Qbar{\overline{Q}}

\define\Qb{Q_b }

\define\Qmt{\tilde{Q}_-}
\define\Qm{Q_-}
\define\Qph{\widehat{Q}_+}
\define\Qpt{\tilde{Q}_+}
\define\Qp{Q_+}

\define\RR{{\Bbb R}}
\define\Rijkl{R_{i\jbar k\bar{\ell }}}

\define\Shat{\hat\Sigma }
\define\Sigbh{\overline{\Shat }}
\define\Sigb{\overline{\Sigma }}
\define\Spin{\operatorname{Spin}}
\define\Spps{S^{\prime\prime*}}
\define\Sps{S^{\prime*}}

\define\Sym{\operatorname{Sym}}
\define\TT{\Bbb{T}}

\define\Wa{W_a }
\define\Wbar{\overline{W}}
\define\Wb{W_b }
\define\What{\hat{W}}
\define\ZZ{{\Bbb Z}}
\define\[{\left[}
\define\]{\right]}
\define\adPP{\operatorname{ad}\scrP}
\define\adP{\operatorname{ad}P}
\define\adj{\operatorname{ad}}
\define\ad{{\dot{a }}}
\define\bdens{|d^nx|}
\define\bd{{\dot{b }}}
\define\bmink{\check{M}}
\define\bres{i ^*}
\define\ccd{\cont{\partial_{cd}}\,|d^3x|\;}
\define\cd{\dot{c}}
\define\chib{\bar{\chi }}
\define\chiup{\raise.5ex\hbox{$\chi$}}
\define\chmu{\check\mu}

\define\cont#1{\iota(#1)}
\define\cp{_\CC}
\define\curv{\operatorname{curv}}
\define\dP{\delta\Phi }
\define\daad{\partial _{a \ad}}
\define\dabd{\partial _{a \bd}}
\define\dab{\partial \mstrut _{a b }}
\define\dad{\bar\partial _{\ad}}
\define\dai{\partial\mstrut  _{a i}}
\define\da{\partial _{a }}
\define\dbad{\partial _{b \ad}}
\define\dbar{{\bar\partial}}
\define\ddabd{\nabla \mstrut _{\!\!a \bd }}
\define\ddab{\nabla \mstrut _{\!\!a b }}
\define\dd{{\dot{d }}}
\define\deln{\delta _{\nabla }}
\define\dens{\bdens\,d^s\theta }

\define\diso{\epsilon}
\define\di{\partial _\mu  }
\define\dj{\partial _\nu }
\define\dlt#1#2{\delta ^{#1}_{#2}}
\define\dm{\partial _- }
\define\dpl{\partial _+ }
\define\dt{\partial _t}
\define\dual{^*}
\define\eQ{\eta ^a Q_a }
\define\eabd{\epsilon ^{\ad \bd }}
\define\eab{\epsilon ^{a b }}
\define\ead{\bar f _{\ad }}
\define\ea{f _{a }}
\define\edd{\dot{e}}
\define\egdd{\epsilon ^{\gd \dd }}
\define\egd{\epsilon ^{c d }}
\define\eij{\epsilon _{ij}}
\define\endo{D}

\define\etabar{\bar{\eta }}
\define\etad{\etabar^{\ad}}
\define\etbd{\etabar^{\bd}}
\define\etbm{\etabar ^-}
\define\etbp{\etabar ^+}

\define\etdd{\etabar^{\dot{d}}}
\define\eted{\etabar^{\dot{e}}}
\define\etgd{\etabar^{\gd}}
\define\etm{\eta ^-}
\define\etp{\eta ^+}
\define\fD#1#2{D^{(#1)}_{#2}}
\define\fDD#1#2{\DD^{(#1)}_{#2}}
\define\fDDb#1#2{\DDbar_{\dot#1(#2)}}
\define\fDb#1#2{\Dbar_{\dot#1(#2)}}
\define\fbar{\bar{f}}

\define\gammatil{\tilde{\gamma}}
\define\gch{\check{\gamma }}
\define\gd{{\dot{c }}}
\define\gmn{g^{\mu \nu }}
\define\grad{\operatorname{grad}}
\define\hXi{\widehat{\Xi }}
\define\htil{\tilde{h}}
\define\hxi{\hat{\xi }}
\define\hyp{hyperk\"ahler}
\define\incl{i }

\define\inv{^{-1}}
\define\jbar{\bar{j}}
\define\lambdab{\bar{\lambda }}

\define\lbar{\bar{\ell }}
\define\lhat{\hat{\lambda }}
\define\lie{\Cal L}
\define\mink#1#2{M^{#1|#2}}
\define\mstrut{^{\vphantom{1*\prime y}}}
\define\nab#1{\nabla\mstrut  _{\!\!#1}}
\define\nabcd{\nu ^{abcd}}
\define\oabd{\omega ^{a\bd}}
\define\oab{\omega ^{ab } }
\define\pK{\partial K}
\define\pR{\phi ^*R}
\define\pad{\bar{\psi}_{\ad}}
\define\pa{\psi _a }
\define\pbd{\bar{\psi}_{\bd}}
\define\pb{\psi _b }
\define\pdd{\bar{\psi}_{\dd}}
\define\pd{\psi _d }
\define\ped{\bar{\psi}_{\dot{e}}}
\define\pe{\psi_e}
\define\pgd{\bar{\psi}_{\gd}}
\define\pg{\psi _c }
\define\phibar{\bar{\phi}}
\define\phib{\bar{\phi}}
\define\phidot{\dot{\phi }}
\define\protag#1 #2{#2\ #1}
\define\prt#1{\partial_#1}
\define\psib{\bar{\psi}}
\define\psidot{\dot{\psi }}
\define\rad{\overline{\rho} ^\ad }
\define\rai{\rho ^{a i}}

\define\ra{\rho ^a }
\define\rbd{\overline{\rho} ^\bd }
\define\rb{\rho ^b }
\define\res#1{\negmedspace\bigm|_{#1}}
\define\scrA{\Cal{A}}
\define\scrE{\Cal E}
\define\scrF{\Cal{F}}
\define\scrLch{\check{\scrL}}
\define\scrL{\Cal L}
\define\scrM{\Cal{M}}
\define\scrP{\Cal{P}}

\define\sdtov{\Gamma}
\define\sigb{\bar{\sigma }}
\define\sqo{\sqrt{-1}}
\define\stov{\tilde{\Gamma}}
\define\sym{\operatorname{sym}}
\define\tQQ#1{\tau_{Q_{#1}}}
\define\tQad{\tau_{\overline Q_\ad }}
\define\tQa{\tau_{\Qa}}
\define\tQbd{\tau_{\overline Q_\bd }}
\define\tQb{\tau_{Q_b }}
\define\tQc{\tau_{Q_c }}
\define\tQ{\tau_{Q}}
\define\tad{\bar\theta ^{\ad}}
\define\tai{\theta ^{a i}}
\define\ta{\theta ^a }
\define\tbd{\bar\theta ^{\bd}}
\define\tb{\theta ^b }
\define\tdens{|d^3x|\,d^2\theta }
\define\teQ{\eta ^a \tau_{Q_a} }
\define\temsquare{\raise3.5pt\hbox{\boxed{ }}}

\define\theprotag#1 #2{#2~#1}

\define\vlo{\tilde{\partial }_{1\dot2}}
\define\vlt{\tilde{\partial }_{2\dot1}}
\define\voll{\operatorname{vol}}
\define\vrt{\operatorname{vert}}
\define\wave{\square}
\define\wch{\check{\omega }}

\define\zhat{\hat{\zeta }} 
\define\zmod#1{\ZZ/#1\ZZ}

\nologo
\redefine\Im{\operatorname{Im}}
\redefine\Re{\operatorname{Re}}
\define\Signature{\operatorname{Signature}}
\define\Vect{\operatorname{VectorFields}}
 
\define\hH{\hat{H}}
\define\hP{\hat{P}}
\define\hQ{\hat{Q}}
\define\scrH{\Cal{H}}
\define\tG{\tilde{\Gamma}}
\redefine\Im{\operatorname{Im}}
\redefine\Re{\operatorname{Re}}

\input epsf
\def\arrowsim{\smash{\mathop{\to}\limits^{\lower1.5pt\hbox{$\scriptstyle\sim$}}}}
\def\Le{{\mathchoice{\,{\scriptstyle\le}\,}
  {\,{\scriptstyle\le}\,}
  {\,{\scriptscriptstyle\le}\,}{\,{\scriptscriptstyle\le}\,}}}
\def\Ge{{\mathchoice{\,{\scriptstyle\ge}\,}
  {\,{\scriptstyle\ge}\,}
  {\,{\scriptscriptstyle\ge}\,}{\,{\scriptscriptstyle\ge}\,}}}

%**end of header

\SeriesTitle{Supersolutions}
\ShortSeriesTitle{Supersolutions}

\Author{Pierre Deligne and Daniel S. Freed}
\ShortAuthor{P. Deligne, D. Freed}

\address School of Mathematics, Institute for Advanced Study, Olden Lane,
Princeton NJ 08540\endaddress 
\email deligne\@math.ias.edu\endemail
 
\address Department of Mathematics, University of Texas, Austin TX
78712\endaddress 
\email dafr\@math.utexas.edu\endemail

	 \abstract  
 We develop classical globally supersymmetric theories.  As much as possible,
we treat various dimensions and various amounts of supersymmetry in a uniform
manner.  We discuss theories both in components and in superspace.
Throughout we emphasize geometric aspects.  The beginning chapters give a
general discussion about supersymmetric field theories; then we move on to
detailed computations of lagrangians, etc\. in specific theories.  An
appendix details our sign conventions.  This text will appear in a two-volume
work {\sl Quantum Fields and Strings: A Course for Mathematicians\/} to be
published soon by the American Mathematical Society.  Some of the
cross-references may be found at {\tt
http://www.math.ias.edu/$\sim$drm/QFT/}.
	 \endabstract

\EndTopInfo 
 \unskip\removelastskip
\rightheadtext{INTRODUCTION}
\Headnn{Introduction}
 This text develops the basic classical supersymmetric field theories in all
dimensions, but without supergravity.  We emphasize at the start that
although we hope to provide a bit of a fresh viewpoint in some places, this
is standard material for the working quantum field theorist (string theorist,
$M$-theorist, \dots )---it was largely developed in the physics literature of
the~1970s and~1980s.  This account began as the solutions to the
Superhomework problems in~\SH{}, but then developed into a larger project.
Unfortunately, we often sink into an unpalatable morass of indices which may
obscure the underlying geometry.\footnote{As Mel Brooks famously said, ``We
mock the thing we are to be.''}  So we hope in this introduction to at least
provide a roadmap to the more conceptual parts of the text.  Certainly some
readers may find the plethora of formulas useful.  Indeed, we have gone to
great lengths\footnote{in units where $c=1$} to ensure their accuracy, though
undoubtedly errors have crept in.  Since there are so many formulas, we have
boxed the ones we thought most crucial or most likely to be referenced.  One
contribution here is a consistent choice of conventions in all dimensions and
with varying amounts of supersymmetry.  Our sign conventions are summarized
in the appendix.  Throughout we freely use notions of supergeometry as
developed in~\LSS{} and notions of classical field theory as developed
in~\CF{}.  Knowledge of spinors~\DS{} is also indispensable.
 
The basic Poincar\'e-invariant classical field theories of a scalar field
(spin~0), spinor field (spin~1/2), and gauge field (spin~1) exist in any
dimension~$n$.  Here we are interested in field theories which are invariant
under a supergroup which extends the Poincar\'e group.  Such a {\it super
Poincar\'e group\/} is determined by a real spin representation~$S^*$ of the
Lorentzian spin group~$\Spin(1,n-1)$ together with a symmetric pairing
$S^*\otimes S^*\to V$ into the vectors, as we explain in~\S{1.1}.  A theory
has {\it minimal supersymmetry\/} if $S^*$~is a minimal (that is,
irreducible) spin representation; otherwise, it is said to have {\it extended
supersymmetry\/}.  We construct theories on Minkowski space which obey the
spin-statistics connection, so that odd symmetries map fields of integral
spin to fields of half-integral spin.  It is convenient to measure the amount
of supersymmetry in terms of~$s=\dim S^*$.  The larger~$s$ is, the greater
the mixing of fields of different spin.  As we consider theories
only\footnote{For a physical explanation of this restriction, see~\DQFT{2.4}}
with fields of spins at most one, we are restricted to small values of~$s$.
Specifically, supersymmetric $\sigma $-models---which only have fields of
spins at most one-half---must have~$s\le 8$ and supersymmetric gauge theories
must have~$s\le 16$.  Because of this restriction the theory has a small set
of general examples and the study of the theory is centered on the examples.
We mostly consider theories with~$s=2,4,8,16$ supersymmetries and these occur
in the maximal dimensions~$n=3,4,6,10$.  ``Maximal'' means, for example, that
only in dimensions~$\le6$ are there real spin representations of
dimension~8. (We also say a bit in~\S{1.3} about theories with a single
supersymmetry and there is more in~\FP{2}, which we recommend as a starting
point before tackling the more complicated theories with more supersymmetry.
Other problems in~\FP{} also deal with formal properties of supersymmetric
field theories.)  By the exceptional isomorphisms among low dimensional Lie
groups, the Lorentzian spin group in dimensions~$n=3,4,6$ is~$SL(2,\FF)$
for~$\FF=\RR,\CC,\HH$.  (There is also a sense in which the spin group in
10~dimensions is~``$SL(2,\OO)$'' for the octonions~$\OO\,$:
see~\DS{6.5--\S6.7}.)  Our treatment emphasizes this exceptional isomorphism,
both in its theoretical and computational aspects.  Throughout we focus on
theories in the maximal dimension, though in some cases (see Chapter~9, for
example) we work out some lower dimensional theories obtained by dimensional
reduction.  (Not everything in lower dimensions comes by dimensional
reduction from the maximal dimension, but we barely mention the exceptions.)

A supersymmetric field theory is first of all an ordinary (lagrangian) field
theory on Minkowski space, described by a set of fields and a lagrangian.  As
usual the Poincar\'e symmetry is manifest.  In addition, there are {\it
nonmanifest\/} odd infinitesimal symmetries corresponding to the odd
generators of the super Poincar\'e algebra.  In this approach one checks by
direct computation that the theory is supersymmetric, i.e., that the odd
transformation of the fields is a symmetry of the lagrangian.  This is the
approach we follow in Chapter~{3} for $\sigma $-models and in Chapter~{6} for
pure Yang-Mills theories.  Although there are heuristic arguments to motivate
the form of the odd symmetries, they are not geometric.  Moreover, in many
examples the bracketing relations of the super Poincar\'e algebra are only
obeyed on-shell.  It is more satisfying to use a {\it superspace
formulation\/} in which the entire super Poincar\'e symmetry is manifest.  In
this approach the spacetime~$M$ of the classical field theory is {\it super
Minkowski space\/}, a supermanifold whose underlying ordinary manifold is
ordinary Minkowski space.  The super Poincar\'e group acts as symmetry
transformations of super Minkowski space.  The fields in the theory, usually
called {\it superfields\/}, are ``functions'' on super Minkowski space.  Then
super Poincar\'e acts on superfields by pullback, and we can write manifestly
supersymmetric lagrangians.  By restricting superfields and their derivatives
to the underlying ordinary Minkowski space, we recover the {\it component
fields\/} of the ordinary field theory on Minkowski space, and by partial
integration we recover the {\it component lagrangian\/}.  These general ideas
are explained in Chapter~{1}.
 
Although superspace formulations have the overriding virtue of exhibiting the
supersymmetry in a manifest way, there are several drawbacks.  First and
foremost, there is not always a superspace formulation!  Unfortunately, that
is the case for theories with the maximal allowed supersymmetry: $\sigma
$-models with 8~supersymmetries and Yang-Mills theories with
16~supersymmetries.  There are good superspace formulations for all theories
with at most 4~supersymmetries and also for pure gauge theories with at most
8~supersymmetries.  Of course, one can use the superspace formulation with
4~supersymmetries even for theories with more supersymmetry; then some of the
supersymmetry is manifest and some is nonmanifest.  Another disadvantage of
superspace formulations is that in many cases the superfields are
constrained.  This complicates the classical field theory computations.
Also, the constraints vary from case to case and add further to the
specificity of the subject.  To a physicist the most damning criticism of the
superspace formulation is that it obscures the physical intuition, which is
based on experience with ordinary fields on Minkowski space.  These drawbacks
are balanced by the manifest geometric realization of odd symmetries,
particularly when applied to the quantum theory.  (For example, the possible
form of quantum corrections to the classical action is clearer in the
superspace formulation.)  So we persist in presenting the superspace point of
view alongside the component approach.
 
Chapters~1 and~2 are preliminary.  Chapter~1 describes features of classical
supersymmetric field theories which go beyond the general ideas of classical
field theory.  In Chapter~2 we describe the coordinates and bases we use to
compute in~3, 4, and 6~dimensions.  As mentioned before, they are based on
the special form of the spin group in those cases.  We list many formulas for
easy future reference.   
 
Our real work begins in Chapter~{3}, where we study supersymmetric $\sigma
$-models in components.  The quantum theories are often discussed in
2~dimensions, but we treat the classical theories in the maximal
dimensions~$n=3,4,6$ (where they have minimal supersymmetry).  In the linear
case we give a uniform description of the theory and a uniform proof of
supersymmetry.  In the nonlinear case the scalar field is a map from ordinary
Minkowski space to a curved Riemannian manifold~$X$.  Supersymmetry
constrains~$X$: in the $n=4$~theory $X$~must be \Kah\ and in the $n=6$~theory
$X$~must by \hyp.  In~\S{3.3} we give a uniform motivation for these
constraints.  In the nonlinear case there are curvature terms in the
lagrangian, and as they vary in the three cases we check supersymmetry on a
case-by-case basis.  In addition to the kinetic terms, there are
supersymmetric potential terms one can add in the $n=3,4$~theories; they are
summarized in~\S{3.4}.
 
Chapters~4 and~5 present the superspace formulation of supersymmetric $\sigma
$-models in $n=3,4$~dimensions, including potential terms.  In Chapter~{4} we
not only derive the component lagrangian and supersymmetry transformations,
but we also discuss the classical field theory computations directly in
superspace, including the computation of the Noether current for the
supersymmetries.  This is the only example we work out in such detail in
superspace.  The superspace lagrangian of the $n=4$ theory is usually given
in terms of a local \Kah\ potential on the target manifold~$X$, and so does
not make sense globally.  In~\S{5.2} we point out a global lagrangian in case
the target manifold is Hodge. 
 
In Chapter~{6} we add gauge fields to the mix.  We begin in~\S{6.1} with pure
supersymmetric Yang-Mills theories in components.  The usual argument which
verifies supersymmetry relies on a ``Fierz identity,'' something we do not
encounter in our treatment.  Instead we rely on a geometric property of
spinors which only holds in dimensions~$n=3,4,6,10$ with the minimal real
spinor representation: the quadratic pairing on spinors takes values on the
light cone.  (In other cases the values are positive timelike or lightlike,
not just lightlike.)  That same property enters again in~\S{6.3}, where we
describe some common features of the superspace formulation of the theory
in~$n=3,4,6$ dimensions.  In each case the basic superfield is a connection
on super Minkowski space constrained to have vanishing curvature along a
canonical odd distribution.  The aforementioned special property of spinors
is used to see that this constraint leads to the correct physical component
fields.  On the other hand, the auxiliary fields\footnote{An {\it auxiliary
field\/} has an algebraic equation of motion and does not correspond to a
physical degree of freedom.}  differ in each case; they are treated in later
chapters.
 
The most general theory without gravity is a gauge theory with matter---a
coupled system consisting of scalar fields, spinor fields, and gauge fields.
The general supersymmetric theories are summarized in {\theprotag{6.33}
{Theorem}}.  Of interest in applications are the nontrivial potential energy
functions which appear; they lead to nontrivial classical moduli spaces.  We
describe the coupled systems in dimensions~$n=3,4,6$ with minimal
supersymmetry.  (Recall that there is no possible $\sigma $-model with
$s=16$~supersymmetries, only the pure Yang-Mills theory described
previously.)  Systems with~$s=2,4,8$ supersymmetries in lower dimensions are
often, but not always, obtained by dimensional reduction from the theories
described here.  For~$n=3,4$ there is a good superspace description from
which the component description may be derived.  This is the subject of
Chapters~7 and~8.  In the $n=4$~case the appearance of the moment map in the
component lagrangian has a geometric explanation, at least if the target
\Kah\ manifold is Hodge (see~\S{8.3}).  In Chapters~7 through~11 we only
describe the Yang-Mills terms in the lagrangian, having treated the $\sigma
$-model terms in Chapters~4 and~5.  Chapter~10 gives the superspace
description of super Yang-Mills theory in~$n=6$ dimensions.  (Recall here
that there is no superspace description of the $\sigma $-model, hence none of
the coupled system.)  For each of the cases~$n=3,4,6$ we prove that the
category\footnote{Because of the local gauge transformations, the collection
of connections (gauge fields) forms a category, not a set.} of constrained
connections is equivalent to the category of component fields.  For the
$n=6$~case the proof is presented in Chapter~{11}.
 
A connection is locally specified by $n$~Lie algebra-valued functions in
$n$~dimensions, and the dimensional reduction to~$n-k$ dimensions still has
$n$~functions: $n-k$~of them transform as a connection and $k$~of them as
scalar fields (see~\S{9.1}).  There are two cases of dimensionally reduced
super Yang-Mills theory which are of particular interest---the reduction of
the $n=4$~theory to 2~dimensions and the reduction of the $n=6$~theory to
4~dimensions.  The former is the subject of Chapter~{9} and the latter the
subject of~\S{10.2}.
 
Finally, in~\S{5.5} and~\S{10.3} we describe theories in $n=4$~dimensions
with extended supersymmetry on the minimal super Minkowski space.  Then four
of the supersymmetries are manifest and the extended supersymmetry is
nonmanifest.   
 
In several places we refer to the $n=3,4,6$~cases with minimal supersymmetry
as the $\FF=\RR,\CC,\HH$ cases.
 
Writing this has been an absolute {\it cauchemar de signes\/}!  Standard
differential geometry has some bad signs, classical field theory has tricky
sign conventions, and odd variables add a whole new level of complication.
Our sign conventions are explained in the previous texts in Part~1 and are
summarized in~\SM{}.

During the preparation of this manuscript we received advice and help from
many people.  Certainly Edward Witten's Superhomework~\SH{} was our starting
point for organizing the material, and Ed was invaluable at many other
times.  Input from our classmates, notably David Kazhdan and John Morgan, is
also reflected here, as is the advice of Joseph Bernstein, Nati Seiberg, and
many others.  However, they should not be held accountable for any errors.

\SubHeadnn{Contents}
 
\def\Page#1{\relax}

\toc
\Chapter{1}{Preliminary Topics }\Page{5}
\Head{1.1}{Super Minkowski spaces and super 
  Poincar\'e groups}\Page{5}
\Head{1.2}{Superfields, component fields, and lagrangians}\Page{10}
\Head{1.3}{A simple example}\Page{15}
\Chapter{2}{Coordinates on Superspace}\Page{17}
\Head{2.1}{$\mink32$, $\mink44$, $\mink6{(8,0)}$ and their complexifications}\Page{17}
\Head{2.2}{Dimensional reduction}\Page{19}
\Head{2.3}{Coordinates on $\mink32$}\Page{20}
\Head{2.4}{Coordinates on $\mink44$}\Page{23}
\Head{2.5}{Coordinates on $\mink6{(8,0)}$}\Page{28}
\Head{2.6}{Low dimensions}\Page{31}
\Chapter{3}{Supersymmetric $\sigma$-Models}\Page{35}
\Head{3.1}{Preliminary remarks on linear algebra}\Page{35}
\Head{3.2}{The free supersymmetric $\sigma$-model}\Page{37}
\Head{3.3}{Nonlinear supersymmetric
   $\sigma$-model}\Page{41}
\Head{3.4}{Supersymmetric potential terms}\Page{47}
\Head{3.5}{Superspace construction}\Page{50}
\Head{3.6}{Dimensional reduction}\Page{50}
\Chapter{4}{The Supersymmetric $\sigma $-model 
   in dimension 3 }\Page{53}
\Head{4.1}{Fields and supersymmetry 
   transformations on~$\mink32$}\Page{53}
\Head{4.2}{The $\sigma $-model action on~$\mink32$}\Page{56}
\Head{4.3}{The potential term on~$\mink32$}\Page{56}
\Head{4.4}{Analysis of the classical theory}\Page{58}
\Head{4.5}{Reduction to~$\mink2{(1,1)}$}\Page{63}
\Chapter{5}{The Supersymmetric $\sigma $-Model 
   in dimension 4 }\Page{65}
\Head{5.1}{Fields and supersymmetry transformations on~$\mink44$}\Page{65}
\Head{5.2}{The $\sigma $-model action on~$\mink44$}\Page{67}
\Head{5.3}{The superpotential term on~$\mink44$}\Page{70}
\Head{5.4}{Analysis of the classical theory}\Page{71}
\Chapter{6}{Supersymmetric Yang-Mills Theories }\Page{73}
\Head{6.1}{The minimal theory in components}\Page{73}
\Head{6.2}{Gauge theories with matter}\Page{77}
\Head{6.3}{Superspace construction}\Page{83}
\Chapter{7}{$N=1$ Yang-Mills Theory in Dimension 3 }\Page{87}
\Head{7.1}{Constrained connections on~$\mink32$}\Page{87}
\Head{7.2}{The Yang-Mills action on~$\mink32$}\Page{91}
\Head{7.3}{Gauge theory with matter on~$\mink32$}\Page{92}
\Chapter{8}{$N=1$ Yang-Mills Theory in Dimension 4 }\Page{95}
\Head{8.1}{Constrained connections on~$\mink44$}\Page{95}
\Head{8.2}{The Yang-Mills action on~$\mink44$}\Page{99}
\Head{8.3}{Gauge theory with matter on~$\mink 44$}\Page{101}
\Chapter{9}{$N=2$ Yang-Mills in Dimension 2 }\Page{105}
\Head{9.1}{Dimensional reduction of bosonic Yang-Mills}\Page{105}
\Head{9.2}{Constrained connections on~$\mink2{(2,2)}$}\Page{107}
\Head{9.3}{The reduced Yang-Mills action}\Page{108}
\Chapter{10}{$N=1$ Yang-Mills in Dimension 6 and $N=2$ 
  Yang-Mills in Dimension 4 }\Page{111}
\Head{10.1}{Constrained connections on~$\mink6{(8,0)}$}\Page{111}
\Head{10.2}{Reduction to $M^{4\vert 8}$}\Page{114}
\Head{10.3}{More theories on $M^{4\vert 4}$ with
   extended supersymmetry}\Page{118} 
\Chapter{11}{The Vector Multiplet on $M^{6\vert
   (8,0)}$}\Page{121}
\Head{11.1}{Complements on $M^{6\vert (8,0}$}\Page{121}
\Head{11.2}{Constrained connections}\Page{122}
\Head{11.3}{An auxiliary Lie algebra}\Page{123}
\Head{11.4}{Components of constrained connections}\Page{126}
\ULecture{Appendix: Sign Manifesto}
\ULecture{References}\Page{129}
\endtoc

 \Chapter1{Preliminary Topics}

 \comment
 lasteqno 1@ 57
 \endcomment

A supersymmetric field theory has as symmetry group a {\it super Poincar\'e
group\/}.  In~\S{1.1} we define these supergroups as symmetry groups of {\it
super Minkowski space\/}.  In many cases supersymmetric theories have a
superspace formulation in terms of fields defined on super Minkowski space.
Passage from the superspace formulation to the formulation in ``components''
on ordinary Minkowski space is the subject of~\S{1.2}.  The simplest example
is in one dimension (time) with one odd variable; this is described
in~\S{1.3}.

\Head{1.1}{Super Minkowski spaces and super Poincar\'e groups}  
 A super Minkowski space~$M$ is constructed as follows.  The starting data
are a Minkowski space~$\bmink$ with vector space of translations~$V$, a
positive cone~$C$ of timelike vectors in~$V$, a real spinorial
representation~$S$ of~$\Spin(V)$, and a symmetric morphism~$\sdtov$ of
representations of~$\Spin(V)$:
  $$ \sdtov\:S^*\otimes S^*\longrightarrow V. \tag{1.1} $$
which is positive definite in the sense that $\Gamma(s^*,s^*)\in \Cbar$ for
$s^*\in S^*$, with\break $\Gamma(s^*,s^*)=0$ only for $s^*=0$.

Minkowski space is an affine space: it has no origin.  By the classification
in~\DS{6.3(i)}, once the representation $S$ is given, positive definite
$\Gamma$ as in (1.1) exist and are unique up to automorphisms of the
representation $S$.  If $S$ is an irreducible real spinorial representation,
nonzero symmetric morphisms $S^*\otimes S^*\to V$ are unique up to a factor,
are positive or negative definite, and are respected by elements of norm $1$
in the field of endomorphisms of $S$ (isomorphic to $\RR$, $\CC$ or $\HH$)
~\DS{6.1}.

Note that so far we have not used the metric on $V$, only its conformal
structure and the choice of a positive cone.

By~\DS{6.2} there is a unique symmetric morphism
  $$ \stov\colon\, S\otimes S\to V\tag{1.2} $$
related to $\Gamma$ by the following formula.  With respect to a basis
$\{e_\mu\}$ of $V$ and $\{f^a\}$ of $S$, write\footnote{We use the usual
summation convention throughout: Repeated indices are summed if one index is
``upstairs'' and one is ``downstairs.''  Also, the Kronecker $\delta $-symbol
has its usual meaning.} $\sdtov(f_a ,f_b ) = \sdtov^\mu _{a b }e\mstrut _\mu
$ and $\stov(f^a ,f^b ) = \stov^{\mu a b }e_\mu $.  Then
  $$ \boxed{\Gamma_{ab}^\mu\,\stov^{\nu bc}+\Gamma_{ab}^\nu\, \stov^{\mu
     bc}=2g^{\mu\nu}\delta_a^c\,\,.} \tag{1.3} $$
If one uses the Minkowski bilinear form $g(\cdot ,\cdot )=(\cdot ,\cdot )$ on
$V$ to convert~\thetag{1.1} and~\thetag{1.2} into morphisms
  $$  \gamma\colon\, V\to\Hom(S^*,S),\qquad \gammatil\colon\,
     V\to\Hom(S,S^*) \tag{1.4} $$ 
i.e., if one defines $\gamma$ and $\gammatil$ by
  $$  \aligned t^*\bigl(\gamma(v)(s^*)\bigr) &=(\Gamma(s^*,t^*),v)\quad
     \text{and}\\ \gammatil(v)(s)(t)
     &=(\stov(s,t),v),\,\, \endaligned \tag{1.5} $$ 
then \thetag{1.3} means that for $v$ in $V$,
  $$  \align \gammatil(v)\gamma(v) &\colon\,S^* \to S^*\quad \text{and}\\
     \gamma(v)\gammatil(v) &\colon\,S\to S \endalign $$
are  multiplication by $(v,v)$.  In other words,
$\gamma$ and $\gammatil$ turn $S\oplus S^*$ into a
module over the Clifford algebra $C(V)$.
By~\DS{4.9.6} the action of $\Spin(V)$ on $S$ and
$S^*$ is induced by this $C(V)$-module structure.
Taking the trace in $\gammatil(v)\gamma(v)=(v,v)$, i.e.,
contracting the indices $a$ and $c$ in \thetag{1.3}, one obtains
  $$  \Gamma_{ab}^\mu\,\stov^{\nu ab}+\Gamma_{ab}^\nu\, \stov^{\mu
     ab}=2\,\dim(S)g^{\mu\nu}\,\,. \tag{1.6} $$ 

The identity \thetag{1.3} also means that if $(v,v)\not=0$, the
inverse of the symmetric bilinear form $(\Gamma(s^*,t^*),v)$
on $S^*$ is the bilinear form $(v,v)^{-1}(\stov(s,t),v)$
on $S$.
For $v$ in $C$, the bilinear form $(\stov(s,t),v)$ is
hence positive definite, and $\stov\colon\, S\otimes
S\to V$ is positive definite.

By the unicity of $\Gamma$ up to automorphisms of $S$, it
follows that if the representation $S$ of $\Spin(V)$ is
isomorphic to its dual, there exists an isomorphism of
representations $\alpha\colon\, S\to S^*$ such that
  $$  \stov(s,t)=\Gamma(\alpha(s),\alpha(t))\,\,. \tag{1.7} $$ 
This isomorphism corresponds to a nondegenerate
$\Spin(V)$-invariant bilinear form~$\epsilon$ on $S$:
  $$  \epsilon(s,t)=\alpha(s)(t)\,\,. \tag{1.8} $$ 
The form $\epsilon(t,s)$ corresponds to the transpose
${}^t\alpha$ of $\alpha$.
One has also
$\stov(s,t)=\Gamma({}^t\alpha(s),{}^t\alpha(t))$.
Indeed, expressing for $(v,v)\not=0$ that the nondegenerate
forms $(v,\stov(s,t))$ and $(v,\Gamma(\alpha(s),\alpha(t)))$ 
have the same inverse, one finds that
$\Gamma(s',t')=\stov(\alpha^*(s'),\alpha^*(t'))$,
with $\alpha^*={}^t\alpha^{-1}$.

In coordinates, \thetag{1.7} reads
  $$ \boxed{\stov^{\mu ab}=\Gamma_{cd}^\mu\,\epsilon^{ac}\epsilon^{bd}\,\,.}
     \tag{1.9} $$

We now assume that $S$ is irreducible, and is isomorphic to its dual.  This
excludes the cases where the dimension $n$ of $V$ is congruent to $2$ or $6$
modulo $8$.  We treat in turn the cases where the field of endomorphisms $Z$
of $S$ is isomorphic to $\RR$ ($n\equiv 1$ or $3\bmod{8}$), $\HH$ ($n\equiv
5$ or $7\bmod{8}$) or $\CC$ ($n\equiv 0$ or $4\bmod{8}$).  Note that
\thetag{1.7} holds up to a factor for any invariant non-zero bilinear form
$\epsilon$.

\subsubhead Case $Z=\RR$\endsubsubhead
 Up to a factor, $S$ admits a unique invariant bilinear form, symmetric for
$n\equiv 1(\bmod8)$ and alternating for $n\equiv 3(\bmod8)$.  The form
$\epsilon$ for which \thetag{1.7} holds is unique up to sign.

\subsubhead Case $Z=\HH$\endsubsubhead
 Up to a factor, $S$ admits a unique invariant bilinear form
such that
  $$ \epsilon(sh,t)=\epsilon(s,t\bar h),\qquad h\in H,\quad s,t\in
     S. \tag{1.10} $$
(We take~$S$ to be a {\it right\/} $\HH$-module.)  After extension of scalars
from $\RR$ to~ $\CC$, $S$~ becomes $S_0\otimes W$, with $S_0$ irreducible and
$W$ of dimension $2$, $\HH$ becomes $\End(W)$ and \thetag{1.10}~means that
$\epsilon=\epsilon_0\otimes\epsilon_W$, for $\epsilon_0$ invariant on $S_0$ and
$\epsilon _W$ the alternating form on $W$.  The form $\epsilon$ for which
\thetag{1.10} holds is symmetric for $n\equiv 5(\bmod8)$ and alternating for
$n\equiv 7(\bmod8)$, as $\epsilon_0$ is respectively alternating and
symmetric~\DS{1.5.1}.  The form $\epsilon$ for which \thetag{1.10} and
\thetag{1.7} holds is unique up to sign.

\subsubhead Case $Z=\CC$\endsubsubhead
 After extension of scalars to $\CC$, $S$ becomes the sum
of two irreducible representations, orthogonal for $n\equiv
0(\bmod8)$ and symplectic for $n\equiv 4(\bmod8)$.
Over $\RR$, the invariant bilinear form will correspondingly be
orthogonal or symplectic, with $\epsilon(zs,t)=\epsilon(s,zt)$.
The ones for which \thetag{1.7} holds form a $U_1$-orbit under
$\epsilon(s,t)\mapsto \epsilon(s,ut)$.

\medskip
 In all cases, $\epsilon$ is symmetric or antisymmetric, so that the
resulting isomorphism $\alpha\colon\, S\to S^*$ is unambiguous, at least up
to sign.  Transporting $\epsilon$ by $\alpha$, we obtain the dual form
$\tilde{\epsilon}$ on $S^*$, which is completely unambiguous.  One
has\footnote{We do not set a convention for raising and lowering a single
index with~$\epsilon $ in the antisymmetric case, since, as was just
explained, this involves an arbitrary sign choice we prefer not to make.
However, there is no ambiguity in raising and lowering an even number of
indices, as for example in~\thetag{1.10}.  Thus $\epsilon _{12} = \epsilon
^{ab}\epsilon _{1a}\epsilon _{2b}=\epsilon ^{21}\epsilon _{12}\epsilon _{21}
= -\epsilon ^{21}=\epsilon ^{12}$.  Beware: Many standard physics texts use a
different convention here.}
  $$ \boxed{\epsilon^{ab}\,\tilde{\epsilon}_{cb}=\delta_c^a\,\,.}
     \tag{1.11} $$

If the dimension is congruent to~2 or~6 $(\bmod 8)$---i.e., for
the range of dimensions of physical interest, in dimensions~
2, 6 and 10---there are super Minkowski spaces based on
unequal number of copies of the half spinor representations
$S^{+}$ and $S^-$.
The invariant  pairing in those dimensions is between
$S^{+}$ and $S^-$, so that $\Gamma$ and $\stov$
encode information not related by a self-duality pairing.

Let $\lie$~be the following super Lie algebra: $\lie^0=V$, $\lie^1=S^*$, the
subalgebra~$\lie^0=V$ is central, and the Lie bracket on~$S^*$ is~$-2\sdtov$.
The corresponding super Lie group~$\exp\lie$ is defined to be a simply
connected super Lie group, given with an isomorphism of~$\lie$ with its Lie
algebra of left-invariant vector fields.  It can be constructed as follows.
As a space it is 
  $$ \exp\lie=V\times \Pi S^*. \tag{1.12} $$
In other words, for any supercommutative ring~$R$ the set of its $R$-points
is the even part of~$\lie_R=\lie\otimes R \cong R\otimes \lie$.  The group
law is derived from the Hausdorff formula, which terminates after the
quadratic term since $[x,[y,z]]=0$ for all $x,y,z\in \lie$:
  $$ (v_1,s^*_1)(v_2,s^*_2) = (v_1+v_2 + \frac
     12[s^*_1,s^*_2],s^*_1+s^*_2),\qquad v_1,v_2\in R^0\otimes V,\quad
     s^*_1,s^*_2\in R^1\otimes S^*. \tag{1.13} $$
The sign rule applied to~$[s^*_1,s^*_2]=-2\sdtov(s^*_1,s^*_2)$ gives
  $$ \frac 12[b_1\otimes t^*_1,b_2\otimes t^*_2] =
     b_1b_2\sdtov(t^*_1,t^*_2),\qquad
      b_1,b_2\in R^1, \quad t^*_1,t^*_2\in S^*.  \tag{1.14} $$

Fix bases~$\{e_\mu \}$ of~$V$ and $\{f^a \}$ of~$S=(S\dual)\dual$.  As
before, write $\sdtov(f_a ,f_b ) = \sdtov^\mu _{a b }e\mstrut _\mu $ and
$\stov(f^a ,f^b ) = \stov^{\mu a b }e_\mu $.  In the super Lie
algebra~$\lie$, with the basis~$\{e_\mu ;f_a\}$, the nontrivial brackets are
  $$ [f_a,f_b] = -2\sdtov^\mu _{ab}e_\mu . \tag{1.15} $$
In the physics literature it is customary to denote~$e_\mu ,f_a$ by~$P_\mu
,Q_a$, and we sometimes follow that convention.  Thus we write 
  $$ \boxed{[Q_a,Q_b] = -2\sdtov^\mu _{ab}P_\mu .} \tag{1.16} $$
The bases of~$V^*,S$ induce a coordinate system~$(x^\mu ,\theta ^a )$
on~$\exp(\lie)=V\times \Pi S^*$: the coordinates of~$P\in (\lie\otimes R)^0$
are $x^\mu (P)=\langle e^\mu ,P \rangle$ and~$\ta(P)=\langle f^a ,P \rangle$.  If
$\{e_\mu \}$~and $\{f_{a }\}$~are the dual bases of~$V$ and~$S^*$,
then $P=e_\mu x^\mu  + f_a
\ta$.  Let $\di$~and $\partial _a $~be the corresponding vector fields.  In
other words,
  $$ \di x^\nu =\delta _\mu ^\nu ,\qquad \partial _a \theta ^b  = \delta
     _a ^b ,\qquad \di\ta=\partial _a x^\mu =0. \tag{1.17} $$
The Lie algebra~$\lie$ is identified with the tangent space to~$\exp\lie$ at
the origin via $e_\mu \mapsto\di\res0$ and $f_a \mapsto\partial _a \res0$.
We shall write~$\Da$ for the left-invariant vector field which is~$\partial
_a $ at~0, and we write~$\tQa$ for the right-invariant vector field which
is~$\partial _a $ at~0.  Then
  $$ \aligned
      \Da = \partial _a -\tb\sdtov^\mu _{ab }\di, \\
      \tQa = \partial _a +\tb\sdtov^\mu _{a b}\di\endaligned
     \tag{1.18} $$
at a point with coordinates~$(x^\mu ,\tb)$.  The vector fields~$\di$ are both
left- and right-invariant.  The bracketing relations are 
  $$ \boxed{\aligned
      [\Da,\Db] &= -2\sdtov^\mu _{a b }\di , \\ 
      [\tQa,\tQb] &= \hphantom{-}2\sdtov^\mu _{a b }\di, \\ 
      [\Da,\tQb] &= \hphantom{-}0.\endaligned} \tag{1.19} $$
The~$D_a$, being left-invariant, have the same brackets as the abstract Lie
algebra~$\scrL$.  (See~\thetag{1.15}.)  The vector fields~$\Da$ and~$\tQb$
commute, since right and left translations commute.

{\it Super Minkowski space\/}~$M$ is the underlying supermanifold of the
super Lie group~$\exp(\lie)$.  We let $\exp(\lie)$~act on the left on~$M$ by
left translations.  Corresponding to~\thetag{1.12} we have a splitting
  $$ M = \bmink\times \Pi S^*. \tag{1.20} $$
It is well-defined since the ambiguity in identifying~$M$ with~$\exp(\lie)$
is translation by an element of~$V$, and such translations preserve the
splitting~\thetag{1.12}.  The {\it super Poincar\'e group\/} is the
semi-direct product $\Spin(V)\ltimes\exp(\lie)$.  It acts on~$M$, and the
structures we will consider on~$M$ are invariant under the super Poincar\'e
group, with one notable exception---the splitting~\thetag{1.20}.  For
example, the left translates of~$S^*=\lie^1\subset \lie$ form a
left-invariant distribution~$\tau $ on~$M$ which is preserved by the action
of the super Poincar\'e group.
 
As we saw, bases~$\{e^\mu \}$ of~$V\dual$ and $\{f^a \}$ of~$S$ give a
coordinate system~$(x^\mu ,\ta)$ on~$M=\exp(\lie)$.  It will often be more
convenient to make computations not in this coordinate system, but rather
using the left-invariant moving frame consisting of the vector
fields~$\Da,\di$ and/or the dual moving coframe consisting of the
1-forms~$\ra, \omega ^\mu$.  We fix the sign in the duality by putting vectors
to the left of forms:
  $$ \langle \Da,\rb  \rangle = \delta ^b _a . \tag{1.21} $$
Also, we will sometimes use {\it complex\/} bases and so complex vector
fields and differential forms.  The complex conjugate of a product is the
product of the complex conjugates in the {\it same\/} order, even for odd
elements.\footnote{See~\SM{5}.}
 
The infinitesimal generators of the action of~$\exp(\lie)$ on~$M$ are the
right-invariant vector fields.  The $\tQa$ defined above are a basis of the
odd right-invariant vector fields. 

A spinor field~$\psi =\psi _{a} f^a $ on~$\bmink$ is a map ~$\psi \:\bmink\to
\Pi S$.  The form~$\stov$ then determines a symmetric (in the super sense)
bilinear form on spinor fields, which we denote with the usual physicists'
notation:
  $$ \boxed{\psi \Dirac\psi = \stov^\mu (\psi ,\partial _\mu \psi ) =
     \stov^{\mu a b } \psi _a \partial _\mu \psi _b .} \tag{1.22} $$
We call~\thetag{1.22} the {\it Dirac form\/}.  One should
understand~`$\Dirac$' in this formula as the name of the bilinear form, which
has been inserted between the arguments.  The Dirac form is symmetric up to
an exact term, so is exactly symmetric when integrated over~$\bmink$
(assuming no contributions at infinity).  A dual spinor field $\lambda
=\lambda ^a f_a $ is a map~$\lambda \:\bmink\to \Pi S^*$, and we
have the (dual) Dirac form 
  $$ \boxed{\lambda \Dirac\lambda =\sdtov^\mu (\lambda ,\partial _\mu \lambda
     ) = \sdtov^\mu _{a b }\lambda ^a \partial _\mu \lambda ^b .}
     \tag{1.23} $$
If there is a duality pairing~\thetag{1.8}, then we can use it to identify
dual spinors with spinors, in which case~\thetag{1.23} is identified
with~\thetag{1.22}. 

We complexify~$V$ and~$S$ to construct complexified super Minkowski
space~$M\cp$.  Inside~$\bmink\cp$ we may consider a real affine subspace~$E$
of Euclidean signature.\footnote{With our sign conventions the metric will be
{\it negative\/} definite; simply change the sign of the metric to obtain a
positive definite metric.}  However, the complex spinors~$S\cp$ generally
will not carry a real structure under the Euclidean spin group~$\Spin(E)$,
and so the corresponding superspace $\bmink\times \Pi S\cp$ is complex in odd
directions.  It is a cs supermanifold~\LSS{4.8}.  The lagrangians we
construct can be rotated to Euclidean space.\footnote{For more details on the
transition from Minkowski space to Euclidean space, see~\CF{7}.}

        \remark{\protag{1.24} {Remark}}
 The super Minkowski space is usually called ``$N=k$~superspace'' if $S$~is
the sum of $k$~irreducible real representations of~$\Spin(V)$.  However, the
usage fluctuates somewhat for $n=\dim(\bmink)$ small.  For example, if
$\dim(\bmink)=2$, then $N=1$~usually means that $S$~is the sum of the two
inequivalent real semi-spinorial representations~$S^+$ and~$S^-$.  It is more
informative to use the notation~`$\mink ns$' to indicate $s$~supersymmetries
in $n$~dimensions, and in dimensions~$n\equiv2 (\bmod4)$ to use a special
notation to indicate the precise spinorial representation:
`$M^{n|(s^+,s^-)}$'~denotes a super Minkowski space of dimension~$n|s^++s^-$
built from a spinorial representation which is a sum of copies of~$S^+$, of
dimension~$s^+$, and a sum of copies of~$S^-$, of dimension~$s^-$.
        \endremark

The connected group of outer automorphisms of the super Poincar\'e group
which fix the Poincar\'e subgroup is called the {\it R-symmetry group\/}.
Dimensional reduction leads to an R-symmetry group (double covering
rotations in the extra spatial dimensions), but it is not the only source of
R-symmetries.

\Head{1.2}{Superfields, component fields, and lagrangians\footnote{Our
formulas in this subsection are meant to be suggestive only.  Precise
versions depend on the particular theory; they appear in subsequent sections.
For example, versions of~\thetag{1.26} for the supersymmetric $\sigma $-model
appear in~\thetag{4.7} and~\thetag{5.7}.  Also, versions of~\thetag{1.29}
are used in the supersymmetric $\sigma $-model to compute~\thetag{4.14}
and~\thetag{5.10}.}}

For many theories we work directly on super Minkowski space~$M$.  In such
situations we say the theory has a ``superspace formulation.''  Fields on~$M$
are called {\it superfields.}  Just as in ordinary nonsupersymmetric field
theory, typical examples are: (i)\ maps $\Phi \:M\to X$ into an ordinary
manifold~$X$; and (ii)\ a connection~$A$ on a principal bundle~$\scrP\to M$.
Very often superfields are constrained.  For example, in the four-dimensional
supersymmetric $\sigma $-model we constrain certain derivatives of~$\Phi $ to
vanish, and in all super gauge theories we constrain certain components of
the curvature to vanish.  The usual elementary facts about functions apply to
superfields.  Thus elementary operations such as composition, sum, and
product produce new superfields.  We can also produce new superfields by
differentiation, for example by the vector fields~$\Da,\tQa$ defined
in~\thetag{1.18}.
 
{\it Component fields\/} are ordinary (even and odd) fields on Minkowski
space~$\bmink$.  We view these as maps from~$\bmink$ into a supermanifold.
Given a superfield~$\Phi $ we define its component fields to be the
restriction to~$\bmink$ of certain derivatives of~$\Phi $.  Let
  $$ \incl \: \bmink\longrightarrow M \tag{1.25} $$
denote the inclusion of Minkowski space into super Minkowski space.  Then for
a map $\Phi \:M\to X$ we typically define a {\it multiplet\/} of component
fields~$(\phi ,\psi ,F)$ with formulas like
  $$ \aligned
      \phi &= \bres\Phi , \\ 
      \psi  &= \bres D\Phi,  \\ 
      F &= \bres DD\Phi .\endaligned \tag{1.26} $$
For specific instances, see~\thetag{4.2} or~\thetag{5.3}.  The definition of
the multiplet is chosen so that the supermanifold of superfields is
isomorphic to the supermanifold of multiplets.  (In gauge theory this is an
equivalence of categories rather than a bijection.)  In~\thetag{1.26} $\phi
$~is a map $\phi \:\bmink\to X$ and $\psi $~is a spinor field on~$\bmink$
with values in~$\phi ^*TX$.  Note that $\phi ,F$~are even, whereas $\psi $~is
odd.  The field~$F$ is {\it auxiliary\/} in the sense that for fundamental
lagrangians no derivatives of~$F$ occur in the lagrangian.  The classical
field equation for~$F$ (keeping the other fields fixed) expresses~$F$ in
terms of the other fields.  In some cases there are no auxiliary fields.  If
$X$~is flat space, then the formulas~\thetag{1.26} are well-defined as
written.  If $X$~is curved, however, then the outermost~$D$ in the definition
of~$F$ acts on a section of the nontrivial bundle~$\Phi ^*TX$, and we use a
covariant derivative to define the action of~$D$.  In the simplest
supersymmetric $\sigma $-models $X$~is Riemannian and we use the pullback of
the Levi-Civita connection.  In other models different connections on~$TX$
(e.g.~with torsion) might be more convenient.  In gauge theories we meet
superfields~$\Lambda$ which are sections of nontrivialized bundles associated
to the principal bundle\footnote{In other places we call a `principal
$G$~bundle' a `$G$-torsor'.}~$\scrP$.  A typical example is the adjoint
bundle.  Since one of the fields in the theory is a connection~$A$
on~$\scrP$, we use it to define derivatives of~$\Lambda $.  Our covariant
definitions of component fields are global---we never choose coordinate
systems or local trivializations.
 
Component fields are defined in terms of the left-invariant vector
fields~$D$.  {\it Lagrangian densities\/} on superspace are also written in
terms of~$D$.  Recall that a lagrangian density~$L$ is a section of $\Dens
M\to \scrF\times M$, where $\scrF$~is the supermanifold of superfields and
$\Dens M$~the density bundle of~$M$, pulled back over the product.  Let $\eta
^a $~be odd parameters. Then the element~$\eta ^aQ_a$ induces an even vector
field~$\xi $ on~$\scrF\times M$.  The corresponding one-parameter group
combines the motion on~$M$ with the {\it inverse\/}\footnote{We use pullback
by the {\it inverse\/} to achieve a {\it left\/} action on functions.  This
introduces a minus sign into the infinitesimal action of vector fields on
functions.  For example, $\di$~acts on functions by {\it minus\/} the
indicated derivative.} pullback on fields.  We now give a general argument
that the lagrangians we construct are invariant under~$\xi$.  First, note
that in our two examples---a $\sigma $-model with field~$\Phi \:M\to X$ and a
gauge theory with connection~$A$ on a principal bundle $\scrP\to M$ over
super Minkowski space---the connection~$\scrA$ we use to define the action
of~$D$ on superfields depends on the field, so is best thought of as living
on~$\scrF\times M$.  For example, in the $\sigma $-model $\scrA$~is the
pullback of the Levi-Civita connection on~$TX$ by the evaluation map $
\scrF\times M\to X$.  In the gauge theory $\scrA$~is the natural ``universal
connection'' on $\scrP\to \scrF\times M$ which is flat in the
$\scrF$~direction.  Now for the $\sigma$-model $\scrA$~is canonically trivial
along the orbits of~$\xi$, since $\xi$~is tangent to the fibers of the
evaluation map.  Hence the actions of~$\xi$ and~$D$ on superfields via the
connection~$\scrA$ do (super)commute.  In the gauge theory case the same
statement is true up to gauge transformations, as the reader may verify with
some computation (cf.~the last remark in~\CF{4.2}).  Therefore, if $L$~is a
gauge-invariant lagrangian density constructed from derivatives of
superfields by the~$D$'s, then $L$~is invariant under~$\xi$.  This is the
statement that ``$L$~is supersymmetric''.
 
Next, we derive the general shape of the formula for the action of~$\eQ$ on
component fields.  We denote a typical component field as
  $$ f=\bres D^r\Phi . \tag{1.27} $$
Implicit in the notation is that $D$~acts by covariant derivative where
necessary.  Now $\eQ $~generates a diffeomorphism~$\exp(\eta^a \tQa)$ of~$M$,
and we define\footnote{Usually, $\hxi$~is denoted as~`$\delta $', but since
we use~`$\delta $' as the differential along~$\scrF$, we avoid this
notation.}~$\hxi f$ to be the change in the component~$f$ under this
diffeomorphism acting on~$\Phi $ by the {\it inverse\/} pullback:
  $$ \hxi f := \frac{D}{Dt}\res{t=0}\bres D^r\bigl(\exp(-t\eta^a
     \tQa)^*\Phi \bigr). \tag{1.28} $$
We write `$D/Dt$' to indicate that we may have to use a covariant derivative
to differentiate in~$t$.  To compute~\thetag{1.28} we use the fact that
$\exp(-t\eta^a \tQa)$ preserves~$D$ and so
  $$ \aligned
      \hxi f&= \frac{D}{Dt}\res{t=0}\bres\exp(-t\eta^a \tQa)^*(D^r \Phi
     ) \\
      &= -\eta^a \bres\tQa D^r\Phi\\
      &= -\eta^a \bres \Da D^r \Phi .\endaligned \tag{1.29} $$
At the last stage we use the fact that $\tQa$~and $\Da$ agree when restricted
to~$\bmink$.  In practice we use~\thetag{1.29} and the commutation relations
among the $D$s to express~$\hxi f$ in terms of other component fields. 
 
With our sign conventions, the infinitesimal symmetry~$P_\mu $ acts on a
component field~$f$ by~$-\partial _\mu f$.
 
The supersymmetry transformations~\thetag{1.29} on components obey the
commutation relations derived from those of the vector fields~$Q$ if there
are no covariant derivatives.  Explicitly, if $\eta _1,\eta _2$~are odd
parameters and $\hxi_i$ the vector field which corresponds to~$\eta ^a_iQ_a$,
then\footnote{In the abstract supersymmetry algebra we have $[\eta
^a_1Q_a,\eta ^b _2Q_b] = -\eta^a _1\eta ^b_2[Q_a,Q_b] = 2\eta ^a_1\eta
^b_2\sdtov^\mu _{ab}P_\mu$.  There are two minus signs which cancel in
passing to~\thetag{1.30}: a minus sign for brackets since we use left
actions, and the minus sign in the action of~$P_\mu $. (See~\SM{5}.)}
  $$ \boxed{[\hxi_1,\hxi_2]f = 2\eta _1^a\eta _2^b\sdtov^\mu _{ab}\partial
     _\mu f .} \tag{1.30} $$
The same is true for nonlinear $\sigma $-models, but in gauge theory these
commutation relations may be altered by curvature terms which act as
infinitesimal gauge transformations.
 
The most straightforward way to define a {\it component lagrangian\/} for the
component fields on~$\bmink$, starting from a lagrangian~$L$ in superspace,
is to integrate~$L$ over the odd variables using the splitting~\thetag{1.20}.
In other words, the splitting~\thetag{1.20} determines a projection 
  $$ \pi \:M\longrightarrow \bmink \tag{1.31} $$
and so an integration 
  $$ \pi _*\:\Dens(M)\longrightarrow \Dens(\bmink) \tag{1.32} $$
on densities.  However, we find better formulas in certain cases by a
procedure which differs from this by an exact term.  First, the metric on the
vector space~$V$ determines a canonical positive density~$\bdens$ on
Minkowski space~$\bmink$; it is invariant under the Poincar\'e group.
Similarly, if we fix a volume form\footnote{Neither the superspace
lagrangian~$L$ nor the component lagrangian function~$\Lch$ depends on this
volume form; only the superspace lagrangian function~$\ell $ in~\thetag{1.33}
and the choice of~$D^s$ in~\thetag{1.34} depend on it.}  on~$S^*$, then
together with the metric on~$V$ this determines a canonical density~$\dens$
on super Minkowski space~$M$; it is invariant under the super Poincar\'e
group.  We define a lagrangian {\it function\/}~$\ell $ on~$M$ by
  $$ L=\dens \,\ell.  \tag{1.33} $$
Then we define the component lagrangian density $\bdens\,\Lch$ on Minkowski
space~$\bmink$ by applying some definite combination of $D$~operators, which
we symbolically denote~`$D^s$', to~$\ell $ and then restricting to~$\bmink$:
  $$ \bdens\,\Lch = \bdens\,(\bres D^s\ell ).\tag{1.34} $$ 
Notice that the component lagrangian {\it function\/}~$\Lch$ is naturally
expressed in terms of component fields.  Often $\bres D^s$~is chosen to agree
with integration over the odd variables in the splitting~\thetag{1.20}, but
in some cases not.  For an example of the former (on~$\mink32$)
see~\thetag{2.40}; for an example of the latter (on~$\mink44$) see the
discussion following~\thetag{2.72}.  In any case we suppose there is a
Poincar\'e invariant differential operator~$\Delta $ on~$\bmink$ such that
  $$ \split
      \pi _*L &= \bdens\,(\bres D^s\ell + \Delta\bres\ell ) \\
      &= \bdens\,(\Lch + \Delta\bres\ell ).\endsplit \tag{1.35} $$
For integration on~$\mink44$ this is formula~\thetag{2.71}.

        \proclaim{\protag{1.36} {Theorem}}
  The component lagrangian~\thetag{1.34} is supersymmetric. 
        \endproclaim

\flushpar
 Here the supersymmetry is {\it nonmanifest\/} so that $\Lie{\hxi}
(\bdens\,\Lch)$~is a nonzero exact form, where $\hxi $~is defined
in~\thetag{1.29}.

        \demo{Proof}
 The verification is easy if the component
lagrangian is defined by ordinary integration~\thetag{1.32}, i.e., if $\Delta
=0$ in~\thetag{1.35}:
  $$ \split
      \Lie{\hxi} \bigl(\bdens\,\Lch \bigr) &= \Lie{\hxi} (\pi _*L) \\ 
      &= \pi _*\bigl(\operatorname{Lie}(-\teQ)L \bigr) \\ 
      &= \pi _*\bigl(d\cont{-\teQ}L \bigr) \\ 
      &= d\Bigl\{ \pi _*\bigl(-\cont{\teQ}L \bigr)\Bigr\},\qquad \text{if
     $\Delta =0$}.\endsplit \tag{1.37} $$
In this computation `Lie'~denotes the Lie derivative, and we use the Cartan
formula $\operatorname{Lie}(\zeta )=d\cont{\zeta } + \cont{\zeta }d$ to
define the {\it integral density\/}\footnote{See \LSS{\S3.9--3.12} for an
explanation of integral forms and integral densities.}~$\cont{\teQ}L$.  Even
if~$\Delta \not= 0$ we use~\thetag{1.35} to rewrite this last expression.
First, use~\thetag{1.31} to lift~$\bdens$ to~$M$, where we view it as a
twisted differential form (twisted by the orientation bundle).  Then
  $$ \split
      d\Bigl\{ \pi _*\bigl(-\cont{\eQ}L \bigr) \Bigr\} &= d\Bigl\{ \pi
     _*\bigl(-\cont{\eta ^a \da + \eta ^a \sdtov^\mu _{a b
     }\tb\di}\,\bdens\,d^s\theta \,\ell \bigr) \Bigr\} \\
      &= d\Bigl\{ \pi _*\bigl(-\cont{\eta ^a \sdtov^\mu _{a b
     }\tb\di}\,\bdens\,d^s\theta \,\ell \bigr) \Bigr\} \\
      &= d\Bigl\{ \bres D^s\bigl(-\cont{\eta ^a \sdtov^\mu _{a b
     }\tb\di}\,\bdens\,\ell \bigr) \\ 
      &\qquad \qquad \qquad \qquad + \Delta \bres\bigl(-\cont{\eta ^a
     \sdtov^\mu _{a b }\tb\di}\,\bdens\,\ell \bigr) \Bigr\} \\
      &= d\Bigl\{ \bres D^s\bigl(-\cont{\eQ}\,\bdens\,\ell \bigr)
     \Bigr\}.\endsplit \tag{1.38} $$
In the penultimate step we use the fact that~$\bres\tb=0$.  Now if~$\Delta
\not= 0$ we write 
  $$ \Delta =\Delta ^\mu \di \tag{1.39} $$
for some constant differential operators~$\Delta ^\mu $.  The
expression~\thetag{1.39} is not in general unique.  Now 
  $$ \split
      \Lie{\hxi} \bigl(\bdens\,\Delta \bres\ell  \bigr) &= \bdens\,\Delta
     \bres(-\teQ\ell ) \\ 
      &= \bdens\,\Delta \bres(-\eta^a \Da\ell ) \\
      &= d\Bigl\{ -\cont{\di}\,\bdens\,\Delta ^\mu \bres(\eta ^a \Da\ell )
     \Bigr\}.\endsplit \tag{1.40} $$
Putting together~\thetag{1.35}, \thetag{1.37}, \thetag{1.38},
and~\thetag{1.40} we have the desired result
  $$ \align
      \Lie{\hxi} \bigl(\bdens\,\Lch \bigr) &= d\Bigl\{ \bres
     D^s\bigl(-\cont{\eQ}\,\bdens\,\ell \bigr) -\cont{\di}\bdens\,\Delta ^\mu
     \bres(\eta ^a \Da\ell ) \Bigr\} \tag{1.41} \\
      &= d\Bigl\{ -\cont{\di}\bdens \bigl(\sdtov^\mu _{a b }\bres D^s(\eta ^a
     \tb\ell ) + \Delta ^\mu \bres(\eta ^a \Da\ell ) \bigr)
     \Bigr\}. \tag{1.42} \endalign $$
        \enddemo

\flushpar
 Equation~\thetag{1.41} gives one contribution to the {\it supercurrent\/},
which is minus the Noether current for the supersymmetry transformation~$\hxi
$.  If $\Delta \not= 0$, then the supercurrent we derive from these formulas
depends on the choice we make in~\thetag{1.39}.
 
We sometimes define (a term in) a lagrangian on~$\bmink$ simply by
restricting a scalar superfield~$S$ on~$M$ to Minkowski space~$\bmink$: 
  $$ \Lch=\bres S. \tag{1.43} $$
Of course, \thetag{1.34}~is of this form with~$S=D^s\ell $.  In general
\thetag{1.43}~is {\it not\/} supersymmetric.  Put differently, only in
special cases does the assumption that $\hxi S=0$ on~$M$ imply that
$\Lie{\hxi} (\bdens\,\Lch)$~is exact.  

For some theories there is no (known) superspace formulation and we work
directly in components.  Then the supersymmetry transformation is defined
directly on the component fields and a direct computation is needed to verify
that a (component) lagrangian is supersymmetric.

\Head{1.3}{A simple example\footnote{Further analysis of this example appears
in~\FP2.}}
 Here we consider a supersymmetric version of a particle moving on a line.
Consider~$\mink11$ with coordinates~$t,\theta $.  The basic odd vector fields
are
  $$ \aligned
      D &= \partial _\theta - \theta \partial _t \\
      \tQ &= \partial _\theta + \theta \partial _t,\endaligned \tag{1.44} $$
with $[D,D]=-2\partial _t$ and $[\tQ,\tQ]=2\partial _t$.  A superfield~$\Phi
\:\mink11\to\RR$ has components 
  $$ \aligned
      \phi &=\bres\Phi , \\ 
      \psi &=\bres D\Phi .\endaligned \tag{1.45} $$
Then for $\hxi$ corresponding to~$\eQ$ we have, according to~\thetag{1.29},
  $$ \aligned
      \hxi \phi &=-\eta \bres D\Phi = -\eta \psi , \\ 
      \hxi \psi &=-\eta \bres D^2\Phi  = \eta\phidot ,\endaligned
     \tag{1.46} $$
where $\phidot=\dt\phi $.  We check the bracketing in the simplest case:
  $$ \split
      (\hxi _1\hxi _2 -\hxi _2\hxi _1)\phi  &= \hxi _1(-\eta _2\psi
     ) - \hxi _2(-\eta _1\psi ) \\ 
      &= -\eta _2\hxi _1\psi  + \eta _1\hxi _2\psi  \\ 
      &= -\eta _2\eta _1\phidot  + \eta _1\eta _2\phidot  \\ 
      &= 2\eta _1\eta _2\phidot .\endsplit \tag{1.47} $$
This is consistent with~\thetag{1.30}.\footnote{It is important to use the
odd parameter~$\eta $ in these computations.  If we write informally
  $$ \Phi (t,\theta ) = \phi (t) + \theta \psi (t), \tag{1.48} $$
then the action of~$-\tQ$ is 
  $$ -\tQ\Phi (t,\theta ) = -\psi (t) - \theta\phidot (t), \tag{1.49}
     $$
from which we might {\it erroneously\/} conclude 
  $$ \aligned
      (-\tQ)\phi  &= -\psi , \\ 
      (-\tQ)\psi &= -\phidot .\endaligned \tag{1.50} $$
Multiplying by~$\eta $ we do {\it not\/} obtain~\thetag{1.46}.  Rather,
instead of~\thetag{1.49} we should write
  $$ (-\eta \tQ)\Phi (t,\theta ) = -\eta \psi (t) + \theta \eta\phidot
     (t) \tag{1.51} $$
and correctly deduce~\thetag{1.46}.  The process of taking components
involves odd derivatives; the odd parameter~$\eta $ protects us from making a
sign mistake.}

A fundamental lagrangian for the field~$\Phi $ is
  $$ L = |dt|\,d\theta \;\Bigl\{-\frac {1}2D\Phi \dt\Phi
     \Bigr\}. \tag{1.52} $$
The component lagrangian~\thetag{1.34} is easy to compute: 
  $$ \Lch = \frac 12(\phidot )^2 +\frac 12 \psi \psidot . \tag{1.53} $$
We evaluate the expression in braces on the right hand side
of~\thetag{1.41}, noting that $\Delta ^i=0$ in this case:
  $$ \split
      \bres D\bigl(-\cont{\eta Q}\,|dt|\;\frac{-1}{2}D\Phi \dt\Phi \bigr) &=
     \frac 12\bres D(\eta \theta D\Phi \dt\Phi ) \\
      &= -\frac 12\eta \psi \phidot.\endsplit \tag{1.54} $$
We check~\thetag{1.41} by computing the left hand side directly
from~\thetag{1.53} using~\thetag{1.46}:
  $$ \split
      \hxi \Lch &= -\bigl(\dt(\eta \psi )\phidot \bigr) + \frac 12 \eta
     \phidot \psidot + \frac 12 \psi \dt(\eta \phidot ) \\
      &= \dt(-\frac 12\eta \psi \phidot ).\endsplit \tag{1.55} $$
To compute the Noether current~$j$ corresponding to the symmetry~$\hxi$, we
first record the variational 1-form computed from~\thetag{1.53}:
  $$ \gamma =\phidot\,\delta \phi +\frac 12\psi \,\delta \psi . \tag{1.56}
     $$
Then the current is 
  $$ \split
      j &= \cont{\hxi}\gamma  - (-\frac 12 \eta \psi \phidot) \\ 
      &= (-\eta \phidot\psi -\frac 12\eta \psi \phidot) - (-\frac 12\eta \psi
     \phidot) \\ 
      &= -\eta \phidot\psi .\endsplit \tag{1.57} $$
The {\it supercharge\/} is defined to be {\it minus\/} the Noether current
of~$Q$, which is~$\phidot\psi $.

\Chapter2{Coordinates on Superspace}

 \comment
 lasteqno 2@119
 \endcomment

 Our task in this chapter is to develop the formalism we need to compute in
our main examples: the super Minkowski spaces~$\mink32$, $\mink44$,
and~$\mink6{(8,0)}$.  Most of the formulas we derive also apply to
differential geometric computations (including spinors) in 3-, 4-, and
6-dimensional ordinary Minkowski space.  The entire treatment is based on the
exceptional isomorphism of Lie groups in low dimensions, as explained
in~\S{2.1}.  After a brief interlude~(\S{2.2}) on dimensional reduction, we
flush out the definitions into concrete formulas (\S\S2.3--2.5).  The final
section~\S{2.6} records some specializations to dimension~2.

\Head{2.1}{$\mink32$, $\mink44$, $\mink6{(8,0)}$ and their complexifications}
 We consider the cases $n=\dim(V)=3,4,6$ with minimal supersymmetry.  The
irreducible real spinorial representation~$S$ of~$\Spin(V)$ is respectively
of dimension~2, 4, and~8, with algebra of endomorphisms~$\endo$ respectively
reduced to~$\RR$ ($n=3$) or isomorphic to~$\CC$ ($n=4$) or~$\HH$ ($n=6$).
Once $\sdtov$~is chosen, $V$~can be identified with the space of
$\endo$-hermitian symmetric sesquilinear forms on~$\Hom_\endo(S,\endo)$, with
the positive cone~$C$ corresponding to positive forms.  Namely, as will be
clear from the complexified description, there is up to a real factor a
unique symmetric morphism of representations $\sdtov\:S^*\otimes S^*\to V$,
and $\sdtov(\bar{a}s,t)=\sdtov(s,at)$ for~$a\in \endo$.  It induces an
isomorphism of the symmetric quotient of~$S^*\otimes _\endo S^*$ with~$V$.
This symmetric quotient is identified with the space of hermitian symmetric
sesquilinear forms on~$\Hom_\endo(S^*,\endo)$ by letting~$s\otimes s$
correspond to the form for which~$\langle \alpha ,\alpha
\rangle=\overline{\alpha (s)}\alpha (s)$.
 
For $V$~of dimension~4 (resp.~6) we choose an isomorphism of~$\endo$
with~$\CC$ (resp.~$\HH$).  On the super Minkowski space~$M$, the odd
distribution~$\tau $ then carries a $\CC$- (resp left~$\HH$-) module
structure, and this structure is respected by the super Poincar\'e group.
 
Over~$\CC$ one has the exceptional isomorphisms $\Spin(3)\cong SL(2)$,
$\Spin(4)\cong SL(2)\times SL(2)$, and $\Spin(6)\cong SL(4)$.  They make it
convenient to describe the complexification~$V_\CC$ of~$V$ starting from that
of~$S$.  The real form~$V$ corresponds to a complex conjugation (antilinear
automorphism) of~$V_\CC$. 
 
After complexification $\HH$~becomes isomorphic to the $2\times 2$~matrix
algebra.  This enables us to fix a 2-dimensional complex vector space~$W$, a
pseudoreal (or quaternionic) structure on~$W$ (that is, an antilinear
automorphism~$j$ of square~$-1$) and an isomorphism of~$\HH$ with the ring of
real endomorphisms of~$W$, where `real' means commuting with~$j$.  We take
these endomorphisms to act on the right.  The model is~$W=\HH$, the complex
structure given by left multiplication by~$I\in \HH$, the pseudoreal
structure given by left multiplication by~$J\in \HH$.  The real endomorphisms
are the right multiplications by elements of~$\HH$.  We fix on~$W$ a
symplectic form~$\epsilon_W $ which is real, i.e., $\epsilon_W (jx,jy) =
\overline{\epsilon_W (x,y)}$, and positive definite, i.e., $\epsilon_W
(x,jx)>0$ for~$x\not= 0$.  Note that $\epsilon_W (x,jx)$~ is real since
$\epsilon_W $~is.  We also fix a basis $\{e',e''=j(e')\}$ of~$W$ for which
$\epsilon_W (e',e'')=1$ and for which the $\HH$-module multiplication
by~$z\in \CC\subset \HH$ is the multiplication by~$z$ on~$e'$ and by
$\bar{z}$ on~$e''$.  In the model~$W=\HH$ introduced above, one may take
$e'=1$, $e''=J$.
 
We now describe, case by case, the complexifications~$S\cp$ and~$V\cp$ of~$S$
and~$V$, the symmetric morphism~$\sdtov\:S\cp^*\otimes S\cp^*\to V\cp$, the
$\endo$-module structure of~$S\cp$, and~$\Spin(V)$.  For~$\mink32$
and~$\mink44$ we give a duality pairing $\diso\:S\cp\otimes S\cp\to \CC$.  In
these cases the form $\stov\:S\cp\otimes S\cp\to V\cp$ is constructed
using~\thetag{1.9}.  (We write the formulas in coordinates below.)
For~$\mink6{(8,0)}$ there is no duality pairing and we specify~$\stov$
explicitly.

\subsubhead $3|2$~case\endsubsubhead
 Here there is no need to complexify to have a nice description, so we
directly describe the real superspace.  Let $S$~be a 2-dimensional real
vector space with volume form~$\epsilon $, viewed as a symplectic form
on~$S$.  Define 
  $$ \gathered
      V = \Sym^2(S^*) \\
      \sdtov\:(s,t)\longmapsto st \endgathered \tag{2.1} $$
There is an isomorphism $SL(S) @>\sim>>\Spin(V)$.
 
\subsubhead $4|4$~case\endsubsubhead  
 Let $S'$ and $S''$~be 2-dimensional complex vector spaces with complex
volume forms~$\epsilon ',\epsilon ''$, viewed as symplectic forms on~$S'$
and~$S''$.  Define
  $$ \gathered
      S\cp = S'\oplus S'' \\
      V\cp =\Sps\otimes \Spps \\
      \sdtov\:(s'_1+s_1'',\; s'_2+s_2'')\longmapsto (s'_1\otimes s''_2) +
     (s'_2\otimes s''_1) \\
      \epsilon =\epsilon '\oplus \epsilon ''\endgathered \tag{2.2} $$
The $\CC$-module structure of~$S\cp$ is multiplication by~$z$ on~$S'$ and
multiplication by~$\bar{z}$ on~$S''$.  There is an isomorphism $SL(S')\times
SL(S'')@>\sim>>\Spin(V\cp)$.  The bigger group 
  $$ \{\langle g,g'  \rangle\:\det g'\det g''=1\}\subset GL(S')\times GL(S'')
     \tag{2.3} $$
acts on~$S'$ and~$S''$ preserving the bilinear form on~$V\cp$.  The new
symmetry in this larger group is called {\it R-symmetry\/}; it is the
action of unit norm scalars in~$\CC$ on~$S\cp$.

\subsubhead $6|(8,0)$~case\endsubsubhead 
 Let $S_0$~be a 4-dimensional complex vector space with a volume
form~$\nu$.  Define
  $$ \gathered
      S\cp = S_0\otimes W \\
      S^*\cp = S_0^*\otimes W^*\\
      V\cp= {\tsize\bigwedge} ^2S_0^* \\
      \sdtov\:(w'\otimes s'_0,w''\otimes s''_0)\longmapsto \epsilon_W
     (w',w'')\;(s'_0\wedge s_0'')\endgathered \tag{2.4} $$
Then $SL(S_0)@>\sim>> \Spin(V\cp)$.  The $\HH$-module structure of~$S\cp$ is
induced from that of~$W$.  The volume form~$\nu $ defines a symmetric pairing
  $$ \nu \:{\tsize\bigwedge} ^2S_0\otimes {\tsize\bigwedge}
     ^2S_0\longrightarrow \CC \tag{2.5} $$
via which we identify~$V\cp$ and~$V\cp^*$.  This pairing is, up to a factor
of~2, the inverse metric.  Then $\stov$~is the composition 
  $$ \stov\:S\cp\otimes S\cp\longrightarrow V\cp^*\longrightarrow V\cp,
     \tag{2.6} $$
where the first arrow is given by a similar formula to that above
for~$\sdtov$ and the second arrow is constructed from the
duality~\thetag{2.5}.  To make~\thetag{1.3} valid, we define the metric~$g$
on~$V\cp$ to be such that the inverse metric~$g\inv $ on~${\tsize\bigwedge}
^2S_0$ is~$\nu /2$.

\vskip\subsubheadskip 
 
In the $6|(8,0)$ and $4|4$~cases we now describe the real spaces~$V$ and~$S$
as fixed points of an antilinear involution of~$V\cp$ and~$S\cp$.  The real
structures must be compatible with~$\sdtov$ as well as with the symmetric
bilinear form~$g$ on~$V\cp$.  The induced form on~$V$ is to have
signature~$(1,n-1)$.  The real description above of the $3|2$~case does give
the required signature~$(1,2)$.
 
\subsubhead $4|4$~case\endsubsubhead 
 Fix an antilinear involution on~$S\cp=S'\oplus S''$ which exchanges~$S'$
and~$S''$ as well as $\epsilon '$ and~$\epsilon ''$.  From this we get real
structures on~$S\cp$ and~$V\cp=\Sps\otimes \Spps$.  The symmetric bilinear
form~$\epsilon '\otimes \epsilon ''$ is real of signature~$(1,3)$.

\subsubhead $6|(8,0)$~case\endsubsubhead 
 A pseudoreal structure~$j$ on~$S_0$ induces real structures
on~$S\cp=S_0\otimes W$ as well as on~$V={\tsize\bigwedge} ^2S_0^*$.  The
vector~$s\wedge j(s)\in V$ is real for any~$s\in S_0$, since 
  $$ j\bigl(s\wedge j(s) \bigr)=j(s) \wedge (-s) = s\wedge j(s). \tag{2.7}
     $$
If the volume form~$\nu $ on~$S_0$ is real, the corresponding quadratic
form on the real form of~$V$ has signature~$(1,5)$ or~$(5,1)$.  We want
$(\nu ,j)$~to give signature~$(1,5)$.  This is the case if, for a basis
of~$S_0$ of the form~$\{s,j(s),t,j(t)\}$, one has $\nu \bigl(s\wedge
j(s)\wedge t\wedge j(t) \bigr)$ real and positive.

\Head{2.2}{Dimensional reduction}
The super Lie algebras~$V^3\oplus S^2$ and~$V^4\oplus S^4$
underlying~$\mink32$ and~$\mink44$ are related as follows.  Any (real)
isomorphism $\alpha \:\CC\otimes_\RR S^2@>\sim>>S^4$ extends to $\alpha
\:V^3\hookrightarrow V^4$.  The image~$\alpha (V^3)$ is the orthogonal
complement to a one-dimensional spacelike subspace~$T\subset V^4$.  This
defines embeddings
  $$ \align
      \mink32&\hookrightarrow \mink44 \tag{2.8} \\
      \mink32&\hookrightarrow \mink34 := \mink44/T. \tag{2.9}\endalign $$
Notice that the left translates of~$S''$ form a two-dimensional integrable
odd distribution~$\Lambda $ on~$\mink34$, and the composition
of~\thetag{2.9} with the quotient map 
  $$ \mink32\hookrightarrow \mink34 \longrightarrow \mink34/\Lambda
     \tag{2.10} $$
is an isomorphism.
 
Similarly, for the super Lie algebras~$V^4\oplus S^4$ and~$V^6\oplus S^8$
underlying~$\mink44$ and~$\mink6{(8,0)}$, any isomorphism $\alpha
\:\HH\otimes _\CC S^4 @>\sim>>S^8$ extends to $\alpha \:V^4\hookrightarrow
V^6$.  The image~$\alpha (V^4)\subset V^6$ is the orthogonal complement to a
two-dimensional spacelike subspace~$T$.  This defines embeddings
  $$ \aligned
      \mink44&\hookrightarrow \mink6{(8,0)} \\ 
      \mink44&\hookrightarrow \mink4{(8,0)} := \mink6{(8,0)}/T.\endaligned \tag{2.11} $$
 
After complexification we obtain the following picture.

\subsubhead $\mink32\hookrightarrow \mink44$\endsubsubhead 
 The map~$\alpha $ complexifies to isomorphisms $\alpha '\:S^2\cp\to S'$ and
$\alpha ''\:S^2\cp\to S''$ which are complex conjugate.  The extension
to~$V^3\cp=\Sym^2(S^2\cp)^*$ is given by $st\mapsto \alpha '(s)\alpha ''(t) +
\alpha '(t)\alpha ''(s)$, with values in~$S^4\cp=S'\otimes S''$.  Identify
$S'\otimes S''\cong S^2\cp\otimes S^2\cp$ via~$(\alpha ',\alpha '')$; then
the subspace~$T\subset \Sps\otimes \Spps$ is identified with the orthogonal
complement~${\tsize\bigwedge} ^2(S^2\cp)^*$ to~$\Sym^2(S^2\cp)^*$.  The
extension $\alpha \:V^3\cp\to V^4\cp$ is compatible with the symmetric
bilinear forms if, for some~$\lambda $, we have
  $$ \alpha '(\epsilon ) = \frac{\lambda }{\sqrt2}\,\epsilon ',\qquad \alpha
     ''(\epsilon ) = \frac{\lambda \inv}{\sqrt2}\,\epsilon ''. \tag{2.12} $$

\subsubhead $\mink44\hookrightarrow \mink6{(8,0)}$\endsubsubhead 
 The complexified map $\alpha \:S^4\cp=S'\oplus S''\to S^8\cp=S_0\otimes W$
is of the form $\alpha 'e'+\alpha ''e''$ for $(\alpha ',\alpha '')\:S'\oplus
S''\to S_0$ an isomorphism.  (Recall that $\{e',e''\}$ is a fixed basis
of~$W$.)  The reality of our starting~$\alpha $ means that the pseudoreal
structure of~$\Hom(S'\oplus S'',S_0)$ transforms~$\alpha '$ into~$\alpha ''$
and $\alpha ''$ into~$-\alpha '$.  The extension to~$V^4\cp=\Sps\otimes
\Spps$ is given by $s'\otimes s''\mapsto \alpha '(s')\wedge \alpha ''(s'')$,
with values in~${\tsize\bigwedge} ^2S_0^*$.  The subspace~$T$
of~${\tsize\bigwedge} ^2S_0^*$ is~${\tsize\bigwedge} ^2\alpha '(S') \oplus
{\tsize\bigwedge} ^2\alpha ''(S'').$ The extension $\alpha \:V^4\cp\to
V^6\cp$ will be compatible with the symmetric bilinear forms if
  $$ \alpha '(\epsilon ') \wedge \alpha ''(\epsilon'') = -\nu
     _0. \tag{2.13} $$

\Head{2.3}{Coordinates on $\mink32$}
 A basis $\{f_a \}_{a =1,2}$ of~$S^*$ gives a basis $\{e_{a
b} =f_a f_b \}$ of~$V=\Sym^2(S^*)$.  Notice that $e_{a
b} =e_{b a }$.  We use on~$V\dual$ not the dual basis, but
rather the basis~$\{f^a f^b \}$ of~$\Sym^2(S)$, in duality
with~$\Sym^2(S^*)$ by
  $$ \langle s^*t^*,st \rangle = \frac 12\bigl(\langle s^*,s \rangle\langle
     t^*,t \rangle + \langle s^*,t \rangle\langle t^*,s \rangle\bigr),\qquad
     s^*,t^*\in S^*,\quad s,t\in S.  \tag{2.14} $$
 
The chosen bases of~$V$ and~$S^*$ give us vector fields~$\da$ and~$\dab$,
while the bases of~$V\dual$ and~$S$ give us coordinates~$y^{a b }$
and~$\theta ^a $.  We have
  $$ \aligned
      \da\tb &= \delta ^b _a \\
      \dab y^{a' b' } &= \frac 12(\dlt{a'}{a}
     \dlt{b'}{b} +\dlt{b'}{a} \dlt{a'}{b}) \\
      \da y^{b c } &= \dab\theta ^{c
     }=0. \endaligned\tag{2.15} $$
The left-invariant vector field~$\Da$ corresponding to~$f_{a }$
in~$S\dual$ is
  $$ \Da = \da - \tb\dab, \tag{2.16} $$
and corresponding to the relation $[f_a ,f_b] =-2f_{a b }$
in the Lie algebra we have 
  $$ [\Da,\Db] = -2\dab. \tag{2.17} $$
The right-invariant vector field~$\tQa$ corresponding to~$f_a $
in~$S\dual$ is
  $$ \tQa = \da + \tb\dab \tag{2.18} $$
with 
  $$ [\tQa,\tQb] = 2\dab. \tag{2.19} $$
 
We use $\{\Da,\dab\}$ as a left-invariant moving frame on~$\mink32$.  Dually,
we use the left-invariant moving coframe $\{\ra,\oab\}$ which has the
following duality pairings:\footnote{Recall the convention~\thetag{1.21} that
we write vector fields to the left of differential forms: $\langle D,df
\rangle = \cont D df = Df$, where $\cont D$ is contraction by the vector
field~$D$.}
  $$ \aligned
      \langle \Da,\rb \rangle &= \dlt b a \\
      \langle \dab,\omega ^{a' b' } \rangle &= \frac
     12(\dlt{a'}{a} \dlt{b'}{b} +\dlt{b'}{a}
     \dlt{a'}{b}) \\
      \langle \Da,\omega ^{b c } \rangle &= \langle \dab,\rho
     ^c
       \rangle = 0. \endaligned\tag{2.20} $$
This gives 
  $$ \aligned
      \ra &= d\ta \\ 
      \oab &= dy^{a b } - (\ta d\tb + \tb d\ta)\endaligned \tag{2.21}
     $$
and 
  $$ \aligned
      d\ra &= 0 \\
      d\oab &= -2\ra\wedge \rb.  \endaligned \tag{2.22} $$
 
The symmetric pairing~$\sdtov$ is 
  $$ \sdtov(f_a,f_b)= e_{ab}=\sdtov^{(a'b')}_{ab}e_{a'b'}= \frac
     12(\dlt{a'}a\dlt{b'}b + \dlt{b'}a\dlt{a'}b)e_{a'b'}.  \tag{2.23} $$
Similarly, 
  $$ \stov(f^a,f^b)=\stov^{(a'b')ab}e_{a'b'} = \frac 12(\epsilon
     ^{a'a}\epsilon ^{b'b}+\epsilon ^{a'b}\epsilon ^{b'a})e_{a'b'},
     \tag{2.24} $$
where $\epsilon $~is the volume form on~$S$.  The metric on Minkowski
space~$\bmink^3$ is the restriction to~$\bmink^3$ of 
  $$ g=2\epsilon _{aa'}\epsilon _{bb'}\;\omega ^{ab}\otimes \omega ^{a'b'},
     \tag{2.25} $$
and the inverse metric is the restriction to~$\bmink^3$ of
  $$ g\inv =\frac{1}{2}\epsilon ^{a a' }\epsilon ^{b b'
     }\;\partial _{a b }\otimes \partial _{a' b'
     }. \tag{2.26} $$
The reader should be mindful of~\thetag{2.20}, which implies 
  $$ g\inv (\omega ^{a b },\omega ^{c d }) = \frac
     14(\epsilon ^{a c }\epsilon ^{b d } + \epsilon
     ^{a d }\epsilon ^{b c }). \tag{2.27} $$
The Dirac form~\thetag{1.22} on a spinor field~$\psi =\psi _a f^a $ is
  $$ \psi \Dirac\psi =-\epsilon ^{a b }\epsilon ^{c d }
     \psi _a \partial _{b c }\psi _d .  \tag{2.28} $$
The Dirac form~\thetag{1.22} on a dual spinor field~$\lambda =\lambda ^a f_a
$ is  
  $$ \lambda \Dirac\lambda = \lambda ^a\partial _{ab}\lambda ^b. \tag{2.29}
     $$
We also introduce the Dirac {\it operator\/}
  $$ (\Dirac\psi )_b  = -\epsilon ^{c d }\partial _{b
     c }\psi _d , \tag{2.30} $$
which maps sections of~$S$ to sections of~$S$.  Then \thetag{2.28}~and
\thetag{2.30}~are compatible in the sense that
  $$ \eab\chi _a (\Dirac\psi )_b =\chi \Dirac\psi  \tag{2.31} $$
for all spinor fields~$\chi ,\psi $.  The wave operator is
  $$ \aligned
      \wave &= \frac 12 \epsilon^{a a '}\epsilon^{b b
     '}\partial _{ab  }\partial _{a 'b' }\\
      &= \partial _{11}\partial _{22} - \partial _{12}^2; \endaligned
     \tag{2.32} $$
the sign is chosen so that the continuation to Euclidean space is the
nonnegative laplace operator.  The reader may check that $\Dirac^2=-\wave$.

{\it Preferred\/} bases are those for which $\epsilon ^{12}= -\epsilon ^{21}
= 1$ and hence\footnote{See~\thetag{1.11}.} $\epsilon _{12} = -\epsilon _{21}
= 1$.  Put differently, $\epsilon (f_1,f_2) = 1$.  In a preferred basis we
introduce standard coordinates~$\{x^\mu\}$ on Minkowski space~$\bmink^3$ by
  $$ \aligned
      y^{11} &= \frac{x^0 + x^1}{2}, \\
      y^{22} &= \frac{x^0-x^1}{2}, \\
      y^{12} &= \frac{x^2}{2}.\endaligned \tag{2.33} $$
The dual vector fields~$\partial _{ab}$ and~$\partial /\partial x^\mu $ are
related by 
  $$ \aligned
      \partial _{11}&= \frac{\partial }{\partial x^0} + \frac{\partial
     }{\partial x^1}, \\ 
      \partial _{22}&= \frac{\partial }{\partial x^0} - \frac{\partial
     }{\partial x^1}, \\ 
      \partial _{12} &= \frac{\partial }{\partial x^2}.\endaligned
     \tag{2.34} $$
The metric~\thetag{2.25} is then the standard metric
  $$ g = (dx^0)^2 - (dx^1)^2 - (dx^2)^2. \tag{2.35} $$
The density~$|d^3x|$ in~\thetag{1.33} is the standard positive density 
  $$ |d^3x| = |dx^0dx^1dx^2|. \tag{2.36} $$
 
It is convenient in our gauge theory computations to work with the Hodge
$*$~operator.  For this we stipulate that $x^0,x^1,x^2$~is a positively
oriented coordinate system.\footnote{The orientation makes it easier to write
intermediate steps in some of our computations, but the final formulas do not
depend on it.}  Then the volume form is the restriction to~$\bmink^3$ of 
  $$ \voll = 4\omega ^{11}\wedge \omega ^{12}\wedge \omega ^{22} \tag{2.37}
     $$
and the $*$~operator acts by 
  $$ \aligned
      *\omega ^{a b } &= \epsilon _{a 'b '}\,\omega
     ^{a a '}\wedge \omega ^{b b '} \\
      *(\omega ^{a a '}\wedge \omega ^{b b '}) &=
     \frac14\,(\eab\omega ^{a 'b '} + \epsilon ^{a
     b '}\omega ^{a 'b } + \epsilon ^{a 'b }\omega
     ^{a b '}+ \epsilon ^{a 'b '}\omega ^{a b
     }).\endaligned \tag{2.38} $$

For integration we introduce the notation 
  $$ \int_{}d^2\theta = -\frac 12\int_{} d\ta\!d\tb  \epsilon _{a b } 
     = \int_{}d\theta ^2\!d\theta ^1. \tag{2.39} $$
Explicit in~\thetag{2.39} is our choice of volume form on spinors.  Using the
identification between integration and differentiation, we write
  $$ \int_{}d^2\theta = \bres\partial _2\partial _1 = -\frac 12 \bres
     \epsilon ^{a b } \Da\Db = -\bres D^2= \frac
     12\bres(D_2D_1 - D_1D_2), \tag{2.40} $$
where we have introduced the notation 
  $$ D^2 = \frac 12\epsilon ^{a b }\Da\Db. \tag{2.41} $$ 
 
The following relations are easy to verify:  
  $$ \align
      \Da\Db &= -(\dab - \epsilon _{a b }D^2) \tag{2.42} \\
      \Da D^2 &= -D^2\Da = -\epsilon ^{b c }\dab\Dg \tag{2.43} \\
      D^2D^2 &= -\wave \tag{2.44} \endalign $$

\Head{2.4}{Coordinates on $\mink44$ }
 We fix a basis $\{f_a \}_{a =1,2}$ of~$\Sps$ and let $\{\ead\}$ be the
complex conjugate basis of~$\Spps$.  As a basis for~$V\cp = \Sps\otimes
\Spps$ we use~$\{e_{a\bd}=f_a \bar{f}_{\bd}\}$.  Those bases and their duals
give us a complex coordinate system $\{y^{a \bd},\ta,\tad\}$, with $\ta$~and
$\tad$~complex conjugate and $y^{a \bd}$ the complex conjugate to~$y^{b
\ad}$.  We write $\da$, $\dad$, and $\dabd$~for the corresponding vector
fields.  Note our convention about indices: If both `$a $'~and `$\ad$'~appear
in the same equation, then~$a =\ad$.  This allows us to write reality
conditions and equations such as~\thetag{2.78} and~\thetag{2.79}.
 
Let $\Da$ and $\Dad$~be the left-invariant complex vector fields corresponding
to~$f_{a }$ and~$\ead$.  One has 
  $$ \aligned
      \Da &= \da - \tbd\dabd \\ 
      \Dad &= \dad - \tb\partial _{b  \ad}.\endaligned \tag{2.45} $$
Corresponding to the relations $[f_a ,f_b ] = [\ead,\bar f_\bd] = 0$
and $[f_a ,\bar f_\bd]= -2e_{a \bd}$ in the Lie algebra, one has
  $$ \aligned
      [\Da,\Db] &= [\Dad,\Dbd]=0 \\ 
      [\Da,\Dbd] &= -2\dabd.\endaligned \tag{2.46} $$
The similar right-invariant complex vector fields are 
  $$ \aligned
      \tQa &= \da + \tbd\dabd \\ 
      \tQad &= \dad + \tb\dbad\endaligned \tag{2.47} $$
with 
  $$ \aligned
      [\tQa,\tQb] &= [\tQad,\tQbd]=0 \\ 
      [\tQa,\tQbd] &= 2\dabd.\endaligned  \tag{2.48} $$
 
We use $\{\Da,\Dad,\dabd\}$ as a left-invariant moving frame and also use the
dual left-invariant moving coframe~$\{\ra,\rad,\oabd\}$.  The nontrivial
duality pairings are
  $$ \aligned
      \langle \Da,\rb  \rangle&=\delta_a ^b  \\ 
      \langle \Dad,\rbd  \rangle&=\delta_{\ad}^{\bd} \\ 
      \langle \dabd,\rho ^{a '\bd '}  \rangle &= \delta _a
     ^{a '}\delta _\bd ^{\bd '}. \endaligned \tag{2.49} $$
One has
  $$ \aligned
      \ra &= d\ta \\
      \rad &= d\tad \\
      \oabd &= dy^{a \bd } - (\tbd d\ta + \ta d\tbd)\endaligned
     \tag{2.50} $$
and, dual to the bracket relations,
  $$ \aligned
      d\ra &= d\rad =0 \\
      d\oabd &= -2\ra\wedge \rbd.  \endaligned  \tag{2.51} $$
 
The symmetric pairing~$\sdtov$ is 
  $$ \sdtov(f_a,\fbar_{\bd})=e_{a\bd}=\sdtov^{(a'\bd')}_{a\bd}e_{a'\bd'}
     =(\dlt{a'}a\dlt{\bd'}\bd) e_{a'\bd'}. \tag{2.52} $$
Similarly, the symmetric pairing~$\stov$ is
  $$ \stov(f^a,\fbar^{\bd})= \stov^{(a'\bd')a\bd} e_{a'\bd'}= (\epsilon
     ^{a'a}\epsilon ^{\bd'\bd})e_{a'\bd'}, \tag{2.53} $$
where $\epsilon $~is the volume form on~$S',S''$.  The metric on Minkowski
space~$\bmink^4$ is the restriction to~$\bmink^4$ of
  $$ g=2\epsilon _{aa'}\epsilon _{\bd\bd'}\;\omega ^{a\bd}\otimes \omega
     ^{a'\bd'}, \tag{2.54} $$
and the inverse metric is the restriction to~$\bmink^4$ of
  $$ g\inv =\frac{1}{2}\epsilon ^{a a' }\epsilon^{\bd\bd'}\;\partial
     _{a \bd}\otimes \partial_{a' \bd'}. \tag{2.55} $$
The Dirac form~\thetag{1.22} pairs a positive chirality spinor field~$\psi
=\psi _a f^a $ with a negative chirality spinor field~$\bar\psi
=\bar\psi _\ad \bar{e}^{\ad}$: 
  $$ \bar\psi \Dirac\psi = \epsilon ^{a b }\epsilon ^{\gd \dd }
     \bar\psi \mstrut _\dd \partial _{a \gd }\psi \mstrut _b
     . \tag{2.56} $$
The Dirac form~\thetag{1.22} on dual spinor fields is   
  $$ \bar\lambda \Dirac\lambda =\bar\lambda ^{\bd}\partial _{a\bd}\lambda
     ^a. \tag{2.57} $$
The Dirac operator on spinor fields exchanges sections of~$S'$ and~$S''$:
  $$ \aligned
      (\Dirac\psi )_{\gd} &= -\eab\partial _{a \gd}\pb \\
      (\Dirac\psib )_{c} &= -\eabd\partial _{c \ad}\pbd\endaligned
     \tag{2.58} $$
The wave operator is 
  $$ \aligned
      \wave &= \frac 12\eab\egdd\partial _{a \gd}\partial _{b \dd}\\
      &= \partial _{1\dot1}\partial _{2\dot2} - \partial _{1\dot2}\partial
     _{2\dot1},\endaligned \tag{2.59} $$
and $\Dirac^2=-\wave$.

{\it Preferred\/} bases are those for which $\epsilon '(f_1,f_2) = \epsilon
''(\bar f_{\dot{1}},\bar f_{\dot{2}}) = 1$.  In a preferred basis we
introduce standard coordinates~$\{x^\mu\}$ on Minkowski space~$V$ by
  $$ \aligned
      y^{1\dot1} &= \frac{x^0 + x^1}{2}, \\ 
      y^{2\dot2} &= \frac{x^0-x^1}{2}, \\ 
      y^{1\dot2} &= \frac{x^2 + ix^3}{2}, \\ 
      y^{2\dot1} &= \frac{x^2 - ix^3}{2}.\endaligned  \tag{2.60} $$
The dual vector fields~$\partial _{a\bd}$ and~$\partial /\partial x^\mu $ are
related by 
  $$ \aligned
      \partial _{1\dot1}&= \frac{\partial }{\partial x^0} + \frac{\partial
     }{\partial x^1}, \\ 
      \partial _{2\dot2}&= \frac{\partial }{\partial x^0} - \frac{\partial
     }{\partial x^1}, \\ 
      \partial _{1\dot2} &= \frac{\partial }{\partial x^2} - i\frac{\partial
       }{\partial x^3}, \\ 
      \partial _{2\dot1} &= \frac{\partial }{\partial x^2} + i\frac{\partial
       }{\partial x^3}.\endaligned \tag{2.61} $$
The metric~\thetag{2.54} is then the standard metric
  $$ g = (dx^0)^2 - (dx^1)^2 - (dx^2)^2 - (dx^3)^2.  \tag{2.62} $$
The density~$|d^4x|$ in~\thetag{1.33} is the standard positive density 
  $$ |d^4x| = |dx^0dx^1dx^2dx^3|. \tag{2.63} $$
 
To define self-duality we need to orient~$\bmink^4$, and so we stipulate that
$x^0,x^1,x^2,x^3$ is an oriented coordinate system.\footnote{The orientation
only serves to choose which of~\thetag{2.64} and~\thetag{2.65} below is
self-dual and which is anti-self-dual.  It does not enter in a fundamental
way in our formulae.}  Then for a symmetric tensor~$F_{a b }$ the
2-form
  $$ \frac 12F_{a b }\epsilon _{\gd\dd}\;\omega ^{a \gd}\wedge
     \omega ^{b \dd} \tag{2.64} $$
is self-dual, and for a symmetric tensor~$\overline{F}_{\gd\dd}$ the 2-form 
  $$ \frac 12\epsilon _{a b }\overline{F}_{\gd\dd}\;\omega ^{a
     \gd}\wedge \omega ^{b \dd} \tag{2.65} $$
is anti-self-dual.  Our convention is that a self-dual 2-form~$\sigma $
satisfies $*\sigma =-\sqrt{-1}\,\sigma $.  Then the analytic continuation to
Euclidean space gives the usual notion of self-duality.  The complex
conjugate of a self-dual form (in Minkowski space) is anti-self-dual.

For the 2-form 
  $$ F = \frac 12 (F_{a b }\epsilon _{\gd\dd} + \epsilon _{a
     b }\overline{F}_{\gd\dd})\; \omega ^{a \gd}\wedge \omega ^{b
     \dd} \tag{2.66} $$
we have the formulas 
  $$ \aligned
      |F|^2 &= \frac 14 (F_{a b }F_{c d }\epsilon ^{a
     c }\epsilon ^{b d } + \Fb_{\ad\bd}\Fb_{\gd\dd}\epsilon
     ^{\ad\gd}\epsilon ^{\bd\dd}), \\
      F\wedge F &= \frac i4 (F_{a b }F_{c d }\epsilon
     ^{a c }\epsilon ^{b d } -
     \Fb_{\ad\bd}\Fb_{\gd\dd}\epsilon ^{\ad\gd}\epsilon
     ^{\bd\dd})\;d^4x,\endaligned \tag{2.67} $$
where $d^4x$~is the standard volume form 
  $$ d^4x = dx^0\wedge dx^1\wedge dx^2\wedge dx^3. \tag{2.68} $$
In general the 2-form~$F$ is complex; then `$|F|^2$' in~\thetag{2.67} is not
a norm, but rather is a quadratic form.  The 2-form~$F$ is real if
$\Fb_{\ad\bd}=\overline{F_{a b }}$.

We mention the fact that the analytic continuation of the $*$~operator on
2-forms in Minkowski space of {\it any\/} dimension is $-\sqrt{-1}$~times the
$*$~operator on 2-forms in Euclidean space.

A complex function $\Phi \:\mink44\to\CC$ is {\it chiral\/} if $\Dad\Phi =0$
and {\it antichiral\/} if\break $\Da\Phi =0$.

Following~\thetag{2.39} we introduce the notations
  $$ \aligned
      \int_{}d^2\bar\theta &= -\frac 12\int_{} d\tad\!d\tbd \epsilon _{\ad
     \bd} = \int_{}d\bar\theta ^{\dot2}\!d\bar\theta ^{\dot1} \\
      \int_{}d^4\theta &= \int_{} d^2\theta d^2\bar{\theta }.
      \endgathered\tag{2.69} $$
We discuss the meaning of $\int_{}d^2\theta $ and $\int_{}d^2\bar{\theta }$
in~\S{5.3}; in any case we compute them using~\thetag{2.40} and the analogous
formula with the conjugate vector fields.  Note that we will use the notation
  $$ \overline{D}^2=\frac 12\epsilon ^{\ad\bd}\Dad\Dbd, \tag{2.70} $$
analogous to~\thetag{2.41}.  For integration over all odd coordinates we find
the formulas
  $$ \align
      \int_{}d^4\theta &= \bres\partial _2\partial
     _1\dbar_{\dot2}\dbar_{\dot1}\\
       &= \frac 12\;\bres\;\{D_1\Dbar_{\dot1}\Dbar_{\dot2} D_2 +
      D_2\Dbar_{\dot2}\Dbar_{\dot1} D_1\} - \wave\bres \tag{2.71}\\
      &= \frac 12\;\bres\left\{ D^2\overline{D}^2 +\overline{D}^2 D^2
     \right\} +\wave\bres.\tag{2.72} \endalign $$
In each of the last two lines the term in braces disagrees with the integral
by a multiple of the wave operator.  In a lagrangian this wave operator gives
an exact term which is usually omitted anyway.  The particular combination of
derivatives in the first term of~\thetag{2.71} has the virtue that it
exactly annihilates {\it chiral\/} and {\it antichiral\/} superfields.  This
is an advantage over the usual integral and leads to nicer formulas.  Hence
we compute component lagrangians~\thetag{1.34} using this improved
expression, which is the usual integral plus the wave operator composed with
restriction to~$V$.  Formula~\thetag{2.72} is useful in some computations.
Both of these integration formulas follow from direct computation.
 
The following relations are easy to verify. 
  $$ \align
      \Da D^2 &= D^2 \Da = \Dad \Dbar^2 = \Dbar^2\Dad=0 \tag{2.73}\\
      \Da\Dbar^2 &= \Dbar^2\Da - 2\epsilon ^{\dot{b }\dot{c
     }}\dabd\Dbar_{\dot c } \tag{2.74} \\ 
       \Dad D^2 &= D^2\Dad - 2\epsilon ^{b c }\partial _{b
     \dot{a} }\Dg \tag{2.75} \\ 
      \Dbar^2 D^2 &= -4\wave\text{\qquad\qquad on chiral superfields}
     \tag{2.76} \\ 
      D^2 \Dbar^2 &= -4\wave\text{\qquad\qquad on antichiral superfields}
     \tag{2.77} \endalign $$ 

Next, we describe the dimensional reduction map $\mink32\hookrightarrow
\mink44$.  Decorate the bases (of spinors and vectors) and coordinates
in~$\mink32$ with a `$\hat{\ }$'.  We take the inclusion to be compatible
with the $\diso$~and $\sdtov$~tensors:
  $$ \gather
      {\aligned
      \hat{f}_a &\longmapsto \frac{\ea + \ead}{\sqrt2}\\
      \hat{e}\mstrut _{a b } &\longmapsto \frac{e _{a
     \bd} + e\mstrut _{b \ad}}{2}.\endaligned} \tag{2.78} \endgather $$
The image of the inclusion on spinors is the subspace where $\ta=\tad$ and
both are real, described by the equations
  $$ \ta=\tad=\frac{\hat{\theta }^a }{\sqrt2}. \tag{2.79} $$
The image of the inclusion on vectors is the subspace
where~$y^{1\dot2}=y^{2\dot1}$, or equivalently where~$x^3=0$, described by
the equations
  $$ \aligned
      y^{1\dot{1}} &= \hat{y}^{11} \\ 
      y^{2\dot{2}} &= \hat{y}^{22} \\ 
      y^{1\dot{2}} = y^{2\dot{1}} &= \hat{y}^{12}.\endaligned \tag{2.80} $$
As an exercise the reader may check that \thetag{2.74}--\thetag{2.75} reduce
to~\thetag{2.42}--\thetag{2.44} under dimensional reduction.

The R-symmetry is generated by an even vector field~$R$ with brackets 
  $$ \aligned
      [R,\Da] &= \hphantom{-}\Da \\ 
      [R,\Dad] &= -\Dad \\ 
      [R,\partial _{a \bd}] &= \hphantom{-}0.\endaligned \tag{2.81} $$
Note that $R$~is purely imaginary: $\overline{R}=-R$.

\Head{2.5}{Coordinates on $\mink6{(8,0)}$}
 A basis $\{f_a \}_{a =1,2,3,4}$ of~$S_0^*$ gives us a basis ${
e^{\prime*}\otimes \ea, e^{\prime\prime*}}\otimes \ea$ of~$S\cp^*$ and a
basis $\{e_{a b } = \ea\wedge f_b \}_{a <b }$ of~$V\cp$.  For the basis
elements of~$S\cp^*$ we use the notation~`$f_{ai}$', where $f_{a1 }=
e^{\prime*}\otimes f_a$ and $f_{a2} =  e^{\prime\prime*}\otimes f_a$.  Let
$y^{a b },\tai$ be the corresponding coordinate system.  For $b <a $ define
$e_{b a } = -e_{a b }$ and~$y^{b a } = -y^{a b }$.  Write $\dai$ and~$\dab$
for the corresponding complex vector fields on~$\mink6{(8,0)}$.
 
Let $\epsilon _{ij}$ denote the symplectic form on the vector space~$W^*$.
We use our standard choice~$\epsilon _{12}=1$.
 
The left-invariant complex vector field~$\Dai$ corresponding to~$e_{a i
}$ is
  $$ \Dai = \dai  - \eij\theta ^{b j}\dab \tag{2.82} $$
with brackets 
  $$ [\Dai,D_{b j}] = -2\eij\dab.\tag{2.83} $$
The similar right-invariant vector fields are 
  $$ \tQQ{a i} = \dai + \eij\theta ^{b j}\dab \tag{2.84} $$
with  
  $$ [\tQQ{a i},\tQQ{b j}] = 2\eij\dab. \tag{2.85} $$

We will use $\{\Dai,\dab\}$ as a left-invariant moving frame with dual
coframe $\{\rai,\oab\}$.  The nonzero duality pairings are 
  $$ \aligned
      \langle \Dai,\rho ^{b j} \rangle &= \delta _a ^b \delta
     _i^j \\
      \langle \dab,\omega ^{c d } \rangle &= \delta ^c _a
     \delta _b^d -\delta _a ^d \delta _b^ c
     .\endaligned \tag{2.86} $$
One has
  $$ \aligned
      \rai &= d\tai\\
      \oab &= dy^{a b } + \eij\theta ^{a i}d\theta ^{b j} -
     \eij\theta ^{b i}d\theta ^{a j}\endaligned \tag{2.87} $$
with differentials
  $$ \aligned
      d\rai &= 0 \\
      d\oab &= 2\eij\,\rai\wedge \rho ^{b j}.\endaligned \tag{2.88} $$
 
The symmetric pairing~$\sdtov$ is 
  $$ \sdtov(f_{ai},f_{bj}) = \epsilon _{ij}e_{ab}=\frac 12
     \sdtov^{(pq)}_{(ai)(bj)}e_{pq} =\frac 12\bigl[\epsilon
     _{ij}(\dlt{p}a\dlt{q}b - \dlt{q}a\dlt{p}b)\bigr]e_{pq}. \tag{2.89} $$
The symmetric pairing~$\stov$ is
  $$ \stov(f^{ai},f^{bj})=\frac 12 \stov^{(pq)(ai)(bj)} e_{pq}= \frac
     12(\epsilon ^{ij}\nu ^{pqab})e_{pq}, \tag{2.90} $$
where $\nu $~is the volume form.  The metric on Minkowski space~$\bmink^6$ is
the restriction to~$\bmink^6$ of
  $$ g=\frac 12\nu _{pqrs}dy^{pq}\otimes dy^{rs} =
     \sum\limits_{{p<q}\atop{r<s}} 2\nu _{pqrs}\omega^{pq}\otimes
     \omega^{rs}, \tag{2.91} $$
and the inverse metric is the restriction to~$\bmink^6$ of
  $$ g\inv =\frac{1}{8}\nu ^{pqrs}\;\partial _{pq}\otimes \partial_{rs }=
     \sum\limits_{{p<q}\atop{r<s}} \frac 12\nu ^{pqrs}\;\partial _{pq}\otimes
     \partial_{rs }. \tag{2.92} $$
The volume form and its inverse are related by~$\nu ^{1234}\nu _{1234}=1$.
The Dirac form~\thetag{1.22} on spinor fields~$\psi =\psi _{a i
}f^{a i}$ is
  $$ \psi \Dirac\psi = \frac 12\nu ^{a bc d } \epsilon ^{ij}\psi _{a i}
     \partial\mstrut _{bc }\psi_{d j} \tag{2.93} $$
On dual spinor fields~$\lambda =\lambda ^{a i}f_{a i}$ we have the
Dirac form~\thetag{1.23} 
  $$ \lambda \Dirac\lambda =\eij\lambda ^{a i}\dab\lambda ^{b j
     }. \tag{2.94} $$

{\it Preferred\/} bases are those for which $\nu ^{1234}=-1$.  We also use
the dual form with~$\nu _{1234}=-1$.  A real structure is given by the
pseudoreal structure on~$S_0$ for which~$f_3=j(f_1)$ and~$f_4=j(f_2)$.  Then
  $$ \aligned
      \overline{D_{11}} &= \hphantom{-}D_{32} \\
      \overline{D_{21}} &= \hphantom{-}D_{42} \\
      \overline{D_{31}} &= -D_{12} \\
      \overline{D_{41}} &= -D_{22}.\endaligned \tag{2.95} $$
In a preferred basis we introduce standard coordinates~$\{x^\mu\}$ on
Minkowski space~$\bmink^6$ by
  $$ \aligned
      y^{12} &= \frac{x^4 + ix^5}{2}, \\ 
      y^{13} &= \frac{x^0 + x^1}{2}, \\ 
      y^{14} &= \frac{x^2 + ix^3}{2}, \\ 
      y^{23} &= \frac{x^2 - ix^3}{2}, \\ 
      y^{24} &= \frac{x^0 - x^1}{2}, \\ 
      y^{34} &= \frac{x^4 - ix^5}{2}.\endaligned \tag{2.96} $$
The dual vector fields~$\partial _{ab}$ and~$\partial /\partial x^\mu $ are
related by 
  $$ \aligned
      \partial _{12}&= \frac{\partial }{\partial x^4} -i \frac{\partial
     }{\partial x^5}, \\
      \partial _{13}&= \frac{\partial }{\partial x^0} + \frac{\partial
     }{\partial x^1}, \\
      \partial _{14} &= \frac{\partial }{\partial x^2} - i\frac{\partial
     }{\partial x^3}, \\
      \partial _{23} &= \frac{\partial }{\partial x^2} + i\frac{\partial
     }{\partial x^3}, \\
      \partial _{24}&= \frac{\partial }{\partial x^0} - \frac{\partial
     }{\partial x^1}, \\
      \partial _{34}&= \frac{\partial }{\partial x^4} +i \frac{\partial
     }{\partial x^5}.\endaligned \tag{2.97} $$
The metric~\thetag{2.91} is then the standard metric
  $$ g = (dx^0)^2 - (dx^1)^2 - (dx^2)^2 - (dx^3)^2- (dx^4)^2- (dx^5)^2.
      \tag{2.98} $$
 
Next, we describe the dimensional reduction map $\mink44\hookrightarrow
\mink6{(8,0)}$.  Denote the bases (of spinors and vectors) and coordinates
in~$\mink6{(8,0)}$ with a `$\tilde{\ }$'.  We take the inclusion maps to be
  $$ \alignedat2
      f_1 &\longmapsto \tilde{f}_{1(1)} &\qquad \qquad e_{1\dot1}&\longmapsto
      \tilde{e}_{13} \\
      f_2 &\longmapsto \tilde{f}_{2(1)} &\qquad \qquad e_{2\dot2}&\longmapsto
      \tilde{e}_{24} \\
      \bar{f}_{\dot1} &\longmapsto \tilde{f}_{3(2)} &\qquad \qquad
     e_{1\dot2}&\longmapsto \tilde{e}_{14} \\
      \bar{f}_{\dot2} &\longmapsto \tilde{f}_{4(2)} &\qquad \qquad
     e_{2\dot1}&\longmapsto \tilde{e}_{23} \endalignedat \tag{2.99} $$
The image of the inclusion on spinors is described by the equations  
  $$ \alignedat2
      \theta ^{1(1)} &= \theta ^1 &\qquad \qquad \theta ^{1(2)} &= 0 \\
      \theta ^{2(1)} &= \theta ^2 &\qquad \qquad \theta ^{2(2)}&=0\\
      \theta ^{3(1)} &= 0 &\qquad \qquad \theta ^{3(2)} &= \bar\theta
     ^{\dot1} \\
      \theta ^{4(1)} &=0 &\qquad \qquad \theta ^{4(2)} &= \bar\theta
     ^{\dot2}. \endalignedat \tag{2.100} $$
The image of the inclusion on vectors is described by the equations
  $$ \alignedat2
      \tilde{y}^{12} &= 0 &\qquad \qquad \tilde{y}^{23} &= y^{2\dot1} \\
      \tilde{y}^{13} &= y^{1\dot1} &\qquad \qquad \tilde{y}^{24} &=
       y^{2\dot2} \\
      \tilde{y}^{14} &= y^{1\dot2} &\qquad \qquad \tilde{y}^{34} &=
     0\endalignedat \tag{2.101} $$
 
There is an R-symmetry group~$SU(2)_R$ which acts on~$\mink6{(8,0)}$.  The action
of a matrix~$S=(S_i^j)$ is 
  $$ D_{a i}\longrightarrow D_{a i'}S^{i'}_i. \tag{2.102} $$
Demanding that this preserve~\thetag{2.83} requires $S\in SL(2;\CC)$; asking
that the real structure also be preserved reduces us to~$SU(2)$.  More
conceptually, this $SU(2)_R=Sp(1)_R$ symmetry group is the group of real
endomorphisms of the complex symplectic vector space~$W$.

There is also a dimensional reduction map $\mink4{8}\hookrightarrow
\mink6{(8,0)}$, described simply by equations~\thetag{2.101}.  Usually
$\mink4{8}$~is described with vector fields $\{D^{(i)}_a ,\Dbar_{\dot{a}
(j)}\}_{{a =1,2}\atop{i,j=1,2}}$ with bracketing~\thetag{2.46} supplemented
by a $\delta $-function:
  $$ [\Da^{(i)},\Dbar_{\ad (j)}] = -2\delta ^{i}_j\daad. \tag{2.103} $$
One possible correspondence between these vector fields and the
$\{\tilde{D}_{ai}\}_{{a =1,2,3,4}\atop{i=1,2}}$ is: 
  $$ \alignedat2
      \tilde{D}_{11} &= \hphantom{-}D_1^{(1)} &\qquad \qquad
     \tilde{D}_{12} &= D^{(2)}_1 \\
      \tilde{D}_{21}&= \hphantom{-}D_2^{(1)} &\qquad \qquad
     \tilde{D}_{22}&= D^{(2)}_2 \\
      \tilde{D}_{31}&= -\overline{D}_{\dot1 (2)} &\qquad \qquad
     \tilde{D}_{32}&= \overline{D}_{\dot1 (1)} \\
      \tilde{D}_{41}&= -\overline{D}_{\dot2 (2)} &\qquad \qquad
     \tilde{D}_{42}&= \overline{D}_{\dot2 (1)} \endalignedat \tag{2.104} $$

A new feature of~$\mink48$ is an enlarged R-symmetry group~$U(2)_R$.
It acts trivially on the even variables.  The representation of~$U(2)_R$ on
each of the pairs $\{D^{(1)}_a ,D^{(2)}_a \}$ of odd vector fields is the
standard one, whereas on each of the pairs
$\{\overline{D}_{\ad(1)},\overline{D}_{\ad(2)}\}$ of odd vector fields
$U(2)_R$~acts by the conjugate to the standard action.  Under the dimensional
reduction $\mink44\hookrightarrow \mink4{(8,0)}$, specified
by~\thetag{2.100}, only the subgroup $U(1)_R\subset U(2)_R$ given
by~\thetag{2.81} survives.  Notice that this subgroup is the induced action
by the double cover in the spin group of rotations in the $x^4$-$x^5$ plane.

\Head{2.6}{Low dimensions}  
 The basic superspace~$\mink11$ in dimension one is built starting from a
one-dimensional odd space~$S$ and letting~$V=S^{\otimes (-2)}$.  In terms of
the obvious left-invariant vector fields~$D$ and~$\partial $ we have
  $$ [D,D] = -2\partial . \tag{2.105} $$
Analogs of other formulas we have developed above are easy to come by.
Dimensional reduction from the higher dimensional spaces sets all spacelike
coordinates of Minkowski space to zero leaving just the time
coordinate~$x^0$.  In this way we also obtain spaces~$\mink12$, $\mink14$,
and~$\mink18$ with more odd variables.  As these spaces are reduced
from~$\mink32$, $\mink44$, and~$\mink6{(8,0)}$, we obtain R-symmetry
groups~$\Spin(2)$, $\Spin(3)$, and~$\Spin(5)$ from rotations in the
additional spatial directions.
 
The basic\footnote{$\mink2{(1,1)}$ is usually termed $N=1$ superspace, though
it is not the minimal possibility---there is a superspace $\mink2{(s^+,s^-)}$
for any nonnegative integers~$s^+,s^-$.} superspace~$\mink2{(1,1)}$ in
dimension two is simply a product
  $$ \mink2{(1,1)} = \mink11\times \mink11. \tag{2.106} $$
(Note that the metric~\thetag{2.109} is not a product, however.)  This
corresponds to the fact that there are real semi-spinors in 2~dimensions
(termed {\it Majorana-Weyl spinors\/} in the physics literature).  For
$S^+,S^-$ real one-dimensional vector spaces, we set $S=S^+\oplus S^-$ and
$V=(S^+)^{\otimes (-2)} \oplus (S^-)^{\otimes (-2)}$ with the obvious
bilinear pairing.  We can see $\mink2{(1,1)}$ as coming from~$\mink32$ by
dimensional reduction: in~$\mink32$ set $y^{12}=0$.  It is customary to
use~`$+$' for the odd index~`1' and~`$-$' for the odd index~`2'.  (These
indices appear beginning with formula~\thetag{2.15}.)  Also, we set
$y^+=y^{11}$ and $y^-=y^{22}$ with similar notation $\partial _+=\partial
_{11}$ and $\partial _-=\partial _{22}$.
 
$N=2$ superspace in 2~dimensions is~$\mink2{(2,2)}$, and we view it as
embedded in~$\mink44$ (dimensional reduction) by
setting~$y^{1\dot2}=y^{2\dot1}=0$, or equivalently~$x^{2}=x^3=0$.  The real
description of~$\mink2{(1,1)}$ makes clear that there is also a real
description of~$\mink2{(2,2)}$, but we use the complex vector fields and
forms induced from~$\mink44$ instead.  Again we use~`$+$' for the odd
index~`1' and~`$-$' for the odd index~`2' (which appear beginning with
formula~\thetag{2.45}.)  The `$+$'~ and`$-$'~on the bosonic coordinates are
the usual lightcone coordinates in two dimensions. Thus the left-invariant
odd vector fields are~$\Dp,\Dbp,\Dm,\Dbm$, and we denote the left-invariant
even vector fields by~$\dpl,\dm$.  The bracketing relations are
  $$ \gathered
       [\Dp,\Dp] = [\Dp,\Dm] = [\Dp,\Dbm] = [\Dbp,\Dbp] =0  \\  
        [\Dm,\Dm] = [\Dbp,\Dbm] = [\Dm,\Dbp] = [\Dbm,\Dbm]=0 \\
      [\Dp,\Dbp] = -2\dpl \\
      [\Dm,\Dbm] = -2\dm.\endgathered \tag{2.107} $$
For convenience we record the nonzero components of~$\sdtov,\stov$: 
  $$ \aligned
      &\sdtov^+_{++}=\sdtov^-_{--}=1 \\ 
      &\stov^{+--}=\stov^{-++}=1.\endaligned \tag{2.108} $$
The formulas in~$\mink44$ all apply, though many simplify.  For example, the
metric~\thetag{2.54}, the inverse metric~\thetag{2.55}, the Dirac
form~\thetag{2.56}, the Dirac operator~\thetag{2.58}, and the wave
operator~\thetag{2.59} reduce to
  $$ \align
      g &= 2(dy^+\otimes dy^- + dy^-\otimes dy^+) \tag{2.109}\\
      g\inv &= \frac 12(\partial _+\otimes \partial _- + \partial _-\otimes
     \partial _+) \tag{2.110} \\
      \bar{\psi}\Dirac\psi &= \psib_+\partial _-\psi _+ + \psib_-\partial
     _+\psi _-  \tag{2.111} \\ 
      (\Dirac\psi )_+ &= -\partial _+\psi _-\qquad (\Dirac\psi )_- =
     \partial _-\psi _+  \tag{2.112} \\ 
      (\Dirac\psib )_+ &= -\partial _+\psib _-\qquad (\Dirac\psib )_- =
     \partial _-\psib _+ \tag{2.113} \\ 
      \wave &= \partial _+\partial _-. \tag{2.114} \endalign $$
The relationship to standard coordinates~$x^0,x^1$ is 
  $$ \aligned
      y^+ &= \frac{x^0+x^1}{2}, \\ 
      y^- &= \frac{x^0-x^1}{2},\endaligned \tag{2.115} $$
and 
  $$ \aligned
      \partial _+ &= \frac{\partial }{\partial x^0} + \frac{\partial
     }{\partial x^1}, \\ 
      \partial _- &= \frac{\partial }{\partial x^0} - \frac{\partial
     }{\partial x^1}.\endaligned \tag{2.116} $$
There is a 2-dimensional space of R-symmetries with basis~$J_+,J_-$ described
by the brackets 
  $$ \alignedat2
      [J_+,D_+] &= \hphantom{-}\Dp &\qquad\quad [J_-,\Dm]&=\hphantom{-}\Dm \\
      [J_+,\Dbp] &= -\Dbp &\qquad\quad [J_-,\Dbm]&=-\Dbm \\
      [J_+,\Dm] &= [J_+,\Dbm] = 0 &\qquad\quad [J_-,\Dp] &= [J_-,\Dbp] =
     0.\endalignedat \tag{2.117} $$
The R-symmetry~\thetag{2.81} in~$\mink44$ induces~$J_+ + J_-$
on~$\mink2{(2,2)}$.  The double cover in the spin group of rotation in the
$x^2$-$x^3$ plane in~$\mink44$ induces the R-symmetry~$J_+ - J_-$
of~$\mink2{(2,2)}$.

As in~$\mink44$, a complex function $\Phi \:\mink2{(2,2)}\to\CC$ is {\it chiral\/}
if 
  $$ \Dbp\Phi =\Dbm\Phi =0; \tag{2.118} $$
then the complex conjugate~$\overline{\Phi }$ is {\it antichiral\/}.  There
is a new possibility as well.  Namely, a complex function $\Sigma
\:\mink2{(2,2)}\to\CC$ is {\it twisted chiral\/} if 
  $$ \Dbp\Sigma  = \Dm\Sigma =0; \tag{2.119} $$ 
then the complex conjugate~$\overline{\Sigma }$ is {\it twisted antichiral}. 
 
Formulas~\thetag{2.71} and~\thetag{2.72} for computing~$\int_{}d^4\theta $
are, of course, valid on~$\mink2{(2,2)}$.  Again we use $\int_{}d^4\theta
+\wave\bres$ to compute component lagrangians when chiral and antichiral
fields are involved.  When twisted chiral and twisted antichiral fields are
involved, it is more convenient to use $\int_{}d^4\theta  -\wave\bres$, which
equals the first term in~\thetag{2.72}.  This has the advantage that the
expression $D^2\Dbar^2 + \Dbar^2D^2$ annihilates both twisted chiral and
twisted antichiral fields.  In a more complicated (term in a) lagrangian
involving both chiral and twisted chiral fields one would have to simply
choose a procedure and then deal with the exact terms which would most likely
appear.

\Chapter3{Supersymmetric $\sigma $-Models}

 \comment
 lasteqno 3@102
 \endcomment

 A scalar field is a map~$\phi \:\bmink^n\to X$ from ordinary $n$-dimensional
Minkowski space to a manifold~$X$.  If $X$~is a linear space, then a theory
with the field~$\phi $ is called a linear $\sigma $-model; if $X$~is curved,
then it is a nonlinear $\sigma $-model.  In supersymmetric $\sigma $-models
there is in addition a spinor field.  After some linear algebra preliminaries
in~\S{3.1}, we begin in~\S{3.2} with the linear case, where we show that a
supersymmetric extension of the ordinary free $\sigma $-model exists in
dimensions~$n=3,4,6$.  We label these three cases according to the
ring~$\FF=\RR,\CC,\HH$ which underlies the spin group~$SL(2,\FF)$.  In the
nonlinear case ($X$~a general Riemannian manifold) supersymmetry imposes
constraints on the target manifold~$X$: For~$\FF=\CC$ it must be \Kah\ and
for $\FF=\HH$ it must be \hyp.  The sufficiency of these constraints is
explained at the beginning of~\S{3.3}, after which we derive the exact form
of the component lagrangian.  In the bosonic case any function~$V\:X\to\RR$
may serve as a potential energy function, but supersymmetry constrains the
form of~$V$.  The constraints are described in~\S{3.4}, though the proofs are
deferred to future chapters.  We make some brief remarks about the superspace
formulation in~\S{3.5} and about the relationship among the different
theories in~\S{3.6}.

\Head{3.1}{Preliminary remarks on linear algebra}
 We discuss simultaneously the fields~$\FF=\RR,\CC,\HH$.  Since $\HH$~is
noncommutative, we must be careful about the order of multiplication.  Let
$\bar{\ }$ denote the standard conjugation on~$\FF$, which is trivial
for~$\FF=\RR$.  For $x,y\in \FF$ we have
  $$ \overline{xy} = \bar{y}\bar{x}. \tag{3.1} $$
The extension of~\thetag{3.1} to superalgebra has a sign, in accordance with
the sign rule.  It is useful to note that:
  $$ \Re(\lambda \mu ) = \Re(\mu \lambda ) = \Re(\bar\lambda \bar\mu ) =
     \Re(\bar\mu \bar\lambda ),\qquad \lambda ,\mu \in \FF. \tag{3.2} $$

Let $V$~be the vector space of translations of~$\bmink^n$ ($n=3,4,6$).  If
$S$~is an irreducible real spinorial representation of~$\Spin(V)$, the
field~$\FF$ of endomorphisms of the representation~$S$ is respectively
isomorphic to~$\RR$, $\CC$, or~$\HH$; the vector space~$S$ is of dimension~2
over~$\FF$; and $\Spin(V)\cong SL_{\FF}(S)$.  We choose~$S$ and a {\it
right\/} vector space structure over~$\FF=\RR,\CC$ or~$\HH$ commuting with
the action of~$\Spin(V)$.  The dual $S^*=\Hom\mstrut _{\FF}(S,\FF)$ is then a
{\it left\/} $\FF$-vector space, and $\Spin(V)\cong SL_\FF(S^*)$ as well.  We
write~$\overline{S}$ for~$S$ with the left vector space structure $\lambda s
:=s\lambdab$ and $\overline{S^*}$ for~$S^*$ with the right vector
space structure $ s^*\lambda  := \lambdab s^*$.  The $\FF$-dual~$S^*$ can be
identified with the $\RR$-dual $\Hom\mstrut _{\RR}(S,\RR)$ by $\alpha \mapsto
\text{real part of the $\FF$-linear form }\alpha $.  If $\{f^a\}$ is a basis
of~$S$ over~$\FF$, then there are induced bases $\{\fbar^{\ad}\}$, $\{f_a\}$,
$\{\fbar_{\ad}\}$ of~$\overline{S}$, $S^*$, $\overline{S^*}$.
 
As explained at the beginning of~\S{2.1} there are invariant symmetric
$\RR$-bilinear forms $\Gamma \:S^*\otimes _{\RR}S^*\to V$ and
$\stov\:S\otimes _\RR S\to V$.  They factor through~$\overline{S^*}\otimes
_\FF S^*$ and ${S}\otimes _{\FF}\overline{S}$, respectively.  In general, if
$A,B,C$ are respectively $\FF$~right, $\FF$~left, $\RR$~vector spaces, then
an $\RR$-linear map $T\:A\otimes _\FF B\to C$ gives rise to an
$(\FF,\FF)$-bilinear form $T_1\:A\otimes _\RR B \to C\otimes _\RR \FF$,
characterized by
  $$ T(a,b) = \Re\bigl(T_1(a,b) \bigr). \tag{3.3} $$
Conversely, the real part of any $(\FF,\FF)$-bilinear~$T_1$ obeys 
  $$ \Re T_1(a\lambda ,b) = \Re T_1(a,b)\lambda  = \Re \lambda T_1(a,b) = \Re
     T_1(a,\lambda b),      \tag{3.4} $$
so factors through $A\otimes _\FF B$.  We apply this construction to~$\sdtov$
and~$\stov$, and denote by~$\sdtov^\mu _{a\bd}$ and~$\stov^{\mu \ad b}$ the
coordinates of~$\sdtov_1$ and~$\stov_1$ in bases~$\{e_\mu \}$ of~$V$ and
$\{f^a\}$ of~$S$, up to a factor: 
  $$ \aligned
      \sdtov(x^af_a,y^bf_b) &= \frac{1}{\kappa ^2}\Re(y^b\sdtov^\mu
     _{b\ad}\overline{x^a})\,e_\mu , \\
      \stov(f^ax_a,f^by_b) &= \phantom{\frac{1}{\kappa
     ^2}}\Re(\overline{y_b}\stov^{\mu \bd a}x_a)\,e_\mu .\endaligned
     \tag{3.5} $$
Here $\kappa $~is the factor 
  $$ \kappa = \cases 1 ,&\FF=\RR;\\\sqrt2,&\FF=\CC,\HH. \endcases
     \tag{3.6} $$
On the left hand side of these equations, $\sdtov, \stov$~denote the real
forms acting on elements of the underlying real vector space.  Thus, for
example, we might write $\sdtov\bigl(\Re(x^af_a),\Re(y^bf_b) \bigr)$ to
distinguish it from~\thetag{2.52}.  The factor of~$\kappa ^2$ is introduced
to make these formulas compatible with~\thetag{2.52} and similar formulas of
Chapter~2.  The symmetry of~$\sdtov$ and~$\stov$ becomes the hermiticity
property
  $$ \aligned
      \overline{\sdtov^\mu _{a\bd}} &= \sdtov^\mu _{b\ad}, \\ 
      \overline{\stov^{\mu \ad b}} &= \stov^{\mu \bd a}.\endaligned
     \tag{3.7} $$
The Clifford relation~\thetag{1.3} with this normalization is
  $$ \sdtov^\mu _{a\bd}\stov^{\nu \bd c} + \sdtov^\nu _{a\bd}\stov^{\mu \bd
     c} = 2g^{\mu \nu }\delta ^c_a \tag{3.8} $$
and its conjugate.  We emphasize that $\sdtov^\mu _{a\bd}, \stov^{\mu \ad b}$
lie in~$\FF$, whereas $g_{\mu \nu },g^{\mu \nu }$ are real.

For the quaternion case~$\FF=\HH$ we can choose bases so that the
pairing~$\Gamma $ is given by
  $$ \aligned
      \Gam011&=1,\quad \Gam111=1, \\ 
      \Gam022&=1,\quad \Gam122=-1, \\ 
      \Gam212&=1,\quad \Gam312=-i,\quad \Gam412=j,\quad
     \Gam512=k,\endaligned \tag{3.9} $$
where $i,j,k$ are the usual quaternions.  Then the pairing~$\stov$ is given
by 
  $$ \aligned
      \Gamt011&=1,\phantom{-}\quad \Gamt111=-1, \\ 
      \Gamt022&=1,\phantom{-}\quad \Gamt122=1, \\ 
      \Gamt221&=-1,\quad \Gamt321=-i,\quad \Gamt421=j,\quad
     \Gamt521=k.\endaligned \tag{3.10} $$
In this basis the metric has the standard form~\thetag{2.98}.  It is useful
to note that the vector~$\Gamma (f_a,f_a)$ is lightlike (has vanishing norm
square) for all~$a$; this property will be important in Chapter~6.  The
pairings for the complex case~$\FF=\CC$ are obtained by omitting the entries
with~$j,k$; those for the real case~$\FF=\RR$ by taking the real part of all
entries.  Then the chosen bases agree with those in~\S{2.3}, \S{2.4}.  Recall
from~\thetag{1.9} that for~$\FF=\RR,\CC$ we have a skew tensor~$\epsilon $,
and the pairings~$\stov$ and~$\sdtov$ are related by
  $$ \stov^{\mu \bd a} = \sdtov^\mu _{c\dd}\epsilon ^{ac}\epsilon
     ^{\bd\dd},\qquad \FF=\RR,\CC. \tag{3.11} $$

 In~\S{3.3} we have use for the following.  Suppose $V_R$~is a right
$\HH$~vector space and $V_L$~a left $\HH$~vector space.  Suppose $\stov $~is
a symmetric real-valued bilinear form on~$V_R$ which factors
through~$V_R\otimes _\HH\overline{V_R}$ and $h$~a symmetric real-valued
bilinear form on~$V_L$ which factors through~$\overline{V_L}\otimes _\HH
V_L$.  Then we claim that there is a well-defined symmetric real bilinear
form~$B$ on the real vector space~$V=V_R\otimes _\HH V_L$ constructed
from~$\stov,h$.  In fact, if $\stov _1,h_1$ are the $\HH$-valued pairings
constructed as in~\thetag{3.3}, then
  $$ B(f \otimes v ,f'\otimes v') = \Re\Bigl\{\stov_1
     (f,\bar{f'})\,h_1(\bar{v},v)\Bigr\},\qquad f,f'\in V_R,\quad v,v'\in
     V_L. \tag{3.12} $$
We easily check that this factors through the quaternionic tensor product
using~\thetag{3.2}.  Relative to bases~$\{f^a\}$ of~$V_R$ and~$v_i$
of~$V_L$, any vector in~$V$ can be written
  $$ f^a\psi ^i_{a}\otimes v_i = f^a\otimes \psi ^i_av_i,\qquad \psi ^i_a\in
     \HH. \tag{3.13} $$
Then the formula for~$B$ is 
  $$ B(f^b\psi ^j_bv_j,f^a\lambda ^i_av^a) = \Re(\overline{\psi^j_b}\stov
     ^{ab}\lambda ^i_ah_{ij}). \tag{3.14} $$

\Head{3.2}{The free supersymmetric $\sigma $-model}
 The bosonic $\sigma $-model lagrangian in $n$-dimensional Minkowski
space~$\bmink^n$ for a scalar field $\phi \:\bmink^n\to X$ with values in a
Riemannian manifold~$X$ is 
  $$ L = \frac 12 |d\phi |^2\;\bdens. \tag{3.15} $$
We look for supersymmetric extensions with various amounts of supersymmetry.
With a single supersymmetry the supersymmetric extension for~$n=1$ is given
in~\thetag{1.53}.  (That lagrangian is in~$\bmink^1$, but in fact it is the
dimensional reduction of a lagrangian in~$\bmink^2$; see~\FP6.)  To construct
models with more supersymmetry we first consider the linear case~$X=\RR^k$
for some~$k$; in the next subsection we treat nonlinear $\sigma $-models.
The linear model is free, and to predict which supersymmetric extensions
exist we invoke an argument from the quantum theory.  Namely, upon
quantization we obtain a representation of the supersymmetry algebra, and the
representation theory shows that models are possible with
$s=2,4,8$~supersymmetries for~$k=1,2,4$.  That is, the smallest model has a
scalar field in~$\RR^k$ for $k=1,2,4$ and a single spinor field.  The maximal
dimension in which such models can occur is~$n=3,4,6$, and the models in
these dimensions have minimal supersymmetry.
 
We refer to the theories in~$n=3,4,6$ according to the field
$\FF=\RR,\CC,\HH$.

To treat the various cases simultaneously, we introduce constants~$\kappa $
in~\thetag{3.6} and
  $$ A = \cases \frac12 ,&\FF=\RR;\\1,&\FF=\CC,\HH.\endcases \tag{3.16} $$
(The constant~$A$ enters the lagrangian~\thetag{3.19} and the
constant~$\kappa $ enters the supersymmetry transformation
laws~\thetag{3.23}, as well as~\thetag{3.5} above.)

The fields in the theory are
  $$ \aligned
      \phi \:&\bmink^n\longrightarrow \FF \\ 
      \psi \:&\bmink^n\longrightarrow \Pi S,\endaligned \tag{3.17} $$
where `$\Pi $'~denotes parity reversal.  (The spinor field~$\psi $ is odd.)

        \proclaim{\protag{3.18} {Theorem}}
 The $\sigma $-model lagrangian 
  $$ \boxed{L = A\Bigl\{ \langle d\phi,d\phib \rangle + \Re(\psi
     \Dirac\bar{\psi }) \Bigr\}\;\bdens} \tag{3.19} $$
is supersymmetric in the three cases~$\FF=\RR,\CC,\HH$. 
        \endproclaim

        \demo{Proof}
 The lagrangian~\thetag{3.19} in more detail is\footnote{The fermion kinetic
term is $A$~times 
  $$ \stov(f^a\pa,f^b\di\pb) = \stov(f^b\di\pb,f^a\pa) = \Re(\pad\stov^{\mu
     \ad b}\di\pb) = \Re(\pa\stov^{\mu \bd a}\di\pbd).  \tag{3.20} $$
}
  $$ \boxed{L = A\Bigl\{ g^{\mu \nu }\partial _\mu \phi\cdot \partial _\nu
     \phib
      + \frac 12\pa \cdot \stov^{\mu \bd a}\cdot \partial _\mu \pbd - \frac
     12 \partial _\mu \pa\cdot\stov^{\mu \bd a}\cdot \pbd\Bigr\}\;\bdens,}
     \tag{3.21} $$
where `$\cdot $'~denotes multiplication in~$\FF$.  We usually omit~`$\cdot$'.
In case $\FF=\RR,\CC$ we can combine the two fermion terms at the cost of an
exact term.  For odd parameters~$\eta^a\in \FF$ we let $\hxi$~denote
the {\it real\/} vector field on the space of fields which corresponds to the
abstract supersymmetry transformation 
  $$ \frac{\kappa ^2}{2}(\eta^a Q_a - \Qad\etad), \tag{3.22}  $$ 
where $\etad = \overline{\eta^a}$.  (Recall the abstract supersymmetry
algebra~\thetag{1.16}.)  We postulate
  $$ \boxed{\aligned
      \hxi\phi &= \kappa \pa\,\eta^a \\
      \hxi\pa&=\kappa \di\phi\,\etbd\, \sdtov^\mu _{a\bd }
     ,\endaligned} \tag{3.23} $$
and of course we have the conjugate equations 
  $$ \aligned
      \hxi\phib &= -\kappa \etad\,\pad \\
      \hxi\pad&=\phantom{-}\kappa \sdtov^\mu _{b\ad}\,\eta ^b \,\di\phib\,.
     \endaligned \tag{3.24} $$
There is a minus sign in passing from~$\hxi\phi $ to~$\hxi\phib$ since we
commute the odd quantities~$\eta ^a$ and~$\pa$.  These formulas are motivated
by Lorentz invariance, parity, and power counting.
 
We must check that $\Lie{\hxi}L$ is exact and that the
transformations~\thetag{3.23} generate the supersymmetry algebra
on-shell.\footnote{By contrast, in one dimension~\thetag{1.47} holds
everywhere, that is, off-shell.  We can achieve off-shell supersymmetry
for~$\FF=\RR, \CC$ by introducing an auxiliary field; see~\S4.1 and~\S5.1.}
The variation of the bosonic term in~$L$ is
  $$ \hxi(Ag^{\mu \nu }\,\di\phi\cdot\dj\phib ) = \kappa Ag^{\mu \nu
     }\{\di\psi_{a}\,\eta^a\,\dj\phib - \di\phi\,\etad\,\dj\pad
     \}. \tag{3.25} $$
The second term in~\thetag{3.21} has variation
  $$ \split
      \hxi(\frac A2\,\pa\cdot\stov^{\mu \bd a}\cdot\di\pbd) &= \frac{\kappa
     A}{2}\{\dj\phi\,\etgd\,\sdtov^\nu _{a\gd}\stov^{\mu \bd a}\,\di\pbd +
     \pa\stov^{\mu \bd a}\sdtov^\nu _{c\bd}\eta^c\di\dj\phibar\} \\
      &=\frac{\kappa A}{2}\{\dj\phi\,\etgd\,\sdtov^\nu _{a\gd}\stov^{\mu \bd
     a}\,\di\pbd - \dj\pa\,\stov^{\mu \bd a}\sdtov^\nu
     _{c\bd}\,\eta^c\,\di\phib \} \\
      &\qquad \qquad \qquad \qquad \qquad \qquad + \dj\bigl\{\frac {\kappa A}
     2\,\pa\,\stov^{\mu \bd a}\sdtov^\nu _{c\bd}\,\eta^c\,\di\phib
     \bigr\}.\endsplit \hskip-16pt\tag{3.26} $$
The variation of the third term in~\thetag{3.21} is the conjugate 
  $$ \multline
      \hxi(-\frac A2\,\di\pa\cdot\stov^{\mu \bd a}\cdot\pbd) =
      \frac{\kappa A}{2}\{-\di\pa\,\stov^{\mu \bd a}\sdtov^\nu
     _{c\bd}\,\eta^c\,\dj\phib + \di\phi\,\etgd\,\sdtov^\nu _{a\gd}\stov^{\mu
     \bd a}\,\dj\pbd\} \\
      \qquad \qquad \qquad \qquad \qquad \qquad - \dj\{\frac {\kappa
     A}2\di\phi\,\etgd\,\sdtov^\nu _{a\gd}\stov^{\mu \bd a}\pbd\}.\qquad
     \qquad \endmultline\tag{3.27} $$
Adding~\thetag{3.26} and~\thetag{3.27} and using the Clifford
relation~\thetag{3.8} we see
  $$ \hxi(\text{fermion term}) = \frac{\kappa A}{2}g^{\mu \nu
     }\{\di\phi\,\etad \,\dj\pad - \di\pa\,\eta^a\,\dj\phib \} + EXACT,
     \tag{3.28} $$
where  
  $$ EXACT = \dj\{\frac{\kappa A}{2}\pa\,\stov^{\mu \bd a}\sdtov^\nu
     _{c\bd}\,\eta^c\,\di\phib \ - \ \frac{\kappa
     A}{2}\di\phi\,\etgd\,\sdtov^\nu _{a\gd}\stov^{\mu \bd
     a}\,\pbd\}. \tag{3.29} $$
The nonexact term in~\thetag{3.28} cancels against~\thetag{3.25}, which
proves that $\hxi$~is a nonmanifest symmetry of~$L$. 
 
We now verify that the supersymmetry algebra holds on-shell.  The vector
field~\thetag{3.22} is $\kappa ^2\Re(\eta^a\Qa)$, and in the abstract
supersymmetry algebra~\thetag{1.16} we have, according to~\thetag{3.5}, 
  $$ \split
      \bigl[ \Re(\eta ^a_1\Qa),\Re(\eta ^b_2\Qb)\bigr] &= - \bigl[ \Re(\eta
     ^a_2\Qa),\Re(\eta ^b_1\Qb)\bigr]  \\ 
      &= 2\,\sdtov\bigl(\Re(\eta _2^a\Qa),\Re(\eta ^b_1\Qb) \bigr) \\ 
      &= \frac{2}{\kappa ^2}\Re(\eta ^b_1\sdtov^\mu _{b\ad}\etad_2)P_\mu  \\ 
      &= \frac{2}{\kappa ^2}\Re(\etbd_1\sdtov^\mu
     _{a\bd}\eta^a_2)P_\mu.\endsplit \tag{3.30} $$
So on a field~$f$ we must verify (see~\thetag{1.30}) 
  $$ [\hxi_1,\hxi_2]f = 2\kappa ^2\Re(\etbd_1\sdtov^\mu _{a\bd}\eta^a_2)\,\di
     f.      \tag{3.31} $$
The check for~$\phi $ is straightforward: 
  $$ \hxi_1\hxi_2\phi  = \hxi_1(\kappa \pa\eta^a_2) = \kappa ^2\di\phi
     \,\etbd_1\sdtov^\mu _{a\bd}\eta^a_2, \tag{3.32} $$
and so 
  $$ \split
      [\hxi_1,\hxi_2]\phi  &= \kappa ^2\di\phi (\etbd_1\sdtov^\mu
     _{a\bd}\eta^a_2 -\etbd_2\sdtov^\mu _{a\bd}\eta^a_1) \\ 
      &= 2\kappa ^2\Re(\etbd_1\sdtov^\mu _{a\bd}\eta^a_2)\,\di\phi ,\endsplit
     \tag{3.33} $$
as desired.  For~$\psi $ we must use the equation of motion~\thetag{3.39},
which appears below.  Using~\thetag{3.10} we write this equation as 
  $$ \aligned
      \prt0\psi _1-\prt1\psi _1 &= \prt2\psi _2 - \prt3\psi _2i + \prt4\psi
     _2j + \prt5\psi _2k, \\ 
      \prt0\psi_2 + \prt1\psi _2 &= \prt2\psi _1 + \prt3\psi _1i - \prt4\psi
     _1j - \prt5\psi _1k.\endaligned \tag{3.34} $$
The action of the supersymmetry transformation on~$\psi $ is 
  $$ \hxi_1\hxi_2\pa = \hxi_1(\kappa \di\phi \,\etbd_2\sdtov^\mu _{a\bd}) =
     \kappa ^2\di\psi _c\,\eta^c_1\etbd_2\sdtov^\mu _{a\bd}. \tag{3.35} $$
So 
  $$ [\hxi_1,\hxi_2]\pa = \kappa ^2\di\psi _c(\eta^c_1\etbd_2 -
     \eta^c_2\etbd_1)\sdtov^\mu _{a\bd}. \tag{3.36} $$
Checking explicitly for~$a=1,2$ using~\thetag{3.9} and~\thetag{3.34} we
find the desired result 
  $$ [\hxi_1,\hxi_2]\psi _a = 2\kappa ^2\Re(\etbd_1\sdtov^\mu
     _{c\bd}\eta^c_2)\,\di \psi _a. \tag{3.37} $$

This completes the proof. 
        \enddemo

We compute the equations of motion and the variational 1-form.  Write the
lagrangian~\thetag{3.21} as~$L=\Lch\;\bdens$.  Then 
  $$ \multline
      \frac 1A\delta \Lch = \delta \phi(-\gmn\di\dj\phib ) +
     (-\gmn\di\dj\phi)\,\delta \phib + \delta \pa(-\stov^{\mu \bd a}\di\pbd) +
     (\di\pa\,\stov^{\mu \bd a})\,\delta \pbd \\
      + \dj\{\gmn\delta \phi\,\di\phib + \gmn\di\phi\,\delta \phib + \frac
     12\delta \pa\,\stov^{\nu \bd a}\,\pbd - \frac 12\pa\,\stov^{\nu
     \bd a}\,\delta \pbd\}.\endmultline \tag{3.38} $$
This leads to the equations of motion 
  $$ \boxed{\aligned
      \gmn\di\dj\phi &=0 \\ 
      \di\pa\stov^{\mu \bd a}&=0\endaligned} \tag{3.39} $$
and the variational 1-form 
  $$ \boxed{\gamma = A(\gmn\delta \phi\,\di\phib + \gmn\di\phi\,\delta \phib
     + \frac 12\delta \pa\,\stov^{\nu \bd a}\,\pbd - \frac 12\pa\,\stov^{\nu
     \bd a}\,\delta \pbd)\,\frac{\partial }{\partial x^\nu }\otimes \bdens.}
     \tag{3.40} $$
In coordinate-free notation the equations of motion are 
  $$ \boxed{\aligned
      \Delta \phi &=0 \\ 
      \Dirac\psi &=0.\endaligned} \tag{3.41} $$
 
The Noether current associated to~$\hxi$ is computed from~\thetag{3.40}
and~\thetag{3.29} to be 
  $$ \multline
      \cont{\hxi}\gamma - (\frac{\kappa A}{2}\pa\,\stov^{\mu \bd a}\sdtov^\nu
     _{c\bd}\,\eta^c\,\di\phib - \frac{\kappa A}{2}\di\phi\,\etgd\,\sdtov^\nu
     _{a\gd}\stov^{\mu \bd a}\,\pbd\ ) \,\frac{\partial }{\partial x^\nu
     }\otimes \bdens\\
      = \kappa A(\pa\,\stov^{\nu \bd a}\sdtov^\mu _{c\bd}\,\eta^c\,\di\phib -
     \di\phi\,\etgd\,\sdtov^\mu _{a\gd}\stov^{\nu \bd a}\pbd
     )\,\frac{\partial }{\partial x^\nu }\otimes
     \bdens. \endmultline\tag{3.42} $$
(We use the Clifford identity~\thetag{3.8} in this computation.)  Thus for
$\FF=\RR,\CC$ the {\it supercurrent\/}, which is {\it minus\/} the Noether
current for~$\Qa$, is
  $$ \boxed{j_a = \frac{2A}{\kappa }\pb\,\stov^{\nu \cd b}\sdtov^\mu
     _{a\cd}\,\di\phib\,\frac{\partial }{\partial x^\nu }\otimes \bdens,
     \qquad \FF=\RR,\CC.}  \tag{3.43} $$
(Recall the factor~$\kappa ^2/2$ in~\thetag{3.22}.)  As a check, in the
case~$\FF=\RR$ this reduces in one dimension with one supersymmetry
to~\thetag{1.57}.  Other formulas from that section also check against those
here.

\Head{3.3}{Nonlinear supersymmetric $\sigma $-model}
 Now consider the supersymmetric $\sigma $-model of the previous subsection
where we replace the target~$X$ by a curved Riemannian manifold.  Then $\psi
$~transforms as an element of the pullback tangent bundle, and it is easy to
see that $\dim X$~is divisible by~$\dim \FF$ if the linearization is to
reduce to the linear $\sigma $-model.  But more is true: To write
formulas~\thetag{3.23} we need to be able to multiply tangent vectors in~$X$
by elements of~$\FF$.  Thus we assume that $TX$~is endowed with a {\it
left\/} $\FF$-structure.  Of course, this is no extra condition
for~$\FF=\RR$.  For $\FF=\CC,\HH$ the $\FF$~multiplication is a tensor field
on~$X$ (which is called an almost complex structure for~$\FF=\CC$).  Let
$T$~denote the covariant derivative of this tensor.  Then $X$~ is {\it
\Kah\/} ($\FF=\CC$) or {\it \hyp\/} ($\FF=\HH$) if and only if $T$~vanishes.
The fields in the theory are now
  $$ \aligned
      \phi &\in C^{\infty}(\bmink^n,X) \\
      \psi &\in C^{\infty}(\bmink^n,\Pi S\otimes _{\FF}\phi ^*TX).\endaligned
     \tag{3.44} $$
The lagrangian~\thetag{3.19} continues to make sense if we use the Riemannian
metric~$h$ on~$TX$ as well as the Levi-Civita covariant derivative~$\nabla $.
But for~$\nabla $ to be well-defined on~$TX$ as an $\FF$-bundle, we need
$T$~to vanish.  Therefore, we now assume that $X$~is \Kah\ for~$\FF=\CC$ and
that $X$~is \hyp\ for~$\FF=\HH$.  For~$\FF=\HH$ we explained above that
if~$\{v_i\}$ is a basis of~$TX$ at some point, then the Riemannian metric is
represented by a quaternionic matrix~$\pmatrix h_{ij} \endpmatrix$.
Using~\thetag{3.14} we write the lagrangian~\thetag{3.21} as
  $$ L = \Re\Bigl\{g^{\mu \nu }\partial _\mu \phi^j\cdot h_{ij}\cdot\partial
     _\nu \phib ^i + \psib_{\bd}^j\cdot \stov^{\mu \bd a}\cdot \nabla _\mu
     \psi _a^i\cdot h_{ij}\Bigr\}\;|d^6x|. \tag{3.45} $$
In~\thetag{3.57} below we abbreviate the second term as~`$\langle
\pbd\stov^{\mu \bd a}\nabla _\mu \pa \rangle$'.  In all cases the
lagrangian~\thetag{3.19} ceases to be supersymmetric if $X$~is not flat.
Rather, we have the following result.  Denote the metric~$h$ on~$X$
as~`$\langle \cdot ,\cdot \rangle$' and let $R$~be the Riemann curvature
tensor.

        \proclaim{\protag{3.46} {Theorem}}
 \rom(i\rom)\ \rom($\FF=\RR$\rom)\ The nonlinear $\sigma $-model lagrangian 
  $$ \boxed{L=\Bigl\{ \frac 12|d\phi |^2 + \frac 12\langle \psi \Diracp\psi
     \rangle + \frac 1{12}\epsilon ^{ab}\epsilon ^{cd}\langle
     \pa,\pR(\pb,\pg)\pd  \rangle\Bigr\}\;|d^3x| } \tag{3.47} $$
is supersymmetric for any Riemannian manifold~$X$. \par
 \rom(ii\rom)\ \rom($\FF=\CC$\rom)\  The nonlinear $\sigma $-model lagrangian
  $$ \boxed{L = \Bigl\{ \langle \overline{d\phi},d\phi \rangle + \Re\langle
     \psib\Diracp\psi \rangle - \frac 14\epsilon ^{ac}\epsilon
     ^{\bd\dd}\langle \pa,\pR(\pg,\pdd)\pbd \rangle\Bigr\}\;|d^4x|}
     \tag{3.48} $$
is supersymmetric if and only if $X$~is \Kah. \par
 \rom(iii\rom)\ \rom($\FF=\HH$\rom)\  The nonlinear $\sigma $-model
lagrangian 
  $$ \boxed{\aligned
      L=\Bigl\{ \frac 1{16}&\htil_{\alpha \beta} \eij\nu ^{abcd}(\dab\phi
     )^{\alpha i}(\partial _{cd}\phi )^{\beta j} + \frac 14\htil_{\alpha
     \beta }\nu ^{abcd}\psi _a^\alpha (\phi ^*\nab{cd})\psi ^\beta _b \\
      &-\frac{1}{24}\nu ^{abcd}\phi ^*\Omega _{\alpha \beta \gamma \delta
     }\psi ^\alpha _a\psi ^\beta _b\psi ^\gamma _c\psi ^\delta
     _d\Bigr\}\;|d^6x|.\endaligned} \tag{3.49} $$
is supersymmetric if and only if $X$~is \hyp. 
        \endproclaim

\flushpar
 In all cases the Dirac form involves the covariant derivative~$\nabla $, and
so it also depends on~$\phi $.  For example, in the case~$\FF=\RR$ we have
  $$ \frac 12\langle \psi \Diracp\psi   \rangle = \frac 12\stov^{\mu
     ab}\langle \pa,(\phi ^*\nabla _\mu )\pb  \rangle. \tag{3.50} $$
Note the use of angle brackets on the left hand side to denote the inner
product on~$TX$.  The supersymmetry transformation laws are the same as for
the linear case.\footnote{Physicists often use a spinor field~$\psi $ which
is not intrinsic, so have different formulas.}  We often omit~`$\phi $'
and~`$\phi ^*$' from the notation.
 
We explain~\thetag{3.49}, where we have switched from the quaternionic
notation used previously to a complex notation.  The complexification of the
tangent bundle of a \hyp\ manifold~$X$ of real dimension~$4n$ can be written
  $$ T\cp X\cong V\otimes W^*, \tag{3.51} $$
where $V$~is a complex vector bundle of complex rank~$2n$ and $W^*$~is the
dual of the fixed complex symplectic vector space described at the beginning
of~\S{2.1}.  Furthermore, $V$~carries a pseudoreal (quaternionic)
structure~$J$ and a skew-symmetric bilinear form~$\htil$ which encodes the
metric.  The complexified Riemann curvature tensor, viewed as a functional on
$(T\cp X)^{\otimes 4}\cong V^{\otimes 4}$, is a totally symmetric
tensor~$\Omega $.  Let $S$~be the real spin representation of~$\Spin(1,5)$.
As described in~\S{2.1}, its complexification is~$S_0\otimes W$ for a
4-dimensional complex vector space~$S_0$.  The spinor field~$\psi $ is a real
odd section of~$ \phi ^*V\otimes_\CC S_0$.  Note there is a trace map
  $$ \phi ^*(T\cp X)\otimes S\cp\cong (\phi ^*V \otimes W^*)\otimes
     (S_0\otimes W)\longrightarrow \phi ^*V \otimes S_0\tag{3.52} $$
which expresses the quaternionic tensor product in~\thetag{3.44}.  (All
tensor products in~\thetag{3.52} are over~$\CC$.)  The pseudoreal structures
on~$S_0$ and~$\phi ^*V$ give a real structure on~$\phi ^*V\otimes S_0$.
Recall that $S_0$~carries a volume form~$\nu $.  Now fix a local framing
of~$V$ (indices~$\alpha ,\beta ,\dots =1,\dots ,2n$), a basis of~$W$
(indices~$i,j,\dots =1,2$), and a basis of~$S_0$ (indices~$a,b,\dots
=1,2,3,4$).  The inner product of two tangent vectors~$\xi _1,\xi _2$ is
  $$ \langle \xi _1,\xi _2 \rangle = \htil_{\alpha \beta }\epsilon _{ij}\xi
     _1^{\alpha i}\xi _2^{\beta j}. \tag{3.53} $$
We will find it useful to lift a spinor field~$\psi $ to an odd section
of~$\phi ^*(T\cp X)\otimes S\cp$ via~\thetag{3.52}:
  $$ \psi ^{\alpha i}_{aj} = \psi ^\alpha _a\delta ^i_j. \tag{3.54} $$
Our normalization of the tensor~$\Omega $ in terms of the Riemann
curvature~$R$ is
  $$ R_{(\alpha i)(\beta j)(\gamma k)(\delta \ell )} = \epsilon _{ij}\epsilon
     _{k\ell }\Omega _{\alpha \beta \gamma \delta }. \tag{3.55} $$
In this notation the supersymmetry variation of the fields~\thetag{3.23} is
  $$ \boxed{ \aligned
      (\hxi\phi )^{\alpha i}&=-\sqrt2 \eta ^{ai}\psi ^\alpha _a \\ 
      \hxi\psi ^\alpha _a&= \phantom{-}\sqrt2\,\eij\eta ^{bj}(\dab\phi
     )^{\alpha i}\endaligned} \tag{3.56} $$
for~$\eta ^{ai}\in \CC$.

        \demo{Proof of \theprotag{3.46} {Theorem}}
 We review the proof of \theprotag{3.18} {Theorem} to determine what changes
when $X$~is curved.  (In this paragraph only we use quaternionic notation for
the $\FF=\HH$~case.)  The variation of the lagrangian~$L$ has a new
contribution from varying the covariant derivative in the fermion kinetic
term, and this cancels the variation of the curvature term.  We do this
computation presently, but first we examine changes in the previous
computations.  For~$\FF=\RR$ they are unaltered except for the substitution
of covariant derivatives.  But for~$\FF=\CC,\HH$ our assumption that $X$~is
\Kah, \hyp\ is crucial.  For in the ``integration by parts''
formula~\thetag{3.26} there is an extra term from differentiating the
$\FF$-structure: 
  $$ \multline
      \hxi_\psi (\frac A2\langle\pa\cdot\stov^{\mu \bd a}\cdot\nab\mu
     \pbd\rangle)
       =\frac{\kappa A}{2}\{\langle\dj\phi\,\etgd\,\sdtov^\nu
     _{a\gd}\stov^{\mu \bd a}\,\nab\mu \pbd \rangle -
     \langle\dj\pa\,\stov^{\mu \bd a}\sdtov^\nu
     _{c\bd}\,\eta^c\,\di\phib\rangle\} \\
       \qquad \qquad \qquad \qquad +\dj\bigl\{\frac {\kappa A}
     2\langle\pa\,\stov^{\mu \bd a}\sdtov^\nu _{c\bd}\,\eta^c\,\di\phib
     \rangle\bigr\} \\
      - \frac{\kappa A}{2}\bigl\{\langle\pa\,T_\nu \,\stov^{\mu \bd
     a}\sdtov^\nu _{c\bd}\,\eta^c\,\di\phib \rangle + \langle\pa\,\stov^{\mu
     \bd a}\sdtov^\nu _{c\bd}\,\eta^c\,T_\nu \,\di\phib \rangle\bigr\}
     .\endmultline \tag{3.57} $$
The subscript on~$\hxi$ indicates that we only vary the fermion; the
variation of~$\phi $ is computed below in~\thetag{3.60}.  We also pick up the
conjugate terms in~\thetag{3.27}.  Our assumption on~$X$ is that $T=0$, so
that this extra term does not appear.  With this assumption the verification
of the supersymmetry algebra also goes through as before.

We turn to the variation of the covariant derivative and the curvature term.
Here it is easier to proceed on a case-by-case basis.  First,
consider~$\FF=\RR$ so that $X$~is any Riemannian manifold.  Let $R$~denote
the Riemann curvature tensor.  Fix a basis~$\{v_i\}$ of~$TX$ at a point.
Then as usual we write 
  $$ R_{ijk\ell } = \langle v_i,R(v_k,v_{\ell })v_j  \rangle. \tag{3.58} $$
The curvature tensor obeys certain symmetries and satisfies the Bianchi
identity:
  $$ \aligned
      &R_{ijk\ell} = -R_{jik\ell } = -R_{ij\ell k} = R_{k\ell ij} \\ 
      &R_{ijk\ell } + R_{jki\ell } + R_{kij\ell } = 0.\endaligned \tag{3.59}
     $$
Now the variation in~$\phi $ of~\thetag{3.50} is 
  $$ \split
      \hxi_\phi \bigl(\frac 12\stov^{\mu ab}\langle \pa,(\phi ^*\nabla _\mu
     )\pb \rangle \bigr) &= \frac 12\stov^{\mu ab}\langle \pa,R(\hxi\phi
     ,\di\phi )\pb \rangle \\
      &= \frac 12\stov^{\mu ab}\eta ^c\langle \pa,R(\pg,\di\phi )\pb
     \rangle\\
      &= -\frac 12\stov^{\mu ab}\eta ^cR_{ijk\ell }\pa^i\pb^j\pg^k(\di\phi)
     ^\ell .\endsplit \tag{3.60} $$
The last minus sign comes from commuting~$\pb$ and~$\pg$.
We claim that this cancels against the variation in~$\psi $ of the curvature
term in the lagrangian~\thetag{3.47}.  That variation is 
  $$ \split
      \hxi_\psi \bigl(\frac 1{12}&\eab\epsilon ^{cd}R_{ijk\ell
     }\pa^i\pd^j\pb^k\pg^\ell \bigr) \\
      &= \frac{1}{12}\eab\epsilon ^{cd}R_{ijk\ell }\eta ^e\Bigl[ \sdtov^\mu
     _{ae}(\di\phi )^i\pd^j\pb^k\pg^\ell - \sdtov^\mu _{de}\pa^i(\di\phi)
     ^j\pb^k\pg^\ell\\
      &\qquad \qquad \qquad \qquad \qquad \qquad \qquad +\sdtov^\mu
     _{be}\pa^i\pd^j(\di\phi )^k\pg^\ell -\sdtov^\mu
     _{ce}\pa^i\pd^j\pb^k(\di\phi )^\ell \Bigr] \\
      &=\frac{1}{12}\eta ^e\sdtov^\mu _{ae} \Bigl[ \epsilon ^{ab}\epsilon
     ^{cd}R_{ijk\ell } - \epsilon ^{db}\epsilon ^{ca}R_{jik\ell } + \epsilon
     ^{da}\epsilon ^{cb}R_{jki\ell } -\epsilon ^{dc}\epsilon ^{ab}R_{jk\ell
     i} \Bigr] (\di\phi )^i\pd^j\pb^k\pg^\ell\\
      &= \frac{1}{12}\eta ^e\sdtov^\mu _{ae}\eab\epsilon
     ^{cd}\Bigl[R_{ijk\ell } + R_{ijk\ell } + R_{kji\ell } + R_{kji\ell
     }\Bigr] (\di\phi )^i\pd^j\pb^k\pg^\ell \\
      &= \frac{1}{6}\eta ^e\sdtov^\mu _{ae}\eab\epsilon ^{cd}(R_{ijk\ell } +
     R_{kji\ell }) (\di\phi )^i\pd^j\pb^k\pg^\ell \\
      &= \frac{1}{6}\sdtov^\mu _{ae}\eab(-\eta ^d\epsilon ^{ec} - \eta
     ^c\epsilon ^{de})(R_{ijk\ell } + R_{kji\ell }) (\di\phi)
     ^i\pd^j\pb^k\pg^\ell \\
      &= \frac{1}{6}(-\stov^{\mu bc}\eta ^d + \stov^{\mu bd} \eta
     ^c)(R_{ijk\ell } + R_{kji\ell }) (\di\phi )^i\pd^j\pb^k\pg^\ell \\
      &= \frac{1}{3}
      \stov^{\mu bd} \eta ^c(R_{ijk\ell } + R_{kji\ell }) (\di\phi)
     ^i\pd^j\pb^k\pg^\ell.\endsplit \hskip-66pt\tag{3.61} $$
In the third to last step we use the fact that $\FF^2$~is two-dimensional,
so that the cyclic sum $\eta ^e\epsilon ^{cd}+ \eta ^d\epsilon ^{ec}+ \eta
^c\epsilon ^{de}$ vanishes.  To compare~\thetag{3.61} to~\thetag{3.60} we
need to move around indices: 
  $$ \hxi_\psi \bigl(\frac 1{12}\eab\epsilon ^{cd}R_{ijk\ell
     }\pa^i\pd^j\pb^k\pg^\ell \bigr) = \frac 13\stov^{\mu ab}\eta ^c(R_{\ell
     ijk} + R_{ijk\ell })\pa^i\pb^j\pg^k(\di\phi )^\ell . \tag{3.62} $$
Using the Bianchi identity~\thetag{3.59} and other symmetries we find 
  $$ \split
      \stov^{\mu ab}\eta ^c R_{\ell ijk}\pa^i\pb^j\pg^k(\di\phi )^\ell &= -
     \stov^{\mu ab}\eta ^c (R_{ij\ell k} + R_{j\ell
     ik})\pa^i\pb^j\pg^k(\di\phi) ^\ell \\
      &= \hphantom{-}\stov^{\mu ab}\eta ^c (R_{ijk \ell} + R_{i\ell
     jk})\pa^i\pb^j\pg^k(\di\phi )^\ell \endsplit \tag{3.63} $$
from which 
  $$ \stov^{\mu ab}\eta ^c R_{\ell ijk}\pa^i\pb^j\pg^k(\di\phi )^\ell =
     \frac{1}{2}\stov^{\mu ab}\eta ^cR_{ijk\ell }\pa^i\pb^j\pg^k(\di\phi )^\ell
     . \tag{3.64} $$
Substituting into~\thetag{3.62} we find 
  $$ \hxi_\psi \bigl(\frac 1{12}\eab\epsilon ^{cd}R_{ijk\ell
     }\pa^i\pd^j\pb^k\pg^\ell \bigr) = \frac 12\stov^{\mu ab}\eta
     ^cR_{ijk\ell }\pa^i\pb^j\pg^k(\di\phi )^\ell  \tag{3.65} $$
which cancels~\thetag{3.60}.

To complete the verification that \thetag{3.47}~is supersymmetric, we check
that the variation in~$\phi $ of the curvature term vanishes: 
  $$ \hxi_\phi \bigl(\eab\epsilon ^{cd}\langle \pa,R(\pb,\pg)\pd  \rangle
     \bigr) = \eab\epsilon ^{cd}\eta ^e\langle \pa,(\nabla
     _{\!\!\pe}R)(\pb,\pg)\pd  \rangle. \tag{3.66} $$
Now the symmetries of~$R$ and the (second) Bianchi identity imply that the
expression $\langle \pa,(\nabla _{\!\!\pe}R)(\pb,\pg)\pd \rangle$ is
symmetric in~$a,d$ and in~$b,c,e$.  In addition, it is symmetric under the
interchange~$a,d\leftrightarrow b,c$.  Altogether this implies that it is
totally symmetric in~$a,b,c,d,e$, whence \thetag{3.66}~vanishes.  This
completes the proof for~$\FF=\RR$.
 
Next, consider~$\FF=\CC$.  Thus $X$~is \Kah\ and we work with the
complexified tangent bundle as usual.  We consider~$\pa$ to be a complex
vector of type~$(1,0)$.  The Riemann curvature tensor is of type~$(1,1)$.  If
$\{v_i\}$~is a basis of type~$(1,0)$ vectors at a point of~$X$, we write
  $$ \Rijkl = \langle v_i,R(v_k,\overline{v_\ell })\overline{v_j}
     \rangle.  \tag{3.67} $$
Then the symmetries of the curvature tensor are
  $$ \Rijkl = R_{k\jbar i\lbar} = R_{i\lbar k\jbar} = -R_{\jbar ik\lbar} =
     -R_{i\jbar\lbar k}.\tag{3.68} $$
Now the variation in~$\phi $ of the fermion kinetic term in~\thetag{3.48} is 
  $$ \split
      \hxi_\phi \bigl(\frac 12\langle \psi&\Dirac\psi \rangle - \frac
     12\langle (\Dirac\psib)\psi \rangle \bigr) \\
      &= \frac 12\stov^{\mu a\bd}\langle \pbd,R(\hxi\phi ,\di\phi )\pa
     \rangle - \frac 12\stov^{\mu a\bd}\langle R(\hxi\phi ,\di\phi )\pbd,\pa
     \rangle \\
      &= \stov^{\mu a\bd}\langle \pa,R(\hxi\phi ,\di\phi )\pbd \rangle \\
      &= -\kappa \stov^{\mu a\bd}\Rijkl\pa^i\pbd^{\jbar}\bigl[\eta
     ^e\pe^k(\di\phib)^{\lbar} - \eted(\di\phi
     )^k\ped^{\lbar}\bigr],\endsplit \tag{3.69} $$
where $\kappa =\sqrt2$.  We claim this cancels against the variation in~$\psi
$ of the curvature term in the lagrangian~\thetag{3.48}.  That variation is 
  $$ \split
      \hxi_\psi \bigl(-\frac 14&\epsilon ^{ac}\epsilon
     ^{\bd\dd}\Rijkl\pa^i\pbd^{\jbar}\pg^k\pdd^{\lbar} \bigr) \\
      &= -\frac 14\kappa \epsilon ^{ac}\epsilon
     ^{\bd\dd}\Rijkl\eted[\sdtov^\mu _{a\edd}(\di\phi )^i\pbd^{\jbar}
     \pg^k\pdd^{\lbar} + \sdtov^\mu _{c\edd}\pa^i\pbd^{\jbar}(\di\phi
     )^k\pdd^{\lbar} ] \\
      &\qquad \qquad + \frac 14\kappa \epsilon ^{ac}\epsilon
     ^{\bd\dd}\Rijkl\eta ^e[\sdtov^\mu
     _{e\bd}\pa^i(\di\phib)^{\jbar}\pg^k\pdd^{\lbar} + \sdtov^\mu
     _{e\dd}\pa^i\pbd^{\jbar}\pg^k(\di\phib)^{\lbar} ] \\
       &= -\frac 12\kappa \sdtov^\mu _{a\edd}\epsilon ^{ac}\epsilon
     ^{\bd\dd}\eted\Rijkl (\di\phi )^i\pbd^{\jbar}\pg^k\pdd^{\lbar} + \frac
     12\kappa \sdtov^\mu _{e\bd}\epsilon ^{ac}\epsilon ^{\bd\dd}\eted\Rijkl
     \pa^i(\di\phib)^{\jbar}\pg^k\pdd^{\lbar} \\
      &= -\frac 12\kappa \Rijkl \left[ -\sdtov^\mu _{a\edd}\epsilon
     ^{ac}(-\epsilon ^{\dd\edd}\etbd - \epsilon ^{\edd\bd}\etdd)(\di\phi
     )^i\pbd^{\jbar} \right.\\
      &\qquad \qquad \qquad \qquad \left. + \sdtov^\mu _{e\bd}(-\epsilon
     ^{cd}\eta ^a - \epsilon ^{ea}\eta ^c) \epsilon
     ^{\bd\dd}\pa^i(\di\phib)^{\jbar}\right] \pg^k\pdd^{\lbar} \\
      &= -\frac 12\kappa \Rijkl\bigl[(-\stov^{\mu c\dd}\etbd+ \stov^{\mu
     c\bd}\etdd )(\di\phi )^i\pbd^{\jbar} \\ 
      &\qquad \qquad \qquad \qquad + (\stov^{\mu c\dd}\eta ^a -
     \stov^{\mu a\dd}\eta ^c)\pa^i(\di\phib)^{\jbar}\bigr]
     \pg^k\pdd^{\lbar}\\
      &= -\kappa \stov^{\mu c\dd}\Rijkl\bigl[\eta ^e\pe^i(\di\phib)^{\jbar} -
     \eted(\di\phi )\ped^{\jbar}\bigr]\pg^k\pdd^{\lbar} \\
      &= \kappa \stov^{\mu a\bd}\Rijkl\pa^i\pbd^{\jbar}\bigl[\eta
     ^e\pe^k(\di\phib)^{\lbar} - \eted(\di\phi
     )^k\ped^{\lbar}\bigr].\endsplit \tag{3.70} $$
This cancels~\thetag{3.69} as claimed. 
 
To complete the proof that \thetag{3.48}~is supersymmetric, we verify that
the variation in~$\phi $ of the curvature term vanishes: 
  $$ \multline
      \hxi_\phi \bigl(\epsilon ^{ac}\epsilon ^{\bd\dd}\langle
     \pa,R(\pg,\pdd)\pbd \rangle \bigr) = \kappa \epsilon ^{ac}\epsilon
     ^{\bd\dd}\eta ^e\langle \pa,(\nabla _{\!\!\psi _e}R)(\pg,\pdd)\pbd \rangle
     \\
      -\kappa \epsilon ^{ac}\epsilon ^{\bd\dd}\etabar^{edd}\langle \pa,(\nabla
     _{\!\!\ped}R)(\pg,\pdd)\pbd \rangle. \endmultline\tag{3.71} $$
The Bianchi identity implies that the factor $\langle \pa,(\nabla _{\!\!\psi
_e}R)(\pg,\pdd)\pbd  \rangle $ in the first term is symmetric in~$e,c$.
Combining with~\thetag{3.68} we see that it is symmetric in~$a,e,c$.  Hence
the first term vanishes.  The argument for the second term is similar.

Finally, consider~$\FF=\HH$.  The variation in~$\phi $ of the fermion kinetic
term in~\thetag{3.49} is 
  $$ \split
      \hxi_\phi \bigl(\frac 14\htil_{\alpha \beta }\nabcd\psi ^\alpha
     _a\nab{cd}\psi ^\beta _b \bigr) &= \hxi_\phi \bigl(\frac
     18\nabcd\epsilon ^{ij}\langle \psi _{ai},\nab{cd}\psi _{bj} \rangle
     \bigr) \\
      &= \frac 18\nabcd\epsilon ^{ij}\langle \psi _{ai},R(\hxi\phi ,\partial
     _{cd}\phi )\psi _{bj} \rangle \\
      &=\frac 18\nabcd\epsilon ^{ij}R_{(\alpha i')(\beta j')(\gamma k)(\delta
     \ell )}\psi ^{\alpha i'}_{ai} \psi ^{\beta j'}_{bj}(\hxi\phi )^{\gamma
     k}(\partial _{cd}\phi )^{\delta \ell } \\
      &=\frac \kappa 4\nabcd\epsilon _{k\ell }\Omega _{\alpha \beta \gamma
     \delta }\eta ^{ek}\psi ^\alpha _a\psi ^\beta _b\psi ^\gamma _e (\partial
     _{cd}\phi )^{\delta \ell },\endsplit \tag{3.72} $$
where $\kappa =\sqrt2$.  This cancels against the variation in~$\psi $ of the
curvature term in~\thetag{3.49}, which is 
  $$ \split
      \hxi_\psi \bigl(-\frac{1}{24}\nabcd &\Omega _{\alpha \beta \gamma
     \delta }\psi ^\alpha _a \psi ^\beta _b\psi ^\gamma _c\psi ^\delta
     _d\bigr) \\
      &= \frac{\kappa }{24}\nabcd\Omega _{\alpha \beta \gamma \delta}\eij\eta
     ^{ej}\Bigl[ (\partial _{ae}\phi )^{\alpha i}\psi ^\beta _b\psi ^\gamma
     _c\psi ^\delta _d - \psi ^\alpha _a(\partial _{be}\phi )^{\beta i}\psi
     ^\gamma _c\psi ^\delta _d \\
      &\qquad \qquad \qquad \qquad \qquad \qquad + \psi ^\alpha _a\psi ^\beta
     _b(\partial _{ce}\phi )^{\gamma i}\psi ^\delta _d - \psi ^\alpha _a\psi
     ^\beta _b\psi ^\gamma _c(\partial _{de}\phi )^{\delta i}\Bigr] \\
      &= -\frac \kappa 6 \eij \Omega _{\alpha \beta \gamma \delta}\nabcd\eta
     ^{ej}\psi ^\beta _b\psi ^\gamma _c\psi ^\delta _d(\partial _{ae}\phi
     )^{\alpha i}.\endsplit \tag{3.73} $$
Now we use the fact that $\dim S_0=4$, so that the cyclic sum in~$a,b,c,d,e$
of $\nabcd\eta ^{ej}$ vanishes.  After relabeling some indices we conclude
that \thetag{3.73}~simplifies to 
  $$ \frac{\kappa }{4}\nabcd\epsilon _{k\ell }\Omega _{\alpha \beta \gamma
     \delta }\eta ^{ek}\psi ^\alpha _a\psi ^\beta _b\psi ^\gamma _e (\partial
     _{cd}\phi )^{\delta \ell }, \tag{3.74} $$
which cancels~\thetag{3.72}. 
 
As in the previous cases, we complete the proof by showing that the variation
in~$\phi $ of the curvature term vanishes.  That variation is a multiple of
  $$ \hxi_\phi \bigl(\nabcd\Omega (\pa,\pb,\pg,\pd) \bigr) = -\nabcd\eta
     ^{ei}(\nab{\psi _{ei}}\Omega )(\pa,\pb,\pg,\pd). \tag{3.75} $$
Now the Bianchi identity and the symmetry of~$\Omega $ imply that $(\nab{\psi
_{ei}}\Omega )(\pa,\pb,\pg,\pd)$~is totally skew in~$a,b,c,d,e$.  Since the
skew-symmetrization of~$\nabcd\eta ^{ei}$ vanishes ($\dim S_0=4$), the entire
expression~\thetag{3.75} vanishes.
        \enddemo

In the nonlinear model the formulas~\thetag{3.40} and~\thetag{3.43} for the
variational 1-form and the supercurrent are replaced by the covariant
versions; that is, we use covariant derivatives in place of ordinary
derivatives.  We do not bother recording them here; see \theprotag{6.33}
{Theorem}.  The equations of motion~\thetag{3.41} are affected in a more
drastic way---the right hand sides are nonzero expressions in the curvature
and its covariant derivative.  They vanish if we set~$\psi =0$.  (In fact,
the terms are quadratic, cubic, and quartic in~$\psi $.)  The interested
reader can work out the precise formulas, some of which appear in~\S{4}
and~\S{5}.

\Head{3.4}{Supersymmetric potential terms}
To the bosonic $\sigma $-model lagrangian~\thetag{3.15} with target~$X$ we can
add a potential term.  Let 
  $$ V\:X\longrightarrow \RR \tag{3.76} $$
be a real-valued function, called the {\it potential energy\/}.  We assume
that $V$~is bounded below.  Then the bosonic lagrangian with potential is
  $$ L = \Bigl\{ \frac 12|d\phi |^2 - \phi ^*V\Bigr\}\;\bdens. \tag{3.77} $$
The moduli space of classical vacua for this theory---that is, the space of
static field configurations of minimal energy---is 
  $$ \Mvac=V\inv (0) \tag{3.78} $$
if we assume that the minimum value of~$V$ occurs at~$0$.

Now we consider supersymmetric extensions of~\thetag{3.77}.  One basic
principle, which follows from the supersymmetry algebra, is that in a
supersymmetric theory the potential energy is nonnegative.  That is manifest
in our formulas below.  Our starting point is the nonlinear $\sigma $-models
developed in the previous section.

We will not carry out a detailed analysis in components, but rather content
ourselves with pointing out a few elementary features.  First, note that the
variation of the potential term is, using~\thetag{3.23},
  $$ \hxi(\phi ^*V ) = -\kappa \eta ^a\phi ^*\bigl(dV(\pa) \bigr). \tag{3.79}
     $$
Thus there must be an additional term in the lagrangian to cancel this.
Power counting---that is, weighting the boson~0, the fermion~$1/2$, and a
spacetime derivative~1---shows that in addition we must change the
supersymmetry transformation law~\thetag{3.23} for the fermion.  The form of
the new transformation is
  $$ \hxi\pa=\kappa \sdtov^\mu _{a\bd}\,\etbd\, \di\phi + \kappa
     \epsilon _{ab}\eta ^b\phi ^*F_a \tag{3.80} $$
for some vector fields~$F_a$ on~$X$, and the additional term in the
lagrangian takes the form 
  $$ \phi ^*A^{ab}(\pa,\pb) \tag{3.81} $$
for some bilinear forms~$A^{ab}$ on~$X$.  Imposing the condition that the
lagrangian be supersymmetric gives relations among~$V,A^{ab},F_a$.  In fact,
one can analyze those conditions in the dimensional reduction of the models
to one spacetime dimension.  Such an analysis leads to the form of the
lagrangians in the following result. 

        \proclaim{\protag{3.82} {Theorem}}
  \rom(i\rom)\ \rom($\FF=\RR$\rom)\ Let $X$~be a Riemannian manifold and
$h\:X\to\RR$ a real function.  Then the lagrangian
  $$ \boxed{\aligned
      L =\Bigl\{\frac 12|d\phi |^2 + \frac 12\langle &\psi \Diracp\psi
     \rangle + \frac 1{12}\epsilon ^{ab}\epsilon ^{cd}\langle
     \pa,\pR(\pb,\pg)\pd \rangle\\
       &-\frac 12\phi ^*|\grad h|^2 - \frac 12\eab\phi ^*(\Hess
     h)(\pa,\pb)\Bigr\}\;|d^3x|.\endaligned } \tag{3.83} $$
is supersymmetric.  The supersymmetry transformation law is
  $$ \boxed{\aligned
      \hxi\phi &= - \eta^a\,\pa \\
      \hxi\pa&=\sdtov^\mu _{ab}\,\eta^b\, \di\phi + \epsilon _{ab}\eta
     ^b\phi ^*\grad h\endaligned} \tag{3.84} $$
 \rom(ii\rom)\ \rom($\FF=\CC$\rom)\  Let $X$~be a \Kah\ manifold and
$W\:X\to\RR$ a holomorphic function.  Then the lagrangian
  $$ \boxed{\aligned
      L = \Bigl\{ \langle d\phib,d\phi \rangle + \frac 12\langle
     \psib\Diracp&\psi \rangle - \frac 12\langle (\Diracp\psib)\psi \rangle -
     \frac 14\epsilon ^{ac}\epsilon ^{\bd\dd}\langle \pa,\pR(\pg,\pdd)\pbd
     \rangle \\
      &-\phi ^*\|\grad W\|^2 - \Re\bigl[\eab\phi ^*(\Hess
     W)(\pa,\pb)\bigr]\Bigr\}\;|d^4x|.\endaligned} \tag{3.85} $$
is supersymmetric.  The supersymmetry transformation law is 
  $$ \boxed{\aligned
      \hxi\phi &= - \sqrt2\,\eta^a\,\pa \\
      \hxi\pa&=\sqrt2\,\sdtov^\mu _{a\bd}\,\etbd\, \di\phi + \sqrt2\,\epsilon
     _{ab}\eta ^b\phi ^*\grad W\endaligned} \tag{3.86} $$
        \endproclaim

\flushpar
 Here $\Hess h$ is the covariant hessian 
  $$ \Hess h = \nabla dh, \tag{3.87} $$
which is a symmetric bilinear form.  Similarly, 
  $$ \Hess W = \nabla \partial W. \tag{3.88} $$
The double bars denote the hermitian norm: 
  $$ \|\grad W\|^2 = \langle \overline{\grad W},\grad W  \rangle. \tag{3.89}
     $$
For $\FF=\RR$ the potential energy is 
  $$ V_\RR = \frac 12|\grad h|^2; \tag{3.90} $$
for $\FF=\CC$ it is 
  $$ V_\CC = \|\grad W\|^2. \tag{3.91} $$
In both cases it is nonnegative.  The function~$W$ is called the {\it
superpotential\/}.  The moduli space of classical vacua is 
  $$ \Mvac^{\RR}=\Crit(h) \tag{3.92} $$
for~$\FF=\RR$ and
  $$ \Mvac^{\CC}=\Crit(W) \tag{3.93} $$
for~$\FF=\CC$.  Here `Crit' denotes the set of critical points.
 
We defer the proof of \theprotag{3.82} {Theorem} to~\S{4} and~\S{5}, where we
give a manifestly supersymmetric version in superspace.  There we also derive
the supercurrent for the model with potential.
 
For the $\FF=\HH$ case there is no way to add a potential energy term and
preserve supersymmetry.  Presumably this follows in a straightforward way
from the component analysis indicated above.  On the other hand, the
dimensional reduction of this 6-dimensional theory to 4~dimensions (and
below) permits a supersymmetric potential energy term---the bosonic potential
energy~$V$ is the norm square of a vector field on~$X$ which preserves the
\hyp\ structure.

\Head{3.5}{Superspace construction}
Consider a superspace~$\mink ns$ in any dimension built out of any real spin
representation, and let $X$~be any Riemannian manifold.  Introduce a scalar
superfield 
  $$ \Phi \:\mink ns\longrightarrow X. \tag{3.94} $$
Let $i\:\bmink^n\hookrightarrow \mink ns$ be the inclusion.  There are
component fields 
  $$ \aligned
      \phi &=i^*\Phi  \\ 
      \pa&=\bres D_a\Phi \endaligned \tag{3.95} $$
of the type we need, but in general higher derivatives of~$\Phi $ lead to
more components.   
 
It is tempting to write a lagrangian 
  $$ L = \Lch\;\bdens \tag{3.96} $$
of the form $\Lch=\bres S$ for some function~$S$ on superspace.  We need
4~fermionic derivatives in superspace to have 2~spacetime derivatives on the
scalar~$\phi $, and the natural invariant expression is 
  $$ \Lch=\bres\Bigl\{ kg_{\mu \nu }\stov^{\mu ac}\stov^{\nu bd}D_aD_b\langle
     D_c\Phi ,D_d\Phi   \rangle\Bigr\}, \tag{3.97} $$
where $k$~is a constant.  In fact, the term 
  $$ \Lch=\bres\Bigl\{ -kg_{\mu \nu }\stov^{\mu ac}\stov^{\nu bd}\langle
     D_aD_c\Phi ,D_bD_d\Phi   \rangle\Bigr\}  \tag{3.98} $$
contains a multiple of the bosonic lagrangian~\thetag{3.15}, and this fixes
the constant~$k$.  It is not hard to see that we also obtain the kinetic term
for the fermion~$\psi $.  In addition---if there are at least
2~supersymmetries---we find terms with no derivatives built out of {\it
auxiliary fields\/}, which are restrictions to~$\bmink^n$ of second
derivatives of~$\Phi $.  But in general this construction fails to give a
supersymmetric lagrangian: there are too many component fields.  We can
eliminate some of them from the supersymmetry transformation laws derived
from~\thetag{1.29}, but there is no guarantee that \thetag{3.97}~is
supersymmetric, nor that the supersymmetry algebra closes on-shell.  In fact,
this construction does work for $s=1$~supersymmetry~\thetag{1.52} and
$s=2$~supersymmetries~(\S{4}).  For $s=1$~there are no auxiliary fields;
for~$s=2$ there is an auxiliary real scalar field. For
$s=4$~supersymmetries~(\S{5}) the superspace model presented here works if we
impose a {\it constraint\/} on~$\Phi $ which eliminates many of the component
fields.  There is then a single auxiliary complex scalar field.  In all of
these cases we express~\thetag{3.97} in a form which is manifestly
supersymmetric.

\Head{3.6}{Dimensional reduction}
In a standard way we can reduce a theory on~$\bmink^n$ to a theory
on~$\bmink^m$ for~$m<n$ by considering fields on~$\bmink^n$ invariant under
an~$(n-m)$-dimensional space of spatial translations.  The scalar field~$\phi $
reduces to a scalar field on~$\bmink^m$, and the spinor field~$\psi $ reduces
to (nonchiral) spinor fields on~$\bmink^m$.  (A single spinor field
on~$\bmink^n$ gives possibly many spinor fields on~$\bmink^m$.  For example,
the 8-component chiral spinor field in the 6-dimensional $\FF=\HH$~model
reduces to 4~right handed and 4~left handed spinor fields in~$\bmink^2$.)
The reduced model has the same number of supersymmetries as the original. 
 
In particular, we can relate the three supersymmetric $\sigma $-models
considered here.  For example, the dimensional reduction of the linear
$\FF=\CC$~model to 3~dimensions gives two copies of the $\FF=\RR$~model.
Similarly, the dimensional reduction of the nonlinear $\FF=\CC$~model with
target a \Kah\ manifold~$X$ gives the nonlinear $\FF=\RR$~model with target
the underlying real manifold.  In this case the fermion~$\hat{\psi}$ of the
3-dimensional model is defined in terms of the fermion~$\psi $ of the
4-dimensional model by 
  $$ \boxed{\hat{\psi }= \frac{\psi +\psib}{\sqrt2}.} \tag{3.99} $$
Then the lagrangian~\thetag{3.48} reduces to twice the
lagrangian~\thetag{3.47}.  For example, if we plug~\thetag{3.99} into the
curvature term in~\thetag{3.47} we have 6~nonzero terms with a factor
of~$1/48$ in front, and twice~$6\cdot 1/48$ is~$1/4$.  The sign also works.
The dimensional reduction of the potential term~\thetag{3.85} in the
$\FF=\CC$~model gives the potential term~\thetag{3.83} with
  $$ \boxed{h = \Re W.} \tag{3.100} $$

To dimensionally reduce the $\FF=\HH$~model to 4~dimensions, we must fix a
\Kah\ structure on the \hyp\ target manifold~$X$.  In fact, there is a
2-sphere of such \Kah\ structures parametrized by the unit imaginary
quaternions acting on the vector space~$W^*$ (cf.~\thetag{3.51}).  We choose
the basis~$\{e_1,e_2\}$ of~$W$ so that $V\otimes \CC\, e^1$ is the bundle of
type~$(1,0)$ tangent vectors to~$X$ and $V\otimes \CC \,e^2$~the bundle of
type~$(0,1)$ tangent vectors to~$X$.  As for the spinor fields, recall
from~\S{2.2} that under dimensional reduction we identify $S_0\cong S'\oplus
S''$, where $S'\oplus S''$ is the complexified spinor representation
of~$\Spin(1,3)$.  Thus the spinor fields of the dimensionally reduced theory
transform in $(S'\oplus S'')\otimes \phi ^*V$.  In indices, if $\hxi^{\alpha
i}$~is a complex tangent vector to~$X$, then we define
  $$ \boxed{\xi ^\alpha =\hxi^{\alpha 1},\qquad \bar\xi ^{\bar\alpha
     }=\hxi^{\alpha 2}.} \tag{3.101} $$
Note $\xi $~is real if $\bar\xi ^{\bar\alpha }$~is the complex conjugate
of~$\xi ^\alpha $.  If $\hat\psi ^\alpha _a$~is a spinor field on~$\bmink^6$,
we define a spinor field~$\psi $ on~$\bmink^4$ by 
  $$ \boxed{\alignedat2
      \psi ^\alpha _1 &=\hat\psi _1^\alpha ,&\qquad \qquad \bar\psi
     ^{\bar\alpha }_{\dot1} &= J^{\bar\alpha }_\beta \hat\psi ^\beta _3, \\ 
      \psi ^\alpha _2 &=\hat\psi _2^\alpha ,&\qquad \qquad \bar\psi
     ^{\bar\alpha }_{\dot2} &= J^{\bar\alpha }_\beta \hat\psi ^\beta
     _4,\endaligned} \tag{3.102} $$
where $J$~is the pseudoreal structure on~$V$.  The reality condition on~$\psi
$ is that $\bar\psi ^{\bar\alpha }_{\dot a}$ is the complex conjugate
of~$\psi ^\alpha _a$.  Armed with these formulas, together with~\thetag{2.99}
and~\thetag{3.55}, we dimensionally reduce~\thetag{3.49} to~$\bmink^4$, i.e.,
we evaluate~\thetag{3.49} on fields~$\phi ,\psi $ which satisfy $\partial
_{12}\phi =\partial _{12}\psi =0$.  A bit of computation shows that we
recover~\thetag{3.48}.

\Chapter4{The Supersymmetric $\sigma $-Model in Dimension 3}

 \comment
 lasteqno 4@ 59
 \endcomment

In this section we discuss the $\sigma $-model with 2~supersymmetries.  We
already described it in components in Chapter~{3}.  Here we give a manifestly
supersymmetric treatment in superspace.  We begin in~\S{4.1} by deriving the
component fields and supersymmetry transformation laws, following the general
principles laid out in~\S{1.2}.  Then in~\S{4.2} we state the superspace
lagrangian and derive the component lagrangian.  We recover the lagrangian of
\theprotag{3.46(i)} {Theorem} except with the addition of an auxiliary field.
One virtue of the auxiliary field is that now the bracketing relations among
the supersymmetry transformations~\thetag{4.14} are precisely those of the
super Poincar\'e algebra; without the auxiliary field we need to impose the
equations of motion to get the correct algebra.  In~\S{4.3} we prove
\theprotag{3.82(i)} {Theorem}, which describes the supersymmetric potential
term.  We carry out the analysis of the classical theory in~\S{4.4}---we
compute the variational 1-form, equations of motion, symplectic structure,
supercurrent---directly in superspace and from that rederive the component
expressions obtained in Chapter~{3}.  We conclude in~\S{4.5} by briefly
considering the dimensional reduction to 2~dimensions, where the Poisson
brackets of the supercharges leads to a central extension of the super
Poincar\'e algebra.

\Head{4.1}{Fields and supersymmetry transformations on~$\mink32$}
\subsubhead Linear Case\endsubsubhead 
 We begin with a real {\it scalar superfield\/} 
  $$ \Phi \:\mink32\longrightarrow \RR. \tag{4.1} $$
Define the component fields as
  $$ \aligned
      \phi &=\bres \Phi  \\ 
      \pa &= \bres\Da\Phi \\ 
      F&=\bres(-D^2)\Phi. \endaligned \tag{4.2} $$
So $\phi $~is a real function on Minkowski space~$\bmink^3$, the field $\psi
=\pa f^a$~is a spinor field on~$\bmink^3$, and $F$~is again a real function
on~$\bmink^3$.  Note that $\phi $~and $F$~are {\it even\/} whereas $\psi $~is
{\it odd\/}.  We will see that the field~$F$ only enters algebraically, and
its equations of motion are algebraic.  Such a field is termed {\it
auxiliary\/} since we can solve for it algebraically in terms of the other
fields. The collection of component fields is termed a {\it multiplet\/}, so
here $(\phi ,\psi ,F)$~is a {\it real scalar multiplet\/}.
 
Let $\hxi$~be the vector field on the space~$\scrF_{(\phi ,\psi ,F)}$ of
component fields induced by the supersymmetry transformation~$\eta ^aQ_a$,
where $\eta ^a$~are odd parameters.  We compute the action of~$\hxi $ on
component fields using~\thetag{1.29}.  For the lowest component~$\phi $ this
is straightforward:
  $$ \hxi \phi  = -\eta ^a \bres \Da\Phi  = -\eta ^a \psi _a
     . \tag{4.3} $$
For the action on~$\pa$ we use~\thetag{2.42}: 
  $$ \split
      \hxi\pa &= -\eta ^b \bres\Db\Da\Phi \\
      &=\eta ^b \bres(\partial _{a b } - \epsilon _{b a
     }D^2)\Phi \\
      &= \eta ^b \dab\phi -\epsilon _{a b }\eta ^b
     F.\endsplit \tag{4.4} $$ 
To compute the action on~$F$ we use~\thetag{2.43}: 
  $$ \split
      \hxi F &= -\eta ^a \bres \Da(-D^2)\Phi  \\ 
      &= -\eta ^a \epsilon ^{b c }\bres\dab\Dg\Phi  \\ 
      &= -\eta ^a \epsilon ^{b c }\dab\pg \\ 
      &= \eta ^a (\Dirac\psi )_a .\endsplit \tag{4.5}
     $$
The reader should check the supersymmetry algebra~\thetag{1.30} and should
also compare~\thetag{4.3} and~\thetag{4.4} to~\thetag{3.23}.

\subsubhead Nonlinear Case\endsubsubhead 
 For the {\it nonlinear\/} $\sigma $-model into a Riemannian manifold~$X$ the
superfield is a map 
  $$ \Phi \:\mink32\longrightarrow X. \tag{4.6} $$
The components are defined by the (covariant) extensions of~\thetag{4.2} using
the Levi-Civita covariant derivative: 
  $$ \boxed{\aligned
      \phi &=\bres \Phi  \\ 
      \pa &= \bres\Da\Phi \\ 
      F&=\bres(-\DD^2)\Phi. \endaligned} \tag{4.7} $$
Now $\phi \:\bmink^3\to X$ is a map into the target manifold, i.e., the field
of the underlying bosonic $\sigma $-model.  Then $\psi =\pa f^a$ is a spinor
field on~$\bmink^3$ with values in~$\phi ^*TX$, i.e., $\psi \:\bmink^3\to\Pi
S\otimes \phi ^*TX$.  (Recall that `$\Pi $'~denotes parity reversal.)  For
pedagogical purposes we use the symbol~`$\DDa$' to denote the covariant
derivative of a section of~$\phi ^*TX$ in the direction~$\Da$ and
`$\ddab$'~for the covariant derivative in the direction~$\dab$.  Then $\DD^2
= \frac 12\epsilon ^{a b }\DDa\DDb$, and $F$~is a scalar with values in~$\phi
^*TX$, i.e., a section of~$\phi ^*TX$ on~$\bmink^3$.

To compute with the nonlinear superfield~$\Phi $ it is useful to recall that
  $$ \DDa\Db\Phi =\DDb\Da\Phi , \tag{4.8} $$
since the Levi-Civita connection is torsionfree.  Also, 
  $$ R(\Da,\Db) = \DDa\DDb + \DDb\DDa + 2\ddab\qquad \text{on $\Phi ^*TX$}
     \tag{4.9} $$
by the definition of the curvature~$R$ of~$\Phi ^*TX$ and the sign rule.  One
must also be careful with the sign rule when applying the Bianchi identity.

Now the analogs of equations~\thetag{2.43} for covariant derivatives are
easily derived: 
  $$ \align
      \DDa\DD^2\Phi &= -\epsilon ^{b c }\bigl(\ddab - \frac 13
     R(\Da\Phi ,\Db\Phi ) \bigr)D_c \Phi, \tag{4.10} \\
      \DD^2\Da\Phi &= \hphantom{-}\epsilon ^{b c }\bigl(\ddab -
     \frac 16 R(\Da\Phi ,\Db\Phi ) \bigr)D_c \Phi. \tag{4.11} \endalign $$
To illustrate the computations, we derive~\thetag{4.11} for~$a =1$:
  $$ \split
      2\DD^2D_1\Phi &= \DD_1\DD_2D_1\Phi - \DD_2\DD_1D_1\Phi \\
      &= \bigl[ R(D_1\Phi ,D_2\Phi )D_1\Phi - \DD_2\DD_1D_1\Phi -2\nabla
     _{12}D_1\Phi \bigr] + \DD_2\partial _{11}\Phi \\
      &= 2D_2\partial _{11}\Phi - 2\nabla _{12}D_1\Phi + R(D_1\Phi ,D_2\Phi
     )D_1\Phi\\ 
      &= 2\nabla _{11}D_2\Phi - 2\nabla _{12}D_1\Phi + R(D_1\Phi ,D_2\Phi
     )D_1\Phi .\endsplit \tag{4.12} $$
At the last stage, and also when we compare~\thetag{4.12} with~\thetag{4.11},
we use the Bianchi identity $R(D_1\Phi ,D_1\Phi )D_2\Phi = -2R(D_1\Phi
,D_2\Phi )D_1\Phi $.
 
Using \thetag{4.10} and \thetag{4.11} it is a routine matter to derive the
transformation laws for the fields in the nonlinear case.  In fact, the only
change comes in~\thetag{4.5}: 
  $$ \split
      \hxi F &= -\eta ^a \bres \DDa(-\DD^2)\Phi \\
      &= -\eta ^a \epsilon ^{b c }\bres\bigl(\ddab - \frac 13
     R(\Da\Phi ,\Db\Phi )\bigr)\Dg\Phi \\
      &= -\eta ^a \epsilon ^{b c }\bigl(\ddab - \frac 13 R(\pa
     ,\pb )\bigr)\pg \\
      &= \hphantom{-}\eta ^a \bigl[ (\Dirac\psi )_a + \frac 13
     \epsilon ^{b c }R(\pa ,\pb )\pg\bigr]
     .\endsplit\tag{4.13} $$

For convenience we collect the transformation formulas for the component
fields:
  $$ \boxed{\aligned
      \hxi \phi &= -\eta ^a \pa \\
      \hxi \pa &= \hphantom{-}\eta ^b (\dab\phi - \epsilon _{a
     b }F) \\
      \hxi F &= \hphantom{-}\eta ^a \bigl[ (\Dirac\psi )_a + 
     \frac 13 \epsilon ^{b c }R(\pa ,\pb
     )\pg\bigr].\endaligned}\tag{4.14} $$

These formulas have easy reductions to~$\mink22$.  As mentioned
after~\thetag{2.106} we simply set $\partial _{12} = \nabla _{12}=0$ and
relabel indices as 1=$+$ and 2=$-$: 
  $$ \boxed{\aligned
      \hxi\phi &= -\eta ^+\psi _+ - \eta ^-\psi _- \\
      \hxi\psi _+ &= \eta ^+\partial _+\phi - \eta ^-F \\
      \hxi\psi _- &= \eta ^-\partial _-\phi + \eta ^+F \\
      \hxi F &= -\eta ^+\nabla _+\psi _- + \eta ^-\nabla _-\psi _+ - \eta
     ^+R(\psi _+,\psi_-)\psi_+ + \eta ^-R(\psi_-,\psi_+)\psi_-.\endaligned}
     \tag{4.15} $$

\Head{4.2}{The $\sigma $-model action on~$\mink32$}
Since the computations in the nonlinear case are no more difficult than in
the linear case, we proceed directly to it.  So our superfield~\thetag{4.6}
is a map $\Phi \:\mink32\to X$ into a Riemannian manifold~$X$.  The
lagrangian density on~$\mink32$ is
  $$ \boxed{L_0 = |d^3x|\,d^2\theta\;\frac 14\epsilon ^{a b }\langle
     \Da\Phi ,\Db\Phi \rangle,} \tag{4.16} $$
where $\langle \cdot ,\cdot \rangle$~is the metric on~$X$ pulled back to a
metric on~$\Phi ^*TX$.  Our goal is to derive the component lagrangian as
defined in~\thetag{1.33}.  Recall that in this case we use integration over
the $\theta $~variables, expressed as differentiation in~\thetag{2.40}.  The
resulting lagrangian  {\it function\/} on Minkowski space is
  $$ \Lch_0 = \int_{}d^2\theta \;\frac 14\epsilon ^{a b }\langle
     \Da\Phi ,\Db\Phi   \rangle. \tag{4.17} $$

We carry out the integration using~\thetag{2.40}, at intermediate stages
using~\thetag{4.11} and the covariant analog of~\thetag{2.42}, and finally
using the definition of the component fields~\thetag{4.7} and
formulas~\thetag{2.26}, \thetag{2.28} to restrict to Minkowski space:
  $$ \split
      4\Lch_0 &= -\eab\bres D^2\langle \Da\Phi ,\Db\Phi \rangle \\
      &= \bres\left\{ -2\eab\langle \Da\Phi ,\DD^2\Db\Phi \rangle +
     \eab\egd\langle \DDg\Da\Phi ,\DDd\Db\Phi \rangle \right\} \\
      &= \bres\left\{ -2\eab\egd\langle \Da\Phi ,\nabla _{b c }\Dd\Phi
     \rangle +\frac 13\eab\egd\langle \Da\Phi ,R(\Db\Phi ,\Dg\Phi )\Dd\Phi
     \rangle \right.\\
      &\qquad \qquad\left.\vphantom{\dfrac 13\eab} + \eab\egd\langle \partial
     _{c a }\Phi -\epsilon _{c a }\DD^2\Phi\; ,\; \partial _{d b }\Phi
     -\epsilon _{d b }\DD^2\Phi \rangle \right\} \\
      &= 2\langle\psi \Dirac\psi\rangle + \frac 13 \eab\egd\langle
     \pa,R(\pb,\pg)\pd \rangle + 2|d\phi |^2 + 2|F|^2.\endsplit \tag{4.18} $$
Of course, here the Dirac form~\thetag{2.28} is defined with a covariant
derivative.  So the final formula is\footnote{We put `$\phi $' and~`$\phi
^*$' into the notation to remind ourselves that these terms contribute to the
equation of motion for~$\phi $.} 
  $$ \boxed{\Lch_0 = \frac 12|d\phi |^2 + \frac 12\langle\psi \Diracp\psi
     \rangle+ \frac{1}{12}\eab\epsilon ^{c d }\langle \pa,\pR(\pb,\pg)\pd
     \rangle + \frac 12 |F|^2.} \tag{4.19} $$
The resulting equations of motion imply that~$F=0$.  Putting~$F=0$
in~\thetag{4.19} we recover~\thetag{3.47}.

\Head{4.3}{The potential term on~$\mink32$}
Let $h\:X\to\RR$ be a real-valued function on~$X$, and set
  $$ \boxed{L_1 = |d^3x|\,d^2\theta \;\Phi ^*(h).} \tag{4.20} $$
The expansion in components is 
  $$ \split
      \Lch_1 &= -\frac 12\bres\eab\Da\Db\Phi ^*(h) \\
      &= -\frac 12\bres\eab\Da\bigl( \cont{\Db\Phi }dh \bigr) \\
      &= -\frac 12\bres\eab\bigl( \cont{\DDa\Db\Phi }dh - \cont{\Db \Phi }\Da
     \Phi ^*dh \bigr) \\
      &= \cont{F}\phi ^*dh - \frac 12\eab \phi ^*(\Hess
     h)(\pa,\pb).\endsplit \tag{4.21} $$
In this expression $\Hess h=\nabla dh$~is the covariant hessian of~$h$, a
symmetric tensor on~$X$.  We include~`$\phi ^*$' in our equations to make
clear the dependence of~$\phi $.  In the total lagrangian function~$\Lch_0 +
\Lch_1$ the field~$F$ still enters algebraically, and its equation of motion
is
  $$ \boxed{F = -\phi ^*\grad h.} \tag{4.22} $$
Substituting into the lagrangian we obtain 
  $$ \boxed{\aligned
      \Lch_0 + \Lch_1 \sim \frac 12 |d\phi |^2 + \frac 12\langle\psi
     &\Diracp\psi\rangle + \frac{1}{12}\eab\epsilon ^{c d }\langle
     \pa,\pR(\pb,\pg)\pd \rangle\\
      &- \frac 12 \phi ^*|\grad h|^2
       - \frac 12 \eab\phi ^*(\Hess h)(\pa,\pb).\endaligned} \tag{4.23} $$
The `$\sim$'~indicates that we have eliminated auxiliary fields.  This is
precisely the lagrangian~\thetag{3.83}, and since the superspace model is
manifestly supersymmetric we have proved \theprotag{3.82(i)} {Theorem}.  The
potential energy for the bosons is
  $$ V= \frac 12 |\grad h|^2. \tag{4.24} $$
It is nonnegative, as we expect in general for a supersymmetric lagrangian.
The bilinear form in~$\psi $ is a mass term for the fermions, which we
abbreviate as
  $$ \phi ^*(\Hess h)(\psi ,\psi ). \tag{4.25} $$
 
A {\it vacuum solution\/} of a classical field theory on Minkowski space is a
field configuration with all fermions set to zero and all scalar fields set
to constants which minimize the energy.  (If there are gauge fields, then
they are chosen to be trivial.)  For $\Lch_0+\Lch_1$~this means that $\psi
\equiv 0$ and $\phi $~is a constant which is a critical point
of~\thetag{4.24}.  Now 
  $$ dV = \langle \grad h, \nabla \grad h \rangle. \tag{4.26} $$
This certainly vanishes at critical points of~$h$.  At such points the
potential energy vanishes and supersymmetry is {\it unbroken\/}.  This means
that the solution is annihilated by the supersymmetry
transformation~\thetag{4.14}.  For any vacuum solution
\thetag{4.14}~simplifies to (using~\thetag{4.22})
  $$ \aligned
      \hxi \phi &=0 \\
      \hxi \pa &= \eta ^b \epsilon _{a b }\phi ^*\grad
     h,\endaligned \tag{4.27} $$
and clearly $\hxi \psi =0$ only at a critical point of~$h$.  Thus at
critical points of~$V$ which are {\it not\/} critical points of~$h$,
supersymmetry is broken.  At such a point $\grad h$~is in the kernel of the
mass form~\thetag{4.25} for the fermions.  In other words, at such a point
there is a massless fermion, the so-called {\it Goldstone fermion\/}.

\Head{4.4}{Analysis of the classical theory}
We first compute the variation of the lagrangian density~$L_0+L_1$ under an
arbitrary {\it even\/} variation~$\delta \Phi $ of the superfield~$\Phi $.
One should interpret~$L_0+L_1$ as living on~$\scrF_\Phi \times \mink32$,
where $\scrF_\Phi $~is the supermanifold of superfields~$\Phi $.  Then
`$\delta $'~is the component of the differential along~$\scrF_\Phi $.  We
have 
  $$ \delta (L_0+L_1) = |d^3x|\,d^2\theta \;\Bigl\{\frac 12 \eab\langle
     \deln \Da\Phi ,\Db\Phi \rangle + \cont{\dP}dh\Bigr\}, \tag{4.28} $$
where $\deln$~is the extension of the differential~$\delta $ using the
covariant derivative.  The first term in braces in~\thetag{4.28} is
  $$ \split
      |d^3x|\,d^2\theta \;\frac 12&\eab\langle \deln \Da\Phi ,\Db\Phi \rangle\\
     &= |d^3x|\,d^2\theta \;\frac 12\eab\langle \DDa\dP,\Db\Phi \rangle \\
      &=|d^3x|\,d^2\theta \;\frac 12\eab\Bigl\{ \Da\langle \dP,\Db\Phi
     \rangle - \langle \dP,\DDa\Db\Phi \rangle\Bigr\} \\
      &= d\Bigl\{\cont{\Da}\bigl(|d^3x|\,d^2\theta \;\frac 12\eab\langle
     \Db\Phi ,\dP\rangle \bigr)\Bigr\} - |d^3x|\,d^2\theta\;\langle
     \dP,\DD^2\Phi \rangle.\endsplit \hskip-60pt\tag{4.29} $$
At the last stage we use the fact that the canonical
density~$|d^3x|\,d^2\theta $ is invariant under~$\Da$ and we use the Cartan
formula for the Lie derivative.  So altogether
  $$ \aligned
      \delta (L_0+L_1) &= d\Bigl\{ \cont{\Da}\bigl(|d^3x|\,d^2\theta
     \;\frac 12\eab\langle \Db\Phi,\dP \rangle \bigr)\Bigr\} \\ 
      &\qquad \qquad \qquad -\,
     |d^3x|\,d^2\theta \;\langle \dP\,,\, \DD^2\Phi - \Phi ^*\grad h
     \rangle.\endaligned \tag{4.30} $$
From this we read off the equation of motion in superspace 
  $$ \boxed{\DD^2\Phi  = \Phi ^*\grad h} \tag{4.31} $$
and the variational 1-form 
  $$ \boxed{\gamma = -\cont{\Da}|d^3x|\,d^2\theta \;\frac 12\eab\langle
     \Db\Phi,\dP \rangle .} \tag{4.32} $$
The appellation `1-form' refers to the fact that $\gamma $~is a 1-form on the
space~$\scrF_\Phi $ of superfields.  With respect to~$\mink32$ it is an
integral density of degree~$-1$, written here as a Berezinian contracted with
a vector field, i.e., $\gamma \in \Omega ^{1,|-1|}(\scrF_\Phi \times
\mink32)$.  The differential ~$\delta \gamma $ is $\omega \in \Omega
^{2,|-1|}(\scrF_\Phi \times \mink32)$:
  $$ \boxed{\omega = \cont{\Da}|d^3x|\,d^2\theta \;\frac 12\eab\langle
     \DDb\dP\wedge\dP   \rangle .} \tag{4.33} $$
The 2-form~$\omega $ is a local version of the symplectic form on the space
of classical solutions~$\scrM\subset \scrF_\Phi $; the global symplectic
2-form on~$\scrM$ is obtained by integrating~$\omega $ over a spacelike
hypersurface of codimension~$1|0$ in~$\mink32$.
 
This completes the analysis in superspace.  Now we expand in components.  We
begin with the equation of motion~\thetag{4.31}.  Restricting to Minkowski
space, we simply recover the equation of motion~\thetag{4.22} for~$F$.  The
equation of motion for~$\psi $ is found by applying~$\bres\DDa$
to~\thetag{4.31} and using~\thetag{4.10}:
  $$ \boxed{(\Diracp\psi )_a =-\epsilon ^{b c }\,\phi ^*\nab{a b
     }\pg = -\frac 13\epsilon ^{b c }\phi ^*R(\pa,\pb)\pg +  \phi
     ^*(\nabla \grad h)(\pa).} \tag{4.34} $$
The equation of motion for~$\phi $ is a complicated expression which
schematically is
  $$ \boxed{
      \wave\phi = -\frac 12\phi ^*\grad |dh|^2 + Q^{(2)}_\phi (\psi )+
     R^{(2)}_\phi (\psi )d\phi + Q^{(4)}_\phi (\psi ), } \tag{4.35} $$
where `$\wave$'~is the covariant wave operator 
  $$ \wave=\nabla _{11}\partial _{22} - \nabla _{12}\partial
     _{12} \tag{4.36} $$
and $Q^{(2)}_\phi $, $R^{(2)}_\phi $, and~$Q^{(4)}_\phi$ are forms in~$\psi $
whose coefficients are made from the pullback of derivatives of~$h$ and
derivatives of the curvature~$R$.  The precise formula is not of interest to
us here.  The {\it bosonic\/} classical equation for~$\phi $, obtained by
setting the fermion~$\psi =0$, is
  $$ \wave\phi = -\phi ^*\grad V, \tag{4.37} $$
where $V$~is the potential energy~\thetag{4.24}.  Newton's law at last!
 
Next we compute $\gch,\wch$ which are obtained from~$\gamma ,\omega $ by
integrating out the~$\theta $'s.  Recall from~\thetag{2.16} that
$\Da=\partial_a -\theta ^c \partial _{a c }$.  Now
in~\thetag{4.32} the term with~$\cont{\da}$ drops out after~$\int_{}d^2\theta
$, so 
  $$ \split
      \gch &= \cont{\partial _{a c }}\,|d^3x|\;\int_{}d^2\theta \;\frac
     12\eab\theta ^c \langle \Db\Phi,\dP \rangle \\
      &= -\cont{\partial _{a c }}\,|d^3x|\;\Bigl\{\frac 14\eab\epsilon ^{d e }
     \bres\Dd D_e \theta ^c \langle \Db\Phi,\dP \rangle\Bigr\} \\
      &= -\cont{\partial _{a c }}\,|d^3x|\; \Bigl\{-\frac 12\eab\egd\bres
     \bigl[\langle \DD_d \Db\Phi,\dP \rangle - \langle\Dd\Phi ,\delta \Db\Phi
     \rangle \bigr]\Bigr\} \\
      &= -\cont{\partial _{a c }}\,|d^3x|\; \Bigl\{\frac
     12\eab\egd\bigl[\langle
      \partial _{d b }\phi,\delta \phi \rangle + \epsilon _{d b }\langle
     F,\delta \phi \rangle+\langle \pb,\delta \pd \rangle
     \bigr]\Bigr\}.\endsplit \tag{4.38} $$
Note that `$\delta \pd$' is a covariant derivative of~$\pd$.
Simplifying further we obtain
  $$ \boxed{\aligned
      \gch &= -\ccd\Bigl\{\frac 12\epsilon ^{bd}\epsilon ^{ac}\bigl[\langle
     \partial _{ab}\phi ,\delta \phi \rangle+ \langle \pa,\delta \pb
     \rangle\bigr]\Bigr\}\\
      &= -\cont{\partial _\mu }\;\,|d^3x|\;\Bigl\{ g^{\mu \nu }\langle \partial
     _\nu \phi ,\delta \phi \rangle + \frac 12 \langle\psi \,\stov ^\mu
     \,\delta \psi\rangle \Bigr\}.\endaligned}\tag{4.39} $$
This last expression is written in terms of the basic bilinear forms: the
inverse metric~$g^{\mu \nu }$ and the paring~$\stov ^\mu$ of~\thetag{1.2}.
It agrees with~\thetag{3.40}.  As expected, no derivatives of fermions and
only one derivative of a boson enter into the formula for~$\gch$, and the
potential terms do not contribute.  We compute~$\wch$ analogously
from~\thetag{4.33}, or by simply differentiating~\thetag{4.39}.  The results
agree: 
  $$ \boxed{\aligned
      \wch &= \ccd\Bigl\{\frac 12\epsilon ^{bd}\epsilon ^{ac}\bigl[\langle
     \partial _{ab}\delta \phi \wedge \delta \phi \rangle+ \langle \delta
     \pa\wedge \delta \pb \rangle + \langle \pa\wedge R_\nabla (\delta \phi
     \wedge \delta \phi )\pb \rangle\bigr]\Bigr\}\\
      &=\cont{\di}\;\,|d^3x|\;\Bigl\{ g^{\mu \nu }\langle \partial _\nu\delta
     \phi \wedge \delta \phi \rangle + \frac 12 \langle\delta \psi \stov ^\mu
     \,\delta \psi\rangle + \frac 12\langle \psi \stov^\mu R_\nabla (\delta
     \phi \wedge \delta \phi )\psi \rangle \Bigr\}.\endaligned} \tag{4.40} $$
We compute the global symplectic form~$\Omega $ on the space of
solutions~$\scrM$ by integrating~$\wch$ over a spacelike hypersurface.  Using
coordinates~$x^0,x^1,x^2$ as in~\thetag{5.8}, with $x^0$~representing time,
and integrating over a constant time, we obtain 
  $$ \boxed{\Omega =\int_{x^0=\text{const}}dx^1dx^2\,\Bigl\{ \langle \delta
     \dot{\phi} \wedge \delta\phi \rangle - \frac 12\langle\delta \psi \stov
     ^0\delta \psi\rangle + \frac 12\langle \psi \stov^0 R_\nabla (\delta
     \phi \wedge \delta \phi )\psi \rangle\Bigr\}.} \tag{4.41} $$
Here `$\dot{\phi }$'~denotes the time derivative of~$\phi $ and $\stov
^0$~is the symmetric form~\thetag{1.2} on spinors evaluated in the time
direction. 
 
We compute the Noether current of the supersymmetry transformation~$\eQ$ in
two ways.  First, we work with superfields---so on~$\scrF_{\Phi }\times
\mink32$---where supersymmetry is manifest.  The supersymmetry
transformation~$\eQ$ induces the vector field
  $$ \xi  = \eta ^a\tQa + \hXi \tag{4.42} $$
on~$\scrF_{\Phi }\times \mink32$, where
  $$ \hXi\Phi =-\eta ^c\tau _{Q_c}\Phi . \tag{4.43} $$
Then 
  $$ J = \cont\xi (L_0 + L_1 + \gamma ). \tag{4.44} $$
Alternatively, we can work in components---so on~$\scrF_{(\phi ,\psi
,F)}\times \bmink^3$---with nonmanifest supersymmetry and compute 
  $$ \Jch = \cont{\hxi}\gch - \alpha , \tag{4.45} $$
where $\alpha $~satisfies 
  $$ \Lie{\hxi}\bigl(|d^3x|\;(\Lch_0 + \Lch_1)\bigr)=d\alpha  \tag{4.46} $$
and may be computed directly from~\thetag{1.42}.  Of course, we pass from~$J$
to~$\Jch$ by integration~\thetag{1.32} over the odd variables. 
 
As a first step in computing~\thetag{4.44} we have 
  $$ \split
      \cont\xi L_0 &= \cont{\eta ^c\tQc}L_0 \\
      &= \cont{\eta ^c\partial _c + \eta ^c\theta ^d\partial
     _{cd}}\,\tdens\;\frac 14\eab\langle \Da\Phi ,\Db\Phi \rangle \\
      &= \cont{\partial _c}\,\tdens\;\frac 14\eab\eta ^c\langle \Da\Phi
     ,\Db\Phi \rangle + \cont{\partial _{cd}}\,\tdens\;\frac 14\eab\eta
     ^c\theta ^d\langle \Da\Phi ,\Db\Phi \rangle.\endsplit
     \hskip-40pt\tag{4.47} $$
Similarly, 
  $$ \cont\xi L_1=\cont{\partial _c}\,\tdens\;\eta ^c\Phi ^*(h) +
     \cont{\partial _{cd}}\,\tdens\;\eta ^c\theta ^d\Phi ^*(h) . \tag{4.48}
     $$
From~\thetag{4.32} we compute the $(0,|\!-\!1|)$ component of~$\cont\xi
\gamma $; the other components play no role. 
  $$ \split
      \bigl(\cont\xi \gamma \bigr)^{0,|-1|} &=
     \cont{\hXi}\,\cont{\Da}\,\tdens\;\frac 12\eab\langle \Db\Phi, \delta \Phi
     \rangle \\
      &= \cont{\Da}\,\tdens\;\Bigl\{\frac 12\eab\langle \Db\Phi ,-\eta
     ^c\tQc\Phi \rangle \Bigr\}\\
      &= \cont{\partial_a}\,\tdens\;\Bigl\{\frac 12\eab\eta ^c\langle \Db\Phi
     ,\tQc\Phi \rangle\Bigr\}\\
      &  \qquad\qquad \qquad +
     \cont{\partial_{ad}}\,\tdens\;\Bigl\{\frac 12\eab\eta ^c\theta ^d\langle
     \Db\Phi ,\tQc\Phi \rangle\Bigr\}.\endsplit \tag{4.49} $$
The $(0,|\!-\!1|)$ component of~$J$ is the sum of~\thetag{4.47},
\thetag{4.48}, and~\thetag{4.49}. 
 
Next, we integrate~$J$ over the odd variables using~\thetag{2.40}.  The terms
with~$\cont{\da}$ drop out; only the terms with~$\cont{\dab}$ survive.
From~\thetag{4.47} we have 
  $$ \split
      \ccd\int d^2&\theta \,\frac 14\eab\eta ^c\theta ^d\langle \Da\Phi
     ,\Db\Phi \rangle\\
      &= \ccd\Bigl\{-\frac 18\eab\epsilon ^{ef}\eta ^c\bres D_eD_f\theta
     ^d\langle \Da\Phi ,\Db\Phi \rangle \Bigr\} \\
      &= \ccd\Bigl\{\frac 12\eab\epsilon ^{de}\eta ^c\bres\langle D_eD_a\Phi
     ,D_b\Phi \rangle \Bigr\} \\
      &= \ccd\Bigl\{\eab\epsilon ^{de}\eta ^c\langle -(\partial _{ae}\phi
     +\epsilon _{ea}F),\pb \rangle \Bigr\} \\
      &= \ccd\Bigl\{-\frac 12\eab\epsilon ^{de}\eta ^c\langle \partial
     _{ae}\phi ,\pb \rangle + \frac 12\epsilon ^{db}\eta ^c\langle F,\pb
     \rangle\Bigr\}.\endsplit \tag{4.50} $$
From~\thetag{4.48} we have 
  $$ \split
      \ccd\int_{}d^2&\theta \,\eta ^c\theta ^d\Phi ^*(h) \\
       &= \ccd\Bigl\{-\frac 12\eab\eta ^c\bres D_aD_b\theta ^d\Phi ^*(h)
     \Bigr\} \\
      &= \ccd\Bigl\{\,-\epsilon^{ad}\eta ^c\bres\langle \Phi ^*\grad
     h,D_a\Phi \rangle \Bigr\} \\
      &= \ccd\Bigl\{ -\epsilon^{ad}\eta ^c\langle \phi ^*\grad h,\pa
     \rangle.\Bigr\}\endsplit \tag{4.51} $$
Finally, from~\thetag{4.49} we compute 
  $$ \split
      \cont{\partial _{cd}}\,&|d^3x|\int_{}d^2\theta \,\frac 12\epsilon
     ^{cb}\eta ^a\theta ^d\langle D_b\Phi ,\tQa\Phi \rangle \\
      &= \ccd\Bigl\{-\frac 14\epsilon ^{cb}\epsilon ^{ef}\eta ^a\bres
     D_eD_f\theta ^d\langle D_b\Phi ,\tQa\Phi \rangle \Bigr\}\\
      &= \ccd\Bigl\{-\frac 12\epsilon ^{cb}\epsilon ^{ef}\eta
     ^a\bres\bigl[\langle D_eD_b\Phi ,\tQa\Phi \rangle - \langle D_b\Phi
     ,D_e\tQa\Phi \rangle \bigr]\Bigr\} \\
      &= \ccd\Bigl\{-\frac 12\epsilon ^{cb}\epsilon ^{ed}\eta
     ^a\bres\bigl[\langle D_eD_b\Phi ,\Da\Phi \rangle + \langle D_b\Phi ,\Da
     D_e\Phi \rangle \bigr]\Bigr\} \\
      &= \ccd\Bigl\{-\frac 12\epsilon ^{ed}(\epsilon ^{cb}\eta ^a + \epsilon
     ^{ac}\eta ^b)\bres\langle D_eD_b\Phi ,D_a\Phi \rangle + \epsilon
     ^{cb}\epsilon ^{ed}\eta ^a\bres\langle \partial _{ea}\Phi ,D_b\Phi
     \rangle\Bigr\} \\
      &= \ccd\Bigl\{\frac 12\epsilon ^{ed}\epsilon ^{ab}\eta ^c\langle
     \partial _{eb}\phi +\epsilon _{ab}F ,\pa \rangle + \epsilon
     ^{cb}\epsilon ^{ed}\eta ^a\bres\langle \partial _{ea}\phi ,\pb
     \rangle\Bigr\}.\endsplit \hskip-50pt\tag{4.52} $$
Summing~\thetag{4.50}, \thetag{4.51}, and~\thetag{4.52} we find after
using~\thetag{4.22} that 
  $$ \Jch = \ccd\Bigl\{\epsilon ^{bc}\epsilon ^{de}\eta ^a\langle \partial
     _{ae}\phi ,\pb  \rangle + \epsilon ^{db}\eta ^c\langle \phi ^*\grad
     h,\pb  \rangle\Bigr\}. \tag{4.53} $$
The supercurrent~$j_a$, which is {\it minus\/} the Noether current of the
supersymmetry transformation~$\Qa$, is 
  $$ \boxed{j_a = \ccd\Bigl\{\epsilon ^{cb}\epsilon ^{de}\langle \partial
     _{ae}\phi ,\pb  \rangle + \epsilon ^{bd}\delta _a^c\langle \phi ^*\grad
     h,\pb  \rangle\Bigr\}.} \tag{4.54} $$
 
As a check, we compute~$\Jch$ directly using~\thetag{4.45}.
Using~\thetag{4.39} we compute the first term of~\thetag{4.45}:
  $$ \split
      \cont{\hxi}\gch &= \frac 12\epsilon ^{ac}\epsilon ^{bd}\cont{\partial
     _{cd}}\,|d^3x|\;\Bigl\{\langle \partial _{ab}\phi,-\eta
     ^e\pe \rangle + \langle \pb,\eta ^e(\partial _{ae}\phi -\epsilon _{ae}F)
     \rangle\Bigr\} \\
      &= \ccd\Bigl\{\frac 12\epsilon ^{ac}(\epsilon ^{db}\eta ^e + \epsilon
     ^{de}\eta ^b)\langle \partial _{ab}\phi ,\pe \rangle - \frac12 \epsilon
     ^{ec}\eta ^d\langle \phi ^*\grad h,\pe \rangle\Bigr\}.\endsplit
     \hskip-20pt\tag{4.55} $$
Note we use~\thetag{4.22}.  Next, we compute~$\alpha $ directly
from~\thetag{1.42}, setting~$\Delta =0$: 
  $$ \split
      \alpha &= \cont{\di}\,|d^3x|\;\sdtov^\mu _{ab}(-\frac 12\epsilon
     ^{ef})\bres D_eD_f\eta ^a\theta ^b\bigl[\frac 14\epsilon ^{gh}\langle
     D_g\Phi ,D_h\Phi\rangle+\Phi ^*(h) \bigr] \\
      &=\ccd\Bigl\{\frac 12\epsilon ^{ef}\eta ^c\bres D_eD_f\theta ^d\bigl[
     \frac 14\eab\langle D_a\Phi ,D_b\Phi \rangle+ \Phi ^*(h)\bigr]\Bigr\} \\
      &= \ccd\Bigl\{\frac 12\epsilon ^{ed}\eta ^c\bres\bigl[\eab\langle
     D_eD_a\Phi ,D_b\Phi \rangle + 2\langle \Phi ^*\grad h,D_e\Phi
     \rangle\bigr]\Bigr\} \\
      &= -\ccd\Bigl\{\frac 12\eab\epsilon ^{ed}\eta ^c\langle \partial
     _{ae}\phi ,\pb \rangle -\frac 12\epsilon ^{ec}\eta ^d\langle \phi
     ^*\grad h,\pe \rangle\bigr]\Bigr\}.\endsplit \tag{4.56} $$
The first term agrees with~\thetag{3.29}.  Combining~\thetag{4.55}
and~\thetag{4.56} we recover~\thetag{4.53}.

\Head{4.5}{Reduction to~$\mink2{(1,1)}$}
We can reduce the model to $n=2$~dimensions either in superspace or in
components.  In both cases we consider the $n=3$~dimensional theory with
fields constrained to be invariant under~$\partial _{12}$.  All the formulas
worked out in $n=3$~dimensions hold, but there are simplifications.  We
already wrote in~\thetag{4.15} the supersymmetry transformations.  We record
here the supercharges~$\Qpt,\Qmt$ obtained by integrating the Noether
current~\thetag{4.54} over space:
  $$ \aligned
      \Qpt &= \int_{x^0 = \text{const}} \Bigl\{ \langle \partial _{+}\phi
     ,\psi _+ \rangle - \langle \phi ^*\grad h,\psi _- \rangle
     \Bigl\}\;|dx^1| \\
      \Qmt &= \int_{x^0 = \text{const}} \Bigl\{ \langle \partial _{-}\phi
     ,\psi _- \rangle + \langle \phi ^*\grad h,\psi _+ \rangle
     \Bigl\}\;|dx^1|. \endaligned \tag{4.57} $$
 
In the abstract supersymmetry algebra~\thetag{1.15} the Lie
bracket~$[\Qp,\Qm]$ of the supersymmetry generators vanishes.  However, the
Poisson bracket~$\{\Qpt,\Qmt\}$ of the Noether charges is nonzero in this
case; it is a locally constant function on the space of classical solutions.
We compute it by acting~$\Qph$ on~$\Qmt$, where $\Qph$~is the Hamiltonian
vector field which corresponds to the function~$\Qpt$; its action on fields
is given in~\thetag{4.15}: 
  $$ \aligned
      \Qph\phi &=-\psi_+ \\ 
      \Qph\psi _+&=\partial _+\phi  \\ 
      \Qph\psi _-&=-\phi ^*\grad h \\ 
      \Qph(\phi ^*\grad h) &= \nabla _+\psi _- + R(\psi _+,\psi _-)\psi
     _+.\endaligned \tag{4.58} $$
In the following computation most terms drop out after we use the equation
of motion~\thetag{4.34} and the Bianchi identity:
  $$ \split
      \frac 12\,\{\Qpt,\Qmt\} &= \frac
     12\int_{x^0=\text{const}}\Bigl\{-\langle \partial _-\phi ,\phi ^*\grad h
     \rangle + \langle \partial _+\phi ,\phi ^*\grad h \rangle\Bigr\}\;|dx^1|
     \\
      &= \int_{x^0=\text{const}} \langle \partial _1\phi ,\phi ^*\grad h
     \rangle \;|dx^1| \\
      &= \int_{x^0=\text{const}} \partial _1(\phi ^*h)\;|dx^1| \\
      &= h\bigl(\phi (\infty ) \bigr) - h\bigl(\phi (-\infty )
     \bigr).\endsplit \tag{4.59} $$
A scalar field~$\phi =\phi (x^1)$ depending on the spatial variable has
finite energy only if it has limits as~$x^1\to\pm\infty $ which are critical
points of~$h$.  The Poisson bracket~\thetag{4.59} measures the difference of
the critical values.  It induces a central extension of the abstract
supersymmetry algebra.

\Chapter5{The Supersymmetric $\sigma $-Model in Dimension 4}

 \comment
 lasteqno 5@ 40
 \endcomment

In this chapter we consider the $\sigma $-model with 4~supersymmetries, which
was described in components in Chapter~{3}.  Here we give a manifestly
supersymmetric treatment in superspace.  In many respects the treatment
parallels that of Chapter~{4}, so we are briefer here.  (One new feature over
theories with 2~supersymmetries is the presence of an R-symmetry.)  Although
the model is well-defined and supersymmetric for any \Kah\ target
manifold~$X$, the superspace model is nicest when $X$~is Hodge.  In that case
we give a global superspace lagrangian (see~\thetag{5.22}).  In these
theories the scalar potential is the norm square of the gradient of a
holomorphic function called the {\it superpotential\/}.  We describe it
in~\S{5.3}, and in particular prove \theprotag{3.82(ii)} {Theorem}.  We do
not give a detailed analysis of the classical theory, but simply summarize
the most important equations in~\S{5.4}.

\Head{5.1}{Fields and supersymmetry transformations on~$\mink44$}
\subsubhead Linear Case\endsubsubhead 
 The simplest superfield for which we can write sensible lagrangians is the
(linear) {\it chiral superfield\/} 
  $$ \Phi \:\mink44\longrightarrow \CC, \tag{5.1} $$
which is required to satisfy 
  $$ \Dad\Phi =0. \tag{5.2} $$
The complex conjugate of a chiral superfield is called an {\it antichiral
multiplet}.  Define the component fields as 
  $$ \boxed{\aligned
      \phi &=\bres \Phi  \\ 
      \pa &= \frac{1}{\sqrt2}\bres\Da\Phi \\ 
      F&=\frac{1}{2}\bres(-D^2)\Phi. \endaligned} \tag{5.3} $$
Component fields for the antichiral superfield~$\Phibar$ are defined by the
conjugate equations to~\thetag{5.3}.  Note that $F$~is a {\it complex\/}
scalar field.
 
Let $\hxi$~be the vector field on the space~$\scrF_{(\phi ,\psi ,F)}$ of
component fields induced by the supersymmetry transformation $ \eta ^a \Qa +
\etabar^{\ad} \Qad$.  Here $\eta ^a ,\etabar^{\ad}$~are independent odd
parameters.  The action on the components of~$\Phi $ is computed by similar
manipulations as in \thetag{4.3}--\thetag{4.5}.  For example,
using~\thetag{2.73} and~\thetag{2.75} we compute
  $$ \split
      \hxi F &= -\bres(\eta ^a \Da + \etabar^{\ad} \Dad)(\frac 12
     D^2)\Phi \\
      &= -\etabar^{\ad}\epsilon ^{b c }\bres \partial _{b
     \ad}\Dg\Phi \\
      &= -\sqrt2\etabar^{\ad}\epsilon ^{b c }\partial_{b
     \ad}\psi_{c } \\
      &= \sqrt2\etabar^{\ad}(\Dirac\psi )_{\ad}.\endsplit \tag{5.4} $$
Collecting the results of similar computations, we display the supersymmetry
transformation laws:
  $$ \boxed{\aligned
      \hxi \phi &= -\sqrt2\eta ^a \pa \\
      \hxi \pa &= \hphantom{-}\sqrt2(\etabar^{\bd}\dabd\phi-\eta ^b \epsilon
     _{a b }F) \\
      \hxi F &= \hphantom{-}\sqrt2\etabar^{\ad}(\Dirac\psi )_{\ad}
     .\endaligned} \tag{5.5} $$
The transformation laws for the components of~$\Phibar$ are obtained by
conjugating the equations in~\thetag{5.5}. 
 
\subsubhead Nonlinear Case\endsubsubhead 
 The nonlinear chiral superfield is a map
  $$ \Phi \:\mink44\longrightarrow X, \tag{5.6} $$
where $X$~is a complex manifold and $\Phi $~is required to pullback
holomorphic functions on~$X$ to linear chiral superfields on~$\mink44$.
Another way to state this is that $\Da\Phi $~is a vector field of
type~$(1,0)$, and so $\Dad\Phibar $~is a vector field of type~$(0,1)$.  Again
the composition of a chiral superfield and a holomorphic function is a chiral
superfield.  The conjugate conditions define a nonlinear antichiral
superfield.  To define the component fields we assume that $X$~is
K\"ahler\footnote{In any case $X$~must be K\"ahler to define the $\sigma
$-model action.} and use the Levi-Civita covariant derivative as
in~\thetag{4.7}:
  $$ \boxed{\aligned
      \phi &=\bres \Phi  \\ 
      \pa &= \frac{1}{\sqrt2}\bres\Da\Phi \\ 
      F&=\frac 12\bres(-\DD^2)\Phi. \endaligned} \tag{5.7} $$
Note that $\pa$ and~$F$ have type~$(1,0)$.  The conjugate fields~$\pad$
and~$\Fb$ have type~$(0,1)$.

Now the analogs of equations~\thetag{2.74} and~\thetag{2.75} for covariant
derivatives of nonlinear chiral superfields are easily derived:
  $$ \align
      \DDa\DDbar^2\Phi &= \DDbar^2\Da\Phi = 0,\tag{5.8} \\
       \DDad \DD^2 \Phi &= - 2\epsilon ^{b c }\Bigl\{\nabla _{b \dot{a} } -
     \frac 14R(\Db\Phi ,\Dad\Phibar )\Bigr\}\Dg\Phi. \tag{5.9} \endalign $$
The conjugate equations to~\thetag{5.8} and \thetag{5.9} give information
about covariant derivatives of nonlinear antichiral superfields.  There is
only one curvature term which alters~\thetag{5.5} in the nonlinear case:
  $$ \boxed{\aligned
      \hxi \phi &= -\sqrt2\eta ^a \pa \\
      \hxi \pa &= \hphantom{-}\sqrt2(\etabar^{\bd}\dabd\phi - \eta ^b
     \epsilon _{a b }F) \\
      \hxi F &= \hphantom{-}\sqrt2\etabar^{\ad}\bigl[ (\Dirac\psi )_{\ad} +
     \frac 14\epsilon ^{bc}R(\pb,\pad)\pg\bigr] .\endaligned}\tag{5.10} $$

\Head{5.2}{The $\sigma $-model action on~$\mink44$}
\subsubhead Linear Case\endsubsubhead 
 Here it is worth computing separately for the linear
superfield~\thetag{5.1}.  The lagrangian density on~$\mink44$ is
  $$ \scrL_0 = |d^4x|\, d^4\theta \;\frac 14\Phibar \Phi. \tag{5.11} $$
Recall from the discussion following~\thetag{2.72} that we compute component
lagrangians by integrating over the~$\theta $'s and adding the wave operator.
Hence by~\thetag{2.72}
  $$ \split
      4\scrLch_0 &= \left\{ \int_{}d^4\theta + \wave\bres \right\} \Phibar
     \Phi \\
      &= \frac 12\bres(D^2\Dbar^2 + \Dbar^2D^2 +4\wave)\Phibar\Phi \\
      &= \frac 12\bres\bigl\{D^2( \Dbar^2\Phibar \cdot \Phi) +
     \Dbar^2(\Phibar \cdot D^2\Phi) +4\wave(\Phibar \Phi)\bigr\} \\
      &= \frac 12 \bres \bigl\{(D^2\Dbar^2+4\wave)\Phibar\cdot \Phi +
     \Phibar\cdot(\Dbar^2D^2+4\wave)\Phi + 8\langle d\Phibar,d\Phi\rangle \\
      &\qquad + 2\, \Dbar^2\Phibar\cdot D^2\Phi
     +\eab\Da\Dbar^2\Phibar\cdot\Db\Phi - \epsilon
     ^{\ad\bd}\Dbd\Phibar\cdot\Dad D^2\Phi \bigr\} \\
      &= \frac 12\bigl\{8\langle d\phibar,d\phi\rangle +8\Fbar F
     +4\eab\epsilon ^{\dot c\dot d}\pb\partial _{a \gd}\pdd + 4\epsilon
     ^{\ad\bd}\egd\pbd\partial _{c \ad}\pd\bigr\}, \endsplit \tag{5.12} $$
where we use~\thetag{2.74}--\thetag{2.77} and the definition~\thetag{5.7} of
component fields in the last step.  Finally, by~\thetag{2.56}
  $$ \boxed{\scrLch_0 = \langle d\phibar,d\phi\rangle + \frac 12(\psib
     \Dirac\psi + \psi \Dirac\psib) + \Fbar F.}  \tag{5.13} $$
The Dirac form in~\thetag{5.13} is real, but up to an exact term we can write
it as the complex form~$\psib\Dirac\psi $: 
  $$ \frac 12(\psib \Dirac\psi + \psi \Dirac\psib)=\psib\Dirac\psi -\frac
     12\partial _\mu (\psib\sdtov^\mu \psi ). \tag{5.14} $$
We drop this exact term in subsequent formulas.  As in $3$~dimensions,
$F$~enters the lagrangian~\thetag{5.13} as an auxiliary field and can be
eliminated through its equation of motion~$F=0$.

\subsubhead Nonlinear Case\endsubsubhead 
 Now we carry out the same computation for the nonlinear chiral
superfield~\thetag{5.6} with values in a K\"ahler manifold~$X$.  The
superfield expression for the lagrangian depends on the choice of a local
K\"ahler potential~$K$, which is a locally defined real-valued function
on~$X$ such that the K\"ahler form is 
  $$ \omega =\sqrt{-1}\,\partial \bar\partial K \tag{5.15} $$ 
and the metric is 
  $$ \langle \cdot ,\cdot \rangle = 2\sqrt{-1}\,\omega =
     2\bar\partial\partial K, \tag{5.16} $$
viewed as a {\it bilinear\/} form on~$ \overline{TX}\otimes TX$.  For
example, on~$\CC$ we have a global K\"ahler potential $K(\bar z, z)= \frac
12\|z\|^2$.  We use double bar notation~$\|\cdot \|$ throughout for the norm
on a hermitian vector space.
 
The lagrangian density on~$\mink44$ is 
  $$ \boxed{\scrL_0 = |d^4x|\,d^4\theta \;\frac 12K(\Phibar,\Phi).}
     \tag{5.17} $$
Although it depends on a choice of K\"ahler potential, the component
lagrangian makes sense globally by the following argument.  Consider a change
$K\to K + f + \bar{f}$ for $f$~a holomorphic function on~$X$.  Of course, it
does not affect the K\"ahler form~\thetag{5.15} and is exactly the ambiguity
in the choice of K\"ahler potential.  Now since we compute the component
lagrangian using the first term of~\thetag{2.71}---the integral over~
$\theta $ plus the wave operator---and since that operation annihilates
chiral and antichiral superfields, such a change in~$K$ does not affect the
component lagrangian.

There is a global formulation of the lagrangian density~$\scrL_0$ on
superspace when $X$~is a {\it Hodge\/} manifold.  A Hodge manifold is a
K\"ahler manifold~$X$ together with a holomorphic hermitian line bundle $L_{\omega }\to
X$ whose curvature is related to the K\"ahler form by
  $$ \curv(L_{\omega }) = \sqrt{-1}\,\omega .  \tag{5.18} $$
Recall that if $s$~is a local holomorphic section of $L_{\omega }\to X$, then 
  $$ \curv(L_{\omega }) = \dbar\partial \log\|s\|^2, \tag{5.19} $$
and so $\log\|s\|^2$~is a local K\"ahler potential.  Consider a chiral
superfield $\Phi \:\mink44\to X$.  Since Minkowski space is contractible,
there is a {\it global\/} lift to a chiral superfield\footnote{The existence
of a global lift depends on the fact that
  $$ H^1(\mink44,\Cal{C}^*)=0, \tag{5.20} $$
where $\Cal{C}^*$ is the sheaf of nonzero chiral superfields.  The
vanishing~\thetag{5.20} is obvious in components.}
  $$ \tilde{\Phi }\:\mink44\longrightarrow \Phi ^*L_{\omega } \tag{5.21} $$
which is a section of $\Phi ^*L_{\omega }\to\mink44$.  For any such lift we write the
lagrangian~\thetag{5.17} globally as 
  $$ \boxed{\scrL_0 = |d^4x|\,d^4\theta \;\frac 12\log\|\tilde{\Phi
     }\|^2.} \tag{5.22} $$
By the argument in the previous paragraph the corresponding component
lagrangian is independent of the choice of lift~$\tilde{\Phi }$.  We remark
that it is not unfamiliar to have an action whose lagrangian density depends
on an auxiliary map:  consider, for example, the usual formulation of the
Wess-Zumino-Witten term in two or four dimensions. 
 
As a preliminary to computing the component lagrangian from~\thetag{5.17},
note that the Cartan formula $\Lie{\eta }=d\cont{\eta }+\cont{\eta }d$ for
the Lie derivative, valid for all vector fields~$\eta $, implies the
following equations when acting on 1-forms:
  $$ \alignat2
      \cont{\zeta }\cont{\zeta '}d &= -\Lie{\zeta }\cont{\zeta '} -\Lie{\zeta
     '}\cont{\zeta }
      + \cont{[\zeta ,\zeta ']},\qquad &&\text{$\zeta ,\zeta '$
     odd;}\tag{5.23} \\
      \cont{\zeta }\cont{\xi}d &= \phantom{-}\Lie{\xi}\cont{\zeta}
     \;-\Lie{\zeta }\cont{\xi} \;+ \cont{[\zeta,\xi]},\qquad &&\text{$\zeta$
     odd, $\xi$ even.}  \tag{5.24} \endalignat $$
We use these formulas several times in the next computation. 
 
We choose~\thetag{2.72} to compute the component lagrangian, though
\thetag{2.71}~also leads to a straightforward computation.  Thus we expand~\thetag{5.17}:
  $$ \split
      \DDbar^2\DD^2\Phi ^*K &= \frac12 \eab\DDbar^2\Da\cont{\Db\Phi }\partial
     K\\
      &= \frac14 \eab\egdd\DDgd(-\DDa\Ddd-2\partial _{a \dd})\cont{\Db\Phi
     }\partial K \\
      &= \frac14 \eab\egdd\DDgd\Bigl\{ \Da\cont{\Ddd\Phibar}\cont{\Db\Phi
     }\dbar\pK + 2\Da\cont{\partial _{b \dd}\Phi }\pK \\
      &\qquad \qquad \qquad \qquad \qquad - 2\cont{\Db\Phi }\cont{\partial
     _{a \dd}\Phi }\dbar\pK -2\Db\cont{\partial _{a \dd}\Phi }\pK \Bigr\} \\
      &= \frac14 \eab\egdd\DDgd\Bigl\{- \frac 12 \Da\langle \Ddd\Phibar
     ,\Db\Phi \rangle - \langle \partial _{a \dd}\Phibar,\Db\Phi \rangle +
     4\Da\cont{\partial _{b \dd}\Phi }\pK \Bigr\} \\
      &= \frac 14 \egdd \Dgd \langle \Ddd\Phibar ,\DD^2\Phi \rangle +
     \eab\egdd \DDgd\Da\cont{\partial _{b \dd}\Phi }\pK.\endsplit
     \hskip-20pt\tag{5.25} $$
The first term is  
  $$ \split
      \frac 14\egdd\Dgd\langle \Ddd\Phibar ,\DD^2\Phi \rangle &= \frac 12
     \langle \DDbar^2\Phibar,\DD^2\Phi \rangle -\frac 14\egdd
     \langle\Ddd\Phibar , \DDgd\DD^2\Phi \rangle \\
      &= \frac 12\eab\egdd \Bigl\{ \langle \Ddd\Phibar,\nabla \mstrut _{\!\!a
     \gd}\Db\Phi \rangle \\
      &\qquad - \frac 14 \langle \Ddd\Phibar ,R(\Da\Phi
     ,\Dgd\Phibar)\Db\Phi \rangle \Bigr\} + \frac 12\langle \DDbar^2\Phibar
     ,\DD^2\Phi\rangle.\endsplit \hskip-30pt\tag{5.26} $$
The second term in~\thetag{5.25} is 
  $$ \split
      \eab\egdd\DDgd\Da\cont{\partial _{b \dd}\Phi }\pK &= \eab\egdd \Bigl\{
     -\DDa\Dgd\cont{\partial _{b \dd}\Phi }\pK -2\partial _{a
     \gd}\cont{\partial _{b \dd}\Phi }\pK \Bigr\} \\
      &= \eab\egdd\Bigl\{ -\frac 12 \Da\langle \Dgd\Phibar , \partial _{b
     \dd}\Phi\rangle - 2\partial _{a \gd}\cont{\partial _{b \dd}\Phi }\pK
     \Bigr\} \\
      &= \eab\egdd\Bigl\{ \frac 12 \langle \Dgd\Phibar ,\nabla \mstrut
     _{\!\!b \dd}\Da\Phi \rangle + \langle \partial _{a \gd}\Phibar ,\partial
     _{b \dd}\Phi\rangle \\
      &\qquad \qquad \qquad \qquad \qquad \qquad \qquad - 2\partial _{a
     \gd}\cont{\partial _{b \dd}\Phi }\pK \Bigr\}.\endsplit \tag{5.27} $$
Combining~\thetag{5.25}--\thetag{5.27} we have 
  $$ \multline
      \frac 12\bigl(\DDbar^2\DD^2 + 2\eab\egdd\partial _{a \gd}\cont{\partial
     _{b \dd}\Phi }\pK \bigr)(\frac 12\Phi ^*K) \\ =\eab\egdd\Bigl\{ \frac
     14\langle \partial _{b \dd}\Phibar,\partial _{a \gd}\Phi \rangle
      + \frac 14\langle \Ddd\Phibar,\nabla\mstrut _{\!\!a \gd}\Db\Phi 
     \rangle \\
      - \frac 1{32}\langle \Ddd\Phibar,R(\Da\Phi ,\Dgd\Phibar)\Db\Phi 
     \rangle\Bigr\} + \frac{1}{8}\langle \DDbar^2\Phibar,\DD^2\Phi 
     \rangle.\endmultline \tag{5.28} $$
Now we add~\thetag{5.28} to its complex conjugate.  Since
  $$ 2\eab\egdd\partial _{a \gd}\cont{\partial
     _{b \dd}\Phi }(\pK + \dbar K) = 2\eab\egdd\partial _{a
     \gd}\partial _{b \dd}(\Phi ^*K) = 4\wave(\Phi ^*K), \tag{5.29} $$
the left hand side restricted to~$\bmink^4$ is the component lagrangian (as
in the second line of~\thetag{5.12}), and so finally
  $$ \boxed{ \scrLch_0 =\langle \overline{d\phi},d\phi \rangle + \langle\psib
     \Diracp\psi\rangle - \frac 14 \eab\egdd\langle \pdd,\pR(\pa,\pgd)\pb
     \rangle + \langle \Fbar,F \rangle.}\tag{5.30} $$
Recall that we have absorbed an exact term~\thetag{5.14} in writing the Dirac
form.  With this understood, the lagrangian~\thetag{5.30} agrees
with~\thetag{3.48} after eliminating the auxiliary field.

We check in~\FP{10(c)} that the lagrangian~\thetag{5.17} on~$\mink44$ reduces
to~\thetag{4.16} in~$\mink32$.

A new aspect in four dimensions is the $U(1)$ R-symmetry~\thetag{2.81}.  We
claim that the $\sigma $-model lagrangian is invariant under~$R$.  First, in
the superspace expression~\thetag{5.17} (or~\thetag{5.22}) we note that
$d^4\theta =d^2\theta \,d^2\bar{\theta }$ is invariant.  So if $R$~acts
trivially on~$\Phi $, then $\scrL_0$ is obviously invariant.  When $R$~acts
nontrivially it is easier to argue in components.  Suppose for simplicity
that $\Phi $~takes values in~$\CC$, and let $R$~acts on~$\Phi $ with
weight~$k$.  Then
  $$ \boxed{\alignedat2
      R\phi &= k\phi, &\qquad \qquad R\phib &= -k\phib \\
      R\psi &=(k+1)\psi, &\qquad \qquad R\psib&=-(k+1)\psib \\
      RF &= (k+1)F, &\qquad \qquad R\Fb &=
     -(k+2)\Fb.\endalignedat} \tag{5.31} $$
Direct inspection of~\thetag{5.30} shows that $\scrLch_0$~is $R$-invariant.

\Head{5.3}{The superpotential term on~$\mink44$}
As a preliminary, note that if $\Lambda $~is any chiral superfield, then
$|d^4x|\int_{}d^2\theta \,\Lambda $~makes sense as a density on Minkowski
space.  Similarly, for $\overline{\Lambda }$~antichiral we can
define~$|d^4x|\int_{}d^2\bar\theta \,\overline{\Lambda }$.  One view is that
$\mink44$~is a split cs-manifold with a complex two dimensional odd tangent
bundle.  In fact, there are two such cs-manifolds, one for each of the
complex structures on the real four dimensional spin representation
(see~\thetag{2.2}).  These cs-manifolds have canonical
densities---denoted~$|d^4x|\,d^2\theta $ (resp.~$|d^4x|\,d^2\bar\theta
$)---and global functions on these cs-manifolds are chiral
(resp.~antichiral) superfields.  To compute these chiral integrals we
use~\thetag{2.40}.
 
The distinction between terms in the action of the form
$\int_{}|d^4x|\,d^4\theta $ and $\int_{}|d^4x|\,d^2\theta $ is important
in~$\mink44$ and also in~$\mink2{(2,2)}$.
 
We introduce a superpotential term, and we may as well work in the nonlinear
$\sigma $-model.  Let $W\:X\to\CC$ be a {\it holomorphic\/} function, and set
  $$ \boxed{\scrL_1 = |d^4x|\,\Re\Bigl\{d^2\theta \; \Phi ^*(W)\Bigr\}.}
     \tag{5.32} $$
It lives in the direct sum of the spaces of chiral and antichiral densities.
The superpotential term in the component action is the integral of~$\scrL_1$:
  $$ \int_{}\scrL_1 = \int_{}|d^4x|\;\frac 12\left\{ \int_{}d^2\theta \;\Phi
     ^*(W) \;+\; \int_{}d^2\bar\theta \;\Phi ^*(\overline W) \right\} =
     \int_{}|d^4x|\;\scrLch_1. \tag{5.33} $$
We compute~$\scrLch_1$ exactly as in~\thetag{4.21}; recalling our
normalizations~\thetag{5.7} for the component fields we obtain 
  $$ \scrLch_1 = \cont{F}\phi ^*\partial W + \cont{\Fb}\phi
     ^*\overline{\partial W} - \frac 12\eab\phi ^*\nabla \partial W(\pa,\pb)
     - \frac 12\eabd\phi ^*\nabla\overline{ \partial
     W}(\pad,\pbd). \tag{5.34} $$
We eliminate the auxiliary fields~$F,\Fb$ from~$\scrLch_0+\scrLch_1$ using
their equations of motion 
  $$ \boxed{\aligned
      F &= -\phi ^*\grad W \\ 
      \Fb &= -\phi ^*\overline{\grad W},\endaligned} \tag{5.35} $$
where now $\overline{\grad W}$~is the vector field of type~$(0,1)$
corresponding to the $(1,0)$-form~$\partial W$ using the hermitian metric
on~$X$.  The total lagrangian with auxiliary fields eliminated is then 
  $$ \boxed{
      \aligned
      \scrLch_0+\scrLch_1 &\sim \langle \overline{d\phi },d\phi \rangle+
     \langle\psib \Diracp\psi\rangle - \frac 14 \eab\egdd\langle
     \pdd,\pR(\pa,\pgd)\pb \rangle\\
      & \qquad -\phi ^*\|\grad W\|^2 - \Bigl( \frac 12\eab\phi
     ^*\nabla \partial W(\pa,\pb) - \frac 12\eabd\phi ^*\nabla\overline{
     \partial W}(\pad,\pbd) \Bigr) .\endaligned} \tag{5.36} $$
This is the lagrangian~\thetag{3.85}, and since the superspace model is
manifestly supersymmetric we have proved \theprotag{3.82(ii)} {Theorem}.  Of
course, the potential energy in~\thetag{5.36} is
  $$ V=\|\grad W\|^2. \tag{5.37} $$
Other aspects of the discussion following~\thetag{4.23} carry over directly. 
 
Finally, we consider the R-symmetry~\thetag{2.81}.  We already noted that it
leaves~$\scrL_0$ and $\scrLch_0$~invariant, even if $\Phi $~is assigned a
nonzero weight.  On the other hand, $d^2\theta $~has weight~$2$ and so if
$\scrL_1$~is to be invariant we must be able to assign~$\Phi $ a weight so
that $\Phi ^*(W)$~has weight~$-2$.  (In general, there are several~$\Phi _i$
with different weights.)  Such a~$W$ is termed {\it quasi-homogeneous\/} in
the physics literature.

\Head{5.4}{Analysis of the classical theory}
We do not give details, having done so exhaustively in~\S{4} for the $\sigma
$-model on~$\mink32$.  Rather, we simply record the formulas 
  $$ \boxed{\gch=\cont{\partial _{c\dd}}\,|d^4x|\;\frac{1}{\sqrt2}\epsilon
     ^{ac}\epsilon ^{\bd\dd}\bigl[ \langle \dabd\phi ,\delta \phib  \rangle +
     \langle \dabd\phib,\delta \phi   \rangle + \langle\pa,\delta \pbd
     \rangle + \langle \pbd,\delta \pa  \rangle\bigr]} \tag{5.38} $$
for the variational 1-form,
  $$ \boxed{\aligned\wch=\cont{\partial
     _{c\dd}}\,|d^4x|\;\frac{1}{\sqrt2}&\epsilon ^{ac}\epsilon
     ^{\bd\dd}\bigl[ \langle \dabd\delta \phi \wedge \delta \phib \rangle +
     \langle \dabd\delta \phib\wedge \delta \phi \rangle \\
      & \qquad \qquad \qquad + \langle\delta \pa\wedge \delta \pbd \rangle +
     \langle \delta \pbd\wedge \delta \pa \rangle\bigr]\endaligned}
     \tag{5.39} $$
for the local symplectic form, and 
  $$ \boxed{j_a=\cont{\partial _{c\dd}}\,|d^4x|\;\Bigl\{\sqrt2\epsilon
     ^{cb}\epsilon ^{\dd\edd} \langle \partial_{a\edd}\phi , \pb \rangle +
     \sqrt2\,\epsilon ^{\bd\dd}\delta ^c_a\langle \phi ^*\grad W,\pbd
     \rangle\Bigr\}} \tag{5.40} $$
for the supercurrent corresponding to the supersymmetry transformation~$\Qa$.

\Chapter6{Supersymmetric Yang-Mills Theories}

 \comment
 lasteqno 6@ 79
 \endcomment

 Now we consider theories with gauge fields (connections).  In the bosonic
case there is a Yang-Mills theory in any dimension.  The supersymmetric
extension involves a {\it dual\/} spinor field as well, and it exists in
dimensions~$n=3,4,6,10$ with minimal supersymmetry.  Our treatment in~\S{6.1}
is based on a special property of spinors which holds only in these cases:
the quadratic form $s^*\mapsto \Gamma (s^*,s^*)$ on~$S^*$ takes values on the
light cone.  In~\S{6.2} we describe the supersymmetric extensions of the most
general bosonic theory, which is variously called a {\it gauge theory with
matter\/} or a {\it gauged $\sigma $-model\/}.  We consider theories in
dimensions~$n=3,4,6$ with minimal supersymmetry.  (The only theory in~$n=10$
is pure Yang-Mills.)  The basic lagrangians, supersymmetry transformations,
and supercurrents are summarized in {\theprotag{6.33} {Theorem}}.  There are
new terms which appear neither in the pure $\sigma $-model nor in the pure
Yang-Mills theory.  There is a superspace formulation of supersymmetric gauge
theories in dimensions~$n=3,4,6$ (with minimal supersymmetry); we describe
some common features in~\S{6.3}.  The condition on the quadratic form has a
geometric significance here as well.  Details of the superspace formulation
occupy the remaining chapters.

\Head{6.1}{The minimal theory in components}
The bosonic Yang-Mills theory is determined by the data: 
  $$ \alignedat2
      &G\qquad &&\text{compact Lie group} \\
      &\langle \cdot ,\cdot \rangle\qquad &&\text{bi-invariant scalar product
     on~$\frak{g}$} \endalignedat \tag{6.1} $$
The lagrangian in $n$-dimensional Minkowski space is 
  $$ L = -\frac 12|F_A|^2 \;|d^{n}x|, \tag{6.2} $$
where $A$~is a connection on a principal $G$~bundle~$P$.  The minimal
supersymmetric extension has a single dual spinor field~$\lambda $ with
values in the adjoint bundle and lagrangian
  $$ \boxed{L = \{ -\frac 12|F_A|^2 + \frac 12\langle\lambda
     \Dirac_A\lambda\rangle \}\;|d^{n}x|.} \tag{6.3} $$
We ask: When is \thetag{6.3}~supersymmetric?  That is, in which dimensions
and for what type of spinor field~$\lambda $ is \thetag{6.3}~supersymmetric?
Consider first the abelian theory, which is free.  Suppose the spinor field
has $s$~components.  Upon quantization,\footnote{The free quantum theory is
only used here to rule out some cases; otherwise the argument is entirely
classical.} the gauge field~$A$ has $n-2$~physical degrees of freedom and the
spinor field has $s/2$~physical degrees of freedom.  With one exception
(explained presently) the number of bosonic and fermionic degrees of freedom
in a supersymmetric theory must be equal, from which $n=2+s/2$.  This is
satisfied only for $n=3,4,6,10$ with minimal supersymmetry $s=2,4,8,16$.

        \proclaim{\protag{6.4} {Theorem}}
 The super Yang-Mills lagrangian~\thetag{6.3} is supersymmetric in
$n=3,4,6,10$ dimensions with minimal supersymmetry. 
        \endproclaim

\flushpar
 There is an additional supersymmetric theory---the exception referred to
above---in $n=2$~dimensions with $s=(1,0)$~supersymmetry; we will not
consider it.  The quantum argument above shows that \thetag{6.3}~cannot be
supersymmetric in other dimensions and with other amounts of supersymmetry.

        \demo{Proof}
 We work in $n$-dimensional Minkowski space and fix a real spin
representation~$S$.  Thus we have pairings~\thetag{1.1}, \thetag{1.2} which
satisfy the Clifford relation~\thetag{1.3}.  The spinor field~$\lambda $
takes values in ~$S^*\otimes \adP$.  The lagrangian~\thetag{6.3} is
  $$ \boxed{L = \bigl(-\frac 14g^{\mu \sigma }g^{\nu \rho }\langle F_{\mu \nu
     },F_{\sigma \rho} \rangle + \frac 12\sdtov ^\sigma _{a b }\langle
     \lambda ^a ,\nabla _\sigma \lambda ^b \rangle \bigr)\;|d^nx|,}
      \tag{6.5} $$
where $\langle \cdot ,\cdot \rangle$~is the invariant inner product on the
gauge algebra, the covariant derivative is~$\nabla_\sigma =\partial _\sigma +
A_\sigma $, and the curvature is  
  $$ F_A = \frac 12F_{\mu \nu }\,dx^\mu \wedge dx^\nu , \tag{6.6} $$
where
  $$ F_{\mu \nu } = \partial _\mu A_\nu - \partial _\nu A_\mu + [A_\mu ,A_\nu
     ]. \tag{6.7} $$
We work with {\it real\/} bases for spinors and vectors.  We postulate the
induced action of the supersymmetry transformation~$\eta^a Q_a $ on the
fields to be
  $$ \boxed{\aligned
      (\hxi A)_\mu &= \eta^a g_{\mu \nu }\sdtov ^\nu _{a  b  }\lambda
     ^b  \\
      \hxi \lambda ^a  &= \frac 12\eta^b\stov^{\mu a  c  }\sdtov
     ^\nu _{c  b  }F_{\mu \nu }.\endaligned } \tag{6.8} $$
Each side of the first equation is a component of a 1-form with values in the
adjoint bundle.  These equations are motivated by Lorentz invariance, gauge
invariance, parity, and power counting.

There are two things we need to check: (i)\;$\hxi$~is a (nonmanifest)
symmetry of~$L$; and (ii)\ the Lie bracket of transformations~\thetag{6.8}
is consistent with~\thetag{1.30}.  We will find the conditions under which
(i)~and (ii)~hold.   
 
First, we investigate~$\Lie{\hxi}L$.  The variation of the curvature is 
  $$ \boxed{\hxi F_{\mu \nu } = \eta^a g_{\nu \sigma }\sdtov ^\sigma _{a b
     }\nabla_\mu \lambda ^b - \eta^a g_{\mu \sigma }\sdtov ^\sigma _{a b
     }\nabla_\nu \lambda ^b ,} \tag{6.9} $$
from which we easily compute 
  $$ \hxi\bigl(-\frac 14 g^{\mu \sigma }g^{\nu \rho }\langle F_{\mu \nu
     },F_{\sigma \rho } \rangle \bigr) = -\eta^a g^{\mu \sigma }\sdtov ^\nu
     _{a  b  }\langle F_{\mu \nu} ,\nabla_\sigma \lambda ^b 
     \rangle. \ \tag{6.10} $$
In the variation of the second term in~\thetag{6.5} we must remember to
include the variation of the covariant derivative~$\nabla_\sigma =\partial
_\sigma + A_\sigma $:  
  $$ \split
      \hxi\bigl(\frac 12 \sdtov ^\sigma _{a b }\langle \lambda ^a
     ,\nabla_\sigma \lambda ^b \rangle \bigr) &= \frac 12\sdtov ^\sigma _{a b
     }\langle \eta ^c \stov^{\mu a d }\sdtov ^\nu _{d c }F_{\mu \nu
     },\nabla_\sigma \lambda ^b \rangle + \frac 12\sdtov ^\sigma _{a b
     }\langle \lambda ^a ,\eta ^c g_{\sigma \nu }\sdtov ^\nu _{c d }[\lambda
     ^d ,\lambda ^b ] \rangle \\
      &\qquad \qquad \qquad \qquad \qquad \qquad \qquad \qquad \qquad \qquad
     \qquad + EXACT\\
      &= \frac 12\eta^a\sdtov ^\sigma _{b d }\stov^{\mu c d }\sdtov ^\nu _{a
     c }\langle F_{\mu \nu },\nabla_\sigma \lambda ^b \rangle - \frac 12 \eta
     ^c g_{\sigma \nu }\sdtov ^\sigma _{a b }\sdtov ^\nu _{c d }\langle
     \lambda ^a ,[\lambda ^d ,\lambda ^b ] \rangle \\
      &\qquad \qquad \qquad \qquad \qquad \qquad \qquad \qquad \qquad \qquad
     \qquad + EXACT\\
      &= \eta^a g^{\mu \sigma }\sdtov ^\nu _{a b }\langle F_{\mu \nu}
     ,\nabla_\sigma \lambda ^b \rangle - \frac 12 \eta ^c g_{\sigma \nu
     }\sdtov ^\sigma _{a b }\sdtov ^\nu _{c d }\langle \lambda ^a ,[\lambda
     ^d ,\lambda ^b ] \rangle \\ 
      &\qquad \qquad \qquad \qquad \qquad \qquad \qquad \qquad \qquad \qquad
     \qquad + EXACT,\endsplit
       \tag{6.11} $$
where  
  $$ EXACT = \partial _\sigma \bigl(\frac 14 \eta ^a \stov^\sigma
     _{bd}\sdtov^{\mu cd}\stov^\nu _{ac}\langle F_{\mu \nu },\lambda ^b
     \rangle\bigr). \tag{6.12} $$
In the last step we use the Clifford relation $\sdtov ^\mu _{b  d 
}\stov^{\sigma c  d  } + \sdtov ^\sigma _{b  d  }\stov^{\mu c 
d  } = 2g^{\mu \sigma }\delta ^c  _b  $ and the Bianchi identity
$\nabla_\sigma F_{\mu \nu } + \nabla_\nu F_{\sigma \mu } + \nabla_\mu F_{\nu \sigma } = 0.$
Now the first term in the last line cancels against~\thetag{6.10}, and the
second term vanishes if and only if the quartic form
  $$ Q_{a  b  c  d  } = g_{\sigma \nu }\bigl(\sdtov ^\sigma
     _{a  b  }\sdtov ^\nu _{c  d  } +  \sdtov ^\sigma
     _{a  d  }\sdtov ^\nu _{b  c  } +  \sdtov ^\sigma
     _{a  c   }\sdtov ^\nu _{d  b  }\bigr) \tag{6.13} $$
vanishes.  There are two conclusions from this computation.  First, if the
gauge algebra is abelian, then $\hxi$~is a symmetry of~$L$ in any dimension
and with any spin representation~$S$.  Second, if the gauge algebra is
nonabelian, then $\hxi$~is a symmetry of~$L$ if and only if the quartic form
  $$ Q\:s^*\longmapsto |\,\sdtov (s^*,s^*)\,|^2,\qquad s^*\in S^*, \
     \tag{6.14} $$
vanishes identically.  This holds for $n=3,4,6,10$ as in those dimensions the
spin group acts transitively on the nonzero spinors.  For $n=3,4,6$ the spin
representation is~$SL(2,\FF)$ acting on~$\FF^2$ for~$\FF=\RR,\CC,\HH$.
For~$n=10$ octonions can be similarly used~\DS{6.5--\S6.7}.

Next, we investigate the Lie algebra generated by the symmetry
transformation~\thetag{6.8}.  Let $\eta^a_1,\eta^b_2$ be odd parameters and
$\hxi_i$ the action~\thetag{6.8} of~$\eta^a_iQ_a  $.  Recall
from~\thetag{1.30} that for any field~$f$ we need to check
  $$ [\hxi_1,\hxi_2]f = 2\eta^a_1\eta^b_2\sdtov ^\sigma _{a  b 
     }\partial _\sigma f \ \tag{6.15} $$
When acting on the spinor field~$f=\lambda $, we replace~$\partial _\sigma $
with the covariant derivative~$\nabla_\sigma $.  When acting on the
connection~$f=A$, we use Lie derivative by the horizontal lift, and it is easy
to compute that the action of~$\partial _\sigma $ on~$A$ has $\mu
^{\text{th}}$~component equal to~$F_{\sigma \mu }$.

We first verify~\thetag{6.15} for the connection field.  Using~\thetag{6.8}
we find:
  $$ \split
      \bigl([\hxi_1,\hxi_2]A \bigr)_\mu &= -\frac 12\eta^a_1\eta^b_2g_{\mu \nu
     }(\stov^{\rho c  d  }\sdtov ^\nu _{b  c  }\sdtov ^\sigma
     _{a  d  } + \stov^{\rho c  d  }\sdtov ^\nu _{a  c 
     }\sdtov ^\sigma _{b  d  }) F_{\rho \sigma } \\
      &= -\frac 12\eta^a_1\eta^b_2g_{\mu \nu }(4g^{\rho \nu }\sdtov ^\sigma
     _{a  b  } ) F_{\rho \sigma} \\
      &=2\eta^a_1\eta^b_2\sdtov ^\sigma _{a  b  }F_{\sigma \mu }.\endsplit
       \tag{6.16} $$
To pass to the second line we added to the first line terms in parentheses
with $\rho $~and $\nu $~exchanged; then we used the Clifford
identity~\thetag{1.3}.  This is the desired result.  To check the commutation
relation~\thetag{6.15} on the spinor field we need to also impose the
equation of motion
  $$ \sdtov ^\sigma _{a  b  }\nabla_\sigma \lambda ^a  =0,\qquad
     \text{for all $b  $}; \ \tag{6.17} $$
that is, we only verify the supersymmetry algebra {\it on-shell\/}.  Then we
compute from~\thetag{6.8}
  $$ \split
      [\hxi_a,\hxi_2]\lambda ^c  &=-\frac 12\eta^a_1\eta^b_2g_{\mu \nu
     }(\stov^{\sigma c  d  }\sdtov ^\nu _{d  b  }\sdtov ^\mu
     _{a  e  } + \stov^{\sigma c  d  }\sdtov ^\nu _{d 
     a  }\sdtov ^\mu _{b  e  } \\ 
      &\qquad \qquad \qquad \qquad \qquad - \stov^{\nu c  d  }\sdtov
     ^\sigma _{d  b  }\sdtov ^\mu _{a  e  } - \stov^{\nu
     c  d  }\sdtov ^\sigma _{d  a  }\sdtov ^\mu _{b 
     e  } )\nabla_\sigma \lambda ^e  \\
      &= \frac 12\eta^a_1\eta^b_2g_{\mu \nu }\stov^{\sigma c  d  }\sdtov ^\nu
     _{d  e  }\sdtov ^\mu _{a  b  }\nabla_\sigma \lambda
     ^e  - \frac 12\eta^a_1\eta^b_2\stov^{\sigma c  d  }Q_{a  b 
     d  e  }\nabla_\sigma \lambda ^e  \\
       &\qquad \qquad \qquad + \frac
     12\eta^a_1\eta^b_2g_{\mu \nu} (\stov^{\nu c  d  }\sdtov ^\sigma _{d 
     b  }\sdtov ^\mu _{a  e  } + \stov^{\nu c  d  }\sdtov
     ^\sigma _{d  a  }\sdtov ^\mu _{b  e  } )\nabla_\sigma
     \lambda ^e  .\endsplit \hskip-50pt\tag{6.18} $$
Let `$LT$'~denote the last term.  Applying the Clifford identity we see that
the symmetric part in~$\nu ,\sigma $ vanishes by the equation of motion, whence
  $$ \split
      LT &= \frac 12LT - \frac 14\eta^a_1\eta^b_2g_{\mu \nu }(\stov^{\sigma c 
     d  }\sdtov ^\nu _{d  b  }\sdtov ^\mu _{a  e  } +
     \stov^{\sigma c  d  }\sdtov ^\nu _{d  a  }\sdtov ^\mu
     _{b  e  } )\nabla_\sigma \lambda ^e  \\
      &= \frac 12LT +\frac 14\eta^a_1\eta^b_2g_{\mu \nu }\stov^{\sigma c  d 
     }\sdtov ^\nu _{d  e  }\sdtov ^\mu _{a  b  }\nabla_\sigma
     \lambda ^e  - \frac 14\eta^a_1\eta^b_2\stov^{\sigma c  d 
     }Q_{a  b  d  e  }\nabla_\sigma \lambda ^e  .\endsplit
       \tag{6.19} $$
Solving for~$LT$ and plugging into~\thetag{6.18} we have 
  $$ \split
      [\hxi_1,\hxi_2]\lambda ^c  &= \eta^a_1\eta^b_2g_{\mu \nu }\stov^{\sigma
     c  d  }\sdtov ^\nu _{d  e  }\sdtov ^\mu _{a  b 
     }\nabla_\sigma \lambda ^e  -\eta^a_1\eta^b_2\stov^{\sigma c  d 
     }Q_{a  b  d  e  }\nabla_\sigma \lambda ^e  \\
      &= 2\eta^a_1\eta^b_2\sdtov ^\sigma _{a  b  }\nabla_\sigma \lambda
     ^c -\eta^a_1\eta^b_2\stov^{\sigma c  d  }Q_{a  b  d 
     e  }\nabla_\sigma \lambda ^e  . \endsplit \tag{6.20} $$
Here we use the equation of motion~\thetag{6.17} to see that the last term in
the first line vanishes when $\sigma $~and $\nu $~are exchanged.  If the
second term vanishes, then this is the desired bracket.  Hence the
supersymmetry algebra is satisfied on-shell if and only if the quartic
form~\thetag{6.14} vanishes.  This applies both to the abelian and nonabelian
theories.
 
This completes the proof that \thetag{6.3}~is supersymmetric.
        \enddemo

\flushpar 
 We give another interpretation of the quartic form below~\thetag{6.69}.

We now compute the classical equations of motion and the variational 1-form.
First, the variation of the curvature~\thetag{6.7} is 
  $$ \delta F_{\mu \nu } = \nabla _\mu \delta A_\nu -\nabla _\nu \delta A_\mu
     . \tag{6.21} $$
Now write the lagrangian~\thetag{6.5} as~$L = \Lch\;\bdens$.  Then a short
computation yields 
  $$ \multline
      \delta \Lch = \langle \delta A_\rho\,,\,g^{\mu \sigma }g^{\nu \rho
     }\nabla _\sigma F_{\mu \nu } - \frac 12\sdtov^\rho _{ab}[\lambda
     ^a,\lambda ^b] \rangle + \langle \delta \lambda ^a\,,\,\sdtov^\sigma
     _{ab}\nabla _\sigma \lambda ^b \rangle \\
      + \partial _\sigma \Bigl\{ -g^{\mu \sigma }g^{\nu \rho }\langle F_{\mu
     \nu },\delta A_\rho \rangle + \frac 12 \sdtov^\sigma _{ab}\langle
     \lambda ^a,\delta \lambda ^b \rangle \Bigr\}.\endmultline \tag{6.22} $$
This leads to the equations of motion 
  $$ \boxed{\aligned
      g^{\mu \sigma }g^{\nu \rho }\nabla _\sigma F_{\mu \nu } &= \frac
     12\sdtov^\rho _{ab}[\lambda ^a,\lambda ^b] \\
      \sdtov^\sigma _{ab}\nabla _\sigma \lambda ^b&=0\endaligned
     }\tag{6.23} $$
and the variational 1-form
  $$ \boxed{\gamma =\bigl(-g^{\mu \sigma }g^{\nu \rho }\langle F_{\mu \nu
     },\delta A_\rho \rangle + \frac 12 \sdtov^\sigma _{ab}\langle \lambda
     ^a,\delta \lambda ^b \rangle \bigr)\;\frac{\partial }{\partial x^\sigma
     }\otimes \bdens.} \tag{6.24} $$
In coordinate-free notation we write the equations of motion~\thetag{6.23} as
  $$ \boxed{\aligned
      -d_A*F_A &= \frac 12[\lambda \,\sdtov\,\lambda ]\;\bdens \\ 
      \Dirac_A\lambda &=0.\endaligned} \tag{6.25} $$

The Noether current associated to~$\hxi$ is computed from~\thetag{6.12}
and~\thetag{6.24} as
  $$ \cont{\hxi}\gamma - \bigl(\frac 14 \eta ^a \sdtov^\sigma _{bd}\stov^{\mu
     cd}\sdtov^\nu _{ac}\langle F_{\mu \nu },\lambda ^b
     \rangle\;\frac{\partial }{\partial x^\sigma }\otimes \bdens\bigr)
      = -\eta ^ag^{\mu \sigma }\sdtov^\nu _{ab}\langle F_{\mu \nu },\lambda
     ^b \rangle\;\frac{\partial }{\partial x^\sigma }\otimes
     \bdens. \tag{6.26} $$
The {\it supercurrent\/}~$j_a$ is {\it minus\/} the Noether current
for~$Q_a$: 
  $$ \boxed{j_a = g^{\mu \sigma }\sdtov^\nu _{ab}\langle F_{\mu \nu
     },\lambda ^b  \rangle \;\frac{\partial }{\partial x^\sigma }\otimes
     \bdens.} \tag{6.27} $$

\Head{6.2}{Gauge theories with matter}  
 The most general bosonic model without gravity\footnote{We include only
scalar fields and gauge fields.  There are also models with $p$-form fields
for~$p\ge2$, for example.  The models discussed here cover most {\it
fundamental\/} (vs.~{\it effective\/}) lagrangians without gravity.} on
Minkowski space~$\bmink^n$ is specified by the following data:
  $$ \alignedat2
      &G\qquad &&\text{compact Lie group} \\
      &\langle \cdot ,\cdot \rangle\qquad &&\text{bi-invariant scalar product
     on~$\frak{g}$} \\
      &X\qquad &&\text{Riemannian manifold on which $G$~acts by isometries}
     \\
      &V\:X\longrightarrow \RR \qquad && \text{potential function invariant
     under $G$}\endalignedat  \tag{6.28} $$
The fields in the model are
  $$ \alignedat2
      &A\qquad &&\text{connection on some principal $G$ bundle
     $P\longrightarrow \bmink^n$} \\
      &\phi &&\text{section of the associated bundle $X^P=P\times
     _GX\longrightarrow \bmink^n$}\endaligned \tag{6.29} $$
The standard fundamental lagrangian is then the ``minimally coupled'' $\sigma
$-model lagrangian~\thetag{3.77} and Yang-Mills lagrangian~\thetag{6.2}: 
  $$ L = \Bigl\{ -\frac 12 |F_A|^2 + \frac 12|d_A\phi |^2 - \phi
     ^*V\Bigr\}\;\bdens. \tag{6.30} $$
The moduli space of classical vacua on~$\bmink^n$ is the quotient of the set
of minima of~$V$---which we assume occurs at~$V=0$---by the action of
constant gauge transformations: 
  $$ \Mvac=V\inv (0)/G. \tag{6.31} $$
 
Now we consider supersymmetric extensions of~\thetag{6.30}.  As in~\S{3} we
work with models in 3, 4, and 6~dimensions with minimal supersymmetry.  As
in~\S{3} we refer to the model in $3,4,6$~dimensions as the
$\FF=\RR,\CC,\HH$~model.  In these dimensions we have already constructed
separately the supersymmetric $\sigma $-model and the supersymmetric gauge
theory.  The next theorem summarizes what happens when we combine them.  In
these models the bosons are given in~\thetag{6.29} and the fermions are
  $$ \alignedat2
      &\psi \qquad && \text{section of $\phi ^*\bigl(P\times _G(TX\otimes
     _{\FF}\Pi \FF^2) \bigr)$}\\
      &\lambda &&\text{section of~$P\times _G(\frak{g}\otimes \Pi{\FF^2}^*)$
     }\endaligned \tag{6.32} $$
In the $\FF=\HH$~case we use complex notation for~$\psi $ as
in~\thetag{3.49}.

A discussion---including an explanation of notation---follows the statement
of the theorem.

        \proclaim{\protag{6.33} {Theorem}}
 Fix
  $$ \alignedat2
      &G\qquad &&\text{compact Lie group} \\
      &\langle \cdot ,\cdot \rangle\qquad &&\text{bi-invariant scalar product
     on~$\frak{g}$} \endalignedat \tag{6.34} $$
\smallskip
 \rom(i\rom)\ \rom($\FF=\RR$\rom)\ Suppose we are given
  $$ \alignedat2
      &X\qquad &&\text{Riemannian manifold with $G$~action by isometries} \\
      &h\:X\longrightarrow \RR\qquad &&\text{$G$-invariant real-valued
     function} \endalignedat \tag{6.35} $$
Then the lagrangian 
  $$ \boxed{\aligned
       L = \Bigl\{ -\frac 12&|F_A|^2 + \frac 12|d_A\phi |^2 + \frac
     12\langle\lambda \Dirac_A\lambda\rangle + \frac 12\langle \psi
     (\Dirac_A)_{\!\phi}\, \psi \rangle - \frac 12\phi ^*|\grad h|^2 \\
      &- \frac 12\eab\phi ^*(\Hess h)(\pa,\pb)+ \langle \lhat^a_\phi ,\psi_a
     \rangle\\
      & \qquad \qquad + \frac 1{12}\epsilon ^{ab}\epsilon
     ^{cd}\langle \pa,\pR(\pb,\pg)\pd \rangle \Bigr\}\;|d^{3}x|.\endaligned}
     \tag{6.36} $$
on~$\bmink^3$ is supersymmetric with the supersymmetry transformation laws 
  $$ \boxed{\aligned
      \hxi\phi &= - \eta^a\,\pa \\
      (\hxi A)_{a b } &= \eta ^c (\epsilon _{a c}\epsilon _{bd} + \epsilon
     _{ad}\epsilon _{bc})\lambda ^d\\
      \hxi \pa &= \hphantom{-}\eta ^b \bigl((d_{A})_{ab}\phi + \epsilon _{a
     b }\phi ^*\grad h\bigr) \\
      \hxi \lambda ^a &= -\eta ^b\epsilon ^{ac} f_{bc }\endaligned}
     \tag{6.37} $$
The potential energy is 
  $$ \boxed{V = \frac 12|\grad h|^2} \tag{6.38} $$
and the moduli space of vacua is 
  $$ \boxed{\Mvac = \Crit(h)/G.} \tag{6.39} $$
The supercurrent~$j_a$, which is minus the Noether current for~$\Qa$, is 
  $$ \boxed{\aligned
      j_a = \Bigl\{ \epsilon ^{cb}\epsilon ^{de}\langle \partial _{ae}\phi
     ,\psi _b \rangle + \epsilon ^{bd} &\delta ^c_a\langle \phi ^*\grad h,\pb
     \rangle + \frac 12\epsilon ^{bd}\langle f_{ba},\lambda ^c
     \rangle\Bigr\}\,\frac{\partial }{\partial y^{cd}}\otimes |d^3x| \\
      &+\frac 12\epsilon ^{cb}\langle f_{cd},\lambda ^d
     \rangle\,\frac{\partial }{\partial y^{ab}}\otimes |d^3x| \endaligned}
     \tag{6.40} $$
\smallskip
 \rom(ii\rom)\ \rom($\FF=\CC$\rom)\ Suppose we are given
  $$ \alignedat2
      &X\qquad &&\text{\Kah\ manifold with $G$~action} \\
      &\mu \:X\longrightarrow \frak{g}^*\qquad &&\text{moment map for
     $G$~action} \\
      &W\:X\longrightarrow \CC\qquad &&\text{$G$-invariant holomorphic
     function} \endalignedat \tag{6.41} $$
Then the lagrangian
  $$ \boxed{\aligned
       L = \Bigl\{ -\frac 12&|F_A|^2 + \langle d_A\phib,d_A\phi \rangle +
     \langle\bar\lambda \Dirac_A\lambda\rangle + \langle
     \psib(\Dirac_A)_{\!\phi}\, \psi \rangle \\
      &-\phi ^*\|\grad W\|^2-2\phi ^*|\mu |^2 - \Re\bigl[\eab\phi ^*(\Hess
     W)(\pa,\pb)\bigr] \\
      & \qquad + \sqrt2\,\bigl(\langle \lhat^a_\phi ,\psi_a \rangle -
      \langle \pad ,\hat{\lambdab}^{\ad}_\phi \rangle\bigr) - \frac
     14\epsilon ^{ac}\epsilon ^{\bd\dd}\langle \pa,\pR(\pg,\pdd)\pbd
     \rangle\Bigr\}\;|d^{4}x|.\endaligned} \tag{6.42} $$
on~$\bmink^4$ is supersymmetric with the supersymmetry transformation laws
  $$ \boxed{\aligned
      \hxi \phi &= -\sqrt2\,\eta ^a \pa \\
      (\hxi A)_{a \bd } &= 2\eta ^c \,\epsilon _{a c }\epsilon
     _{\bd\dd}\,\lambdab^{\dd} + 2\etgd\,\epsilon _{ad}\epsilon
     _{\bd\gd}\,\lambda ^{d }\\
      \hxi \pa &= \sqrt2\,\bigl[\etabar^{\bd}(d_{A})_{a\bd}\phi
     + \eta ^b \epsilon _{a b }\phi ^*\grad W\bigr] \\
      \hxi \lambda ^a &= -\eta ^b \epsilon ^{ac}F_{ bc} -2 \eta ^a\sqo\phi
     ^*\chmu \endaligned } \tag{6.43} $$
The potential energy is 
  $$ \boxed{V = \|\grad W\|^2 + 2|\mu |^2} \tag{6.44} $$
and the moduli space of vacua is 
  $$ \boxed{\Mvac = \mu \inv (0)\cap\Crit(W)\bigm/G.}  \tag{6.45} $$
The supercurrent~$j_a$, which is minus the Noether current for~$\Qa$, is  
  $$ \boxed{\aligned
      j_a = \Bigl\{ \sqrt2\epsilon ^{cb}\epsilon ^{\dot d\dot e}\langle
     \partial _{a\dot e}&\phi ,\pb \rangle + \sqrt2\,\epsilon ^{\bd\dd}\delta
     ^c_a\langle \phi ^*\!\grad W,\pbd \rangle + \frac 12\epsilon
     ^{bc}\langle F_{ba},\lambdab^{\dd}\rangle \Bigr\}\, \frac{\partial
     }{\partial y^{c\dd}}\otimes |d^4x| \\
      &+ \Bigl\{\frac 12\epsilon ^{\cd\bd}\langle
     \Fbar_{\cd\dd},\lambdab^{\dd} \rangle - 2
      \sqo\,\phi ^*\!\mu (\lambdab^{\bd}) \Bigr\}\, \frac{\partial }{\partial
     y^{a\bd}}\otimes |d^4x|\endaligned }\tag{6.46} $$
\smallskip
 \rom(iii\rom)\ \rom($\FF=\HH$\rom)\ Suppose we are given
  $$ \alignedat2
      &X\qquad &&\text{\hyp\ manifold with $G$~action} \\
      &\mu_{\HH} \:X\longrightarrow \frak{g}^*\otimes \Im\HH\qquad
     &&\text{\hyp\ moment map} \endalignedat \tag{6.47} $$
Then the lagrangian
  $$ \boxed{\aligned
       L = \Bigl\{ -\frac 12&|F_A|^2 + \frac 1{16}\htil_{\alpha \beta} \eij\nu
     ^{abcd}(\dab\phi )^{\alpha i}(\partial _{cd}\phi )^{\beta j} + \frac
     12\langle\lambda \Dirac_A\lambda\rangle \\
      + &\frac 14h_{\alpha \beta }\nu ^{abcd}\psi _a^\alpha (\phi
     ^*\nab{cd})\psi ^\beta _b
      -2\phi ^*|\mu _{\HH}|^2 +\sqrt2\,\htil_{\alpha \beta }\epsilon
     _{ij}(\lhat_\phi ^{aj})^{\alpha i}\psi ^\beta _a\\
      &- \frac{1}{24}\nu ^{abcd}\phi ^*\Omega _{\alpha \beta \gamma \delta
     }\psi ^\alpha _a\psi ^\beta _b\psi ^\gamma _c\psi ^\delta
     _d\Bigr\}\;|d^{6}x|.\endaligned}
      \tag{6.48} $$
on~$\bmink^6$ is supersymmetric with the supersymmetry transformation laws
  $$ \boxed{\aligned
      (\hxi\phi )^{\alpha i} &= -\sqrt2\,\eta ^{ai}\psi ^\alpha _a \\
      (\hxi A)_{pq} &= 2\eta ^{ai}\nu _{pqab}\epsilon _{ij}\lambda ^{bj} \\
      \hxi\psi ^\alpha _a &= \sqrt2\,\eij\eta ^{bj}\bigl((d_{A})_{ab}\phi
     \bigr)^{\alpha i}\\
      \hxi\lambda ^{ai} &= -\eta ^{bi}f^a_b - 2\eta ^{aj}\sqo\,\phi ^*(\chmu
     _{\HH})^i_j.\endaligned}\tag{6.49} $$
The potential energy is
  $$ \boxed{V = 2|\mu _{\HH}|^2}  \tag{6.50} $$
and the moduli space of vacua is the \hyp\ quotient
  $$ \boxed{\Mvac = \mu _{\HH}\inv (0)\bigm/G.}  \tag{6.51} $$
The supercurrent~$j_{ai}$, which is minus the Noether current for~$Q_{ai}$,
is 
  $$ \boxed{\aligned
      j_{ai}= \Bigl\{ \frac{1}{\sqrt2}\nu ^{bcde}&\langle \partial _{ac}\phi
     ,\psi _{bi} \rangle + \frac 12\epsilon _{ij}\langle f^e_a,\lambda ^{dj}
     \rangle\Bigr\} \,\frac{\partial }{\partial y^{de}}\otimes |d^6x| \\
      &-\Bigl\{ \frac 12\epsilon _{ij}\langle f^c_b,\lambda ^{bj} \rangle +
     2\sqo\,\epsilon _{jk}\langle \phi ^*(\chmu_{\HH})^j_i,\lambda ^{ck}
     \rangle\Bigr\} \, \frac{\partial }{\partial y^{ac}}\otimes
     |d^6x|. \endaligned}\tag{6.52} $$
        \endproclaim
\flushpar
 The proof for~$\FF=\RR$ is given in~\S{7.3}, and the proof\footnote{See
\ASH4 for an example, which in particular serves as a check of the constants
in~\thetag{6.42} and~\thetag{6.43}.} for~$\FF=\CC$ (at least when $X$~is
Hodge) in~\S{8.3}.  In both cases we give a manifestly supersymmetric
formulation is superspace.  We do not give a proof for~$\FF=\HH$, a case in
which there is no known superspace formulation for an arbitrary \hyp\
manifold~$X$, though we have done enough of the computations in components
that the reader should have no problem completing the proof directly in
components.  The reduction of~\thetag{6.48} to 4~dimensions is discussed in
~\S{10.3}.
 
In ~\thetag{6.37} the Hodge dual of the curvature~$F_A$ is 
  $$ *F_A = f_{a b }dy^{a b }. \tag{6.53} $$
In~\thetag{6.43} the curvature~$F_A$ has been decomposed into its self-dual
and anti-self-dual parts:
  $$ F_A = \frac 12 (F_{a b }\epsilon _{\gd\dd} +
     \Fb_{\gd\dd}\epsilon _{a b }) \;dy^{a\gd}\wedge dy^{b\dd}. \tag{6.54} $$
In~\thetag{6.49} the curvature 2-form 
  $$ F_A = \frac 18\,F_{(ab)(cd)}\,dy^{ab}\wedge dy^{cd} \tag{6.55} $$
has been rewritten in terms of a traceless $4\times 4$~matrix~$f$: 
  $$ f^a_b = \frac 14\nu ^{acde}F_{(bc)(de)}. \tag{6.56} $$
We explain this further in~\S{10.1}.

For $\FF=\CC$ the scalar product on~$\frak{g}$ identifies $\frak{g}^*\cong
\frak{g}$.  We let 
  $$ \chmu\:X\to\frak{g} \tag{6.57} $$
denote the composition of the moment map with this isomorphism; for $\FF=\HH$
we have $\chmu_{\HH}\:X\to\frak{g}\otimes \Im\HH$.  In the last equation
in~\thetag{6.49} we identify~$\Im\HH$ with traceless hermitian $2\times
2$~matrices:
  $$ aI+bJ+cK \leftrightarrow \pmatrix a\sqo&-b+c\sqo \\ b + c\sqo &-a\sqo 
     \endpmatrix. \tag{6.58} $$

In all three cases there are new terms of the  form $\langle \lhat_\phi ,\psi
\rangle$ which appear in the lagrangian.  Here $\lhat_\phi$~is the odd vector
field on~$P\times _GX$ corresponding to~$\lambda $,  and the inner product is
the Riemannian  metric on~$X$.  For~$\FF=\RR$, in the  linear  case $X$~is an
orthogonal  vector  space      and  $G$~acts linearly   by     an  orthogonal
representation~$\rho \:G\to O(X)$. Then
  $$ \lhat_\phi =  \dot{\rho }(\lambda )\phi, \tag{6.59} $$
where $\dot{\rho }\:\frak{g}\to\frak{o}(X)$ is the infinitesimal action.
This term can be predicted by supersymmetry:  the variation in~$A$
of~$\langle \psi \Dirac_A\psi   \rangle$ is canceled by the variation
in~$\phi$  of this term.  For~$\FF=\CC$ note that $\pa$~is a vector field of
type~$(1,0)$ on~$X$, so only the $(0,1)$-component of~$\lhat_\phi ^a$ and its
conjugate enter~\thetag{6.42}.  In the linear case $G$~acts by a unitary
representation $\rho \:G\to U(X)$ on a hermitian vector space~$X$.  Then the
$(1,0)$-component of~$\lhat_\phi $ is 
  $$ (\lhat_\phi )^{1,0} = \dot\rho \cp(\lambda )\phi , \tag{6.60} $$
where $\dot\rho \cp\:\frak{g}\cp\to\frak{g}\frak{l}\cp(X)$ is the
complexification of the differential to~$\rho $.  To write a formula for the
$(0,1)$-component we identify $\overline{X}\cong X^*$ via the hermitian
metric; then 
  $$ (\lhat_\phi )^{0,1} = - \dot\rho \cp(\lambda )^*\phi ^*\in X^*\cong
     \overline{X} \tag{6.61} $$
at the point~$\phi ^*\in X^*$.  (As a real manifold, $X^*\cong
\overline{X}\cong X$ so we can identify points of~$X$ with points of~$X^*$.)
Note that when $\lambda $~is real, then $\dot\rho \cp(\lambda )$~is
skew-hermitian and \thetag{6.61}~is the complex conjugate of~\thetag{6.60},
i.e., $\lhat_\phi $~is real.  Also,\footnote{We caution the reader that
`$\lambdab$'~denotes the conjugate in the Lie algebra.  For example, if
$G=\TT$~is the circle group, then $\frak{g}\cong \sqo\,\RR$ and
$\frak{g}\cp\cong \CC$ with conjugation {\it minus\/} the usual complex
conjugation in~$\CC$.  See \ASH4 for a sample lagrangian in this case.} 
  $$ \hat{\bar\lambda } =  \overline{\lhat}, \tag{6.62} $$
which also holds in the nonlinear case.  This ensures that \thetag{6.42}~is
real.  

The quotient in~\thetag{6.45} is the symplectic (\Kah) quotient, which can
often be identified with the complex quotient of~$\Crit(W)$ by the
complexification of~$G$.  The quotient in~\thetag{6.51} is the \hyp\
quotient.  (It was in this context that the \hyp\ quotient was discovered.)
For~$\FF=\CC$ the abelian group~$\bigl(\frak{g}/[\frak{g},\frak{g}] \bigr)^*$
of infinitesimal characters acts by translations on the space of moment maps.
Theories with different choices of moment map are different.  Notice that
when $X$~is a point we must still choose a moment map, which is now simply an
infinitesimal character.  The resulting term~$-\bigl(2|\mu |^2\bigr)$
is usually called a {\it Fayet-Iliopoulos term.\/}  For~$\FF=\HH$ the moment
map is also ambiguous, this time up to $\bigl(\frak{g}/[\frak{g},\frak{g}]
\bigr)^*\otimes \Im\HH$.  For $X$~a point this leads to a Fayet-Iliopoulos
term. 
 
The action of R-symmetries for $\FF=\CC$ and~$\FF=\HH$ is summarized
in~\thetag{5.31}, \thetag{8.24}, and~\thetag{10.21}. 
 
We have not written the equations of motion for these general lagrangians; it
seems better to work out the detailed form on a case-by-case basis.  However,
we indicate here some general features.  For the scalar field~$\phi $ we find
an equation of the form 
  $$ \wave_A\phi =-\phi ^*\grad V + Q^{(2)}_\phi (\psi ,\lambda ) +
     R^{(2)}_\phi (\psi )d\phi + Q^{(4)}(\psi ). \tag{6.63} $$
Here $\wave_A = d_A^*d_A$ is the wave operator on sections of
$X^P\to\bmink^n$.  The coefficients of the quadratic form~$Q^{(2)}_\phi $
depend on the pullback of the hessian of~$h$ or~$W$ (for $\FF=\RR$ or
$\FF=\CC$) and also on the action of~$\frak{g}$ on~$X$; the coefficients of
the quadratic form~$R^{(2)}_\phi $ depend on the pullback of the Riemann
curvature tensor; and the coefficients of the quartic form~$Q^{(4)}_\phi $
depend on the pullback of derivatives of the Riemann curvature tensor.  Now
the action of~$\frak{g}$ on~$X$ induces a map from~$\frak{g}$ to vertical
vector fields on~$X^P$, or equivalently a $\frak{g}^*$-valued vector
field~$\zhat$ on~$X^P$.  This vector field enters into some equations of
motion, for example in the equation
  $$ \Dirac_A\lambda =-\langle \zhat_\phi ,\psi   \rangle. \tag{6.64} $$
Here~$\zhat_\phi $ means the pullback of~$\zhat$ by the section~$\phi $.

\Head{6.3}{Superspace construction}
We consider only pure gauge theory.  As usual we fix data~\thetag{6.1}.
Consider a superspace~$\mink ns$ in any dimension built out of any real spin
representation.  Suppose $\scrP\to\mink ns$ is a principal $G$~bundle.
Let~$\scrA$ be a connection on~$\scrP$.  If $\scrA$~is unconstrained, then
there are too many component fields to obtain a sensible theory, even with
$n,s$~small.  Hence we impose a constraint.  Namely, we require that the
curvature~$F_{\scrA}$ be constrained to vanish along the left-invariant odd
distribution~$\tau $ on~$\mink ns$: 
  $$ \cont{D_b}\cont{D_a}F_{\scrA} = 0. \tag{6.65} $$
In that case the ``lowest'' piece of the curvature is in the even-odd
direction.  Let $\DDa$~be the horizontal lift of~$\Da$ and~$\nab\mu $ the
horizontal lift of~$\di$.  Define
  $$ \boxed{\Lambda ^a = \frac 1n\stov^{\mu ab}\cont{D_b}\cont{\di}F_{\scrA}=
     \frac 1n\stov^{\mu ab}[\DDb ,\nab\mu].} \tag{6.66} $$
(The constant is chosen with hindsight.)  Also, we identify $G$-invariant
vertical vector fields with sections of the adjoint bundle; see~\CF{4.2}.)
This is a ``projection'' of the lowest nonvanishing part of the curvature;
$\Lambda ^a$~is an odd section of the adjoint bundle $\adP=\scrP\times
_G\frak{g}\to\mink ns$.  There are component fields
  $$ \alignedat2
      A&=\bres\scrA\qquad &&\text{connection on $P=\bres\scrP\longrightarrow
     \bmink^n$} \\
      \lambda^a&=\bres\Lambda ^a &&\text{dual spinor field on~$\bmink^n$ with
     values in~$\adP$}\endaligned \tag{6.67} $$
and possibly additional component fields as well.  To see what other
component fields there are, we examine consequences of the Bianchi identity.
 
First, we apply Bianchi to~$D_a,D_b,D_c$ to learn that 
  $$ \text{The symmetrization of $\sdtov^\mu _{ab}[\DDg,\nab\mu ]$ in $a,b,c$
     vanishes.}  \tag{6.68} $$
To interpret this, consider the sequence of maps
  $$ \CD
      \Sym^3(S^*) @>{\Sym^3(\sdtov\otimes \operatorname{id})}>>V\otimes S^*
     @>g\Gamma >> S \\
      \vspace{6pt}
      f_af_bf_c @>>> \sym(\sdtov^\mu _{ab}e_\mu \otimes f_c) \\
      \vspace{-2pt}
      @. e_\mu \otimes f_c @>>> g_{\mu \nu }\sdtov^\nu
     _{cd}f^d\endCD \tag{6.69} $$
The composition vanishes if and only if for all~$s^*$ in~$S^*$, $\Gamma
(s^*,s^*)\otimes s^*$ maps to~0 in~$S$, i.e., if and only if for all $t^*\in
S^*$,  
  $$ \langle \text{image of }\Gamma (s^*,s^*)\otimes s^*,t^* \rangle =
     \langle \Gamma (s^*,s^*),\Gamma (s^*,t^*) \rangle \tag{6.70} $$
vanishes.  The last expression is obtained by polarization from$\langle
\Gamma (s^*,s^*),\Gamma (s^*,s^*) \rangle$, and the composition vanishes if
and only if the quadratic form~\thetag{6.14} does.  As we have seen, this
occurs for $n=\dim V=3,4,6,10$ with minimal supersymmetry $s=\dim
S=2,4,8,16$.  Furthermore, in these cases the sequence~\thetag{6.69} is
exact.  Assume from now on that we are in one of these situations.  Now the
dual to~\thetag{6.69} is
  $$ S^* \longrightarrow V^*\otimes S\longrightarrow \Sym^3(S). \tag{6.71} $$
The even-odd piece of the curvature can be identified with an element
of\break $\Omega ^0(\mink ns,\adPP)\otimes V^*\otimes S$;then
\thetag{6.68}~asserts that its image in $\Omega ^0(\mink ns,\adPP)\otimes
\Sym^3(S)$ vanishes.  By the exactness of~\thetag{6.71} we can then identify
the even-odd piece of the curvature with an element $\Lambda \in \Omega
^0(\mink ns,\adPP)\otimes S^*$, as defined in~\thetag{6.66}.  In fact, the
Clifford identity implies that
  $$ \aligned
      V^*\otimes S&\longrightarrow S^* \\ 
      e^\mu \otimes f^a&\longmapsto \frac 1n\stov^{\mu ab}f_b\endaligned
     \tag{6.72} $$
is a splitting of~\thetag{6.71}, whence the formula~\thetag{6.66}
for~$\Lambda $.  To summarize, in the four cases we are considering the
even-odd curvature can be expressed in terms of~$\Lambda $: 
  $$ \boxed{[\DDa,\nab\mu ]= g_{\mu \nu }\sdtov^\nu _{ab}\Lambda ^b.}
     \tag{6.73} $$
 
Next, we apply Bianchi to~$D_a,D_b,\di$.  After some manipulation
with~\thetag{6.73} and the Clifford identity we find 
  $$ -2\stov^{\mu ac}\sdtov^\nu _{cb}\scrF_{\mu \nu }= g_{\mu \nu }\stov^{\mu
     ac}\sdtov^\nu _{bd}\DDg\Lambda ^d + n\DDb\Lambda ^a, \tag{6.74} $$
where the even-even part of the curvature is 
  $$ \scrF_{\mu \nu }= \cont{\dj}\cont{\di}F_{\scrA} = -[\nab\mu ,\nab\nu ].
     \tag{6.75} $$
(The funny sign is due to the fact that we take the bracket as vector fields
on~$P$, not as operators on an associated bundle; cf.~\CF{4.2}.)  The first
term on the right hand side of~\thetag{6.74} depends on the particular case,
and its precise form determines the structure of the auxiliary fields.  In
$n=3,4,6$~dimensions we will find a formula of the form
  $$ \boxed{\DDb\Lambda ^a = -\frac 12\stov^{\mu ac}\sdtov^\nu
     _{cb}\scrF_{\mu \nu }+ \scrE,} \tag{6.76} $$
where $\bres\scrE$~consists of auxiliary fields.  (Compare with the second
equation in~\thetag{6.8}.)
 
We do not learn anything new from the remaining cases of the Bianchi
identity.  Applied to~$D_a,\di,\dj$ we obtain the Jacobi identity for three
vector fields~$\Lambda ^a,\di,\dj$ on~$\scrP$.  (Here we identify~$\Lambda
^a$ with a $G$-invariant vertical vector field.  Upon restriction
to~$\bmink^n$ this equation gives the supersymmetry transformation
law~\thetag{6.9} for the curvature.)  Bianchi applied to~$\di,\dj,\partial
_\rho $ restricts on~$\bmink^n$ to the usual Bianchi identity for~$A$. 
 
Set 
  $$ F_{\mu \nu } = \bres\scrF_{\mu \nu }; \tag{6.77} $$
then 
  $$ F_A = \frac 12F_{\mu \nu }dx^\mu \wedge dx^\nu  \tag{6.78} $$
is the curvature of~$A$.
 
From the general formula~\thetag{1.29} and equations~\thetag{6.73}
and~\thetag{6.76} we recover the supersymmetry transformation
law~\thetag{6.8}, at least if we set~$\scrE=0$.  For example, to compute the
action of~$\hxi$ on the connection~$A$, we lift the vector field~$\eta ^aD_a$
on~$\mink ns$ to the horizontal vector field~$-\eta ^a\DDa$ on~$\scrP$ and
apply the covariant derivative.  From~\thetag{6.73} the $\mu
^{\text{th}}$~component of the answer is
  $$ \split
      (\hxi A)_\mu &= \cont{\di}\bres\Lie{-\eta ^a\DDa}A \\
      &= \bres\cont{\nab\mu}\Lie{-\eta ^a\DDa}A \\
      &=\bres\Bigl[ \Lie{-\eta ^a\DDa}\cont{\nab\mu}A - \iota\bigl([-\eta
     ^a\DDa,\nab\mu] \bigr)A\Bigr] \\
      &= \bres\eta ^a[\DDa,\nab\mu] \\
      &=\eta ^ag_{\mu \nu }\sdtov^\nu _{ab}\lambda ^b,\endsplit \tag{6.79} $$
which agrees with~\thetag{6.8}.  The precise formula for the variation of the
fermion~$\lambda $, which includes the auxiliary fields, must be analyzed on
a case-by-case basis.
 
In~\S{7}, \S{8}, and~\S{10} we construct the superspace models for the
cases~$n=3,4$, and~6 (previously referred to as~$\FF=\RR,\CC$, and~$\HH$).
Supersymmetry is manifest in the superspace formulation, and we determine the
auxiliary fields by computing~$\scrE$ in~\thetag{6.76}.  In addition,
for~$n=3,4$ we give a superspace formulation of the gauge theory coupled to
matter, and so prove \theprotag{6.33(i)} {Theorem} and \theprotag{6.33(ii)}
{Theorem}.  There is no known superspace formulation for the
$n=10$~dimensional theory. 
 
There are also superspace models for dimensional reductions of the $n=3,4,6$
theories; particular cases are treated in~\S{9} and~\S{10}.  We remark here
that \thetag{6.65}~is no longer the correct curvature constraint in
superspace; some components of the odd-odd curvature are allowed to be
nonzero.

\Chapter7{$N=1$ Yang-Mills Theory in Dimension 3}

 \comment
 lasteqno 7@ 30
 \endcomment

 For the scalar superfield of Chapter~{4} and the chiral superfield of
Chapter~{5} it is easy to see that the supermanifold of superfields is
diffeomorphic to the supermanifold of component fields.  But for constrained
connections the argument is more complicated due to gauge symmetry.  The
precise statement is the main task of~\S{7.1}.  In~\S{7.2} and~\S{7.3} we
describe the superspace formulation of the Yang-Mills lagrangian and of the
lagrangian including matter.  There are no surprises, though we do explain
where the new term in the component lagrangian~\thetag{6.36} (coupling the
two sorts of spinors) arises from the superspace point of view.

\Head{7.1}{Constrained connections on~$\mink32$}  
 We specialize the discussion in~\S{6.3} to~$\mink32$.  Our first task is to
evaluate~\thetag{6.74}.  To that end we compute using the formulas
of~\S{2.3}:
  $$ \split
      g_{\mu \nu }\stov^{\mu ac}\sdtov^\nu _{bd}&= g_{\mu \nu }\epsilon
     ^{aa'}\epsilon ^{cc'}\sdtov^\mu _{a'c'}\sdtov^\nu _{bd} \\
      &= \frac 12(\epsilon _{a'b}\epsilon _{c'd}+\epsilon _{a'd}\epsilon
     _{c'b})(\epsilon ^{aa'}\epsilon ^{cc'}+ \epsilon ^{ac'}\epsilon ^{ca'})
     \\
      &= \delta ^a_b\delta ^c_d + \delta ^a_d\delta ^c_b.\endsplit
     \tag{7.1} $$
Plugging into~\thetag{6.74} we obtain 
  $$ -2\stov^{\mu ac}\sdtov^\nu _{cb}\scrF_{\mu \nu }= 4\DDb\Lambda ^a +
     \delta ^a_b\DDg\Lambda ^c. \tag{7.2} $$
Setting~$a=b$ and summing we find 
  $$ 6\DDg\Lambda ^c = -2\epsilon ^{aa'}\epsilon ^{cc'}\sdtov^\mu
     _{a'c'}\sdtov^\nu _{ac}\scrF_{\mu \nu}=0, \tag{7.3} $$
since $\epsilon $~and $\scrF$~are skew-symmetric.  Thus \thetag{7.2}~becomes
  $$ \DDb\Lambda ^a = -\frac 12\stov^{\mu ac}\sdtov^\nu _{cb}\scrF_{\mu \nu
     }. \tag{7.4} $$
This implies~$\scrE=0$ in~\thetag{6.76}, which means that there are no
auxiliary fields.  

The multiplet~$\{A,\lambda\}$ of component fields is called the {\it vector
multiplet\/}, and from the vector multiplet we can reconstruct the
constrained connection~$\scrA$.  The precise statement is the following.
 
        \proclaim{\protag{7.5} {Proposition}}
 The category of connections~$\scrA$ on~$\mink32$ whose curvature vanishes on
the odd distribution~$\tau $ is equivalent to the category of
pairs~$\{A,\lambda\}$ consisting of a connection~$A$ on~$\bmink^3$ and a
dual spinor field~$\lambda $ with values in the adjoint bundle.
        \endproclaim

We first comment on the meaning of \theprotag{7.5} {Proposition}.  As odd
fields are involved, it is crucial to work over a base $S$: for any
supermanifold $S$, what \theprotag{7.5} {Proposition} claims is that the
functor which attaches the component fields to a constrained connection is an
equivalence between the following categories: (a) $G$-torsors on $M^{3\mid 2}
\times S$ provided with a relative connection $\scrA$ satisfying the curvature
constraint; (b) $G$-torsors on $\check M^3\times S$ provided with a relative
connection $A$ and an adjoint bundle valued dual spinor field $\lambda$.
Note that in (b) we have $\check M^3 \times S$, not the reduced space of
$M^{3\mid 2}\times S$.  As usual, in the proof we will keep $S$ silent.

If $\Cal P$ is a $G$-torsor with a connection $\Cal A$ on $M^{3\mid 2}$
(i.e. on $M^{3\mid 2}\times S$), any automorphism of $(\Cal P,\Cal A)$ which
induces the identity on the restriction $P$ of $\Cal P$ to $\check
M^3$(i.e. $\check M^3\times S$) is the identity.  Because of this, $(\Cal P,
\Cal A)$ can be viewed as a structure on $P$.  Concretely, once some choices
have been made on $\check M^3$ (see the proof of \theprotag{7.5}
{Proposition} for an example), if $(\Cal P,\Cal A)$ is a $G$-torsor with
connection on $M^{3\mid 2}$ with restriction $P$ to $\check M^3$, and if $s$
is a trivialization of $P$, the connection gives rise to a canonical
extension of $s$ to a trivialization of $\Cal P$.  Given $s$, then $\Cal A$
becomes a $1$-form~$\alpha$ with values in the Lie algebra $\frak{g}$ obeying
suitable constraints.  What \theprotag{7.5} {Proposition} claims is that the
map from the space of those $\alpha$'s to the space of $(A,\lambda)$ ($A$ a
connection on $P$, and $\lambda$ a $\frak{g}$-valued dual spinor field) is
bijective.  In other words, for $P$ a $G$-torsor on $\check M^3$ the map from
(a) the set of isomorphism classes of $G$-torsors $\Cal P$ on $M^{3\mid 2}$,
given with a constrained connection $\Cal A$ and an isomorphism
$P\overset\sim\to{\rightarrow}\Cal P\mid \check M^3$, to (b) the set of
$(A,\lambda)$, is bijective.  In (a), the objects considered have no
non-trivial isomorphisms.  Because of this, the question is local, and one may
assume that $P$ is the trivial $G$-torsor.

        \demo{Proof}
 The data on $\scrP$ of a constrained connection amounts to that
of a partial connection in the direction of the odd
distribution $\tau$: a $G$-invariant lift to $\scrP$ not of all
vector fields, but only of those in $\tau$.
If $\DD_a$ (resp. $\nabla_{\!ab}$) is the horizontal lift
of $D_a$ (resp. $\partial_{ab}$), the vanishing along $\tau$
of $F$ means that
  $$  \nabla_{\!ab}=-\frac12[\DD_{a},\DD_{b}].\tag{7.6}$$
If only the $\DD_{a}$ are given, the corresponding
constrained connection is obtained by defining $\nabla_{\!ab}$
by this formula.

The functor implicit in \theprotag{7.5} {Proposition} attaches to
$(\scrP,\nabla)$: the restriction to $\bmink^3$ of $\scrP$ (denoted $P$),
that of $[\DD_{a},\DD_{b}]$, and that of $[[\DD_{a},\DD_{b}],\DD_{c}]$;
respectively: a $G$-torsor on $\bmink^3$, a connection on it, and a section
of the adjoint bundle with vanishing symmetrization in $a$, $b$, $c$.
(cf. \thetag{6.68}).

We now choose a coordinate system as in~\S2.3, giving rise to the basis
$D_1$, $D_2$ of $\tau$.  We will use $D_1$, $D_2$ to define a retraction
$q\colon\, \mink32\to\bmink^3$, and, for $\scrP$ a torsor with connection on
$\mink32$, with restriction $P$ to $\bmink^3$, an isomorphism of torsors
$q^*P\to \scrP$ (partial gauge fixing).

We first explain an analog, in classical differential
geometry of what we will do.
Suppose $N$ is of codimension $2$ in a variety $M$, and that
$D_1$, $D_2$ are vector fields which, on $N$,
span the normal bundle $T_M/T_N$.
In a neighborhood of $N$, the flow $\exp(tD_1)$ generates
from $N$ a codimension $1$ subvariety $M_1$, with a
retraction $q_1\colon\, M\to N$ constant on the flow lines.
Repeating the process with $M_1$ and $D_2$, we obtain in a
neighborhood of $N$ a retraction $q_2\colon\, M\to M_1$,
constant on the flow lines of $D_2$:

  $$ \vbox{\epsfxsize=1.5in\epsfbox{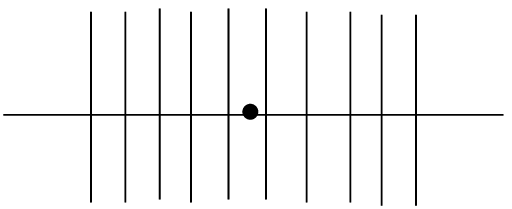}} $$

\noindent If $\scrP$ is a torsor on~$M$ with connection, with restriction $P$
to $N$, integrating the connection on the flow lines of $D_1$, we obtain an
isomorphism $\alpha_1\colon\, q_1^*P\to \scrP$ on $M_1$.  Doing the same with
$D_2$, we obtain $\alpha_2\colon\, q_2^*(\scrP\vert M_1)\arrowsim \scrP$.

In our case of codimension $0\vert 2$, the situation is different in that \
(a) the vector field $D_1$ does not integrate to a foliation with leaves
$\RR^{1\vert 0}$: the obstruction is $D_1^2$ mod $D_1$. \ (b) On $\RR^{0\vert
1}$, a connection $\nabla$ does not need to be flat: curvature obstruction
$\nabla_{\!\theta}^2$.

It however remains true that for $N\subset \mink32$ a subvariety and $D$ an
odd vector field transversal to $N$, the vector field $D$, in fact just its
restriction to $N$, generates from $N$ an embedding $N\times\RR^{0\vert
1}\hookrightarrow \mink32$.  A function $f$ on $\mink32$ vanishes on the
image $M_1$ if and only if $f$ and $Df$ vanish on $N$, and the retraction
$q_1\colon\, M_1\to N$ is such that for any function $u$ on $N$,
$Dq_1^*(u)=0$.  If $\scrP$ is a torsor with connection on $\mink32$, with
restriction $P$ to $N$, we have on $M_1$ a unique isomorphism of torsors
$q_1^*P\to \scrP\vert M_1$ which is the identity on $N$ and such that on $N$
the liftings of $D$ agree.  One should beware that if $D$ is given as a
vector field on $\mink32$, it does not need to be tangent to $M_1$.  This
tangency to $M_1$ holds only on $N$.

We apply this twice, to $\bmink^3\subset \mink32$ and $D_1$, then to $M_1$
and $D_2$, obtaining $q\colon\, \mink32\to M_1\to\bmink^3$ and
$\alpha\colon\, q^*P\to \scrP$.  The construction of $\alpha$ uses only the
connection in the directions of $D_1$ and $D_2$: for $\scrP$ a torsor with a
connection along $\tau$, we obtain $\alpha\colon q^*P\to \scrP$, and the data
of $\alpha$ is equivalent to that of part of the connection, viz. $\DD_1$ on
$\bmink^3$ and $\DD_2$ on $M_1$.  It remains to check that to complete this
partial data to that of a connection $\nabla$ along $\tau$ amounts to giving
$[\DD_{a},\DD_{b}]$ on $\bmink^3$ (a connection on $P$) and
$[[\DD_{a},\DD_{b}],\DD_{c}]$ on $\bmink^3$ (a section on $\bmink^3$ of
$\adP$ with vanishing symmetrization in $a$, $b$, $c$, i.e., the data of
just $[[\DD_{1},\DD_{1}],\DD_{2}]$ and $[[\DD_{2},\DD_{2}],\DD_{1}]$).

Let us choose a local trivialization of $P$, hence of $\scrP$.
The connection along $\tau$ is given by a section $a$ of
$\tau^*$ with values in the Lie algebra, i.e., by
$a_1=\left<D_1,a\right>$ and $a_2=\left<D_2,a\right>$.
That the connection is partially given by $\alpha\colon
q^*P\to \scrP$ amounts to $a_1=0$ on $\bmink^3$ and $a_2=0$ on
$M_1$, i.e. to $a_1=a_2=D_1a_2=0$ on $\bmink^3$.

For the simplicity of notations, let us assume that $G$ is $GL(n)$, i.e.,
that we are considering a vector bundle.  We have then
  $$  \alignedat2 &[\DD_{1},\DD_{1}]=-2\partial_{11}+2D_1a_1
     &\quad&\text{on}\quad \bmink^3\\ &[\DD_{1},\DD_{2}]
     =-2\partial_{12}+D_2a_1 &\quad&\text{on}\quad \bmink^3\\
     &[\DD_{2},\DD_{2}]=-2\partial_{22}+2D_2a_1 &\quad&\text{on}\quad
     \bmink^3\,\,, \endalignedat \tag{7.7} $$
so the data of the first order jet of $a_1$ and $a_2$ along
$\bmink^3$ amounts to the data of the connection on $P$.
The remaining freedom: the addition of $\theta_1\theta_2 b_1$ 
to $a_1$ and $\theta_1\theta_2 b_2$ to $a_2$
amounts to the data of $[[\DD_{i},\DD_{j}],\DD_{k}]$
on $\bmink^3$.
Indeed, by $a_i\mapsto a_i+\theta_1\theta_2 b_i$,
$[[\DD_{1},\DD_{1}],\DD_2]$ on $\bmink^3$ is increased
by $-2b_1$ and $[[\DD_{2},\DD_{2}],\DD_{1}]$ by
$-2b_2$.
        \enddemo

It is easy in this case to reformulate a constrained connection in terms of
an {\it unconstrained\/} field.  Namely, take the unconstrained field to be a
partial connection on~$\scrP$ along the odd distribution~$\tau $.  Then if
$\DDa$ is the horizontal lift of~$\Da$, define the horizontal lift~$\ddab$
of~$\dab$ to be 
  $$ \ddab = -\frac 12[\DDa,\DDb]. \tag{7.6} $$
In other words, the constraint that the odd-odd components of the curvature
vanish uniquely determines the connection~$\scrA$ from the restriction
of~$\scrA$ to~$\tau $.

It is convenient to lower an index on the spinor field and
define\footnote{This goes against our general wish not to raise or lower a
single index using a skew-symmetric tensor, since there is a sign ambiguity.
However, defining~$\Lambda _a$ facilitates comparison with the physics
literature.} 
  $$ \boxed{\Lambda _a = \Lambda ^b\epsilon _{ba} = \frac 13\epsilon
     ^{bc}[\ddab,\DDg].} \tag{7.8} $$
Equation~\thetag{6.73} specializes to 
  $$ [\DD_a,\Lambda _{bc}] = \epsilon _{ab}\Lambda _c + \epsilon _{ac}\Lambda
     _b. \tag{7.9} $$
Note that \thetag{7.3}~is now
  $$ \epsilon ^{ab}\DDa\Lambda _b =0. \tag{7.10} $$
We note a useful consequence of~\thetag{7.10}:
   $$ \DD^2\Lambda _a  = -\epsilon ^{b c }\ddab \Lambda _c
     . \tag{7.11} $$

It is convenient to set 
  $$ f_{ac}= \frac 12\epsilon ^{bd}F_{(ab)(cd)}, \tag{7.12} $$
which is symmetric in~$a,c$.  Then 
  $$ *F_A = f_{a b }dy^{a b } \tag{7.13} $$
is the Hodge dual of the curvature  (c.f.~\thetag{2.38}).  

In these variables the component fields are
  $$ \boxed{\aligned
      A&=\bres\scrA \\ 
      \lambda _a&=\bres\Lambda _a\endaligned} \tag{7.14} $$
and from~\thetag{7.3} and~\thetag{7.10} we have
  $$ \boxed{f_{a b } = \bres\DDa\Lambda _b =\bres\DDb\Lambda
     _a .} \tag{7.15} $$
The supersymmetry transformation laws~\thetag{6.8} specialize to
  $$ \boxed{\aligned
      (\hxi A)_{a b } &= -\eta ^c (\epsilon _{a c
     }\,\lambda _b + \epsilon _{b c }\,\lambda _a )\\
      \hxi \lambda _a &= -\eta ^b f_{a b} \endaligned} \tag{7.16} $$
and \thetag{6.9}~specializes to 
  $$ \boxed{\hxi f_{a b} = \hphantom{-}\eta ^c \bigl[ \nab{c a }\lambda _b
     +\epsilon _{a c }(\Dirac\lambda )_b \bigr].} \tag{7.17} $$

\Head{7.2}{The Yang-Mills action on~$\mink32$}
Assume that $G$~is compact.  Fix a bi-invariant inner product~$\langle \cdot
,\cdot \rangle$ on~$\frak{g}$.  For a constrained connection~$\scrA$
on~$\mink32$ the basic lagrangian density is
  $$ \boxed{L_2 = |d^3x|\,d^2\theta \;\Bigl\{-\frac {1}4\epsilon ^{a b
     }\langle \Lambda _a ,\Lb \rangle\Bigr\}.} \tag{7.18} $$
We compute the component lagrangian function~$\Lch_2$ using~\thetag{2.40}: 
  $$ \split
      -4\Lch_2 &= -\eab\bres D^2\langle \La,\Lb \rangle \\
      &= -\eab\bres\Bigl\{ 2\langle \DD^2\La,\Lb \rangle + \epsilon ^{c
     d }\langle \DDd\La,\DDg\Lb \rangle\Bigr\} \\
      &= 2\eab\epsilon ^{c d }\langle \nab{a c }\lambda
     _d ,\lambda _b \rangle - \eab\epsilon ^{c d }\langle
     f_{d a },f_{c b } \rangle.\endsplit \tag{7.19} $$
In the last step we use~\thetag{7.15}, \thetag{7.11}, and~\thetag{7.14}.  So
from~\thetag{2.27}, \thetag{2.28}, and the fact that $|\!*\!F_A|^2=|F_A|^2$
we conclude
  $$ \boxed{\Lch_2 = -\frac 12|F_A|^2 + \frac 12\langle \lambda
     \Dirac_A\lambda \rangle. } \tag{7.20} $$
This is the super Yang-Mills lagrangian~\thetag{6.3}.  Thus we do have a
superspace formulation of this lagrangian in 3~dimensions.
 
Next, we begin the analysis of the superspace lagrangian directly in
superspace.  This is a redundant exercise, since we already have the
component expressions in~\thetag{6.22} and the equations which follow.  But
it is instructive to see how the computation looks in superspace.

We compute the superspace equation of motion by varying the {\it
unconstrained\/} partial connection~$\DDa$.  Notice that a variation~$\delta
\DDa$ is a $G$-invariant vertical vector field, so corresponds to a
section~$\delta \scrA_a $ of the adjoint bundle.  From~\thetag{7.8},
\thetag{6.66}, and~\thetag{7.6} we compute
  $$ \delta \Lambda_a = -\frac 16 \epsilon ^{c d } \Bigl\{
     [[\delta \DDa,\DDg],\DDd] + [[\DDa,\delta \DDg],\DDd] -2[\nab{a
     c },\delta \DDd]\Bigr\}. \tag{7.21} $$
We use the invariance of~$\langle \cdot ,\cdot \rangle$, \thetag{7.10},
and~\thetag{7.11} to find
  $$ \split
      \delta L_2 &= |d^3x|\,d^2\theta \;\frac {-1}2\eab\langle \delta
     \Lambda_a ,\Lambda_b \rangle \\
      &= |d^3x|\,d^2\theta \; \frac {1}{12} \eab\epsilon ^{c d
     }\Bigl\{ \langle \delta \DDa,[\DDg,[\DDd,\Lambda_b ]] \rangle +
     \langle \delta \DDg,[\DDa,[\DDd,\Lambda_b ]] \rangle\\
      &\qquad \qquad \qquad \qquad + 2\langle \delta \DDd,[\nab{a c
     },\Lambda_b ] \rangle \Bigr\} + {EXACT} \\
      &= |d^3x|\,d^2\theta \; \frac {-1}2\eab\epsilon ^{c d
     }\langle \delta \scrA_a ,\nab{b c }\Lambda _d \rangle
     + {EXACT}.\endsplit \tag{7.22} $$
So the superspace equation of motion is 
  $$ \boxed{\epsilon ^{c d }\nab{b c }\Lambda _d =0\qquad \text{for all
     $b$}.} \tag{7.23} $$
 
The exact term is~$-d\gamma $ for the variational 1-form~$\gamma $.  We do
not compute the precise expression, since it is unwieldy and not useful.
One interesting feature emerges quickly, though.  Namely, $\gamma $~contains
a term proportional to 
  $$ \epsilon ^{ab}\epsilon ^{cd}\cont{D_d}\;|d^3x|\,d^2\theta \;D_c\langle
     \delta \scrA_a,\Lambda _b  \rangle, \tag{7.24} $$
and so as a differential 1-form on the space of fields $\gamma $~is {\it
not\/} linear over functions.  (In other words, $\delta \scrA_a $ is
differentiated.)  
 
We extract from~\thetag{7.23} the equations of motion in components.
Restricting~\thetag{7.23} to~$\bmink^3$ we find
  $$ \boxed{\Dirac_A\lambda = 0.} \tag{7.25} $$
This agrees with~\thetag{6.25}.  To obtain the equation of motion for~$A$, we
either apply~$\bres\DDe$ to~\thetag{7.23} or specialize~\thetag{6.23}.  In
either case we obtain
  $$ \boxed{\epsilon ^{cd}\nab{ad}f_{bc} = [\lambda _a,\lambda _b]\qquad
     \text{for all $a,b$}.} \tag{7.26} $$
This agrees with~\thetag{6.23}.  Finally, the supercurrent~\thetag{6.27}
specializes to 
  $$ \boxed{j_a = \frac 12\epsilon ^{cb}\epsilon ^{de}\langle f_{cd},\lambda
     _e \rangle\;\frac{\partial }{\partial y^{ab}}\otimes |d^3y| + \frac
     12\epsilon ^{cd}\epsilon ^{be}\langle f_{ca},\lambda _e
     \rangle\;\frac{\partial }{\partial y^{db}}\otimes |d^3y| .} \tag{7.27} $$
Since we didn't compute the exact term in~\thetag{7.22}, we do not give a
superspace computation of the supercurrent.

\Head{7.3}{Gauge theory with matter on~$\mink32$}
In this subsection we prove \theprotag{6.33(i)} {Theorem}.  Suppose we are
given the data in~\thetag{6.35}: a Riemannian manifold~$X$ on which $G$~acts
by isometries and a $G$-invariant function $h\:X\to\RR$.  Fix a $G$~bundle
$\scrP\to\mink32$.  Then we can form the associated bundle
$X^\scrP\to\mink32$ whose typical fiber is~$X$.  The superfields in the
theory are a constrained connection~$\scrA$ on~$\scrP$ and a section~$\Phi $
of~$X^{\scrP}$.  We view~$\Phi $ as an equivariant map
  $$ \Phi \:\scrP\longrightarrow X. \tag{7.28} $$
Then the formulas in~\S{4} make sense provided we interpret the vector
fields~$\Da,\dab$ as the horizontal lifts to~$\scrP$ of the indicated vector
fields on~$\mink32$.  For example, the component fields are still defined
by~\thetag{4.7}, but now $\pa=\bres\DDa\Phi $ where $\DDa$~acts by the
covariant derivative associated to~$\scrA$.  Also, in the definition of the
auxiliary field $F=\frac
12\bres(\DD_2\DD_1-\DD_1\DD_2)\Phi $ the outer covariant derivatives are
constructed from both~$\scrA$ and the Levi-Civita connection on~$X$.  

The superspace lagrangian for the theory is~$L_0+L_1+L_2$, where the
individual terms are defined in~\thetag{4.16}, \thetag{4.20},
and~\thetag{7.18}.  The only possible changes to the computation of the
component lagrangian are additional curvature terms from the
connection~$\scrA$.  Recall that $\scrA$~is constrained to be flat on the odd
distribution~$\tau $, and this constraint eliminates many such possible
terms.  With this in mind we examine the computation~\thetag{4.18}.  There is
a change in the formula~\thetag{4.11} for~$\DD^2\DD_b\Phi $, which is used to
pass from the second line of~\thetag{4.18} to the third line.
Examining~\thetag{4.12} we see that there is a new term from the even-odd
curvature of~$\scrA$ in passing to the last line.  Hence~\thetag{4.11} is
replaced by\footnote{Recall a tricky sign in~\CF{4.2}: An equivariant
vertical vector field is identified with a section of the adjoint bundle, and
the action on a section of an associated bundle is by {\it minus\/} the
directional derivative.}
  $$ \DD^2\DDa\Phi = \epsilon ^{b c }\bigl(\ddab -
     \frac 16 R(\DDa\Phi ,\DDb\Phi ) \bigr)\DDg \Phi + 2(\Lambda _a)_\Phi ,
     \tag{7.29} $$
where $(\Lambda _a)_{\Phi }$~is the vector field on~$X^\scrP$ induced by the
action of~$\Lambda _a$ on~$X$.  The new term in~\thetag{7.29} contributes an
extra term to~\thetag{4.18}, and the new contribution to the component
lagrangian is
  $$ -\epsilon ^{ab}\langle \psi _a ,(\lhat _b)_\phi \rangle = -\langle
     \pa,\lhat^a_\phi \rangle= \langle \lhat^a_\phi,\pa \rangle. \tag{7.30} $$
The computation~\thetag{4.21} with the potential term remains unchanged.
Altogether, we find the total component lagrangian to be~\thetag{6.36}, and
this completes the proof of \theprotag{6.33(i)} {Theorem}.

\Chapter8{$N=1$ Yang-Mills Theory in Dimension 4}

 \comment
 lasteqno 8@ 51
 \endcomment

 Here we complete the discussion initiated in~\S{6.3} of the supersymmetric
formulation of super Yang-Mills theory in dimension~4.  This chapter
parallels the previous one.  One new feature in~\S{8.1} is the appearance of
an auxiliary field as well as the existence of an R-symmetry.  Another new
feature appears in~\S{8.2}, where the coupling constant of the Yang-Mills
lagrangian is complex.  Its imaginary part is proportional to the usual
coupling constant, whereas its real part is the coefficient of a Chern-Weil
form.  (If this term is present, then the theory is not invariant under
orientation-reversing symmetries.)  In~\S{8.3} we describe the superspace
approach to gauge theories with matter in case the target manifold of the
scalar field is Hodge.  Then there is a geometric explanation for the
appearance of the moment map in the component lagrangian~\thetag{6.42}.

\Head{8.1}{Constrained connections on~$\mink44$}
We specialize the discussion in~\S{6.3} to~$\mink44$.  As explained
in~\S{2.4} we work in a complex basis.  Our first task is to
evaluate~\thetag{6.74}.  To that end we compute 
  $$ \split
      g_{\mu \nu }\stov^{\mu a\cd}\sdtov^\nu _{b\dd}&= g_{\mu \nu }\epsilon
     ^{aa'}\epsilon ^{\cd\cd'}\sdtov^\mu _{a'\cd'}\sdtov^\nu _{b\dd} \\
      &= \frac 12\epsilon _{a'b}\epsilon _{\cd'\dd}\epsilon ^{aa'}\epsilon
     ^{\cd \cd'} \\
      &= 2\delta ^a_b\delta ^{\cd}_{\dd}\endsplit \tag{8.1} $$
Plugging into~\thetag{6.74} we obtain 
  $$ \DDb\Lambda ^a = -\frac 12\stov^{\mu ac}\sdtov^\nu _{cb}\scrF_{\mu \nu
     }- \frac 12\delta ^a_b\DDgd\Lbar^{\cd}. \tag{8.2} $$
Thus there is an auxiliary field in this theory.  Setting~$a=b$ and summing,
the first term on the right hand side of~\thetag{8.2} vanishes, and so
  $$ \DDg\Lambda ^c = -\DDgd\Lbar ^{\cd}. \tag{8.3} $$
Define the (auxiliary) field 
  $$ E = -\frac{\sqo}{2}\DDg\Lambda ^c. \tag{8.4} $$
Equation~\thetag{8.3} asserts that $E$~is real:
  $$ E = \overline{E}. \tag{8.5} $$
From~\thetag{6.74} we also learn that
  $$ \DDbd\Lambda ^a=0; \tag{8.6} $$
in other words, $\Lambda ^a$~is a chiral superfield. 
 
Let $A,\lambda ,E$ be the restrictions of $\scrA,\Lambda ,E$ to~$\bmink^4$.
(It should cause no confusion that the restriction of~$E$ to~$\bmink^4$ is
also denoted~`$E$'.)  The multiplet~$\{A,\lambda ,E\}$ of component fields is
called the {\it vector multiplet\/}, and from the vector multiplet we can
reconstruct the constrained connection~$\scrA$.  The precise statement is the
following.

        \proclaim{\protag{8.7} {Proposition}}
  The category of connections~$\scrA$ on~$\mink44$ whose curvature vanishes
on the odd distribution~$\tau $ is equivalent to the category of
triples~$\{A,\lambda,E\}$ consisting of a connection~$A$ on~$\bmink^4$, a
dual real spinor field~$\lambda $ with values in the adjoint bundle, and a
section~$E$ of the adjoint bundle.
        \endproclaim

\flushpar 
 For an elucidation of what \theprotag{8.7} {Proposition} claims, we refer to
the comments following \theprotag{7.5} {Proposition}.  The following proof
contains a description of an unconstrained field equivalent to a constrained
connection.

        \demo{Proof}
 We first consider the complex analog of \theprotag{8.7} {Proposition}.
For the complex super Minkowski space $M\cp=M^{4\mid 4}_{\Bbb C}$, instead of a
complex structure on the distribution $\tau$, we get a decomposition $\tau =
\tau'\oplus \tau''$, and the distributions $\tau'$ and $\tau''$ are
integrable.  Dividing by the corresponding foliations, we obtain
  $$ M''\overset \text{pr}''\to{\!\!\!\leftarrow} M\cp \overset
     \text{pr}'\to{\!\!\!\rightarrow} M', \tag{8.8} $$
with $\tau'$ (resp $\tau''$) being the relative tangent bundle
of $M\cp/M''$ (resp $M\cp/M'$).  Let $\check M\cp\subset M\cp$ be the
ordinary Minkowski space, with images $\check M'$ in $M'$ and
$\check M''$ in $M''$.  

Put $M_1 = (\text{pr}^{\prime\prime})\inv (\check M'')$ and $M_2: = (\text{pr}^{\prime})\inv (\check M')$.  
The projections $\text{pr}'$ and $\text{pr}''$ induce isomorphisms
  $$ M_1\overset\sim\to{\rightarrow} M' \tag{8.9} $$
and
  $$ M_2\overset\sim\to{\rightarrow}M''.  \tag{8.10} $$

 \midinsert
 \bigskip
 \centerline{
  \epsfxsize=2in
  \epsfysize=2in
 \epsffile{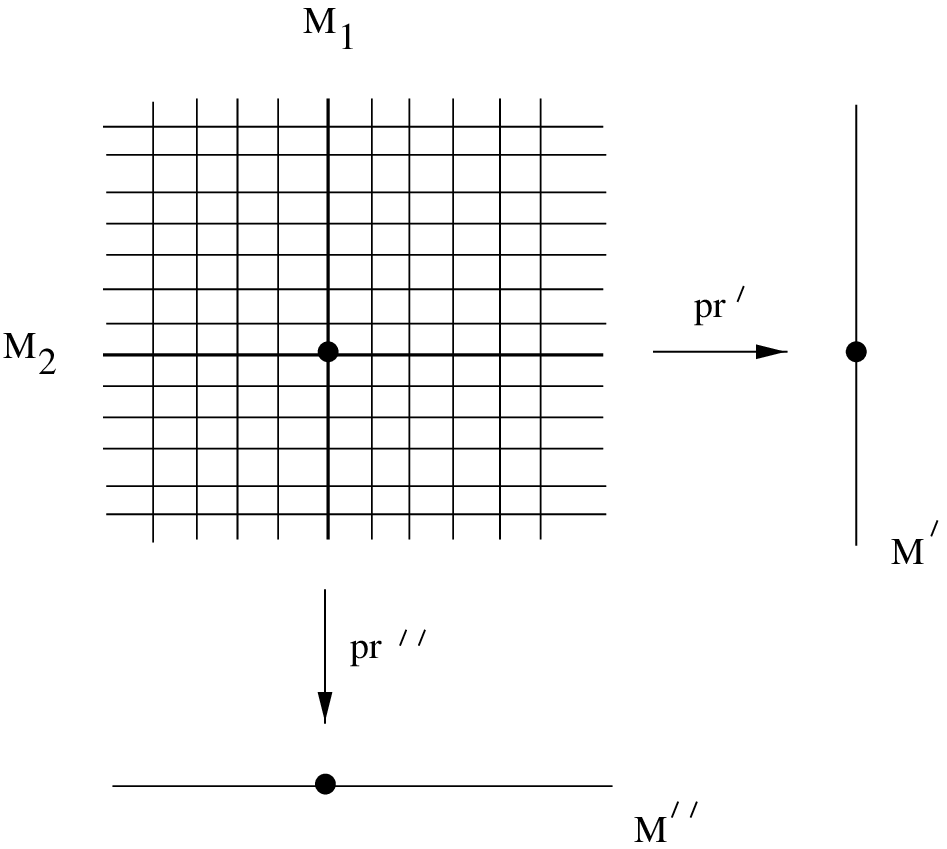}}
% \nobreak
% \botcaption{Figure~: }
% \endcaption
 \bigskip
 \endinsert

If $P$ is a $G$-torsor on $M\cp$, a constrained connection on $\Cal P$ is
determined by its restriction to $\tau$ and, arguing as in the proof of
\theprotag{7.5} {Proposition}, one sees that constrained connections
correspond one to one to connections in the direction of $\tau$, which are
integrable in the directions of $\tau'$ and $\tau''$.  An integrable
connection in the direction of $\tau''$ (resp $\tau'$) amounts to a way to
descend $\Cal P$ from $M\cp$ to $M'$ (resp $M''$).  A constrained connection
hence amounts to the data of $\Cal P'$ on $M'$ and $\Cal P''$ on $M''$
descending $\Cal P$:
  $$ \Cal P\overset\sim\to{\rightarrow} (\text{pr}')\inv (\Cal P')  \tag{8.11} $$
and
  $$  \Cal P\overset\sim\to{\rightarrow} (\text{pr}'')\inv (\Cal P''). \tag{8.12} $$
In the $\tau'$ (resp.~$\tau''$) direction, the connection is
given by~\thetag{8.12} (resp.~\thetag{8.11}).

Let $P$ be the restriction of $\Cal P$ to $\check M\cp$, and let us choose a
trivialization of $P$.  From \thetag{8.12} (resp \thetag{8.11}) we get a
trivialization of $\Cal P$ on $M_1$ (resp $M_2$): the only one extending the
given trivialization and compatible with the connection along $\tau'$ (resp
$\tau''$).  From \thetag{8.9} (resp~\thetag{8.10}) and \thetag{8.11}
(resp~\thetag{8.12}), we then get trivializations $s'$ and $s''$ of $\Cal P'$
and $\Cal P''$.

We keep noting $s'$ and $s''$ their inverse images in $\Cal P$ and let
$g\:M\cp\to G$ be the map such that $s'g = s''$.  The map $g$ to $G$, trivial
on $M_1$ and $M_2$, is an unconstrained field equivalent to the constrained
connection.  It can be written $g = \exp (U)$, with $U$ a map to $\frak{g} =
\operatorname{Lie}(G)$ vanishing on $M_1$ and $M_2$.  In terms of $g$, the
data $(\Cal P,\Cal A)$ is the trivial torsor, with the trivial connection in
the direction of $\tau''$ and, in the direction of $\tau'$, the connection
for which $g$ is horizontal (given by the 1-form $-g^{-1}dg$).

Let us use coordinates as in \S{2.4}.  In those coordinates,
$M_1$ is defined by $\bar\theta^{\dot b} = 0$, and $M_2$ by
$\theta^a = 0$.  Let $J$ be the ideal of functions 
vanishing on $M_1$ and $M_2$, and $I$ be the ideal generated by all
$\theta$ and $\bar\theta$.  Then,
$(J\cap I^n)/J\cap I^{n+1}$ vanishes for $n\not= 2,3,4$ and is spanned
respectively by
  $$ \align &n=2: \text{ the }\theta^a\bar\theta^{\dot b}\\ &n = 3: \text{
     the } \theta^1\theta^2\bar\theta^{\dot b} \text{ and the
     }\theta^a\bar\theta^{\dot 1}\bar\theta^{\dot 2}\\ &n = 4:\;
     \theta^1\theta^2\bar\theta^{\dot 1}\bar\theta^{\dot 2}, \endalign $$
the coefficients being functions on $\check M\cp$.  We have 
  $$ \aligned
      \Cal D_a&= D_a,\\ 
       \Cal D_{\dot a} &= D_a-g^{-1}D_ag.\endaligned \tag{8.13} $$
It is then easy to check that $U\pmod I^3$ gives the connection on $P$; that
changing $U$ by a section of $(J\cap I^3)\otimes \frak{g}$, taken modulo
$J\cap I^4$, gives $\lambda$; and changing $U$ by a section of $(J\cap
I^4)\otimes \frak{g}$ gives $E$.

We now consider the real case.  The arguments used in the complex case work
in just the same way for the cs case, where super Minkowski space is
complexified just in the odd direction.  As explained in the comments after
\theprotag{7.5} {Proposition}, we have to prove that some map is a bijection.
This map is obtained from the similar map in the cs case by taking the fixed
points of complex conjugation, and \theprotag{8.7} {Proposition} follows.
        \enddemo
\flushpar
 In the physics literature the field $U$ is referred to as the vector
superfield in ``Wess-Zumino gauge''.
 
It is convenient to lower an index on the spinor field and define the chiral
superfield\footnote{The symbol `$W$' is standard in the physics literature.}
  $$ \boxed{W_a = \Lambda ^b\epsilon _{ba} = \frac 14\epsilon
     ^{\bd\cd}[\ddabd,\DDgd]} \tag{8.14} $$
together with the conjugate antichiral superfield~$\overline{W}_{\ad}$.
Equation~\thetag{6.73} specializes to 
  $$ [\DDa,\nab{b\cd}] = 2\epsilon _{ab}\Wbar_{\cd}. \tag{8.15} $$
The auxiliary field~$E$ is
  $$ E = -\frac{\sqo}{2}\epsilon ^{ab}\bres\DDa W_b. \tag{8.16} $$
The reality condition~\thetag{8.5} (equivalently~\thetag{8.3}) is 
  $$ \eab\DDa\Wb + \eabd\DDad\Wbar_{\bd} =0 \tag{8.17} $$
This has the useful consequence
  $$ \DD^2\Wa = -2\epsilon ^{\bd\gd}\ddabd\Wbar_{\gd}. \tag{8.18} $$

It is convenient to write the curvature as
  $$ F_A := \frac 12 (F_{a b }\epsilon _{\gd\dd} +
     \Fb_{\gd\dd}\epsilon _{a b }) \;dy^{a\gd}\wedge dy^{b\dd}, \tag{8.19} $$
where $F_{ab}$~is symmetric in~$a,b$.  Our convention is that the curvature
is real---$\Fb_{\ad\bd} = \overline{F_{a b }}$---for a compact gauge
group~$G$.  {\it Warning}: The involution on the complexified Lie
algebra~$\frak{g}\cp$ is such that the real subalgebra is the Lie algebra of
the compact group.  Thus for $G=\TT$ the circle group the real subalgebra
in~$\frak{g}\cp=\CC$ is~$i\RR\subset \CC$.

In these variables the component fields are 
  $$ \boxed{\aligned
      A &= \bres\scrA \\ 
      \lambda _a &= \bres W_a,\endaligned} \tag{8.20} $$
and from~\thetag{8.2} and~\thetag{8.16} we have 
  $$ \boxed{F_{ab} + \sqo\,\epsilon _{ab}E = \bres\DDa W_b.} \tag{8.21} $$
Let $\hxi$~be the vector field on the space of component fields corresponding
to the supersymmetry transformation $\eta ^aQ_a + \etad\Qad$.  We compute it
from the general formula~\thetag{1.29} and the other formulas in this
section.  We also can check the result against~\thetag{6.8} and~\thetag{6.9}
if we set the auxiliary field~$E$ to zero.  The result is: 
  $$ \boxed{\aligned
      (\hxi A)_{a \bd } &= -2\eta ^c \,\epsilon _{a c }\,\lambdab_{\bd} -
     2\etgd\,\epsilon _{\bd\gd}\,\lambda _{a }\\
      \hxi \lambda _a &= -\eta ^b F_{a b} + \eta ^b \epsilon _{a b }\sqo E \\
      \hxi E &= \hphantom{-}\eta ^a\sqo (\Dirac\lambdab)_a -
     \etabar^{\ad}\sqo\,(\Dirac\lambda )_{\ad}\endaligned} \tag{8.22} $$
and 
  $$ \boxed{\hxi F_{a b} = \etgd\bigl\{ \nab{a \gd}\lambda _b + \nab{b
     \gd}\lambda _a \bigr\} +\eta ^c \bigl\{ \epsilon _{ac
     }(\Dirac\lambdab)_b+\epsilon _{bc }(\Dirac\lambdab)_{a } \bigr\}.}
     \tag{8.23} $$

Under the R-symmetry~\thetag{2.81} the component fields transform by
(see~\thetag{8.14} and~\thetag{8.16})
  $$ \boxed{\aligned
      [R,A_{a \bd}] &= 0 \\
      [R,\lambda _a ] &= -\lambda _a \\
      [R,\lambdab _{\ad} ] &= \lambdab _{\ad} \\
      [R,E] &=0.\endaligned} \tag{8.24} $$

\Head{8.2}{The Yang-Mills action on~$\mink44$}
We assume $G$~compact with $\langle \cdot ,\cdot \rangle$ a bi-invariant inner
product on~$\frak{g}$.  Extend~$\langle \cdot ,\cdot \rangle$ to be a {\it
bilinear\/} form on~$\frak{g}\cp$.  Fix~$\tau \in \CC$.  Then for a
constrained connection on~$\mink44$ the basic lagrangian density is 
  $$ \boxed{\aligned
      \scrL_2 &= |d^4x|\;\Bigl\{-\frac {1}{16\pi }\Im\Bigl(d^2\theta\,\tau\,
     \epsilon ^{a b }\langle W _a ,W_b \rangle\Bigr)\Bigr\} \\
      &= |d^4x|\;\frac{1}{2i}\Bigl\{ d^2\theta \,\frac{-\tau }{16\pi
     }\eab\langle W_a ,W_b \rangle + d^2\bar{\theta}
     \,\frac{\bar{\tau}}{16\pi }\eabd\langle \Wbar_{\ad} ,\Wbar_{\bd} \rangle
     \Bigr\}.\endaligned} \tag{8.25} $$
Notice that the two terms in~$\scrL_2$ are defined on different
cs-manifolds.  We compute the contribution of the first to the component
lagrangian using~\thetag{8.20} and \thetag{8.21}:
  $$ \split
      \frac{\tau }{16\pi }\eab\bres D^2\langle W_a ,W_b \rangle &=
     \frac{\tau }{16\pi }\eab\bres\Bigl\{ 2\langle \DD^2W_a ,W_b
     \rangle + \egd\langle \DDd W_a ,\DDg W_b \rangle\Bigr\} \\
      &= -\frac{\tau }{4\pi }\eab\egdd\bres\langle \nab{a
     \gd}\Wbar_{\dd},\Wb \rangle \\ 
      &\qquad \qquad \qquad + \frac{\tau }{16\pi }\eab\egd\langle
     F_{d a }+\epsilon _{d a }\sqo E,F_{c b
     }+\epsilon _{c b }\sqo E \rangle \\
      &= \frac{\tau }{4\pi }\langle \lambdab\;\Dirac_A\;\lambda \rangle -
     \frac{\tau }{16\pi }\eab\egd\langle F_{a c },F_{b d }
     \rangle + \frac{\tau }{8\pi }|E|^2.\endsplit \tag{8.26} $$
The contribution of the second term in~\thetag{8.25} is similar and we
combine them using~\thetag{2.68}.  We fix the standard orientation
on~$\bmink^4$ to write the result in terms of differential forms:
  $$ d^4x\;\scrLch_2 = d^4x\;\Bigl\{ -\frac{\Im\tau }{8\pi }|F_A|^2 +
     \frac{\Im\tau }{4\pi }\langle \lambdab\;\Dirac_A\;\lambda   \rangle +
     \frac{\Im\tau }{8\pi }|E|^2 \Bigr\} + \frac{\Re\tau }{8\pi }\langle
     F\wedge F  \rangle. \tag{8.27} $$
Introduce the notation 
  $$ \boxed{\tau =\frac{\theta }{2\pi } + \frac{4\pi i}{g^2}.} \tag{8.28} $$
Then 
  $$ \boxed{ d^4x\;\scrLch_2 = d^4x\;\frac{1}{g^2}\,\bigl\{ -\frac 12 |F_A|^2
     + \langle \lambdab\;\Dirac_A\;\lambda   \rangle + \frac{1}{2}|E|^2
     \bigr\} + \frac{\theta }{16\pi ^2}\langle F_A\wedge F_A  \rangle.}
     \tag{8.29} $$
This is essentially the super Yang-Mills lagrangian~\thetag{6.3}.  We have
written the second term in complex notation and dropped an exact term along
the way.  Note $E$~enters as an auxiliary field; its equation of motion
is~$E=0$.  The last term is {\it topological\/}; the integral is a
characteristic number (for connections of finite action).  Note that the
first term makes sense as a density whereas to integrate the topological term
we need an orientation on~$\bmink^4$.  Another way to express~\thetag{8.29}
is to write $F=F^+ + F^-$ in terms of its self-dual and anti-self-dual pieces
(see~\thetag{2.64}).  Then the lagrangian function is
  $$ \boxed{\scrLch_2 = -\frac{\tau }{4\pi }|F_A^+|^2 + \frac{\bar{\tau
     }}{4\pi }|F_A^-|^2 + \frac{\Im\tau }{4\pi }\langle
     \lambdab\;\Dirac_A\lambda \rangle + \frac{\Im\tau }{8\pi }|E|^2.}
     \tag{8.30} $$
The orientation on~$\bmink^4$ enters here in distinguishing self-dual from
anti-self-dual.
 
We are following standard physics usage in our choice of constants.  The
closed 4-form $\langle F_A\wedge F_A \rangle$~represents a real
characteristic class of $P\to\bmink^4$; if the bilinear form~$\langle \cdot
,\cdot \rangle$ on~$\frak{g}$ is such that the real characteristic class
$\frac{1}{16\pi ^2}\langle F_A\wedge F_A \rangle$ comes from a class
in~$H^4(BG;\ZZ)$, then $\theta $~and $\theta +2\pi $~yield the same quantum
theory.
 
Next, we consider the action of the R-symmetry~\thetag{2.81}.  Now $d^2\theta
$~has weight~2 under~$R$ and from~\thetag{8.14} we see that $\Wa$~has
weight~$-1$.  It follows that \thetag{8.25}~is invariant under~$R$.  One can
also use~\thetag{8.24} to check that \thetag{8.29}~is invariant under the
R-symmetry. We remark that even though the classical theory has a $U(1)$
R-symmetry, in the quantum theory this is broken to a finite subgroup by an
anomaly (which is computed using the index theorem).

We do not analyze this theory in superspace, but rather simply report the
equations of motion:
  $$ \boxed{\aligned
      \epsilon ^{cd}\nab{d\bd}F_{ac} + \epsilon ^{\cd\dd}\nab{a\dd}F_{\bd\cd}
     &= [\lambda _a,\lambdab_{\bd}] \\ 
      \epsilon ^{ab}\nab{a\cd}\lambda _b&=0 \\ 
      E &=0\endaligned} \tag{8.31} $$
and the supercurrent~$j_a$ which corresponds to the supersymmetry
transformation~$Q_a$:
  $$ \boxed{\aligned
      j_a = \frac 12\epsilon ^{cd}\epsilon ^{\bd\dot e}&\langle
     F_{ca},\lambdab_{\dot e} \rangle\;\frac{\partial }{\partial
     y^{d\bd}}\otimes |d^4y| \\
      &+ \Bigl\{\frac 12\epsilon ^{\cd\bd}\epsilon ^{\dd\dot e}\langle
     \Fbar_{\cd\dd},\lambdab_{\dot e} \rangle -\epsilon ^{\bd\cd}\langle
     \sqo\, E,\lambdab_{\cd} \rangle \Bigr\}\;\frac{\partial }{\partial
     y^{a\bd}}\otimes |d^4y|.\endaligned} \tag{8.32} $$
In this formula we take $g=1,\,\theta =0$.

\Head{8.3}{Gauge theory with matter on~$\mink 44$}
In this subsection we prove \theprotag{6.33(ii)} {Theorem}.  Suppose we are
given the data in~\thetag{6.41}, but with the additional proviso that $X$~be
Hodge.  This assumption is only for exposition, as the computation we review
is local and done for arbitrary \Kah\ manifolds.  Let $L_\omega\to X$ be the Hodge
line bundle, a hermitian line bundle with curvature~$\sqo\omega $.  In this
theory a compact Lie group~$G$ acts by isometries on~$X$, and we assume given
a lift of the $G$~action to~$L_\omega$.  This is roughly equivalent to choosing a
moment map
  $$ \mu \:X\longrightarrow \frak{g}^* \tag{8.33} $$
for the action of~$G$ on~$X$, and in fact defines a moment map by the formula
  $$ \mu (\zeta) = -\sqo \vrt_1(\Hat{\Hat\zeta }),\qquad \zeta \in
     \frak{g}. \tag{8.34} $$
Here $ \mu (\zeta) $~is the function on~$X$ obtained by pairing~\thetag{8.33}
with~$\zeta $; the vector field~$\Hat{\Hat\zeta }$ on~$L_\omega$ is defined
by the lifted action; and `$\vrt_1$'~indicates the vertical part with respect
to the connection on~$L_\omega$, taken at points of unit norm.  It is easy to
check that $\mu $~is a moment map.  In other words
  $$ d \mu(\zeta) = -\cont{\hat\zeta }\omega , \tag{8.35} $$
where $\hat\zeta $~is the vector field on~$X$ induced by the $G$~action.  In
case $X$~is K\"ahler but not Hodge, we must choose a moment map.  We also fix
a $G$-invariant holomorphic function $W\:X\to\CC$.
 
Fix a $G$~bundle $\scrP\to\mink44$.  The superfields in the theory are a
constrained connection~$\scrA$ on~$\scrP$ and a chiral section~$\Phi $ of the
associated bundle~$X^{\scrP}$, which we view as an equivariant chiral map 
  $$ \Phi \:\scrP\longrightarrow X. \tag{8.36} $$
As in~\thetag{5.21} there is a global lift to an equivariant chiral map 
  $$ \Phitil\:\scrP\longrightarrow \Phi ^*L_\omega. \tag{8.37} $$
Then the total superspace lagrangian for the theory is~$\scrL_0 + \scrL_1 +
\scrL_2$, where the individual terms are defined in~\thetag{5.22},
\thetag{5.32}, and~\thetag{8.25}.  We take the coupling constants in the
gauge theory to be~$g=1$ and~$\theta =0$.  Our task is to work out the
expansion in component fields.  Note that component fields are defined
by~\thetag{5.7}, though now the vector fields which enter the definitions are
the $\DDa$ on~$\scrP$---the horizontal lifts of the~$\Da$ defined by the
connection~$\scrA$.
 
We re-examine the computation beginning in~\thetag{5.25}.  Since this uses
the alternative form~\thetag{5.17} of the lagrangian in terms of a local
\Kah\ potential~$K$, which we assume is $G$-invariant, our first job is to
express a moment map~$\mu $ in terms of~$K$.  Let $\zhat$~be the real
vector field on~$X$ which corresponds to~$\zeta \in \frak{g}$.  Then an easy
computation from~\thetag{8.35} shows
  $$ \langle \chmu,\zeta   \rangle= \mu(\zeta) = -\sqo\,\zhat^{1,0}\cdot K =
     \sqo\,\zhat^{0,1}\cdot K \tag{8.38} $$
defines a moment map, where $\zhat =\zhat ^{1,0}+\zhat ^{0,1}$ is expressed
as a sum of its $(1,0)$~and $(0,1)$~components and $\chmu\:X\to\frak{g}$ is
obtained from~$\mu $ using the inner product on~$\frak{g}$.
 
There are several new terms in~\thetag{5.25}--\thetag{5.27} from the nonzero
bracket~\thetag{8.15}.  First, in the third line of~\thetag{5.25} there is a
new contribution from the bracket term in~\thetag{5.24}:
  $$ \split
      \bres\biggl(\frac 14\epsilon ^{ab}\epsilon
     ^{\cd\dd}&\DDgd\Bigl\{2\cont{[\DDb,\nab{a\dd}]}\Phi ^*\partial
     K\Bigr\}\biggr) \\
      &= -2\bres \epsilon ^{\cd\dd}\DDgd\,\cont{\Wbar_{\dd}}\Phi ^*\partial K
     \\
      &= -2\bres \epsilon ^{\cd\dd}\iota \bigl([\DDgd,\Wbar_{\dd}] \bigr)\Phi
     ^*\partial K + 2\bres\epsilon
     ^{\cd\dd}\cont{\DDgd}\cont{\Wbar_{\dd}}\Phi ^*\dbar\partial K \\
      &= 4\bres\sqo\,\iota \bigl(-\hat{E}_\Phi \bigr)\partial K +
     2\bres\epsilon ^{\cd\dd}\cont{\DDgd\Phi }
     \iota\bigl(-(\widehat{\Wbar}_\Phi )_{\dd} \bigr)\dbar\partial K \\
      &= 4\langle \phi ^*\chmu,E \rangle + \sqrt2\,\epsilon ^{\cd\dd}\langle
     \pgd,(\hat{\lambdab}_\phi )_{\dd} \rangle ,\endsplit \tag{8.39} $$
where $(\widehat{\Wbar}_{\dd})_\Phi $ is the vector field on~$X^{\scrP}$
induced by~$\Wbar_{\dd}$ and similarly for~$E$.  In the third line we
use~\thetag{5.23} and the fact that $\cont{\DDgd }\Phi^*\partial K$~is a
$G$-invariant function on~$\scrP$, so its derivative in the fiber
direction~$\Wbar_{\dd}$ vanishes; in the fourth the definition~\thetag{8.16}
of~$E$ and the fact that vertical vectors on~$\scrP$ act with a minus sign
(cf.~the footnote in~\S{7.3}); and in the fifth \thetag{8.38}
and~\thetag{5.16}, together with an extra minus sign due to the odd
variables.\footnote{For even vectors~$\xi _1,\bar\xi_2$ of type~$(1,0),(0,1)$
respectively, equation~\thetag{5.16} asserts
  $$ 2\cont{\xi _1}\cont{\bar\xi _2}\dbar\partial K = -2\cont{\bar\xi
     _2}\cont{\xi _1}\dbar\partial K = \langle \bar\xi _2,\xi _1
     \rangle. \tag{8.40} $$
Now substitute $\xi _i=\eta _i\psi _i$, where $\zeta _i$~are odd vectors and
$\eta _i$~are auxiliary odd parameters.}
From~\thetag{5.28} and~\thetag{5.29} we see that the results of
\thetag{5.25}--\thetag{5.27} enter the component lagrangian with a
factor~$1/4$ and are added to their complex conjugates.  Hence~\thetag{8.39}
leads to a contribution
  $$ 2\langle \phi ^*\chmu,E \rangle + \frac{\sqrt2}{4}\Bigl(\epsilon
     ^{cd}\langle (\lhat_\phi )_c,\psi _d \rangle + \epsilon ^{\cd\dd}\langle
     \pgd,(\hat{\lambdab}_\phi )_{\dd} \rangle \Bigr) \tag{8.41} $$
to the component lagrangian.  Also, in the second line of~\thetag{5.27}
there is a new term
  $$ \split
      \epsilon ^{ab}\epsilon ^{\cd\dd}\Bigr\{-\DDa\,\cont{[\DDgd,\nab{b\dd}]
     }\Phi^*\partial K\Bigr\} &= -4\eab\DDa\,\cont{W_b }\Phi^*\partial K,
     \endsplit \tag{8.42} $$
which by a similar computation contributes 
  $$ -4\langle \phi ^*\chmu,E \rangle - \frac{\sqrt2}{2}\Bigl(\epsilon
     ^{cd}\langle (\lhat_\phi )_c,\psi _d \rangle + \epsilon ^{\cd\dd}\langle
     \pgd,(\hat{\lambdab}_\phi )_{\dd} \rangle \Bigr) \tag{8.43} $$
to the component lagrangian.  

As in~\S{7.3} there is also a new term due to a revision of~\thetag{5.9}:
  $$ \DDad \DD^2 \Phi = - 2\epsilon ^{b c
     }\Bigl\{\nabla _{b \dot{a} }\Phi - \frac 14R(\DDb\Phi ,\DDad\Phibar
     )\Bigr\}\DDg\Phi - 4(\widehat{\Wbar}_{\ad})_{\Phi }, \tag{8.44} $$
This enters the computation
in~\thetag{5.26}, where we pick up
  $$ \frac 14\epsilon ^{\cd\dd}\langle \DDdd\Phibar
     ,4(\widehat{\Wbar}_{\cd})_{\Phi } \rangle = -\epsilon ^{\cd\dd}\langle
     \DDgd\Phibar , (\widehat{\Wbar}_{\dd})_{\Phi } \rangle. \tag{8.45} $$
Another new term enters into the last line of~\thetag{5.27} from
commuting~$\DDa$ past~$\nab{b\dd}$: 
  $$ \eab\epsilon ^{\cd\dd}\Bigl\{\frac 12\langle \DDgd\Phibar
     ,[\DDa,\nab{b\dd}]\Phi \rangle\Bigr\} = -2\epsilon ^{\cd\dd}\langle
     \DDgd\Phibar , (\widehat{\Wbar}_{\dd})_{\Phi } \rangle. \tag{8.46} $$
Taking into account the factor~$1/4$ in~\thetag{5.28}, and
adding~\thetag{8.45} and~\thetag{8.46} to their complex conjugates, we obtain
a contribution to the component lagrangian
  $$ -\frac{3\sqrt2}{4} \Bigl(\epsilon ^{cd}\langle (\lhat_\phi )_c,\psi _d
     \rangle + \epsilon ^{\cd\dd}\langle \pgd,(\hat{\lambdab}_\phi )_{\dd}
     \rangle \Bigr) . \tag{8.47} $$
Combining~\thetag{8.41}, \thetag{8.43}, and~\thetag{8.47} we obtain the total
new contribution to the component lagrangian: 
  $$ -2\langle \phi ^*\chmu,E \rangle - \sqrt2\Bigl(\epsilon ^{cd}\langle
     (\lhat_\phi )_c,\psi _d \rangle + \epsilon ^{\cd\dd}\langle
     \pgd,(\hat{\lambdab}_\phi )_{\dd} \rangle \Bigr) \tag{8.48} $$
The second term appears in~\thetag{6.42}, after using~\thetag{8.14}.
 
To recover the potential energy term in~\thetag{6.42} we must eliminate the
auxiliary field~$E$.  The terms involving~$E$ in~\thetag{8.29}
and~\thetag{8.48} are:
  $$ \frac 12 |E|^2 -2\langle \phi ^*\chmu ,E  \rangle. \tag{8.49} $$
The classical equation for~$E$ is then
  $$ \boxed{E=2\phi ^*\chmu }, \tag{8.50} $$
and eliminating~$E$ we obtain the potential term
  $$ -2\phi ^*|\mu |^2. \tag{8.51} $$
This completes the proof of \theprotag{6.33(ii)} {Theorem}.  Note the
substitution ~\thetag{8.50} which gives the last equation in~\thetag{6.43}
from the second equation in~\thetag{8.22}.

\Chapter9{$N=2$ Yang-Mills in Dimension 2}

 \comment
 lasteqno 9@ 29
 \endcomment

The theory we study in this chapter is the reduction to two dimensions of the
basic four-dimensional supersymmetric gauge theory studied in~\S{8}.  Thus we
begin in~\S{9.1} with some general remarks about dimensional reduction of
ordinary bosonic Yang-Mills.  The crucial observation is that in the
dimensionally reduced theory we have not only a connection, but also some
scalar fields and a nontrivial potential energy.  The dimensional reduction
of a constrained connection on~$\mink44$, which is described in~\S{9.2}, may
be considered as a connection on~$\mink24$ with constraints.  But the
constraints now allow some nonzero curvature along the odd distribution,
unlike the constraint condition in the maximal dimension~4.  Recall also that
the dimensionally reduced theory has a bigger R-symmetry group.  In~\S{9.3}
we derive a superspace expression for the Yang-Mills lagrangian.

\Head{9.1}{Dimensional reduction of bosonic Yang-Mills}
Consider pure Yang-Mills theory on Minkowski space~$\bmink^n$ in
$n$~dimensions with gauge group~$G$.  Fix a bi-invariant inner
product~$\langle \cdot ,\cdot \rangle$ on~$\frak{g}$.  The field is a
connection~$A$ on a principal $G$~bundle $P\to\bmink^n$, and the lagrangian
is
  $$ L = |d^nx|\;\frac {-1}2 |F_A|^2. \tag{9.1} $$
Let $x^0,x^1,\dots ,x^{n-1}$ be standard coordinates on~$\bmink^n$ so that
the metric is 
  $$ g = (dx^0)^2 - (dx^1)^2 -\dots -(dx^{n-1})^2. \tag{9.2} $$
Let~$T$ be the group of translations in the direction of the last $k$~basis
vectors.  Let $p$~be the quotient map $p\:\bmink^n\to\bmink^n/T$.  The
quotient~$\bmink^n/T$ is a Minkowski space of dimension~$n-k$.  If
$\bmink^{n-k}\subset \bmink^n$ is defined by~$x^{n-k}=\dots =x^{n-1} =0$, the
map~$p$ identifies~$\bmink^{n-k}$ with~$\bmink^n/T$. 
 
In the dimensional reduction of pure Yang-Mills theory, one considers: (a)\ a
principal $G$~bundle~$P$ on $\bmink^n/T\cong \bmink^{n-k}$, and (b)\ a
$T$-invariant connection~$A$ on its pullback to~$\bmink^n$.  Let $a$~be the
restriction of~$A$ to~$\bmink^{n-k}$.  Being $T$-invariant, the
connection~$A$ can be written as 
  $$ A = p^*(a) + \sum\limits_{n-k}^{n-1} p^*(\phi _i)dx^i, \tag{9.3} $$
the $\phi _i$~being sections of the adjoint bundle. 
 
The lagrangian density~$\scrL_n$ on~$\bmink^n$ for~$A$ is $T$-invariant.  In
the reduced theory, the lagrangian density~$\scrL_{n-k}$ on~$\bmink^{n-k}$ is
obtained by restricting~$\scrL_n$ to~$\bmink^{n-k}$ and contracting it
against~$\partial _{n-k},\dots ,\partial _{n-1}$:  for $\Lambda $~a lattice
in~$T$,  
  $$ \int_{\bmink^{n-k}}\scrL_{n-k} = \voll(T/\Lambda )\inv
     \int_{\bmink^n/\Lambda }\scrL_n. \tag{9.4} $$

The curvatures ~$F_A$ of~$A$ and $F_a$ of~$a$ are related by 
  $$ F_A = p^*(F_a) +  \nabla _a(\phi _i)dx^i + \sum\limits_{i<j}[\phi
     _i,\phi _j]\,dx^i\wedge dx^j, \tag{9.5} $$
giving for the lagrangian density of the reduced theory 
  $$ \scrL_{n-k} = |d^{n-k}x|\;\Bigl\{ -\frac 12 |F_a|^2 + \frac
     12\sum\limits_{n-k}^{n-1}|\nabla _a(\phi _i)|^2 + \frac
     12\sum\limits_{i<j}|[\phi _i,\phi _j]|^2\Bigr\}. \tag{9.6} $$
The field content of the reduced theory is a connection and $k$~sections of
the adjoint bundle.

Most relevant to us is the reduction to $n-2$~dimensions.  Thus consider a
connection~$A$ invariant by translations in the $x^{n-1},x^{n-2}$ directions
and write 
  $$ A = a + \phi _1dx^{n-2} + \phi _2dx^{n-1}. \tag{9.7} $$
A short computation gives for the reduced lagrangian
  $$ L = |d^{n-2}x|\;\bigl\{ -\frac 12|F_a|^2 + \langle d_a\sigb,d_a\sigma
     \rangle - \frac 12 \|\,[\sigma ,\sigb]\,\|^2\bigr\},
     \tag{9.8} $$
where in the last line we set 
  $$ \sigma = \frac{\phi _1 + \sqo\phi _2}{\sqrt2}. \tag{9.9} $$
Thus $\sigma $~is a section of the {\it complexified\/} adjoint bundle.
Therefore, the field content of the reduction of pure Yang-Mills by two
dimensions is a connection and a {\it complex\/} scalar field in the adjoint
representation.  There is a nonzero potential energy term for the scalar.
 
Consider the reduction of the bosonic part of the 4-dimensional
supersymmetric Yang-Mills lagrangian~\thetag{8.29} to 2~dimensions.  The
kinetic term for the connection reduces as in~\thetag{9.8}.  The auxiliary
field is a scalar, so the reduction is also a scalar.  The topological term
reduces nontrivially, and the total reduced bosonic lagrangian is 
  $$ |d^2x|\;\frac 1{g^2}\Bigl\{-\frac 12|F_a|^2 + \langle d_a\sigb,d_a\sigma
     \rangle- \frac 12 \|\,[\sigma ,\sigb]\,\|^2+\frac 12|E|^2\Bigr\} -
     \sqo\frac{\theta }{8\pi ^2}\langle F_a,[\sigma ,\sigb]
     \rangle. \tag{9.10} $$

\Head{9.2}{Constrained connections on~$\mink2{(2,2)}$}
We use the notation established in~\thetag{2.107}--\thetag{2.119}
for~$\mink2{(2,2)}$.   
 
Let $\scrP\to\mink44$ be a principal bundle with structure group~$G$.  We
suppose that the vector fields~$\partial _{1\dot2},\partial _{2\dot1}$
on~$\mink44$ have been lifted to commuting complex conjugate vector
fields~$\vlo,\vlt$ on~$\scrP$.  Let $\scrA$~be a constrained connection
on~$\scrP$ which is invariant under these translations.
All of the equations in~\S{8.1} remain valid for~$\scrA$, and we use them
freely.  The translation invariance is the equation 
  $$ [\vlo,\DDa] = [\vlt,\DDa] = [\vlo,\DDad] = [\vlt,\DDad]=0.  \tag{9.11} $$
Then the vertical vector field 
  $$ {\Sigma } = \frac1{\sqrt2}(\nab{2\dot1} - \vlt) \tag{9.12} $$
is also translation invariant and corresponds to a section~$\Sigma $ of the
complexified adjoint bundle.  It is the basic invariant of a translationally
invariant constrained connection.  The Bianchi identity relates~$\Sigma $ to
the superfields considered in~\S{8.1}.  First, we consider Bianchi
for~$D_2,D_2,\Dbar_{\dot1}$ and~$D_2,\Dbar_{\dot1},\Dbar_{\dot1}$ to deduce
  $$ \DDbar_{\dot1}\Sigma  = \DD_2\Sigma =0. \tag{9.13} $$
In other words, the restriction of~$\Sigma$ to~$\mink2{(2,2)}$ is {\it
twisted chiral.\/}  Then Bianchi for~$D_1,\Dbar_{\dot1},\Dbar_{\dot2}$ and
Bianchi for~$D_2,\Dbar_{\dot1},\Dbar_{\dot2}$ imply
  $$ \aligned
      \sqrt2W_1 &= -\DDbar_{\dot1}\Sigb \\
      \sqrt2W_2 &= \hphantom{-}\DDbar_{\dot2}\Sigma .\endaligned
     \tag{9.14} $$
Using~\thetag{9.14} and the results of~\S{8.1} we can express the
curvature~$F_{\scrA}$ completely in terms of~$\Sigma $. 
 
Restrict the connection~$\scrA$ and all derived superfields to
$\mink2{(2,2)}\hookrightarrow \mink44$.  Let $i\:\bmink^2\hookrightarrow \mink2{(2,2)}$
denote the inclusion, as usual.  The component fields of the two-dimensional
reduced theory are
  $$ \boxed{\alignedat2
      A &= \bres\scrA\\
      \sigma &= \bres\Sigma \\
      \lambda _+ &= \bres W_+ &&= -\frac 1{\sqrt2}\bres\DDbp\Sigb \\
      \lambda _- &= \bres W_- &&= \hphantom{-}\frac
     1{\sqrt2}\bres\DDbm\Sigma \endaligned} \tag{9.15} $$
We also have 
  $$ \boxed{f_A + [\sigma ,\sigb] + \sqo E = \bres \DD_+W_- = \frac
     1{\sqrt2}\bres\DDp\DDbm\Sigma,   }  \tag{9.16} $$
where the curvature of~$A$ is given in~\thetag{8.19} as
  $$ F_A = 2f_A \;dy^+\wedge dy^-. \tag{9.17} $$
The scalar field~$\sigma $ is a section of the complexified adjoint bundle;
it is the same field which appears in the previous subsection.  So too are
$f_A,E$~sections of the adjoint bundle, but both $f_A$~and $E$~are real.

From the point of view of~$\mink2{(2,2)}$, we can think of~$\scrA$ as a
connection whose curvature is constrained to vanish along the odd
distribution {\it except\/} for the $D_2,\Dbar_{\dot1}$ direction (and its
complex conjugate), where 
  $$ [\DD_2,\DDbar_{\dot1}] = -2\sqrt2\,\Sigma . \tag{9.18} $$

There is an important distinction between the abelian and nonabelian cases.
In the abelian case the superfield~$\Sigma $ has values in the constant
vector space~$\frak{g}$ and is unconstrained, except for the requirement that
it be twisted chiral.  Up to gauge equivalence the connection can be
reconstructed from~$\Sigma $.  In the nonabelian case, on the other hand, the
superfield~$\Sigma $ is a section of the adjoint bundle, and the twisted
chirality condition~\thetag{9.13} depends on the connection~$\scrA$.
 
We can compute the transformation laws of the component fields under $\hxi
=\eta ^a \tQa + \etad\tQad$ directly, or by dimensional reduction
from~\thetag{8.22}.  The result is 
  $$ \boxed{\aligned
      (\hxi A)_+ &= -2\eta ^-\lambdab_+ - 2\etabar^-\lambda _+ \\
      (\hxi A)_- &= \phantom{-}2\etp\lambdab_- + 2\etbp\lambda _- \\
      \hxi \sigma &= \phantom{-}\sqrt2\etp\lambdab_+ - \sqrt2\etbm\lambda _-
     \\
      \hxi \lambda _+ &= -\etm(f_A+ [\sigma ,\sigb] - \sqo E)
     -{\sqrt2}\,{\etp}\nabla _+\sigb \\
      \hxi \lambda _- &= -\etp(f_A+ [\sigma ,\sigb] + \sqo E)
     +{\sqrt2}\,{\etm}\nabla _-\sigma \\
      \hxi f_A &= \phantom{-}\etm\nabla _-\lambdab_+ + \etp\nabla
     _+\lambdab_- + \etbp\nabla _+\lambda _- + \etbm\nabla _-\lambda _+ \\
      \hxi E &= -\sqo (-\etp\nabla _+\lambdab_- + \etm\nabla _-\lambdab_+ +
     \etbp\nabla _+\lambda _- - \etbm\nabla _-\lambda _+).\endaligned}
     \tag{9.19} $$
 
From~\thetag{9.12} it is easy to see that the R-symmetries~$J_+, J_-$
(see~\thetag{2.117}) act on~$\Sigma $ by
  $$ \boxed{[J_{\pm},\Sigma ] = \mp\Sigma .} \tag{9.20} $$
From this and~\thetag{9.15} we deduce the action on the component fields: 
  $$ \boxed{\aligned
      [J_\pm,A] = [J_{\pm},f_A] = [J_\pm,E] &= [J_+,\lambda _+] = [J_-,\lambda
     _-] = 0 \\
      [J_\pm,\sigma ] &= \mp\sigma \\
      [J_\pm,\sigb ] &= \pm\sigb \\
      [J_+,\lambda _-] &= -\lambda _- \\
      [J_-,\lambda _+] &= -\lambda _+.\endaligned}\tag{9.21} $$

\Head{9.3}{The reduced Yang-Mills action}
We rewrite the action~\thetag{8.25} in terms of~$\Sigma $.  First,
using~\thetag{9.14} and~\thetag{9.13} we have 
  $$ \aligned
      \eab\langle \Wa,\Wb  \rangle &= 2\langle W_+,W_-  \rangle \\ 
      &= -\langle \DDbp\Sigb,\DDbm\Sigma   \rangle \\ 
      &= -\Dbp\Dbm\langle \Sigb, \Sigma  \rangle \\ 
      &= -\Dbar^2\|\Sigma \|^2.\endaligned \tag{9.22} $$
Let $\scrLch'_2$ denote the dimensionally reduced component lagrangian
function.  From the second line of~\thetag{8.25} we deduce 
  $$ \aligned
      \scrLch_2' &= \frac{1}{2\sqo}\Bigl\{\int_{}d^2\theta \, \frac{-\tau
     }{16\pi }\eab\langle \Wa,\Wb \rangle + \int_{}d^2\bar{\theta
     }\,\frac{\bar{\tau }}{16\pi } \eabd\langle \Wbar_{\ad},\Wbar_{\bd}
     \rangle\Bigr\} \\
      &= \frac{1}{2\sqo}\bres\Bigl\{ -\frac{\tau }{16\pi }D^2\Dbar^2\|\Sigma
     \|^2 + \frac{\bar{\tau }}{16\pi }\Dbar^2D^2\|\Sigma \|^2 \Bigr\} \\
      &= -\frac{\Im\tau }{4\pi }\bres\frac 12(D^2\Dbar^2 +
     \Dbar^2D^2)\|\Sigma \|^2 - \frac{\Re\tau }{16\pi \sqo}\bres\frac
     12(D^2\Dbar^2 - \Dbar^2D^2)\|\Sigma \|^2 .\endaligned \tag{9.23} $$
Recall from~\S{2.6} that for terms in a lagrangian in~$\mink2{(2,2)}$
involving twisted chiral fields only we compute component lagrangians using
exactly the combination of derivatives which occurs in the first term
of~\thetag{9.23}.  The second term is the topological term in~\thetag{9.10};
it does not seem to reduce to an integral~$\int_{}d^4\theta $.  So
for~$\Re\tau =0$ the $N=2$~superspace Yang-Mills lagrangian
in~$\mink2{(2,2)}$ is
  $$ \boxed{\scrL'_2 = |d^2y|\,d^4\theta \;\frac{-1}{4g^2}\|\Sigma \|^2.}
     \tag{9.24} $$
The corresponding component lagrangian is the dimensional reduction
of~\thetag{8.29}, which we compute either directly or with the aid
of~\thetag{9.8}.  After eliminating the auxiliary field~$E$ we obtain
  $$ \boxed{\aligned
      \scrLch'_2 \sim |d^2y|\;\frac{1}{g^2}\Bigl\{ -\frac 12|&F_A|^2 + \langle
     d_A\sigb,d_A\sigma \rangle + \langle \lambdab\,\Dirac_A\,\lambda \rangle
     \\
      &- \frac 12\|\,[\sigma ,\sigb]\,\|^2 + \sqrt2\,\bigl(\langle
     \lambdab^-,[\sigb ,\lambda ^+] \rangle + \langle \lambdab^+,[\sigma
     ,\lambda ^-] \rangle \bigr) \Bigr\}.\endaligned} \tag{9.25} $$
 
The (classical) theory is invariant under the R-symmetries~$J_+,J_-$, which
are defined in~\thetag{2.117}.  This follows either in superspace
from~\thetag{9.20} and~\thetag{9.24} or in components from~\thetag{9.21}
and~\thetag{9.25}.
 
Another type of topological term may be defined for an invariant trace 
  $$ \langle \cdot   \rangle\:\frak{g}\longrightarrow \RR, \tag{9.26} $$
and it has a superspace expression.  Define 
  $$ \int_{}d\bar\theta ^-d\theta ^+\,\ell  = i^*D_+\Dbar_-\ell . \tag{9.27}
     $$
The topological term is then 
  $$ \boxed{\scrL_3' = |d^2x|\,\frac{1}{\sqrt2}\Re\Bigl( d\bar\theta
     ^-d\theta ^+\,\langle \Sigma   \rangle\Bigr).} \tag{9.28} $$
The component expression is easily seen from~\thetag{9.16} to be 
  $$ |d^2x|\,\scrLch_3' = |d^2x|\,\Bigl\{ \langle f_A  \rangle +
     \sqo\langle E  \rangle\Bigr\}. \tag{9.29} $$

\Chapter{10}{$N=1$ Yang-Mills in Dimension 6 and $N=2$ Yang-Mills in
Dimension 4} 
\ShortLectureName{$N=1$ Yang-Mills in~6d and $N=2$ Yang-Mills in 4d}

 \comment
 lasteqno 10@ 56
 \endcomment

In~\S{10.1} we complete the discussion of the superspace formulation of super
Yang-Mills theory in 6~dimensions.  As in Chapter~7 and Chapter~8, the main
point is to work out the structure of the auxiliary fields and the
R-symmetries.  The proof that the category of component fields (including
the auxiliary fields) is equivalent to the category of superfields is
deferred to the next chapter.  We did not succeed in finding a manifestly
supersymmetric formula for the lagrangian in this case.  Section~10.2 is
parallel to Chapter~9: We reduce the 6-dimensional pure gauge theory to
4~dimensions, where it has twice the minimal amount of supersymmetry.
Neither theories in 4~dimensions with 8~supersymmetries which include matter
fields nor theories in 4~dimensions with 16~supersymmetries have superspace
formulations which exhibit all of the supersymmetry manifestly.  But they can
all be written on~$\mink44$, where 4~of the supersymmetries are manifest.  We
do this in~\S{10.3}.

\Head{10.1}{Constrained connections on~$\mink6{(8,0)}$}
 We specialize the discussion in~\S{6.3} to~$\mink6{(8,0)}$.  Our first task
is to evaluate~\thetag{6.74}.  To that end we compute using
\thetag{2.89}--\thetag{2.91}: 
  $$ \aligned
      g\bigl(\stov(f^{ai},f^{ck}),\sdtov(f_{bj},f_{d\ell }) \bigr) &= \frac
     12\epsilon ^{ik}\nu ^{pqac}\epsilon _{j\ell } \;g(e_{pq},e_{bd})\\
           &= \epsilon ^{ik}\epsilon _{j\ell }\nu ^{pqac}\nu _{pqbd}\\
      &= 2(\delta ^a_b\delta ^c_d - \delta ^a_d\delta ^c_b)\epsilon
     ^{ik}\epsilon _{j\ell }. \endaligned\tag{10.1} $$
The curvature term in~\thetag{6.74} may be expanded using 
  $$ \aligned
      \stov(f^{ai},f^{ck})\otimes \sdtov(f_{ck},f_{bj}) &= \frac12\epsilon
     ^{ik}\nu ^{pqac}\epsilon _{kj}\;e_{pq}\otimes e_{cb}\\ &= -\frac
     12\delta ^i_j\,\nu ^{pqac}\;e_{pq}\otimes e_{cb},\endaligned \tag{10.2} $$
so that the left hand side of~\thetag{6.74} is 
  $$ -2(-\frac 12\delta ^i_j\nu ^{pqac})\scrF_{(pq)(cb)} = \delta ^i_j\,\nu
     ^{pqac}\scrF_{(pq)(cb)} = \delta ^i_j\,\nu
     ^{acpq}\scrF_{(bc)(pq)}. \tag{10.3} $$
For $a=b$ this vanishes, since $\nu ^{acpq}$~ is symmetric
in~$(ac),(pq)$ whereas $\scrF_{(ac)(pq)}$~ is skew.  Now evaluate~\thetag{6.74}
for~$i=j=1$, then~$i=j=2$, and sum over~$a=b$ to learn 
  $$ \DD_{ai}\Lambda ^{ai} = 0. \tag{10.4} $$ 
Define 
  $$ E^i_j = -\frac{\sqo}{4}\DD_{aj}\Lambda ^{ai}. \tag{10.5} $$ 
Then with a bit of work we derive 
  $$ \DD_{bj}\Lambda ^{ai} = -\frac 12\stov^{\mu (ai)(ck)}\sdtov^\nu
     _{(ck)(bj)}\scrF_{\mu \nu } + \delta ^a_b\,{\sqo}\,E^i_j, \tag{10.6} $$
which has the desired form~\thetag{6.76}.
 
To proceed further we remark on some linear algebra.  For $S_0$ of dimension~4,
the natural map 
  $$ {\tsize\bigwedge} ^2({\tsize\bigwedge} ^2S_0) @>>> {\tsize\bigwedge}
     ^2S_0\otimes {\tsize\bigwedge} ^2S_0 @>>> {\tsize\bigwedge} ^3S_0\otimes
     S_0, \tag{10.7} $$
which is $u\wedge v\mapsto v\otimes v-v\otimes u$ followed by $(x\wedge
y)\otimes (z\wedge t)\mapsto(x\wedge y\wedge z)\otimes t - (x\wedge y\wedge
t)\otimes z$, gives rise to a short exact sequence
  $$ 0 @>>> {\tsize\bigwedge} ^2({\tsize\bigwedge} ^2S_0) @>>>
     {\tsize\bigwedge} ^3S_0\otimes S_0@>>>{\tsize\bigwedge} ^4S_0 @>>>
     0. \tag{10.8} $$
For $S_0$~given with a volume form, ${\tsize\bigwedge} ^3S_0$~is the dual
of~$S_0$, the map to~${\tsize\bigwedge} ^4S_0$ becomes the evaluation map,
and we get
  $$ {\tsize\bigwedge} ^2({\tsize\bigwedge} ^2S_0)\cong \{\text{trace zero
     endomorphisms of $S_0$}\}. \tag{10.9} $$
This makes it convenient to define
  $$ f_b ^a :=\frac 14\nu^{a c d e }\scrF_{(b c)(d e)}. \tag{10.10} $$
By~\thetag{10.9} the tensor~$f$ is traceless:
  $$ f^a _a =0. \tag{10.11} $$
Also,
  $$ \scrF_{(a b) (c d) } = \nu _{e b c d }f_a ^e + \nu _{a e c d }f^e _b
     . \tag{10.12} $$
Then using~\thetag{10.3} we see that \thetag{10.6} simplifies to
  $$ \boxed{\DD_{b j}\Lambda ^{a i} = \delta ^i_j\,f^a_b 
     + \delta ^a_b\,\sqo\, E^i_j}. \tag{10.13} $$
We need one additional formula 
  $$ \DD_{ai}E^j_k = \sqo\,(\epsilon _{ik}\nab{ab}\Lambda ^{bj} - \delta
     ^j_i\epsilon _{k\ell }\nab{ab}\Lambda ^{b\ell }), \tag{10.14} $$
which is equivalent to~\thetag{11.30} and will be explained in Chapter~{11}. 

Now we define the component fields corresponding to a constrained
connection.  First, we have a connection~$A$ on the restriction
$P\to\bmink^6$ of~$\scrP$ to Minkowski space.  The odd part of the curvature
restricts to a dual spinor field 
  $$ \boxed{\bres \Lambda ^{a i} = \lambda ^{a i}.} \tag{10.15} $$
The restriction of~$E^i_j$ to~$\bmink^6$, which we continue to
denote~`$E^i_j$', is an auxiliary field.  Using \thetag{2.95}, \thetag{2.97},
and \thetag{6.66} we learn that
  $$ \overline{E^i_j} = E^j_i, \tag{10.16} $$
By~\thetag{10.4} we see that $E$~is traceless: 
  $$ E^i_i = 0. \tag{10.17} $$
Thus $\left(E^i_j\right)$ has three independent real components; the matrix
$\left(\sqo\,E^i_j\right)$ is traceless skew-Hermitian (with values in the
adjoint bundle).  It transforms in the adjoint representation of the
$Sp(1)_R$~symmetry~\thetag{2.102}.  In other words, the auxiliary fields
comprise an imaginary quaternion.  The six-dimensional {\it vector
multiplet\/} is the collection of component fields $\{A,\lambda ^{a
i},E^i_j\}$.

        \proclaim{\protag{10.18} {Proposition}}
   The category of connections~$\scrA$ on~$\mink6{(8,0)}$ whose curvature
vanishes on the odd distribution~$\tau $ is equivalent to the category of
triples~$\{A,\lambda,E\}$ consisting of a connection~$A$ on~$\bmink^6$, a
dual spinor field~$\lambda $ with values in the adjoint bundle, and an
imaginary quaternion scalar field~$E$ with values in the adjoint bundle.
        \endproclaim

\flushpar 
 The proof of~\theprotag{10.18} {Proposition} is presented in Chapter~11.

We derive the supersymmetry transformation laws for the components as in
previous sections.  Denote the restriction of~$f^a_b$ to~$\bmink^6$
by~`$f^a_b$'.  Then the supersymmetry transformation~$\eta ^{ai}Q_{ai}$
induces the following transformation on component fields:
  $$ \boxed{ \aligned
      (\hxi A)_{pq} &= 2\eta ^{ai}\nu _{pqab}\epsilon _{ij}\lambda ^{bj}\\
      \hxi \lambda ^{a i} &= -\eta ^{bi}f^a_b - \eta ^{aj}{\sqo}\,E^i_j\\
      \hxi E^i_j &= \eta ^{ai}\sqo\,(\Dirac\lambda )_{aj} - \eta
     ^{ak}\sqo\,\epsilon _{jk}\epsilon ^{i\ell }(\Dirac\lambda )_{a\ell
     }.\endaligned} \tag{10.19} $$
and 
  $$ \boxed{\hxi f^a_b = 2\eta ^{ai}(\Dirac \lambda )_{bi} - 2\eta
     ^{ci}\epsilon _{ij}\nab{bc}\lambda ^{aj} - \delta ^a_b\eta ^{ci}(\Dirac
     \lambda )_{ci}.} \tag{10.20} $$

Recall from~\thetag{2.102} that the R-symmetry group is~$SU(2)_R$.  The
matrix $S=(S^i_j)$ acts on the component fields by 
  $$ \boxed{\alignedat2
      &(SA) &&= A \\
      &(S\lambda )^{ai}&&= (S\inv )^i_{i'}\lambda ^{ai'} \\
      &(SE)^i_j &&= (S\inv )^i_{i'}E^{i'}_{j'}S^{j'}_j.\endaligned}
     \tag{10.21} $$

\Head{10.2}{Reduction to~$\mink48$}
 The story here is analogous to that in Chapter~{9}, where we reduce
Yang-Mills on~$\mink44$ to~$\mink2{(2,2)}$.  For a worked out example,
see~\ASH8. 
 
We consider a constrained connection~$\scrA$ on a $G$-bundle
$\scrP\to\mink6{(8,0)}$ which is ``constant'' in the $\partial _{12},\partial
_{34}$ directions.  That is, $\partial _{12}$~and $\partial _{34}$~are lifted
to commuting complex conjugate vector fields~$\tilde\partial _{12}$
and~$\tilde\partial _{34}$ on~$\scrP$, and the horizontal vector
fields~$\DD_{ai}$ on~$\scrP$ satisfy 
  $$ [\tilde\partial _{12},\DD_{ai}] = [\tilde\partial _{34},\DD_{ai}] =
     0. \tag{10.22} $$
Set 
  $$ \Sigma =\frac{1}{\sqrt2}(\nab{34} - \tilde\partial _{34}). \tag{10.23}
     $$
Then applying Bianchi to~$D_{ai},D_{31},D_{42}$ we learn 
  $$ \DD_{3i} \Sigma =\DD_{4i}\Sigma =0. \tag{10.24} $$
From Bianchi we also find, after some tedious computation, a formula
for~$\Lambda ^{ai}$ in terms of derivatives of~$\Sigma $: 
  $$ \sqrt2\,\Lambda ^{ai} = \cases J^a_b\epsilon ^{ij}\DD_{bj}\Sigma
     ,&a=1,2;\\J^a_b\epsilon ^{ij}\DD_{bj}\Sigb,&a=3,4.\endcases \tag{10.25} $$
Here $J^a_b$~is the matrix of the pseudoreal structure~$j$ described
before~\thetag{2.95}.  Also, from~ \thetag{10.6} we learn that $\DD_{bj}\Lambda
^{ai}$ ~is independent of~$a,b$ if~$i\not= j$.  It follows that 
  $$ \aligned
      \DD_{11}\DD_{21}\Sigma &=\DD_{31}\DD_{41}\Sigb \\ 
      \DD_{12}\DD_{22}\Sigma &=\DD_{32}\DD_{42}\Sigb .\endaligned \tag{10.26} $$
 
Let $j\:\mink48\hookrightarrow \mink6{(8,0)}$ be the inclusion, as described
in the last paragraph of~\S2.5.  We follow the notation introduced there and
also use formula~\thetag{2.104}.  Now write $\tilde\scrA,\tilde
D_{ai},\tilde\Sigma ,\tilde\Lambda ^{ai},\tilde E^i_j$ for the fields
on~$\mink6{(8,0)}$ to distinguish them from their restrictions to~$\mink48$.
Set
  $$ \Sigma =j^*\tilde\Sigma . \tag{10.27} $$
Then using~\thetag{2.104}, equation~\thetag{10.24} asserts that $\Sigma $~is
a {\it chiral\/} superfield on~$\mink48$ in the sense that 
  $$ \DDbar_{\ad(i)}\Sigma =0,\qquad a=1,2,\quad i=1,2. \tag{10.28} $$
Formula~\thetag{10.25} expresses $\Lambda ^{ai}=j^*\tilde\Lambda ^{ai}$ in
terms of~$\Sigma $.  We record the result in full: 
  $$ \alignedat2
      \sqrt2\,\Lambda ^{11}&=-\DD^{(2)}_2\Sigma &\qquad \qquad \sqrt2\,\Lambda
     ^{12} &= \phantom{-}\DD^{(1)}_2\Sigma \\
      \sqrt2\,\Lambda ^{21}&=\phantom{-}\DD^{(2)}_1\Sigma &\qquad \qquad
     \sqrt2\,\Lambda ^{22} &= -\DD^{(1)}_1\Sigma \\
      \sqrt2\,\Lambda ^{31}&=-\DDbar_{\dot2(1)}\Sigb &\qquad \qquad
     \sqrt2\,\Lambda ^{32} &= {-}\DDbar_{\dot2(2)}\Sigb \\
      \sqrt2\,\Lambda ^{41}&=\phantom{-}\DDbar_{\dot1(1)}\Sigb &\qquad \qquad
     \sqrt2\,\Lambda ^{42} &= \phantom{-}\DDbar_{\dot1(2)}\Sigb\endaligned
     \tag{10.29} $$
We also work out the auxiliary fields~$E^i_j = j^*\tilde E^i_j$ using the
definition~\thetag{10.5} together with~\thetag{10.16}, \thetag{10.29},
and~\thetag{10.26}:  
  $$ \aligned
      E^1_1 &= -\frac{\sqo}{2\sqrt2}\bigl(\DD^{(1)}_2 \DD^{(2)}_1\Sigma -
     \DD^{(1)}_1\DD^{(2)}_2\Sigma \bigr) \\
      E^1_2 &= -\frac{\sqo}{\sqrt2}\DD_2^{(2)}\DD^{(2)}_1\Sigma \\
      &= - \frac{\sqo}{\sqrt2} \DDbar_{\dot2(1)}\DDbar_{\dot1(1)}\Sigb \\
      E^2_1 &= -\frac{\sqo}{\sqrt2}\DD_1^{(1)}\DD^{(1)}_2\Sigma \\
      &= - \frac{\sqo}{\sqrt2} \DDbar_{\dot1(2)}\DDbar_{\dot2(2)}\Sigb \\
      E^2_2 &= -E^1_1.\endaligned \tag{10.30} $$
 
From the point of view of~$\mink48$, we can think of~$\scrA$ as a connection
whose curvature is constrained to vanish along the odd distribution {\it
except\/} for the nontrivial curvature 
  $$ [\DDbar_{\dot1(2)},\DDbar_{\dot2(1)}] =
     -[\DDbar_{\dot2(2)},\DDbar_{\dot1(1)}] = 2\sqrt2\,\Sigma  \tag{10.31} $$
and the complex conjugate.

As in Chapter~{9} there is a sharp distinction between the abelian and
nonabelian cases---in the abelian case $\Sigma $~takes values in the constant
vector space~$\frak{g}$, the only constraint being~\thetag{10.28}; whereas in
the nonabelian case $\Sigma $~is a section of~$\adPP$ and the
constraint~\thetag{10.28} involves the connection~$\scrA$, which therefore
cannot be derived solely from~$\Sigma $. 
 
Formulas~\thetag{10.29} and~\thetag{10.30} are more understandable if we
restrict to~$\mink44\hookrightarrow \mink48$.  (The inclusion
$\mink44\hookrightarrow \mink6{(8,0)}$ is discussed towards the end
of~\S2.5.)  Let $\Ahat,\Shat$ denote the restrictions of~$\scrA, \Sigma$
to~$\mink44$.  Then $\Ahat$~is a vector superfield (cf.~\S{8.1})---a
constrained connection on the restriction~$\Phat$ of~$\scrP$
to~$\mink44$---and \thetag{10.28}~implies that $\Shat$~is a chiral superfield
(cf.~\S{5.1}) with values in~$\adj\Phat$.  The pair~$(\Ahat,\Shat)$ is called
an {\it $N=2$~vector multiplet\/} on~$\mink44$.  We claim, but do not write
out a proof, that the category of constrained connections~$\scrA$
on~$\mink6{(8,0)}$ which satisfy~\thetag{10.22} is equivalent to the category
of pairs~$(\Ahat,\Shat)$ on~$\mink44$.
      
Let $i\:\bmink^4\hookrightarrow \mink48$ be the inclusion.  Motivated by
formulas~\thetag{5.7}, and~\thetag{8.20}, \thetag{8.4} we define the
component fields of the superfields on~$\mink48$ to be 
  $$ \boxed{\alignedat2
      A &= \bres\scrA \\
      \sigma &=\bres\Sigma \\
      \lambda _a &= \bres\epsilon _{ba}\Lambda ^{b1} &&=
     -\frac{1}{\sqrt2}\,\bres \DD^{(2)}_a\Sigma \\
      \chi _a &= \bres\epsilon _{ba}\Lambda ^{b2} &&= \frac{1}{\sqrt2}\,\bres
     \DD^{(1)}_a\Sigma \\
      E &= \bres E^1_1 &&= -\frac{\sqo}{2\sqrt2}\,\bres\bigl(\DD^{(1)}_2
     \DD^{(2)}_1\Sigma - \DD^{(1)}_1\DD^{(2)}_2\Sigma \bigr) \\
      F &= -\frac{\sqo}{\sqrt2}\,\bres E^2_1 &&= \frac12\,\bres
     \DD_2^{(1)}\DD^{(1)}_1\Sigma. \endaligned} \tag{10.32} $$
The indices~$a,b$ in these formulas run over~$\{1,2\}$.  Also, we
use~\thetag{10.29}, \thetag{10.30}.
Following~\thetag{8.14} we define the dual spinor field 
  $$ \lambda ^a = \bres\Lambda ^{a1}= -\frac{1}{\sqrt2}\,\bres\epsilon
     ^{ab}\DD^{(2)}_b\Sigma . \tag{10.33} $$
These component fields group together into $N=1$ supermultiplets: $(A,\lambda
,E)$~is an $N=1$~vector multiplet (\S{8.1}) and $(\sigma ,\chi ,F)$~is an
$N=1$~chiral multiplet (\S{5.1}).  The collection $(A,\sigma ,\lambda ,\chi,
E ,F)$ of all component fields is called an {\it $N=2$~vector multiplet\/}.
The reader should check that~\thetag{10.32} is consistent with~\thetag{5.7}
and~\thetag{8.20}.  The only nontrivial compatibility check is to see that
$\tilde\Lambda ^{a1}$ ($a=1,2$) on~$\mink6{(8,0)}$ restricts on~$\mink44$ to
be the superfield~$\hat\Lambda ^a$, with both $\tilde\Lambda $
and~$\hat\Lambda $ defined by~\thetag{6.66}; we leave this check to the
reader.  Note that on $\mink44$ we have
  $$ \What_a = -\frac{1}{\sqrt2}\,\bres\fDD2a\Shat . \tag{10.34} $$

On $\mink44$ only half of the supersymmetry of the vector
multiplet~$(\hat\scrA,\hat\Sigma )$ is manifest.  For the rest we write the
vector field~$\hxi '$ on the space of $N=1$
superfields~$\scrF_{(\Ahat,\Shat)}$ which corresponds to the supersymmetry
transformation 
  $$ \eta ^{\prime a}Q^{(2)}_a + \etabar^{\prime
     \ad}\Qbar_{\ad(2)}. \tag{10.35} $$
From the formulas in this section we find
  $$ \boxed{\alignedat2
      &(\hxi'\Ahat)_a &&=2\sqrt2\,\eta ^{\prime b}\epsilon _{ba}\Sigbh \\ 
      &(\hxi'\Ahat)_{a\bd} &&=\sqrt2\bigl(\eta ^{\prime c}\epsilon _{ca}\DDbd
     \Sigbh + \etabar^{\prime \cd}\epsilon _{\cd\bd}\DD_a\Shat \bigr) \\ 
      &\phantom{(}\hxi'\Shat &&=\sqrt2\,\eta ^{\prime a}\What_a\endalignedat}
     \tag{10.36} $$
The penultimate line of~\thetag{6.79} is used to derive the first two of
these formulas.

Next, we write all of the supersymmetry transformations on the component
fields on~$\bmink^4$.  Let $\hxi$~be the vector field on the
space~$\scrF_{(A,\sigma ,\lambda ,\chi ,E,F)}$ of component fields induced by
the supersymmetry transformation
  $$ \eta ^aQ^{(1)}_a + \etad\Qbar_{\ad(1)} + \eta ^{\prime a}Q_a^{(2)} +
     \etabar^{\prime\ad}\Qbar_{\ad(2)}. \tag{10.37} $$
Then from~\thetag{10.19} (or~\thetag{8.22} together with~\thetag{10.36} and
the other formulas of this section) we deduce
  $$ \boxed{\aligned
      (\hxi A)_{a\bd} &= \hphantom{-}2\eta ^c\epsilon _{ca}\lambdab_{\bd} +
     2\etabar^{\cd}\epsilon _{\cd\bd}\lambda _{a} + 2\eta ^{\prime c}\epsilon
     _{ca}\chib_{\bd} + 2\etabar^{\prime\cd}\epsilon _{\cd\bd}\chi_{a} \\
      \hxi\sigma &= -\sqrt2\,\eta ^a\chi _a + \sqrt2\,\eta ^{\prime a}\lambda
     _a\\
      \hxi\lambda _a&=-\eta ^bF_{ab} + \eta ^b\epsilon _{ab}\sqo\,E +
     \sqrt2\,\eta ^{\prime b}\epsilon _{ab}\Fbar -
     \sqrt2\,\etabar^{\prime\bd}\nab{a\bd}\sigma \\
      \hxi\chi _a&= -\sqrt2\,\eta ^{b}\epsilon _{ab}F +
     \sqrt2\,\etabar^{\bd}\nab{a\bd}\sigma -\eta ^{\prime b}F_{ab} - \eta
     ^{\prime b}\epsilon _{ab}\sqo\,E \\
      \hxi E&=\eta ^a\sqo\,(\Dirac\lambdab)_{a} -
     \etabar^{\ad}\sqo\,(\Dirac\lambda )_{\ad} - \eta
     ^{\prime a}\sqo\,(\Dirac\chib)_a +
     \etabar^{\prime\ad}\sqo\,(\Dirac\lambda )_{\ad}\\
      \hxi F&= \sqrt2\,\eta ^{\prime a}(\Dirac\lambdab)_a -
     \sqrt2\,\etabar^{\ad}(\Dirac\chi )_{\ad} \endaligned }\tag{10.38} $$
 
The R-symmetry group of the dimensionally reduced theory is~$U(2)$ (see the
end of~\S{2.5}).  The matrix $S=(S^i_j)$ acts on the component fields by 
  $$ \boxed{\aligned
      S\:A &\longmapsto A \\
      S\: \sigma &\longmapsto (\det S)\inv \sigma \\
      S\:\pmatrix \lambda _a\\\chi _a \endpmatrix &\longmapsto S\inv \pmatrix
     \lambda _a\\\chi _a \endpmatrix \\
      S\:\pmatrix E&-\sqo\,\sqrt2\,\Fbar\\\sqo\,\sqrt2\,F&-E \endpmatrix
     &\longmapsto S\inv \pmatrix E&-\sqo\,\sqrt2\,\Fbar\\\sqo\,\sqrt2\,F&-E
     \endpmatrix S.\endaligned} \tag{10.39} $$
We remark that it is useful to check~\thetag{10.38} using
equations~\thetag{5.5}, \thetag{8.22}, and the R-symmetry~\thetag{10.39}
for $S=\left(\smallmatrix 0&1\\1&0 \endsmallmatrix\right)$.

Finally, we write the $N=2$~Yang-Mills lagrangian in four dimensions in a
manifestly supersymmetric way on~$\mink48$ .  As with~$\mink44$ (see the
beginning of~\S{5.3}) there is a ``chiral odd integration'' on~$\mink48$.
Namely, if $\Upsilon $~is any superfield on~$\mink48$ which is chiral in the
sense that $\Dbar_{\ad(i)}\Upsilon =0$, then
  $$ |d^4x|\;\int_{}d^4\theta \,\Upsilon =|d^4x|\;\bres
     \fD12\fD11\fD22\fD21\Upsilon \tag{10.40} $$
is a well-defined density on~$\bmink^4$.  Given a connection~$\scrA$
on~$\mink48$ of the type we have been considering, with $\Sigma $~defined as
above, set
  $$ \boxed{\scrL = |d^4x|\,\Im\Bigl\{d^4\theta \,\frac{\tau }{32\pi }\langle
     \Sigma ,\Sigma \rangle\Bigr\}.} \tag{10.41} $$
We claim that $\scrL$~is a superspace lagrangian which reproduces in
components the dimensional reduction of Yang-Mills that we seek.  To verify
this we transform it into a lagrangian on~$N=1$ superspace~$\mink44$.
Namely, the component lagrangian function~$\scrLch$ may be written as  
  $$ \aligned
      \scrLch&=\Im\Bigl\{ \frac{\tau }{32\pi }\bres\fD12\fD11\fD22\fD21\langle
     \Sigma ,\Sigma   \rangle\Bigr\} \\ 
      &= \Im\Bigl\{ \frac{\tau }{16\pi }\bres\fD12\fD11\bigl(-\langle
     \fDD21\Sigma ,\fDD22\Sigma   \rangle + \langle \fDD22\fDD21\Sigma ,\Sigma
     \rangle \bigr)\Bigr\} \\ 
      &= \Im\Bigl\{\frac{\tau }{16\pi }\bres\fD12\fD11\bigl(-2\langle W_1,W_2
     \rangle + \langle \fDDb21\fDDb11\Sigb,\Sigma   \rangle \bigr)\Bigr\} \\ 
      &= \bres\fD12\fD11\Im\bigl\{-\frac{\tau \epsilon ^{ab}}{16\pi }\langle
     W_a,W_b  \rangle\bigr\} + \frac{\Im\tau }{16\pi
     }\bres\fD12\fD11\fDb21\fDb11\langle \Sigb,\Sigma   \rangle \\ 
      &= \int_{}d^2\theta \,\Im\bigl\{-\frac{\tau \epsilon ^{ab}}{16\pi
     }\langle W_a,W_b  \rangle\bigr\} + \frac{\Im\tau }{4\pi }\int_{}d^4\theta
     \frac 14\langle \Sigb,\Sigma   \rangle.\endaligned \tag{10.42} $$
In the third equation we use~\thetag{10.34} and~\thetag{10.26}
(cf.~\thetag{10.30}); in the fourth we use the fact that $\Sigma $~is chiral;
and in the fifth we use notions of integration on~$\mink44$.  For the
$d^4\theta $~term we use~\thetag{2.71}, so that the last equality holds only
up to exact terms, which we ignore.  The component expansion of the first
term is~\thetag{8.29}.  The component expansion of the second
is~\thetag{5.13}, but we must remember the extra terms explained in~\S{8.3}.
Recall that for this we eliminate the auxiliary field~$E$
using~\thetag{8.50}: 
  $$ \boxed{E = \sqo\,[\sigma,\bar\sigma ].} \tag{10.43} $$
Set 
  $$ \tau  = \frac{\theta }{2\pi } + \frac{4\pi \sqo}{g^2}. \tag{10.44} $$
The component lagrangian is then a special case of~\thetag{6.42}: 
  $$ \boxed{\aligned
      d^4x\;\scrLch = d^4x\;\frac{1}{g^2}\Bigl\{ \!&-\frac 12|F_A|^2 + \langle
     d_A\bar\sigma ,d_A\sigma \rangle + \langle\bar\lambda
     \Dirac_A\lambda\rangle + \langle \chib\Dirac_A\chi \rangle 
      -\frac 12|\,[\sigma ,\bar\sigma ]\,|^2 \\ 
       &-\! \sqrt2\,\bigl(\langle
     \bar\sigma ,[\lambda ^a,\chi _a] \rangle +
      \langle [\lambdab^{\ad},\chib_{\ad}],\sigma \rangle\bigr) + \langle
     \Fbar,F \rangle\Bigr\}
         + \frac{\theta }{16\pi ^2}\langle F_A\wedge F_A
     \rangle.\endaligned} \tag{10.45} $$

The $N=2$~superspace expression~\thetag{10.41} is convenient for studying
quantum corrections to the classical theory (which are explained
in~\DQFT{18}).  Define the {\it holomorphic prepotential\/}
  $$ \scrF(\Sigma ) = \frac{\tau }{2}\langle \Sigma ,\Sigma
     \rangle. \tag{10.46} $$
Then we rewrite~\thetag{10.41} as 
  $$ \boxed{\scrL = |d^4x|\,\Im\Bigl\{d^4\theta \,\frac{\tau }{16\pi
     }\scrF(\Sigma )\Bigr\}.} \tag{10.47} $$
Using~\thetag{5.17} we see that the last term of~\thetag{10.42} corresponds
to the \Kah\ potential 
  $$ K(\Sigb,\Sigma ) = \frac 12\Im\bigl(\Sigb\,\frac{\partial \scrF}{\partial
     \Sigma } \bigr). \tag{10.48} $$
We can replace the classical prepotential~$\scrF$ by any holomorphic function
of~$\Sigma $ and still obtain a supersymmetric theory; the
relation~\thetag{10.48} between the prepotential~$\scrF$ and the \Kah\
potential~$K$ is unchanged.  The computation of the quantum effective action
is then the determination of the function~$\scrF$.  \Kah\ metrics of the
form~\thetag{10.48} are called {\it special\/} by physicists, and indeed they
exhibit many special properties.\footnote{For a mathematical treatment of
special K\"ahler geometry, see:
 D. S. Freed, {\it Special K\"ahler manifolds}, Commun. Math. Phys., 
to appear, {\tt hep-th/9712042}.
}

\Head{10.3}{More theories on~$\mink44$ with extended supersymmetry} 
 We record without proof the form of the general $N=2$~model in 4~dimensions
written on the $N=1$ superspace~$\mink44$.  Recall~\thetag{6.27} that the
scalars take values in a \hyp\ manifold~$X$.  As in~\S{3.6} we fix a \Kah\
structure on~$X$.  Then the rest of the hyperk\"ahler structure is
encoded in a holomorphic symplectic form.  The moment map~$\mu _{\HH}$ of the
$G$~action on~$X$ is 
  $$ \mu _{\HH} = (\mu _{\RR},\mu _{\text{hol}}), \tag{10.49} $$
where $\mu _{\RR}\:X\to\frak{g}^*$ is the moment map for the K\"ahler
structure and $\mu _{\text{hol}}\:X\to\frak{g}_{\CC}^*$ the complex moment
map of the holomorphic symplectic structure.  The fields in the model are a
constrained connection~$\scrA$ on some principal bundle $\scrP\to\mink44$ and
a chiral superfield~$\Phi $ with values in the associated fiber
bundle~$\tilde X^{\scrP}$, where
  $$ \tilde X = X\times \frak{g}_{\CC}. \tag{10.50} $$
Note that the $\frak{g}_{\CC}$-valued scalars are part of the $N=2$~vector
multiplet, as explained in~\S{10.2}.  The total K\"ahler moment map is
  $$ \mu (\phi ,\sigma ) = \mu _{\RR}(\phi ) + \frac {\sqo}2[\sigma ,\bar\sigma
     ]. \tag{10.51} $$
There is a nonzero superpotential 
  $$ W = \sqrt2\;\tilde\mu _{\text{hol}} \tag{10.52} $$
coming from the transpose of the holomorphic moment map: 
  $$ \tilde\mu _{\text{hol}}\:X\times \frak{g}_{\CC}\longrightarrow
     \CC. \tag{10.53} $$
We leave the reader to verify that the component lagrangian~\thetag{6.42} for
these data is the dimensional reduction of~\thetag{6.48}.  As mentioned at
the end of~\S3.4, if $\zeta $~ is a vector field on~$X$ which preserves the
\hyp\ structure, then we can use~$\zeta $ to add a term to the
superpotential.  Namely, if $\eta $~is the holomorphic symplectic form on~$X$
for the particular choice of complex structure we have made, we add to~$W$
the holomorphic function~$W_0$ which satisfies
  $$ \partial W_0 = -\cont{\zeta }\eta . \tag{10.54} $$
In other words, $W_0$~is a holomorphic Hamiltonian function for the
infinitesimal symmetry~$\zeta $. 
 
The only $N=4$~model in 4~dimensions is pure Yang-Mills theory, which is
reduced from 10~dimensions.  In the description on~$\mink44$ there is a
constrained connection~$\scrA$ and three chiral superfields~$\Sigma _1,\Sigma
_2,\Sigma _3$ with values in the complex adjoint bundle.  The moment map is 
  $$ \mu (\sigma _1,\sigma _2,\sigma _3) = \frac{\sqo}{2}\Bigl\{ [\sigma
     _1,\bar\sigma _1] + [\sigma _2,\bar\sigma _2] + [\sigma _3,\bar\sigma
     _3] \Bigr\}, \tag{10.55} $$
and there is a superpotential 
  $$ W(\sigma _1,\sigma _2,\sigma _3) = \sqrt2\,\langle \sigma _1,[\sigma
     _2,\sigma _3]  \rangle. \tag{10.56} $$
Note that the $U(1)$ R-symmetry~\thetag{2.81} is broken by the
superpotential (cf.~the end of~\S{5.3}).  This model does have an $SU(4)$
R-symmetry; the $SU(3)$ subgroup which permutes the chiral superfields is
manifest in this formulation.\footnote{In components the six {\it real\/}
scalar fields are permuted by~$SU(4)\cong \Spin(6)$.}

\Chapter{11}{The Vector Multiplet on~$\mink6{(8,0)}$}

 \comment
 lasteqno 11@ 37
 \endcomment

We prove \theprotag{10.18} {Proposition}, following the treatment in Koller
(see~References). 

\Head{11.1}{Complements on $M^{6\vert(8,0)}$}

As in \S2.1 we write the relevant spinor representation $S$, complexified, as
$S_\CC=S_0\otimes W$.  The real structure comes from pseudoreal structures
on $S_0$ and $W$.

The following lemma gives a convenient description,
compatible with the action of $GL(S_0)\times GL(W)$, of
the symmetric and exterior powers of $S_\CC$.
For $\rho$ a representation of the symmetric group $S_n$, we
note $\Schur_\rho(U)$ the corresponding Schur functor
  $$ \Schur_\rho(U):=\Hom_{S_n}(\rho,U^{\otimes n}). \tag{11.1} $$
One has
  $$  \bigoplus\Schur_\rho(V)\otimes\rho\arrowsim V^{\otimes n},
     \tag{11.2} $$
the sum running over representatives of the isomorphism
classes of irreducible representations of $S_n$.
The map \thetag{11.2} is an isomorphism of representations of
$GL(V)\times S_n$.

        \proclaim{\protag{11.3} {Lemma}}
 Let $U$ and $V$ be vector spaces.  One has
  $$  \align &\bigoplus\Schur_\rho(U)\otimes\Schur_{\rho\upvee}(V) \arrowsim
     \Sym^n(U\otimes V)\tag{11.4}\\
     &\bigoplus\Schur_\rho(U)\otimes\Schur_{\rho\upvee\otimes\sgn}(V)
     \arrowsim {\tsize\bigwedge} ^{n}(U\otimes V)\tag{11.5} \endalign $$
        \endproclaim

\flushpar
 Note that $\rho$ and $\rho\upvee$ are isomorphic.  In \thetag{11.5} $\sgn$
is the sign character of $S_n$.

\demo{Proof}
 Apply \thetag{11.2} to $U$ and $V$, and take the coinvariants (resp.
co-anti-invariants) by $S_n$.
\enddemo

On $M^{6\vert(8,0)}$ the tangent bundle is a direct sum $V\oplus\tau$.  As
vector bundles the structural odd distribution $\tau$ is the constant bundle
$S^*$, parity changed, and $V$ is the  constant bundle real part of
${\tsize\bigwedge}^{2}S_0$.  In particular, each $s$ in $S_\CC^*$ defines a
complex odd vector field $D_s$ in $\tau_\CC$.  The bases $\{f_a\}$ of $S_0^*$
and $\{e_i\}$ of $W^*$ define a moving frame (see \S2.5) consisting of the
$D_{ai}$ and $\partial_{ab}$ $(a<b)$.  For $\partial_{ba}:=-\partial_{ab}$
one has $D_{ai}=D_{f_a\otimes e_i}$ and
  $$  [D_{ai},D_{bj}]=-2\epsilon_{ij}\partial_{ab}.  \tag{11.6} $$

For $x$ in $\Sym^2(S_\CC^*)$ define $D(x)$ by
  $$  D(s't')=[D_{s'},D_{t'}].  \tag{11.7} $$
By the lemma,
  $$  \Sym^2(S_\CC^*)={\tsize\bigwedge}^{2}S_0^*\otimes
     {\tsize\bigwedge}^{2}W^*\,\,\oplus\,\,\Sym^2(S_0^*)\otimes
     \Sym^2(W^*),\tag{11.8} $$
and \thetag{11.6} gives that $D(x)$ vanishes for $x$ in the second factor.
By polarization in $w$ this amounts to saying that for $w$ in $W^*$, we have
$[D_{s_1\otimes w},D_{s_2\otimes w}]=0$.  For $w\not=0$ fixed in $W^*$, the
$D_{s\otimes w}$ ($s\in S_0^*)$ span an integrable distribution $\tau_w$; its
leaves are affine spaces with vector space of translations $S_0\otimes w$,
parity changed.  Of course, $\tau_w$ is a complex distribution so that it
makes sense to speak of its leaves only after complexification, at least in
the odd direction (cs spaces).  One can as well say that for $s\not=0$
fixed in $S_0^*$, the $D_{s\otimes w}$ ($w\in W^*$) span an integrable
distribution $\tau_s$, with leaves affine spaces with vector space of
translations $s\otimes W^*$.

The intersection of $s\otimes W^*$ and of $S_0^*\otimes w$ is
the line spanned by $s\otimes w$.
This line also gives rise to an integrable distribution, with
leaves affine spaces of dimension~ $0\vert 1$, intersections
of a leaf of $\tau_s$ and a leaf of $\tau_w$.

We can now use the volume forms $\nu$ and $\epsilon$ to define
differential operators
  $$ 
\align
&D_w^4 :=\nu^{1234}D_{f_1\otimes w}D_{f_2\otimes w}
D_{f_3\otimes w}D_{f_4\otimes w}\tag{11.9}\\
&D_s^2 :=\epsilon^{12}D_{s\otimes e_1}D_{s\otimes e_2}.\tag{11.10}
\endalign
  $$ 
They can be viewed as integrals on the leaves of $\tau_w$
(resp. $\tau_s$).

        \proclaim{\protag{11.11} {Lemma}}
 $D_w^4D_s^2=D_s^2D_w^4=0$. 
        \endproclaim

        \demo{Proof}
  It suffices to check it for $w$ and $s$ not zero; if we take a basis with
$f_1=s$ and $e_1=w$, the lemma reduces to $D_{s\otimes w}^2=0$.
        \enddemo

\subsubhead Variant\endsubsubhead
 $D_w^4$ produces a function constant along the leaves of $\tau_w$.  The
operator $D_s^2$ is right divisible by $D_{s\otimes w}$.  As $s\otimes w$ is
in $\tau_w$, the operator~$D_s^2$~annuls the image of $D_w^4$.  Permuting the
roles of $S_0^*$ and $W^*$, one similarly proves that $D_w^4D_s^2=0$.

\Head{11.2}{Constrained connections}

Let $G$ be a Lie group, with Lie algebra $\frak{g}$.  We think to connections
$\nabla$ on a $G$-torsor $\scrP$ as follows: For each representation $R$ of
$G$, the principal bundle $\scrP$ gives rise to a vector bundle $R^\scrP$,
and $\nabla$ gives connections on the $R^\scrP$, functorially in $R$ and
compatibly with the tensor product.

From this point of view, the curvature $F$ is the $\frak{g}^\scrP$-valued
$2$-form such that for each $R$, the function $-i_Xi_YF$ acts on $R^\scrP$
(by the action of $\frak{g}^\scrP$ on $R^\scrP$ deduced from the action of
$\frak{g}$ on $R$) by $[\nabla_X,\nabla_Y]-\nabla_{[X,Y]}$.  The formula
  $$ -\cont X\cont YF=[\nabla_X,\nabla_Y]-\nabla_{[X,Y]} \tag{11.12} $$
differs by a sign from the one which holds when $\nabla_X$ is viewed as a
horizontal vector field on $\scrP$ lifting $X$.  (See~\thetag{6.75} for an
example of the latter formula.)

A constrained connection is one whose curvature vanishes in the
$\tau$-direction.  Let us write $\DD_s$ for $\nabla_{\DD_s}$ and $\DD_{ai}$
for $\DD_{f_a\otimes e_i}$.  The constraint is that, in parallel to~
\thetag{11.6},
  $$ 
[\DD_{ai},\DD_{bj}]=-2\epsilon_{ij}\nabla_{ab}.
\tag{11.13}
  $$ 

A constrained connection is uniquely determined by its
restriction to $\tau$.
For this restriction, here is what the constraint becomes.
For $x$ in $\Sym^2(S_\CC^*)$, define~ $\DD(x)$ by
$\DD(s't')=[\DD_{s'},\DD_{t'}]$; the constraint is that
$\DD(x)$ vanishes for $x$ in the second factor \thetag{11.8} of
$\Sym^2(S_\CC)$.
This amounts to saying that for $w$ in $W^*$, the connection
is integrable along the leaves of $\tau_w$.
It also amounts to saying that it is integrable along the
leaves of $\tau_s$ ($s\in S_0^*$).
If we define $\DD_w^4$ and $\DD_s^4$ by \thetag{11.9} and \thetag{11.10},
with $D$ replaced by $\DD$, we still have (in any
$V^\scrP$)
  $$ 
\DD_w^4\DD_s^2=\DD_s^2\DD_w^4=0,
\tag{11.14}
  $$ 
with the same proof.

\Head{11.3}{An auxiliary Lie algebra}

Let $\scrL=\bigoplus\limits_{\nu =0}^{\infty}\scrL^\nu $ be the graded super
Lie algebra (with parity $=$ degree modulo~ $2$) generated in degree $1$ by
$S_0^*\otimes W^*$, and with relations given by
  $$ \Sym^2(S_0^*)\otimes\Sym^2(W^*)\subset \Sym^2(S_0^*\otimes
     W^*). \tag{11.15} $$
For any constrained connection, the $\DD(s)$ obey the relations of $\scrL$:
for any representation $R$ of $G$, the Lie algebra $\scrL$ acts on the local
sections of $R^\scrP$.  In degree one, we get the $\DD(s)$; in degree two,
the $\nabla_{ab}$: the connection in the direction of $V$; in degree three,
the components $F_{ab,ck}=[\nabla_{ab},\DD_{ck}]$ of the curvature, in the
moving frame of~\S2.5; in degree $\Ge 3$ the action of elements of $\scrL$ is
given by sections of $\frak{g}^\scrP$, iterated covariant derivatives of the
components of the curvature.

As a preliminary to a description in component fields of constrained
connections, we now analyze $\scrL$ in degree $\Le 5$.  In degree $1$, we
have a basis $\DD_{ai}$.  In degree $2$ we have $\scrL^2\cong
{\tsize\bigwedge}^{2}S_0^*$, with basis the $\nabla_{ab}$ ($a<b$) defined by
\thetag{11.13}.

        \proclaim{\protag{11.16} {Lemma}}
 The iterated bracket $[[\DD_{ai},\DD_{bj}],\DD_{ck}]
=-2\epsilon_{ij}[\nabla_{ab},\DD_{ck}]$ is antisymmetric in $(a,b,c)$.
        \endproclaim

\demo{Proof}
By \thetag{11.13}, the iterated bracket $B$ is antisymmetric in $(a,b)$
and in $(i,j)$.
As $i$, $j$, $k$ take only two values, two of them are equal.
We consider the cases $i=j$, $j=k$ and $i=k$ in turn:

\hskip25pt\vtop{\roster
\item"$i=j$:"
one has $B=0$;\par
\item"$i=k$:"
$B=[\DD_{ai},[\DD_{bj},\DD_{ck}]]$, which is antisymmetric in
$(b,c)$;
\item"$j=k$:"
$B$ is similarly antisymmetric in $(a,c)$.
\endroster}
\medskip 
\noindent
 In all cases, antisymmetry in $(a,b,c)$ follows.

By the lemma, $[\nabla_{ab},\DD_{ck}]$ defines a morphism
${\tsize\bigwedge}^{3}S_0^*\otimes W^*\to\scrL^3$.  As $S_0^*$ is of
dimension $4$, the volume form of $S_0$ identifies
${\tsize\bigwedge}^{3}S_0^*$ to the dual $S_0$ of $S_0^*$.  Hence we have a
map 
  $$ \align &\Lambda\colon S_0\otimes W\longrightarrow \scrL^3,\tag{11.17}\\
     \intertext{where we take} &[\DD_{ak},\nab{bc}]=2\epsilon
     _{kj}\nu_{abcd}\Lambda^{dj}.  \tag{11.18} \endalign $$
We find it convenient to lower the index on~$\Lambda $ by 
  $$ \Lambda ^d_k = \epsilon _{kj}\Lambda ^{dj}, \tag{11.19} $$
and so 
  $$ [\DD_{ak},\nab{bc}] = 2\nu _{abcd}\Lambda ^d_k. \tag{11.20} $$

For any $x$ in $S_0\otimes W^*$, one easily checks that
$\epsilon_{ij}\nu_{abcd}x^d_k$, averaged by cyclic permutations of
$(ac,\,bj,\,ck)$, vanishes.  This gives the Jacobi identity required to check
that \thetag{11.17} is an isomorphism.  \enddemo

\subhead Degree $4$ \endsubhead  
 The bracket $[\quad]\colon\,\,\scrL^2\otimes \scrL^2\to\scrL^4$ defines a
map 
  $$ \aligned
      {\tsize\bigwedge}^{2}{\tsize\bigwedge}^{2}S_0^*&\longrightarrow
     \scrL^4\\
      (f_a\wedge f_b)\wedge(f_c\wedge f_d)&\longmapsto
     [\nabla_{ab},\nabla_{cd}].\endaligned  \tag{11.21} $$
The iterated exterior power ${\tsize\bigwedge}^{2}{\tsize\bigwedge}^{2}S_0^*$
is the orthogonal Lie algebra  of ${\tsize\bigwedge}^{2}S_0^*$, provided with
the  symmetric  bilinear   form  $h_{ab,cd}= \nu_{abcd}$.     The exceptional
isomorphism $SL(S_0^*)\arrowsim \Spin({\tsize\bigwedge}^{2}S_0^*)$ identifies
it  with the Lie algebra  of traceless endomorphisms of  $S_0^*$.  The map is
$f_a^b\mapsto F_{ab,cd}$, with
  $$  F_{ab,cd}=-\nu_{abef}(\delta_c^ef_d^f-\delta_d^e f_c^f), \tag{11.22} $$
or, moving up the vector index $ab$ by $h^{ab,cd}=\nu^{abcd}$ and multiplying
by a suitable factor, we have
  $$  F_{cd}^{ab}=\text{ antisymmetrization in $(a,b)$ and in $(c,d)$ of
     $\delta_c^af_d^b$}. \tag{11.23} $$
Note that $F$, given by this formula, is antisymmetric in $(ab,cd)$ if and
only if $f$ is traceless.  In particular, we obtain a traceless $f_b^a$ in
$\scrL^4$ with
  $$ [\nabla_{ab},\nabla_{cd}]=-\nu_{abef}\left(\delta_c^e f_d^f-\delta_d^e
     f_c^f\right).\tag{11.24} $$

The degree $4$ part of $\scrL$ is generated by the
$[\DD_{ai},\Lambda^{b}_j]$.  They are related to the
$F_{ab,cd}:=[\nabla_{ab},\nabla_{cd}]$ by
  $$ \alignedat2
      -2\epsilon_{k\ell} F_{ab,cd} &=[\nabla_{ab},[\DD_{ck},\DD_{d\ell}]]\\
     &=[[\nabla_{ab},\DD_{ck}],\DD_{d\ell}] \qquad &&\text{ symmetrized in
     $(ck,d\ell)$}\\ &=2\nu_{abce}[\DD_{d\ell}, \Lambda^{e}_k]\qquad &&\text{
     symmetrized in $(ck,d\ell)$}\\
     &=2\nu_{abef}\delta_c^e[\DD_{d\ell},\Lambda^{f}_k] \qquad &&\text{
     symmetrized in $(ck,d\ell)$}.  \endaligned\tag{11.25} $$
Let us extract from this formula the parts symmetric and antisymmetric in
$(k,\ell )$.  Antisymmetrization gives
  $$ \alignat2
       -4\epsilon_{k\ell}F_{ab,cd} &=
     \nu_{abef}\delta_c^e[\DD_{d\ell},\Lambda^{f}_k]\quad
     &&\text{antisymmetrized in $(c,d)$, $(k,\ell)$, }\tag{11.26}\\
      \intertext{in other words,}
     4\epsilon_{k\ell}f_b^a &= \DD_{b\ell}\Lambda^{a}_k\quad
     &&\text{antisymmetrized in $(k,\ell)$.}\tag{11.27} \endalignat $$
Symmetrization in~$(k,\ell )$ gives the vanishing of the symmetrization in
$(c,d)$, $(k,\ell)$ and antisymmetrization in $(a,b)$ of
$\delta_c^a\DD_{d\ell}\Lambda^{b}_k$.  Let us fix $k$, $\ell$ and let $D_d^b$
be the symmetrization of $\DD_{d\ell}\Lambda^{b}_k$ in $(k,\ell)$.  For
$a=c=1$, $b=2$, $d=1$ we get $D_1^2=0$.  This holding in any basis, $D_d^b$
is a multiple of $\delta_a^b$.  Hence for some $\scrE_{ij}$, we
have\footnote{$\scrE_{ij}$ is related to the~$E^i_j$ of Chapter~{10} by 
  $$ \scrE_{ij} = 2\sqo\,\epsilon _{ik}E^k_j. \tag{11.28} $$
}
  $$ \delta_b^a\scrE_{ij}=\DD_{bj}\Lambda^{a}_i\qquad \text{symmetrized in
     $(i,j)$}.\tag{11.29} $$

The morphism from ${\tsize\bigwedge}^{2}{\tsize\bigwedge}^{2} S_0^*\oplus
\Sym^2(W^*)$ to $\scrL^4$ is an isomorphism.  Indeed, if we define $\scrL^4$
as being ${\tsize\bigwedge}^{2}{\tsize\bigwedge}^{2}S_0^*\oplus\Sym^2(W^*)$
and the bracket by \thetag{11.27}, \thetag{11.29}, we get a Lie algebra.  The
only new Jacobi identity to check is \thetag{11.25}, from which
\thetag{11.27} and \thetag{11.29} were derived.

\subhead Degree $\Ge 5$ \endsubhead  
 From the forthcoming description (\theprotag{11.33} {Proposition}) of
constrained connections in components, one can deduce that the part of degree
$\Ge 2$ of the super Lie algebra $\scrL$ is free, with generators the
$\nabla_{ab}$, $\Lambda^{a}_i$ and $\scrE_{ij}$.  To prepare for
\theprotag{11.33} {Proposition}, we show here that it is generated by those
elements.  If $\scrL_1$ is the subalgebra generated by the $\nabla_{ab}$,
$\Lambda ^{a}_i$ and~$\scrE_{ij}$, it suffices to check that a $\DD_{ai}$,
applied to any of the generators, is again in $\scrL_1$.  It only remains to
show that the $[\DD_{ai},\scrE_{jk}]$ are linear combinations of the
$[\nabla_{cd},\Lambda^{e}_\ell]$.  The formula
  $$ \DD_{ai}\scrE_{jk}=C\epsilon_{ij}\nabla_{ab}\Lambda ^{b}_k\qquad
     \text{symmetrized in $(j,k)$} \tag{11.30} $$
for some constant~$C$ is the only possible formula compatible with the
grading, the actions of $SL(S_0^*)$, $SL(W)$, and with the symmetry in
$(j,k)$.  (Further computation shows~$C=-2$.)  Here, we will content
ourselves to check that, modulo the $[\nabla,\Lambda]$, all
  $$  [\DD_{ai},[\DD_{bj},[\DD_{ck},[\DD_{d\ell}, \DD_{em}]]]]=4\epsilon_{\ell
     m} \nu_{cdef}[\DD_{ai},[\DD_{bj}, \Lambda^{f}_k]] \tag{11.31} $$
vanish.

If we antisymmetrize \thetag{11.31} in $(j,k)$, we get by \thetag{11.27} a
$[\DD,[\nabla,\nabla]]$, which Jacobi turns into a $[\nabla,[\nabla,\DD]]$,
i.e., a $[\nabla,\Lambda]$.  Modulo the $[\nabla,\Lambda]$, this gives the
symmetry of \thetag{11.31} in $(j,k)$ and, by \thetag{11.29}, its
antisymmetry in $(b,c,d,e)$.

If a quantity $x_{ijk}$ is symmetric in $(j,k)$, to prove its vanishing, it
suffices to check the vanishing of its symmetrization $y_{ijk}$ in $(i,j)$.
Indeed, the symmetrization of $y$ in $(j,k)$ gives $x_{ijk}$ back, up to a
multiple of the complete symmetrization (of $x$ or $y$, this amounts to the
same):
  $$  y_{ijk}+y_{jki}=x_{ijk}+(x_{ikj}+x_{jik}+x_{kij}). \tag{11.32} $$

To prove the vanishing of \thetag{11.31} modulo the
$[\nabla,\Lambda]$, it hence suffices to consider its
symmetrization $\thetag{11.31}^S$ in $(i,j)$.
In $\thetag{11.31}^S$, if we symmetrize in $(a,b)$, we get $0$: a
$[[\DD_{ai},\DD_{bj}]\ldots]$ symmetrized in $(i,j)$.
We conclude that $\thetag{11.31}^S$ is antisymmetric in $(a,b)$.
Modulo the $[\nabla,\Lambda]$, it is also antisymmetric in
$(b,c,d,e)$, and as $\dim\,S_0^*=4<5$, its vanishing
follows.

\Head{11.4}{Components of constrained connections}

In classical differential geometry, if $N$ is a subvariety
of $M$ and $(\scrP,\nabla)$ a torsor with connection on $M$,
the restriction of $(\scrP,\nabla)$ to a formal neighborhood
of $N$ is determined up to unique isomorphism by \ (a) the
restriction $(P,\nabla)$ of $(\scrP,\nabla)$ to $N$, and \
(b) the value on $N$ of all iterated covariant derivatives
of the components of the curvature of $P$.
The same holds in the super case.
In addition, if $N$ is of codimension $0\vert s$ in $M$, the
formal neighborhood of $N$ in $M$ is just $M$ itself.

If the codimension is $0\vert 1$, the value on $N$ of the
curvature is enough.
Indeed, take local coordinates $(\theta^0,x^i)$, where the
$x^i$ can be even or odd, for which $N$ is defined by $\theta^0=0$.
A trivialization of $P$ on $N$ extends uniquely to a
trivialization $s$ of $\scrP$ for which $\nabla_{\!\!0}\, s=0$ on
$N$.
Given a trivialization of $P$, the connection on $P$ is given
by a $\frak{g}$-valued $1$-form $\alpha$, while $(\scrP,\nabla)$
is given by a $\frak{g}$-valued $1$-form $\beta$ on $M$, such
that $\left<\partial_0,\beta\right>=0$ on $N$.
Let us write $\beta=d\theta^0\,\theta^0 b+\sum dx^i(c^i+\theta^0
d^i)$, with $b$, $c$, $d$ functions of the $x^i$.
Then, $\alpha$ gives the $c^i$, the components $F_{0,i}$ of
the curvature, restricted to $N$, gives the $d^i$ and
$F_{00}$ restricted to $N$ gives $b$.

To handle the case of codimension $0\vert s$, one chooses $N\subset
N_1\subset\ldots\subset M$, each of codimension $0\vert 1$
in the next, and one uses induction on $s$: on $N_i$, after
recovering $(\scrP,\nabla)\vert N_i$, one recovers from the
data on $N$ the data on $N_i$ of all covariant derivatives
of the components of the curvature, restricted to $N_i$.

Let $(\scrP,\nabla)$ be a constrained $G$-torsor on $M^{6\vert(8,0)}$.  From
\S11.3, we obtain from it a $\frak{g}^\scrP$-valued spinor $\Lambda^{a}_i$,
and $\frak{g}^\scrP$-valued auxiliary fields $\scrE_{ij}$ By restriction to
Minkowski space $\bmink^6$, we obtain a torsor with connection $(P,\nabla)$ on
$M^6$, $\frak{g}^P$-valued spinors $\lambda^a_i$ and $\frak{g}^P$-valued
auxiliary fields $E_{ij}$.  By \S11.3, all covariant derivatives of the
components of $\scrP$, restricted to $M^6$, can be expressed from $\nabla$,
$\lambda$ and $\scrE$.  It follows that $(\scrP,\nabla)$ is determined up to
unique isomorphism by $(P,\nabla,\lambda,E)$.  More precisely, one has the
following rephrasing of \theprotag{10.18} {Proposition}.

        \proclaim{\protag{11.33} {Proposition}}
 The functor $(\scrP,\nabla)\mapsto (P,\nabla,\lambda,E)$ is
an equivalence of categories.
        \endproclaim

\flushpar
 For a spelling out of what this means, we refer to the text following
\theprotag{7.5} {Proposition}.  We will check \theprotag{11.33} {Proposition}
in the complex setting; the real case results from it as in the proof of
\theprotag{8.7} {Proposition}.

        \demo{Proof}
 We have to check that any $(P,\nabla,\lambda,E)$ comes from some constrained
$(\scrP,\nabla)$.  The flat $P$ with $\lambda,E=0$ corresponds to $\scrP$
flat.  We now construct vector fields on the space of connections, tangent to
the constrained ones, to get from there to any $(P,\nabla,\lambda,E)$.
Rather, we work with the space of connections along $\tau$.  A deformation of
a connection $\DD$ in the $\tau$ direction is given by
$\frak{g}^\scrP$-valued $\delta\DD_{ai}$.  It is tangent to constrained
connections if the symmetrization in $(a,b)$ and $(i,j)$ of $[\delta
\DD_{ai},\DD_{bj}]$ vanishes.

We will need the following operators acting on
$\frak{g}^\scrP$:
  $$  \alignedat2 &\DD_{ab}=\tfrac{1}{2!}\epsilon^{ij}\DD_{ai}\DD_{bj}, &\quad
     &\text{symmetrized in $(a,b)$}\\
     &\DD_{ijk\ell}=\tfrac{1}{4!}\epsilon^{abcd}\DD_{ai}\DD_{bj}
     \DD_{ck}\DD_{d\ell}, &\quad &\topfoldedtext{symmetrized in
     $(i,j,k,\ell).$} \endalignedat \tag{11.34} $$
They polarize the $\DD_s^2$ and $\DD_w^4$ of \thetag{11.14}, and
from \thetag{11.14} we get that
  $$  \DD_{ab}\DD_{ijk\ell}=\DD_{ijk\ell}\DD_{ab}=0 \tag{11.35} $$
(for their action on sections of $\frak{g}^\scrP$).

Fix $\frak{g}^P$-valued fields $V^{ij}$ and consider
  $$  \delta\DD_{ai}:=\epsilon^{jj'}\DD_{aj}\DD_{ij'k\ell}
V^{k\ell}. \tag{11.36} $$
If we start from a constrained $\DD$, we have
  $$ \alignedat2
      [\DD_{ai},\delta&\DD_{bj}]\qquad &&\text{symmetrized in $(a,b)$,
     $(i,j)$}\\
      =&\epsilon^{kk'}\DD_{ai}\DD_{bk}\DD_{jk'\ell m}V^{\ell m}\qquad
     &&\text{symmetrized in $(a,b)$} \endalignedat \tag{11.37} $$
and the $(a,b)$-symmetrization of $\DD_{ai}\DD_{bk}$ is alternating in
$(i,k)$: it is $2\epsilon_{ik}\DD_{ab}$.  We get
  $$  \ldots =-2\DD_{ab}\DD_{ji\ell m}V^{\ell m}=0. $$
The $\delta\DD_{ai}$ is the promised vector field, tangent
to the constrained $\DD$.

Let $(\theta)$ be the ideal defining $M^6$ in $M^{6\vert (8,0)}$.  To
conclude, it remains to check that:

\vtop{\roster
\item"(a)"
for $V$ in $(\theta)^6$, the variation of $\nabla_{ab}$ on
$\breve{M}$ depends only on $V$ modulo $(\theta)^7$, and the
$\nabla_{ab}$ are changed freely;
\item"(b)"
for $V$ in $(\theta)^7$, the variation of $\lambda^a_i$
depends only on $V$ modulo $(\theta)^8$, and the $\lambda^a_i$
are changed freely;
\item"(c)"
for $V$ in $(\theta)^8$, the auxiliary fields $E$ are
changed freely.
\endroster}
\medskip\flushpar
 The variation of $[\DD_{ai},\DD_{bj}]$ (resp.
$[\DD_{ai},[\DD_{bj},\DD_{ck}]]$, resp. $[\DD_{ai},
[\DD_{bj},[\DD_{ck},\DD_{d\ell}]]]$) is given by $6$ (resp. $7$, resp. $8$)
operators $\DD$ applied to $V$.  For $V$ in $(\theta)^6$ (resp. $(\theta)^7$,
resp.  $(\theta)^8$), this variation, restricted to $M^6$, depends only on
$V$ modulo the next power of $(\theta)$ and, to compute it, one may replace
each $\DD_{ai}$ by $\partial_{ai}$.  The resulting simplified computation is
left to the reader.
        \enddemo

   \topmatter
   \lecturelabel{Appendix}
   \lecture{}
   \lecturename Sign Manifesto\endlecturename
   \endtopmatter
\rightheadtext{APPENDIX. SIGN MANIFESTO}

\CenteredTagsOnSplits

\define\itm{\removelastskip\medskip\begingroup\it
	\narrower\narrower\noindent
	\parindent0pt\llap{$\bullet$\enspace }} 
\define\enditm{\par\endgroup\medskip}

\refstyle{A}
\widestnumber\key{SSSSS}   % for widest bibliography name

 \comment
 lasteqno A@ 41
 \endcomment

 \Head{1}{Standard mathematical conventions}

 \itm We apply the sign rule relentlessly. 
 \enditm

\flushpar
 This means that when passing from ordinary algebra to $\ZZ/2$-graded, or
super, algebra we pick up a sign $(-1)^{|a|\,|b|}$ when permuting homogeneous
elements~$a,b$ of parity~$|a|,|b|$.  Structure maps (multiplications, Lie
brackets, inner products, \dots) are even.
 
For example, consider a graded complex vector space $V=V^0\oplus V^1$.  A
hermitian inner product~$\langle \cdot ,\cdot   \rangle$ satisfies, among
other properties, 
  $$ \langle v_1,v_2 \rangle = (-1)^{|v_1|\,|v_2|}\overline{\langle v_2,v_1
     \rangle},\qquad v_1,v_2\in V\text{\ homogeneous}. \tag{A.1} $$
and from the evenness of the inner product it follows that $V^0$~is
orthogonal to~$V^1$.  From~\thetag{A.1} we deduce that $\langle v,v \rangle$~is
real for $v$~even and pure imaginary for $v$~odd.  The adjoint~$T^*$ of a
homogeneous linear operator $T\:V\to V$ is characterized by
  $$ \langle Tv_1,v_2  \rangle = (-1)^{|T|\,|v_1|}\langle v_1,T^*v_2
     \rangle. \tag{A.2} $$
Skew-adjoint operators form a super Lie algebra. 
 
\itm Symmetry groups act on the left.\enditm
 
\flushpar
 For example, if $\frak{g}$~is a Lie algebra, then an action of~$\frak{g}$ on
a vector space~$V$ is a {\it homomorphism\/} $\frak{g}\to\End(V)$.  Brackets
are preserved.  On the other hand, an action of~$\frak{g}$ on a manifold~$M$
is an {\it antihomomorphism\/} $\frak{g}\to\Vect(M)$.  The reversal of sign
comes from the rule $\xi f = \frac{d}{dt}\exp(t\xi )^*f$ and the fact that
$f\mapsto u^*f$ is a {\it right\/} action of diffeomorphisms~($u$) on
functions~($f$).

 \Head{2}{Choices}

 \itm A hermitian inner product on a complex vector space~$V$ is conjugate
linear in the {\it first\/} variable: 
  $$ \langle \lambda _1v_1, \lambda _2v_2 \rangle = \overline{\lambda
_1}\lambda _2\langle v_1,v_2  \rangle,\qquad \lambda _i\in \CC,\quad v_i\in
     V. \tag{A.3} $$
 \enditm

 \itm If $V=V^0\oplus V^1$ is a super Hilbert space, then
  $$ -i\langle v,v  \rangle\ge0,\qquad v\in V^1. \tag{A.4} $$
 \enditm 
 
 \itm We pass from self-adjoint operators to skew-adjoint operators using
multiplication by~$-i$: 
  $$ \text{$T$~self-adjoint} \longleftrightarrow \text{$-iT$
     skew-adjoint}. \tag{A.5} $$
 \enditm 
 
 \itm The Lorentz metric~$g$ on $n$~dimensional Minkowski space has
signature~$(1,n-1)$: 
  $$ \Signature(g) = +---\cdots \tag{A.6} $$
 \enditm 
 
 \itm The quantum {\it hamiltonian\/}~$\hH$ is minus the operator which
corresponds to infinitesimal time translation:
  $$ \hH = -\hP_0. \tag{A.7} $$
 \enditm

 \Head{3}{Rationale}

The first choice~\thetag{A.3} is not the usual one in mathematics, but it has
its merits.  For example, since linear operators act on the left, it makes
sense to have the commuting scalar multiplication act on the right.  In
mathematics we do follow this convention for modules over noncommutative
rings.  With right scalar multiplication \thetag{A.3}~reads: $\langle
v_1\lambda _1,v_2\lambda _2 \rangle = \overline{\lambda _1}\langle v_1,v_2
\rangle\lambda _2$.  We do not adopt this convention for scalar
multiplication, but do adopt~\thetag{A.3}.  Physicists like~\thetag{A.3} in view
of Dirac's notation~`$\langle v_1|T|v_2 \rangle$' for~`$\langle
v_1,Tv_2\rangle$'.  Comment: In computations it is often more convenient and
safer to work with a bilinear form rather than a sesquilinear form, and so to
write the sesquilinear inner product as~`$\langle\overline{v_1},v_2\rangle$'.
 
From a mathematical point of view it is more natural to quantize with {\it
skew\/}-adjoint operators, since they form a Lie algebra.  We use~\thetag{A.5}
to convert to {\it self\/}-adjoint operators, whose real eigenvalues
correspond to physical measurements.

The sign choice in~\thetag{A.6} leads to the usual bosonic
lagrangian~\thetag{A.20} with a plus sign in front of the kinetic energy. 
 
In~\thetag{A.7} we assume that time translation is a symmetry of a quantum
theory, so that the infinitesimal generator is represented by a self-adjoint
operator~$\hat P_0$ on the quantum (super)Hilbert space.  The minus sign
gives the standard answer for the hamiltonian of a classical free particle.

 \Head{4}{Notation}

Throughout~$i=\sqrt{-1}$.
 
Let $M$~denote $n$~dimensional affine Minkowski space with associated vector
space of translations~$V$ and future timelike cone~$C\subset V$.  We fix
linear coordinates~$x^0,\dots ,x^{n-1}$ with respect to which the metric is
  $$ g=g_{\mu \nu }\,dx^\mu \otimes dx^n = (dx^0)^2 - \dots
     -(dx^{n-1})^2, \tag{A.8} $$
and the cone is 
  $$ C=\{x: \langle x,x \rangle\ge0 \text{ and } x^0\ge0\}. \tag{A.9} $$
Let $\{e_ \mu \}$ be the corresponding basis of~$V$ and $\partial _\mu $~the
corresponding vector field on~$M$.  The standard density on~$M$ is
  $$ \dens = |dx^0\dots dx^{n-1}|. \tag{A.10} $$

Let $S$~be a real spin representation.  Fix a basis~$\{f^a\}$ of~$S$ and dual
basis~$\{f_a\}$ of~$S^*$.  Then there are symmetric pairings
  $$ \alignedat2
      &\Gamma \:S^*\otimes S^*&&\longrightarrow V \\ 
      &\tG\:S\otimes S&&\longrightarrow V.\endaligned \tag{A.11} $$
We write 
  $$ \alignedat2
      &\Gamma (f_a,f_b) &&= \Gamma ^\mu _{ab}e_\mu , \\ 
      &\tG(f^a,f^b)&&=\tG^{\mu ab}e_\mu ,\endaligned \tag{A.12} $$
where as usual we sum over repeated indices if one is upstairs and the other
is downstairs.  We raise and lower indices using the metric.  The
pairings~\thetag{A.11} are assumed to satisfy the Clifford relation
  $$ \tG^{\mu ab}\Gamma ^\nu _{bc} + \tG^{\nu ab}\Gamma ^\mu _{bc} =2g^{\mu
     \nu }\delta ^a_c \tag{A.13} $$
and the positivity condition  
  $$ \Gamma (s^*,s^*)\in C\qquad \text{for all $s^*\in S^*$}. \tag{A.14} $$
For~$v\in C^o$, the form $\langle v,\Gamma (s^*,s^*) \rangle$ is then
positive definite.  From~\thetag{A.13} and~\thetag{A.14} it follows that
  $$ \tG (s,s)\in C\qquad \text{for all $s\in S$}. \tag{A.15} $$

In a classical field theory we work with a space of fields~$\scrF$, where
$f\in \scrF$ is some sort of function on~$M$.  An infinitesimal symmetry is a
vector field~$\xi $ on~$\scrF$ which preserves the lagrangian in a certain
sense.  Corresponding to~$\xi $ is a Noether current~$J_\xi $, which is a
twisted $(n-1)$-form on~$M$.  The Noether charge~$Q_\xi $ is the integral
of~$J_\xi $ over a time slice.  We usually consider the current and charge
only on the space of classical solutions~$\scrM$, which carries a closed 2-form~$\omega $.  The infinitesimal symmetry and Noether charge are related by
  $$ dQ_\xi =-\iota (\xi )\omega.  \tag{A.16} $$
For $(\scrM,\omega )$ symplectic, \thetag{A.16}~can be rewritten 
  $$ \xi f = \{Q_\xi ,f\} \tag{A.17} $$
for $f$~a function on~$\scrM$.
 
Quantization is, in principle, a map 
  $$ Q\longmapsto \hQ \tag{A.18} $$
from functions on~$\scrM$ to operators on a complex Hilbert space~$\Cal{H}$.
We assume 
  $$ \hat{\overline{Q}} = \hQ^*, \tag{A.19} $$
so that real functions map to self-adjoint operators. 
 
Let $P_\mu $~be infinitesimal translation in the Poincar\'e algebra and
$Q_a$~the odd generator of the supersymmetry algebra.\footnote{The notational
conflict between the supersymmetry generator and the Noether charge is too
ingrained to correct.}  Let $\hP_\mu ,\hQ_a$ be the corresponding quantum
operators.

 \Head{5}{Consequences of~\S2 on other signs}

\itm The kinetic lagrangian for a scalar field $\phi \:M\to\RR$ is 
  $$ L = \frac 12|d\phi |^2\;\dens = \frac 12g^{\mu \nu }\partial _\mu \phi
     \partial _\nu \phi \;\dens. \tag{A.20} $$
\enditm

\flushpar
 The sign of this term is a consequence of~\thetag{A.6}; it is the main
rationale for preferring~\thetag{A.6} over the other choice. 
 
\itm Suppose $V=V^0\oplus V^1$ is a graded hermitian vector space and $T$~an
odd skew-adjoint operator.  Then 
  $$ i[T,T]\ge0. \tag{A.21} $$
\enditm 
 
\flushpar 
 Observe from~\thetag{A.2}, \thetag{A.3} and~\thetag{A.4} that an odd skew-adjoint
operator has eigenvalues on the line~$i^{-1/2}\RR\subset \CC$.
 
\itm The bracket in the supersymmetry algebra is 
  $$ [Q_a,Q_b] = -2\Gamma ^\mu _{ab}P_\mu . \tag{A.22} $$
\enditm 
 
\flushpar 
 Because we use left group actions, upon quantization we expect a
homomorphism from the supersymmetry algebra to {\it skew\/}-adjoint
operators.  Using~\thetag{A.5} we see that the sign in \thetag{A.22}~leads to 
  $$ [-i\hQ_a,-i\hQ_b] = -2\Gamma ^\mu _{ab}(-i\hP_\mu ). \tag{A.23} $$
Setting~$a=b$ we see from~\thetag{A.21} that $-2\Gamma ^\mu _{aa}\hP_\mu >0$
for all~$a$.  From~\thetag{A.14} we see that $\Gamma ^\mu _{aa}P_\mu $ has
nonnegative norm in~$V$.  Except possibly in dimension~2, the positive cone
generated by~$\{\Gamma ^\mu _{aa}P_\mu \}_a$ includes~$P_0$, and so the sign
choice in~\thetag{A.22} renders the hamiltonian nonnegative (rather than
nonpositive), in view of~\thetag{A.7}.
 
\itm The vector field~$\hxi _{P_\mu }$ on~$\scrF$ corresponding to
infinitesimal translation~$P_\mu $ is
  $$ \hxi _{P_\mu }f = -\partial _\mu f,\qquad f\in \scrF. \tag{A.24} $$
\enditm 
 
\flushpar 
 This follows since a diffeomorphism $\varphi \:M\to M$ acts on functions
by~$(\varphi \inv )^*$.  

\itm If $\frak{g}$~is a Lie algebra of infinitesimal symmetries, then the
vector fields~$\xi _\lambda \;(\lambda \in \frak{g})$ on the space of
fields~$\scrF$ satisfy 
  $$ [\xi _{\lambda _1},\xi _{\lambda _2}] = -\xi _{[\lambda _1,\lambda
     _2]}. \tag{A.25} $$
\llap{$\bullet$\enspace }The Noether currents satisfy\footnote{The bracketing
operation on Noether currents is defined in \CF{2.6}.
} 
  $$ \{j _{\lambda _1},j _{\lambda _2}\} = -j _{[\lambda _1,\lambda
     _2]}. \tag{A.26} $$
\llap{$\bullet$\enspace }The Noether charges satisfy
  $$ \{Q _{\lambda _1},Q _{\lambda _2}\} = -Q _{[\lambda _1,\lambda
     _2]}. \tag{A.27} $$
\llap{$\bullet$\enspace }The quantum operators satisfy 
  $$ [-i\hQ _{\lambda _1},-i\hQ _{\lambda _2}] = -i\hQ _{[\lambda _1,\lambda
     _2]}.  \tag{A.28} $$
\enditm 
 
\flushpar 
 Equations~\thetag{A.25}--\thetag{A.27} follow from the fact that
$\frak{g}\to\Vect(\scrF)$ is an antihomorphism and the standard equations for
Poisson brackets which follow from~\thetag{A.16}.  Equation~\thetag{A.28} says
that $\frak{g}\to\End(\scrH)$ is a homomorphism to skew-adjoint operators,
where we use~\thetag{A.5}. 
 
\itm The {\it self\/}-adjoint quantum operators~$\hQ_1,\hQ_2$ which
correspond to classical functions~$Q_1,Q_2$ on~$\scrM$ satisfy 
  $$ [\hQ_1,\hQ_2] = -i\hbar\{Q_1,Q_2\}{\hat{\ }}\quad \text{modulo $O
     (\hbar^2)$}. \tag{A.29} $$
\enditm 
 
\flushpar 
 To the extent that \thetag{A.29}~holds exactly, it says that the map
  $$ Q\longmapsto \frac{-i}{\hbar}\hQ \tag{A.30} $$
to skew-adjoint operators is an antihomorphism.  The sign in~\thetag{A.30} is
dictated by~\thetag{A.5}.  The desire to have an {\it anti\/}homomorphism is
dictated by~\thetag{A.27} and~\thetag{A.28}, and this determines the sign
in~\thetag{A.29}.   
 
\itm The Schr\"odinger equation for the evolution of a state~$\psi $ is 
  $$ \frac{\partial \psi }{\partial t} = \frac{-i}{\hbar}\hH \psi . \tag{A.31}
     $$
Evolution through time~$t$ for a static hamiltonian~$\hH$
is~$e^{-it\hH/\hbar}$. 
\enditm 
 
\flushpar
 The sign follows from~\thetag{A.7} and~\thetag{A.5}.

\itm Let $\eta _1^a,\eta _2^b$ be odd parameters and $\hxi _i$~the even vector
field on~$\scrF$ corresponding to~$\eta ^a_iQ_a$.  Then  
  $$ [\hxi _1,\hxi _2]f = 2\eta ^a_1\eta ^b_2\Gamma ^\mu _{ab} \partial _\mu
     f,\qquad f\in \scrF. \tag{A.32} $$
\enditm 
 
\flushpar 
 To see this, observe from~\thetag{A.22} that in the abstract supersymmetry
algebra we have 
  $$ [\eta ^a_1Q_a,\eta ^b_2Q_b] = 2\eta ^a_1\eta ^b_2\Gamma ^\mu _{ab}P_\mu
     . \tag{A.33} $$
Then \thetag{A.32}~follows from~\thetag{A.25} and~\thetag{A.24}; the minus signs in
these two equations cancel. 
 
\itm If $\psi _1,\psi _2$ are complex classical odd quantities, then complex
conjugation satisfies 
  $$ \overline{\psi _1\psi _2} = \overline{\psi _1}\;\overline{\psi _2}.
     \tag{A.34} $$
\enditm

\flushpar
 This is a consequence of the sign rule if we assume that $\psi
\mapsto\overline{\psi }$ is a {\it $*$-operation\/} and $\psi
_1$~(super)commutes with~$\psi _2$.  (A $*$~operation satisfies
$(ab)^*=(-1)^{|a|\,|b|}\,b^*a^*$.)  Notice that the classical
statement~\thetag{A.34} is consistent with the quantum statement~\thetag{A.19},
since the adjoint operation on linear operators is also a $*$~operation.
Notice that the product of real commuting odd quantities is real.

\itm The kinetic lagrangian for a spinor field $\psi \:M\to S$ is 
  $$ L = \frac 12\psi \Dirac\psi\;\dens = \frac 12\tG^{\mu ab}\psi _a\partial
     _\mu \psi _b\;\dens. \tag{A.35} $$
\llap{$\bullet$\enspace }The kinetic lagrangian for a dual spinor field
$\lambda \:M\to S^*$ is
  $$ L = \frac 12\lambda \Dirac\lambda\;\dens = \frac 12 \Gamma^\mu
     _{ab}\lambda ^a\partial _\mu \lambda ^b\;\dens. \tag{A.36} $$
\enditm 
 
\flushpar 
 The spinor fields are odd.  In view of~\thetag{A.34}, the lagrangians
\thetag{A.35}~and \thetag{A.36}~are real, as they must be in Minkowski space.  It
is easiest to check the sign in classical mechanics~($n=1$).  Then
from~\thetag{A.35} we deduce\footnote{See \FP{2} for a derivation.} the
classical Poisson bracket
  $$ \{\psi ,\psi\} = -1. \tag{A.37} $$
Upon quantization we know the corresponding operators satisfy~\thetag{A.29}.
The sign in~\thetag{A.37} is compatible with~\thetag{A.21}, and this means that
the sign in~\thetag{A.35} is correct. 
 
\itm The energy-momentum tensor is {\it minus\/} the Noether current
of~$P_\mu $.  The supercurrent is {\it minus\/} the Noether current
of~$Q_a$. 
\enditm 
 
\flushpar 
 This is a definition and follows by superPoincar\'e invariance from the
definition~\thetag{A.7} of the hamiltonian.  It means that the charges computed
from the energy-momentum tensor are energy and {\it minus\/} momentum.

 \Head{6}{Differential forms}

\itm When computing with differential forms on superspace, we use a bigraded
point of view.\footnote{See the appendix to \LSS{1}.} \enditm

\flushpar
 Objects have a ``cohomological'' degree, corresponding to a classical (that
is, non-super) degree, and a parity.  The permutation of objects of
parity~$p_1,p_2$ and cohomological degree~$d_1,d_2$ introduces two signs: a
classical sign~$(-1)^{d_1d_2}$ and an additional factor~$(-1)^{p_1p_2}$.

 \itm 
 On $\RR^{p|q}$ with coordinates~$t^1,\dots t^p,\theta ^1,\dots \theta ^q$ we
have the following table of parities and cohomological degrees:
 \enditm

 \centerline{\eightpoint
$$
\vbox{\offinterlineskip
\hrule
\halign{\vrule# &\quad\hfill #\hfill
  &\quad\vrule# &\quad\hfill #\hfill
  &\quad\vrule# &\quad\hfill #\hfill
  &\quad\vrule# &\quad\hfill #\hfill
  &\quad\vrule# &\quad\vrule# \cr
height6pt &\omit &&\omit &&\omit &&\omit  &\cr
& quantity && type  && parity ($\ZZ/2\ZZ$) && coh deg ($\ZZ$) &\cr
height6pt &\omit  &&\omit &&\omit&&\omit &\cr
\noalign{\hrule height 1.5pt depth 0pt}
height3pt &\omit   &&\omit &&\omit &&\omit &\cr
& $t^\mu \phantom{\partial /}$ && even coordinate && 0 && 0  &\cr
height3pt &\omit  &&\omit &&\omit &&\omit  &\cr
\noalign{\hrule}
height3pt &\omit   &&\omit &&\omit &&\omit &\cr
& $\theta ^a \phantom{\partial /}$ && odd coordinate && 1 && 0  &\cr
height3pt &\omit  &&\omit &&\omit &&\omit  &\cr
\noalign{\hrule}
height3pt &\omit   &&\omit &&\omit &&\omit &\cr
& $\partial /\partial t^\mu $, $\iota (\partial /\partial t^\mu) $ && even
vector field && 0 && $-1$  &\cr 
height3pt &\omit  &&\omit &&\omit &&\omit  &\cr
\noalign{\hrule}
height3pt &\omit   &&\omit &&\omit &&\omit &\cr
& $\partial /\partial \theta ^a $, $\iota(\partial /\partial \theta ^a )$ &&
odd vector field && $1$ && $-1$  &\cr 
height3pt &\omit  &&\omit &&\omit &&\omit  &\cr
\noalign{\hrule}
height3pt &\omit   &&\omit &&\omit &&\omit &\cr
& $\Lie {\xi } $ && Lie derivative && $p(\xi )$ && 0  &\cr 
height3pt &\omit  &&\omit &&\omit &&\omit  &\cr
\noalign{\hrule}
height3pt &\omit   &&\omit &&\omit &&\omit &\cr
& $dt^\mu, \epsilon (dt^\mu ) $ && even 1-form && 0 && 1  &\cr
height3pt &\omit  &&\omit &&\omit &&\omit  &\cr
\noalign{\hrule}
height3pt &\omit   &&\omit &&\omit &&\omit &\cr
& $d\theta ^a, \epsilon (d\theta ^a ) $ && odd 1-form && 1 && 1  &\cr
height3pt &\omit  &&\omit &&\omit &&\omit  &\cr
\noalign{\hrule}
height3pt &\omit   &&\omit &&\omit &&\omit &\cr
& $d^pt\,d^q\theta $ && berezinian && $q \pmod2$ && $p$  &\cr
height3pt &\omit  &&\omit &&\omit &&\omit  &\cr
\noalign{\hrule}
height3pt &\omit   &&\omit &&\omit &&\omit &\cr
& $|d^pt|\,d^q\theta $ && density && $q \pmod2$ && 0  &\cr
height3pt &\omit  &&\omit &&\omit &&\omit  &\cr
\noalign{\hrule}
height3pt &\omit   &&\omit &&\omit &&\omit &\cr
& $\cont{\xi _r}\dots \cont{\xi _1}\;d^pt\,d^q\theta $ && integral form &&
$\sum p(\xi _i)+q\pmod2$ && $p-r$  &\cr 
height3pt &\omit  &&\omit &&\omit &&\omit  &\cr
\noalign{\hrule}
height3pt &\omit   &&\omit &&\omit &&\omit &\cr
& $\cont{\xi _r}\dots \cont{\xi _1}\;|d^pt|\,d^q\theta $ && integral density &&
$\sum p(\xi _i)+q\pmod2$ && $-r$  &\cr 
height3pt &\omit  &&\omit &&\omit &&\omit  &\cr
\noalign{\hrule}
}
\hrule}
$$
}

\itm For $X$~a \rom(super\rom)manifold the canonical pairing of vectors and
1-forms is written with the vector on the left.  \enditm
 
\flushpar
 Therefore, by the sign rule, for a tangent vector~$\xi $ and a
1-form~$\alpha $ there is a sign\footnote{See \LSS{3.3}.} when passing from
the canonical pairing~$\iota (\xi )\alpha $ to $(-1)^{p(\xi )p(\alpha
)}\alpha (\xi )$, where $p$~is the parity.

\itm For integration over odd variables we have  
  $$ \gathered
      \int_{}d\theta \;\theta =1\\
      \int_{}d\theta ^2\!d\theta ^1\;\theta ^1\!\theta ^2
       =\int_{}d\theta ^2\left(\int_{}d\theta ^1\;\theta ^1\right)\theta
     ^2=1.\endgathered $$
\enditm

\itm For any vector field~$\xi $ we have the Cartan formula 
  $$ \Lie\xi =[d,\cont\xi ]. \tag{A.38} $$
 \enditm

 \flushpar
 Both sides of~\thetag{A.38} act on differential or integral forms.

 \Head{7}{Miscellaneous signs}

\itm Let $X$~be a smooth manifold, $\xi $~a vector field on~$X$, $\varphi
_t$~the one-parameter group of diffeomorphisms generated, and $T$~a tensor
field.  Then
  $$ \Lie{\xi }T = \frac{d}{dt}\res{t=0}\varphi _t^* T =
     \frac{d}{dt}\res{t=0}(\varphi _{-t})_*T. \tag{A.39} $$
 \enditm

 \itm 
 On a K\"ahler manifold~$X$ the Riemannian metric~$g$, K\"ahler form~$\omega
$, complex structure~$J$, and a local K\"ahler potential~$K$ are related by
the equations
  $$ \aligned
      \omega (\xi _1,\xi _2) &= g(J\xi _1,\xi _2)\\
      \omega &=i\,\partial \bar\partial K.\endaligned \tag{A.40} $$
 \enditm

 \itm  
 Suppose $f$~is a (suitable) function on a vector space~$V$ of dimension~$n$.
Then $f$~and its Fourier transform~$\hat{f}$ on~$V^*$ are related by
  $$ \aligned
      \hat{f}(k) &= \frac{1}{(2\pi )^{n/2}} \int_{V} e^{-i\langle k,x
     \rangle} f(x)\;\dens, \\ 
      f(x) &= \frac{1}{(2\pi )^{n/2}} \int_{V^*} e^{+i\langle k,x
     \rangle}\hat{f} (k)\;|d^nk|.\endaligned \tag{A.41} $$
 \enditm

 \itm  
 Suppose $\frak{g}$~is the Lie algebra of a real Lie group~$G$.  Then the
complexified Lie algebra~$\frak{g}\cp=\frak{g}\otimes \CC$ carries a
conjugation whose set of real points is~$\frak{g}$.  
 \enditm
 
 \flushpar
 For example, starting with the unitary group~$G=U(n)$, we obtain the
conjugation $A\mapsto -A^*$ on complex $n\times n$~matrices.  ($A^*$~is the
conjugate transpose of~$A$.)

   \topmatter
   \lecturelabel{References}
   \lecture{}
   \lecturename\endlecturename 
   \endtopmatter
   \rightheadtext{REFERENCES} 

\parindent0pt\parskip6pt

Most of the material discussed here was developed in the physics literature
in the '70s and ~'80s.  It is hopeless for us to reference all of the
original papers.  Our more modest goal is to give the reader an entry into
the literature and to point out a few more specialized references we used.

Some standard physics books on the subject (alphabetical by author) are:

\medskip 

\ref 
\by P. G. O.  Freund 
\book{Introduction to supersymmetry}
\bookinfo{Cambridge Monographs on Mathematical Physics}
\publ{Cambridge University Press}
\publaddr{Cambridge}
\yr{1986}
\endref\smallskip

\ref 
\by{S. J. Gates, M. T. Grisaru, M. Ro{\v{c}}ek, W. Siegel}
\book{Superspace,
One thousand and one lessons in supersymmetry}
\bookinfo{Frontiers in Physics}
\vol{58}
\publ{Benjamin/Cummings Publishing Co. Inc.}
\publaddr{Reading, Mass.}
\yr{1983}
\endref\smallskip
 
\ref 
\by{M. F. Sohnius}
\paper{Introducing supersymmetry}
\jour{Phys. Rep.}
\vol{128}
\yr{1985}
\pages{39--204} 
\endref\smallskip

\ref 
\by{J. Wess, J. Bagger}
\book{Supersymmetry and supergravity}
\bookinfo{Princeton Series in Physics}
\publ{Princeton University Press}
\publaddr{Princeton, NJ}
\yr{1992}
\endref 

\ref 
\by{P. West}
\book{Introduction to supersymmetry and supergravity}
\publ{World Scientific Publishing Co., Inc.}
\publaddr Teaneck, NJ
\yr{1990} 
\endref

\medskip 
 
An expository introduction to some of this material may be found in 
 
\medskip 
\ref 
\by D. S. Freed 
\book Five Lectures on Supersymmetry
\publ American Mathematical Society 
\yr 1999 
\toappear
\endref
\medskip 

The superspace formulation of 6-dimensional Yang-Mills (Chapters~10 and~11)
does not seem to be widely known, but it can be found in the literature.
See, for example:
 
\medskip 
\ref 
\by P. S. Howe, G. Sierra, P. K. Townsend 
\paper Supersymmetry in six dimensions 
\jour Nucl. Phys. B
\vol 221 
\yr 1983 
\pages 331--348
\endref\smallskip 
 
\ref 
\by J. Koller
\paper A six-dimensional superspace approach to extended superfields
\jour Nucl. Phys. B
\vol 222  
\yr 1983
\pages 319--337
\endref 
\medskip

\enddocument